\newcommand{\kms}{{km~s$^{-1}$}}

\newcommand{\halp}{H$\alpha$}

\newcommand{\oiii}{[O{\scshape iii}]}
\newcommand{\hone}{H{\scshape i}}
\newcommand{\none}{N{\scshape i}}

\newcommand{\gaz}{\theta}
\newcommand{\pa}{\phi_0}

\newcommand{\vsys}{V_{\rm sys}}

\newcommand{\vrot}{V_{\rm rot}}

\newcommand{\slos}{\sigma_{\rm LOS}}

\newcommand{\sinst}{\sigma_{\rm inst}}

\newcommand{\sigr}{\sigma_R}
\newcommand{\sigp}{\sigma_{\gaz}}
\newcommand{\sigz}{\sigma_z}

\newcommand{\sddisk}{\Sigma_{\rm dyn}}

\newcommand{\mls}{\Upsilon_\ast}

\newcommand{\arot}{V_{\rm arot}}
\newcommand{\hr}{h_{\rm R}}

\newcommand{\hs}{h_{\rm \sigma}}
\newcommand{\vmax}{V_{\rm max}}

\newcommand{\itf}{i_{\rm TF}}

\newcommand{\arcdeg}{\mbox{$^\circ$}}
\newcommand{\jhk}{$J\!H\!K$}
\newcommand{\rs}{r_{\rm s}}
\newcommand{\rbulge}{R_{\rm bulge}}
\newcommand{\hsl}{h_{\sigma{\rm,LOS}}}
\newcommand{\hsz}{h_{\sigma{\rm,z}}}
\newcommand{\vmaxoiii}{V_{\rm max}^{\rm OIII}}
\newcommand{\voiii}{V_{2.2h_{\rm R}}^{\rm OIII}}
\newcommand{\vmaxstar}{V_{\rm max}^{\rm star}}

\documentclass[traditabstract]{aa} 
\usepackage{txfonts}
\usepackage{graphicx}
\usepackage{lscape}
\usepackage{subfigure}
\usepackage{rotating}
\usepackage{caption}
\usepackage[title,titletoc]{appendix}       
\usepackage[spanish,dutch,english]{babel}
\usepackage{natbib}
\usepackage{hyperref}

\bibpunct{(}{)}{;}{a}{}{,}

\defcitealias{bershady2010a}{Paper~I}
\defcitealias{bershady2010b}{Paper~II}
\defcitealias{westfall2011a}{Paper~III}
\defcitealias{westfall2011b}{Paper~IV}
\defcitealias{bershady2011}{Paper~V}		

\begin{document}

\title{The DiskMass Survey. VI. Gas and stellar kinematics in spiral galaxies from 
PPak\thanks{Based on observations collected at the Centro Astron\'omico Hispano Alem\'an
(CAHA) at Calar Alto, operated jointly by the Max-Planck Institut f\"ur Astronomie and
the Instituto de Astrof\'isica de Andaluc\'ia (CSIC)} integral-field spectroscopy}

\author{Thomas P. K. Martinsson\inst{1,}\inst{2}
   \and Marc A. W. Verheijen\inst{1}
   \and Kyle B. Westfall\inst{1,}\thanks{National Science Foundation (USA) International
   Research Fellow}
   \and \\Matthew A. Bershady\inst{3}
   \and Andrew Schechtman-Rook\inst{3}
   \and \\David R. Andersen\inst{4}
   \and Rob A. Swaters\inst{5}}

\institute{
  Kapteyn Astronomical Institute, University of Groningen, PO Box 800, 9700 AV Groningen, 
  The Netherlands\\
  \email{verheyen@astro.rug.nl; westfall@astro.rug.nl}
\and
  Leiden Observatory, Leiden University, PO Box 9513, 2300 RA Leiden, The Netherlands\\
  \email{martinsson@strw.leidenuniv.nl}
\and
  Department of Astronomy, University of Wisconsin, 475 N. Charter St., Madison, WI 53706,
  USA\\
  \email{mab@astro.wisc.edu; andrew@astro.wisc.edu}
\and
  NRC Herzberg Institute of Astrophysics, 5071 West Saanich Road, Victoria, 
  British Columbia, Canada V9E 2E7\\
  \email{david.andersen@nrc-cnrc.gc.ca}
\and
  National Optical Astronomy Observatory, 950 North Cherry Ave., Tucson, AZ 85719, USA\\
  \email{swaters@noao.edu}
}

\date{Received 7 October 2012 / Accepted 9 May 2013}

\abstract{
We present ionized-gas (\oiii$\lambda$5007\AA) and stellar kinematics (velocities and
velocity dispersions) for 30 nearly face-on spiral galaxies out to as many as three
$K$-band disk scale lengths ($\hr$). These data have been derived from PPak 
Integral-Field-Unit spectroscopy from $4980-5370$\AA\ observed at a mean resolution of
$\lambda/\Delta\lambda$$=$$7700$ ($\sigma_{\rm inst}$$=$17~\kms). These data are a
fundamental product of our survey and will be used in companion papers to, e.g., derive
the detailed (baryonic$+$dark) mass budget of each galaxy in our sample. Our presentation
provides a comprehensive description of the observing strategy and data reduction,
including a robust measurement and removal of shift, scale, and rotation effects in the
data due to instrumental flexure. Using an in-plane coordinate system determined by
fitting circular-speed curves to our velocity fields, we derive azimuthally averaged
rotation curves and line-of-sight velocity dispersion ($\slos$) and luminosity profiles
for both the stars and \oiii-emitting gas. Along with a clear presentation of the data,
we demonstrate: (1) The \oiii\ and stellar rotation curves exhibit a clear signature of
asymmetric drift with a rotation difference that is 11\% of the maximum rotation speed of
the galaxy disk, comparable to measurements in the solar neighborhood in the Milky Way.
(2) The e-folding length of the stellar velocity dispersion ($\hs$) is $2\hr$ on average,
as expected for a disk with a constant scale height and mass-to-light ratio, with a
scatter that is notably smaller for massive, high-surface-brightness disks in the most
luminous galaxies.  (3) At radii larger than $1.5\hr$, $\slos$ tends to decline slower
than the best-fitting exponential function, which may be due to an increase in the disk
mass-to-light ratio, disk flaring, or disk heating by the dark-matter halo.  (4) A strong
correlation exists between the central vertical stellar velocity dispersion of the disks
($\sigma_{\rm z,0}$) and their circular rotational speed at 2.2$\hr$ ($\voiii$), with a
zero point indicating that galaxy disks are submaximal. Moreover, weak but consistent
correlations exist between $\sigma_{\rm z,0}/\voiii$ and global galaxy properties such
that disks with a fainter central surface brightness in bluer and less luminous galaxies
of later morphological types are kinematically colder with respect to their rotational
velocities.
}               

   \keywords{techniques: imaging spectroscopy --
             galaxies: spiral --
             galaxies: structure --
             galaxies: kinematics and dynamics --
             galaxies: fundamental parameters}

   \titlerunning{DMS-VI. Gas and stellar kinematics from PPak IFU spectroscopy}
   \authorrunning{Martinsson et al.}
   \maketitle

\section{Introduction}
\label{sec:intro}

Kinematic measurements of disk galaxies have been used for decades in the study
of the distribution of dark matter.  In fact, although ``missing mass'' was
first reported in clusters of galaxies \citep{zwicky1933}, the ongoing quest of
finding and explaining dark matter only began in earnest when it became clear
that the rotation curves of spiral galaxies remain constant (``flat'') out to
the largest measured radius, which can be well beyond the optical disk when
using \hone\ data \citep[early observations include][]{roberts1966, rogstad1972,
bosma1978, bosma1981a, bosma1981b, rubin1980}.  Within the framework of
Newtonian gravity, these observations cannot be explained by the implied mass
distribution of the visible stars and gas alone, requiring that a significant
dark component be present as well.

Although the existence of dark-matter halos is firmly established (again within
the context of Newtonian dynamics), a robust measure of the density profile of
these halos and the relative amounts of dark and luminous matter in real
galaxies is highly uncertain.  This is primarily caused by the uncertainty in
the mass associated with the visible stars, or the stellar mass-to-light ratio
($\mls$).  Even when a parameterized density distribution is assumed for the
dark-matter halo (such as the two-parameter pseudo-isothermal sphere
\citep[e.g.][]{ostriker1979} or NFW halo \citep{NFW1997}, one cannot
unambiguously decompose the mass contributions to the observed rotation curve
without determining $\mls$.  For a wide range of $\mls$, equally acceptable
decompositions can be constructed with a corresponding wide range of
dark-matter-density profiles \citep[e.g.][]{albada1985,verheyen1997,dutton2005};
while this is true in general, a poignant illustration of this fact is provided
by \citet[][hereafter \citetalias{bershady2010a}]{bershady2010a} using \hone\
data for NGC~3198.

This ``disk-halo degeneracy'' is most commonly circumvented by adopting the
so-called ``maximum-disk hypothesis'' \citep{AlbadaSancisi1986}, which assumes
that the visible baryons contribute maximally to the gravitational potential
regardless of the corresponding $\mls$.  The maximum-disk hypothesis is
equivalent to minimizing the amount of dark matter needed to explain the
observed rotation curve.  When invoking the presence of a dark-matter halo to
explain the outer parts of observed rotation curves, application of the
maximum-disk hypothesis results in the baryonic disk contributing 85$\pm$10\% to
the rotational velocity at 2.2 disk scale lengths ($\hr$) in order to avoid a
hollow halo \citep{sackett1997}.  Although ad hoc, the maximum-disk
hypothesis can explain the shapes of the inner rotation curves very well.  This
has provided empirical evidence that the dynamical mass distribution of disk
galaxies follows the light distribution in the central regions
\citep[e.g.][]{sancisi2004,noordermeer2007,swaters2011,fraternali2011}; however,
such results only provide circumstantial evidence in validating the maximum-disk
hypothesis.  A robust validation of the maximum-disk hypothesis requires one to
break the disk-halo degeneracy, thereby deriving meaningful density profiles of
the dark-matter halos from rotation curve decompositions.  To do so, one must
obtain independent measures of $\mls$.

Although a flourishing industry exists for deriving stellar masses by matching
photometric measurements to the predictions of stellar population synthesis
(SPS) models \citep[e.g.][]{belldejong2001, kauffmann2003, zibetti2009}, these
models are hampered by significant uncertainties in the stellar initial mass
function, the star-formation history, the effect of late phases of stellar
evolution, and the chemical-enrichment history of a galaxy. Alternatively,
stellar dynamics can be used to determine the dynamical mass surface density of galaxy
disks ($\sddisk$), which provides $\mls$ when combined with other data that
quantifies the gas mass \citep[e.g.,][hereafter
\citetalias{westfall2011b}]{westfall2011b}.  For the Milky Way, the vertical
density distribution and motions of stars in the solar neighborhood
\citep[e.g.][]{kuijken1991} can be used to determine the restoring force to the
Galactic mid-plane and, thus, the Galactic $\sddisk$.  The essence of this
concept can be extended to external galaxies as well \citep[e.g.][]{kruit1984},
and this idea is at the heart of the DiskMass Survey
\citepalias[DMS;][]{bershady2010a}.

Although many surveys exist that include spiral galaxies, few have measured
stellar kinematics in the low-surface-brightness outer regions of their disks.
The main reason for this has been the observational difficulties of obtaining
data with the high signal-to-noise ratio ($S/N$) and high spectral resolution
needed to measure stellar velocity dispersions in galaxy disks, which are
typically $\sim$20 \kms\ in the outer part.  Previous studies
\citep[e.g.,][]{kruit1986, bottema1993, kregel2005} have used long-slit
spectroscopy to measure the stellar velocity dispersions in disks; however,
these observations only reached out to $1-2\hr$ and suffered from large uncertainties
in the vertical density distribution, significant projection effects
due to high inclination, and rather low $S/N$ of the acquired spectra.
Nevertheless, these studies concluded that the disks of the galaxies in their
samples only contribute $\sim60\pm15$\% to their total rotation velocity, significantly
less than the value found by applying the maximum-disk hypothesis.

With the advent of Integral-Field-Unit (IFU) spectroscopy, the study of galaxy
kinematics has taken a significant leap forward.  Instead of only observing along
one-dimensional long-slits, it has now become possible to obtain spectra with
two-dimensional coverage on the sky.  This allows the construction of three-dimensional
data cubes with a spectrum at every position on the sky, a capability achieved in
radio astronomy for decades, although by the use of different techniques. Additionally,
IFU spectroscopy has the potential to boost the $S/N$ of the data by averaging spectra
from adjacent positions in the galaxy disks.

The DMS employs the two custom-built IFUs SparsePak \citep{bershady2004,bershady2005}
and PPak \citep{verheyen2004,kelz2006}, both with a field-of-view in excess of one
arcminute, a spectral resolution of $\lambda/\Delta\lambda=8000-12000$, and
light-collecting fibers (spaxels) that are $2\farcs7-4\farcs5$ in diameter.
The main goal of the survey is to determine $\sddisk$ in $\sim$40 disk galaxies
thereby breaking the disk-halo degeneracy and directly constraining $\mls$ and the
density profiles of their dark-matter halos.  Our strategy includes measuring the gas
and stellar kinematics in the galaxy disks, particular the vertical stellar velocity
dispersion ($\sigz$), which provides a direct estimate of $\sddisk$
\citepalias[][]{bershady2010a} when assuming a reasonable vertical mass distribution.

Existing surveys employing IFU spectroscopy are unsuitable for these analyses.
For instance, \citet{ganda2006} used SAURON on the 4.2m WHT telescope at La Palma
to obtain IFU spectroscopy of the inner regions of 18 late-type galaxies.  Due
to its limited field-of-view ($33\arcsec \times 41\arcsec$), spectral resolution
($\lambda/\Delta\lambda=1200$ or 250 \kms), and surface-brightness sensitivity
($0\farcs94$ spaxels collecting relatively little light), stellar kinematic
observations with SAURON were limited to the inner, high surface-brightness
regions of galaxies ($R_{\rm max}<1\hr$ for 14 out of 18 galaxies) where often a
bulge or bar dominates the light.  Other spectroscopic surveys including disk
galaxies, such as the PINGS survey \citep{rosales-ortega2010} or the CALIFA
survey \citep{sanchez2012}, also lack the high spectral resolution needed in
order to measure velocity dispersions in the outskirts of galaxy disks.

Here, we present gas and stellar kinematic data obtained for thirty nearly face-on
spiral galaxies using the PMAS fiber package (PPak).  We describe, in detail, the
observational strategy, data reduction and analysis methods.  Some preliminary
results regarding the stellar and gas kinematics are presented as well.
Ultimately, these PPak data will allow us to determine $\sddisk$ out to as far as
$3\hr$, to break the disk-halo degeneracy, to calibrate the mass scale of SPS models,
and to determine the dark-matter distribution in spiral galaxies with unprecedented
accuracy.

We start with a description of our galaxy sample in Sect.~\ref{sec:Sample},
followed by a description of our re-analysis of near-infrared 2MASS images in
Sect.~\ref{sec:2MASSphot}.  Sections~\ref{sec:Setup} and \ref{sec:Observations}
provide brief overviews of the instrumental setup and observational strategy,
respectively.  Due to our original reduction methodology, and because of the
importance of these data in future papers, Sect.~\ref{sec:Reduction} describes
the data reduction in great detail; additional detail is provided by
\citet{martinsson2011}.
Section~\ref{sec:Analysis} describes our derivation of the line-of-sight
kinematics of the stars and gas, while the derivation of $\sigz$, the core data
product of the DMS project, is presented in Sect.~\ref{sec:sigmaz}.
Finally, Sect.~\ref{sec:Summary} summarizes this paper.  Throughout this
paper, we adopt $H_0=73$\kms~Mpc$^{-1}$ for the Hubble parameter.

\section{The PPak Sample}
\label{sec:Sample}

The sample selection and observational strategy of the DMS has been described in
detail in \citetalias{bershady2010a}.  In short, we selected a parent sample of
1661 galaxies from the Uppsala General Catalogue of Galaxies
\citep[UGC;][]{nilson1973} that adhere to the following criteria: (a) a
blue-band diameter of $1\arcmin < D_{25} < 3\farcm5$ to appropriately match the
field-of-view of our custom-built IFUs; (b) a blue-band minor-to-major axis
ratio of $b/a \geqslant 0.75$ to ensure a nearly face-on orientation; (c) a
Galactic extinction below $A^g_B<0.25$ magnitudes; and (d) a numerical Hubble
type of $T\geqslant0$ to select disk-dominated systems.  We culled this sample
of galaxies with dominant bars, significant asymmetries, large bulges, possibly
interacting companions, and overcrowding of nearby bright stars via visual
inspection of optical images and luminosity profiles resulting in 231 galaxies
in our ``Phase-A'' sample.  We measured H$\alpha$ velocity fields for 146
galaxies in the Phase-A sample using SparsePak, which we call the H$\alpha$ sample.
Of these 146, we selected 40 galaxies for follow-up stellar-absorption-line spectroscopy;
this ``Phase-B'' sample excludes galaxies in the H$\alpha$ sample that are too
face-on, kinematically lopsided, or significantly warped, and those that
exhibited strong streaming motions.  The 30 galaxies in the Phase-B sample
observed using PPak are presented in this paper, which we refer to as the PPak
sample.
The distributions of galaxy properties in the PPak and Phase-A samples agree well, with
similar color and surface-brightness ranges. However, the PPak sample lacks the
least luminous galaxies and the few early-type galaxies (E \& S0) present in the
Phase-A sample. The PPak sample also has a higher fraction of Sc galaxies.
See \citetalias{bershady2010a} (especially Figs.~10--14 and Table~3) for a more detailed
comparison between the different samples.

Table~\ref{tab:UGC} lists salient properties of the galaxies in this
sample.  In summary, the PPak sample contains galaxies that range in Hubble type
from Sa to Im (22 of them are Sc or later), in absolute $K$-band magnitude
($M_K$) from -25.4 to -21.0, in $B-K$ color from 2.7 to 4.2, and in
central disk $K$-band surface brightness ($\mu_{0,K}$) from 15.8 to 19.7
mag~arcsec$^{-2}$.

%
\begin{table*}
\centering
\caption{\label{tab:UGC}
Properties of the galaxies in the PPak sample}
%
{\tiny 
\renewcommand{\tabcolsep}{0.6mm}
\begin{tabular}{|rrrrlcrcccrccrc|}
\hline
\multicolumn{1}{|c}{UGC}             & 
\multicolumn{1}{c}{R.A.}             & 
\multicolumn{1}{c}{Dec.}             & 
\multicolumn{1}{c}{Dist.}            &   
\multicolumn{1}{c}{Type}             & 
\multicolumn{1}{c}{$B$}              &   
\multicolumn{1}{c}{$K$}              & 
\multicolumn{1}{c}{A$_B^{\rm g}$}    & 
\multicolumn{1}{c}{A$_K^{\rm g}$}    & 
\multicolumn{1}{c}{$B-K$}            &
\multicolumn{1}{c}{$\hr$}            & 
\multicolumn{1}{c}{$\mu_{0,K}$}      & 
\multicolumn{1}{c}{$\mu_{0,K}^{i}$}  & 
\multicolumn{1}{c}{$\rbulge$}        & 
\multicolumn{1}{c|}{B/D}             \\ 
\multicolumn{1}{|c}{}                &   
\multicolumn{1}{c}{(J2000)}          & 
\multicolumn{1}{c}{(J2000)}          & 
\multicolumn{1}{c}{(Mpc)}            & 
\multicolumn{1}{c}{}                 &   
\multicolumn{1}{c}{(mag)}            & 
\multicolumn{1}{c}{(mag)}            &   
\multicolumn{1}{c}{(mag)}            &   
\multicolumn{1}{c}{(mag)}            & 
\multicolumn{1}{c}{(mag)}            &
\multicolumn{1}{c}{(arcsec)}         &
\multicolumn{1}{c}{(mag/arcsec$^2$)} &
\multicolumn{1}{c}{(mag/arcsec$^2$)} &
\multicolumn{1}{c}{(arcsec)}         &
\multicolumn{1}{c|}{}                \\
\multicolumn{1}{|c}{(1)}             &   
\multicolumn{1}{c}{(2)}              & 
\multicolumn{1}{c}{(3)}              & 
\multicolumn{1}{c}{(4)}              & 
\multicolumn{1}{c}{(5)}              &   
\multicolumn{1}{c}{(6)}              & 
\multicolumn{1}{c}{(7)}              &   
\multicolumn{1}{c}{(8)}              &   
\multicolumn{1}{c}{(9)}              & 
\multicolumn{1}{c}{(10)}             &
\multicolumn{1}{c}{(11)}             &
\multicolumn{1}{c}{(12)}             &
\multicolumn{1}{c}{(13)}             &
\multicolumn{1}{c}{(14)}             &
\multicolumn{1}{c|}{(15)}            \\
%
\hline  
  448 & 00:42:22.06 & 29:38:30.1 &  65.3 $\pm$ 2.1 & SABc      & 14.0 $\pm$ 0.2 & 10.13 $\pm$ 0.07 & 0.258 & 0.022 & 3.6 $\pm$ 0.2 & 12.19 $\pm$ 0.27 & 17.70 $\pm$ 0.04 & 17.84 $\pm$ 0.04 & 13.6 & 0.32 \\
  463 & 00:43:32.39 & 14:20:33.2 &  59.6 $\pm$ 2.1 & SABc      & 13.3 $\pm$ 0.2 &  9.38 $\pm$ 0.05 & 0.392 & 0.033 & 3.6 $\pm$ 0.2 & 13.08 $\pm$ 0.47 & 16.79 $\pm$ 0.05 & 16.95 $\pm$ 0.05 & 6.4  & 0.06 \\
 1081 & 01:30:46.63 & 21:26:25.5 &  41.8 $\pm$ 2.1 & SBc       & 13.5 $\pm$ 0.2 & 10.05 $\pm$ 0.08 & 0.252 & 0.021 & 3.2 $\pm$ 0.2 & 15.07 $\pm$ 0.62 & 17.72 $\pm$ 0.09 & 17.82 $\pm$ 0.09 & 6.5  & 0.05 \\
 1087 & 01:31:26.63 & 14:16:39.4 &  59.6 $\pm$ 2.1 & Sc        & 14.3 $\pm$ 0.3 & 10.68 $\pm$ 0.11 & 0.233 & 0.020 & 3.4 $\pm$ 0.3 & 11.19 $\pm$ 0.34 & 17.95 $\pm$ 0.06 & 18.05 $\pm$ 0.06 & 4.9  & 0.04 \\
 1529 & 02:02:31.02 & 11:05:34.8 &  61.6 $\pm$ 2.1 & Sc        & 14.1 $\pm$ 0.3 &  9.98 $\pm$ 0.08 & 0.374 & 0.032 & 3.8 $\pm$ 0.3 & 11.93 $\pm$ 0.26 & 17.17 $\pm$ 0.03 & 17.45 $\pm$ 0.03 & 6.9  & 0.04 \\
 1635 & 02:08:27.78 & 06:23:42.3 &  46.6 $\pm$ 2.1 & Sbc       & 14.1 $\pm$ 0.3 & 10.40 $\pm$ 0.09 & 0.237 & 0.020 & 3.5 $\pm$ 0.3 & 12.92 $\pm$ 0.30 & 17.84 $\pm$ 0.03 & 17.92 $\pm$ 0.03 & 9.0  & 0.07 \\
 1862 & 02:24:24.80 &$-$02:09:45.3& 18.4 $\pm$ 2.1 & SABcd$^1$ & 13.7 $\pm$ 0.1 & 10.37 $\pm$ 0.11 & 0.137 & 0.012 & 3.2 $\pm$ 0.2 & 15.73 $\pm$ 1.14 & 18.10 $\pm$ 0.13 & 18.31 $\pm$ 0.13 & --   & 0.00 \\
 1908 & 02:26:37.27 & 12:09:18.9 & 110.0 $\pm$ 2.1 & SBc$^2$   & 14.1 $\pm$ 0.2 & 10.06 $\pm$ 0.08 & 0.366 & 0.031 & 3.7 $\pm$ 0.2 &  9.12 $\pm$ 0.30 & 16.83 $\pm$ 0.06 & 16.95 $\pm$ 0.06 & 5.8  & 0.09 \\
 3091 & 04:33:56.17 & 01:06:48.8 &  73.8 $\pm$ 2.2 & SABd      & 14.6 $\pm$ 0.3 & 11.25 $\pm$ 0.17 & 0.361 & 0.031 & 3.0 $\pm$ 0.3 & 10.05 $\pm$ 0.46 & 18.26 $\pm$ 0.07 & 18.38 $\pm$ 0.07 & --   & 0.00 \\
 3140 & 04:42:54.93 & 00:37:06.9 &  62.1 $\pm$ 2.1 & Sc        & 13.4 $\pm$ 0.2 &  9.45 $\pm$ 0.06 & 0.349 & 0.030 & 3.6 $\pm$ 0.2 & 11.67 $\pm$ 0.50 & 17.00 $\pm$ 0.10 & 17.04 $\pm$ 0.10 & 5.9  & 0.11 \\
 3701 & 07:11:42.59 & 72:10:11.5 &  43.2 $\pm$ 2.1 & Scd       & 14.4 $\pm$ 0.4 & 11.23 $\pm$ 0.19 & 0.275 & 0.023 & 2.9 $\pm$ 0.4 & 16.96 $\pm$ 1.59 & 19.28 $\pm$ 0.16 & 19.41 $\pm$ 0.16 & 9.0  & 0.05 \\
 3997 & 07:44:38.74 & 40:21:58.9 &  83.1 $\pm$ 2.1 & Im        &       $-$      & 11.56 $\pm$ 0.21 & 0.230 & 0.020 &      $-$      & 13.74 $\pm$ 1.21 & 19.14 $\pm$ 0.16 & 19.29 $\pm$ 0.16 & 5.0  & 0.02 \\
 4036 & 07:51:54.79 & 73:00:56.8 &  50.9 $\pm$ 2.1 & SABbc     & 12.8 $\pm$ 0.2 &  9.56 $\pm$ 0.05 & 0.116 & 0.010 & 3.1 $\pm$ 0.2 & 17.51 $\pm$ 1.41 & 17.54 $\pm$ 0.14 & 17.60 $\pm$ 0.14 & 6.9  & 0.04 \\
 4107 & 07:57:01.87 & 49:34:02.5 &  51.1 $\pm$ 2.1 & Sc        & 13.7 $\pm$ 0.2 & 10.16 $\pm$ 0.07 & 0.183 & 0.016 & 3.4 $\pm$ 0.2 & 12.92 $\pm$ 0.46 & 17.57 $\pm$ 0.06 & 17.68 $\pm$ 0.06 & 4.9  & 0.04 \\
 4256 & 08:10:15.18 & 33:57:23.9 &  74.8 $\pm$ 2.1 & SABc      & 12.9 $\pm$ 0.2 &  9.54 $\pm$ 0.05 & 0.231 & 0.020 & 3.2 $\pm$ 0.2 & 12.92 $\pm$ 0.30 & 17.11 $\pm$ 0.04 & 17.20 $\pm$ 0.04 & 5.7  & 0.08 \\
 4368 & 08:22:44.96 & 24:17:48.9 &  56.4 $\pm$ 2.1 & Scd       & 13.5 $\pm$ 0.3 & 10.45 $\pm$ 0.09 & 0.162 & 0.014 & 2.9 $\pm$ 0.3 & 11.67 $\pm$ 0.87 & 17.39 $\pm$ 0.15 & 17.79 $\pm$ 0.15 & 8.2  & 0.07 \\
 4380 & 08:24:31.87 & 54:51:14.0 & 105.0 $\pm$ 2.1 & Scd       & 14.5 $\pm$ 0.2 & 11.09 $\pm$ 0.13 & 0.258 & 0.022 & 3.2 $\pm$ 0.2 &  9.78 $\pm$ 0.35 & 18.11 $\pm$ 0.07 & 18.18 $\pm$ 0.07 & 4.8  & 0.03 \\
 4458 & 08:32:11.25 & 22:33:37.8 &  68.4 $\pm$ 2.1 & Sa        & 13.2 $\pm$ 0.2 &  8.86 $\pm$ 0.05 & 0.150 & 0.013 & 4.2 $\pm$ 0.2 & 27.14 $\pm$ 0.67 & 18.55 $\pm$ 0.04 & 18.79 $\pm$ 0.04 & 35.0 & 0.72 \\
 4555 & 08:44:08.27 & 34:43:02.1 &  61.8 $\pm$ 2.1 & SABbc     & 13.2 $\pm$ 0.2 & 10.09 $\pm$ 0.06 & 0.137 & 0.012 & 3.0 $\pm$ 0.2 & 13.57 $\pm$ 0.50 & 17.42 $\pm$ 0.07 & 17.72 $\pm$ 0.07 & 6.7  & 0.04 \\
 4622 & 08:50:20.19 & 41:17:21.9 & 178.2 $\pm$ 2.2 & Scd       & 14.6 $\pm$ 0.3 & 11.21 $\pm$ 0.12 & 0.125 & 0.011 & 3.3 $\pm$ 0.3 &  8.75 $\pm$ 0.56 & 17.87 $\pm$ 0.12 & 18.05 $\pm$ 0.12 & 8.3  & 0.16 \\
 6903 & 11:55:36.94 & 01:14:13.8 &  31.3 $\pm$ 2.2 & SBcd      & 13.1 $\pm$ 0.3 &  9.84 $\pm$ 0.13 & 0.090 & 0.008 & 3.2 $\pm$ 0.3 & 27.83 $\pm$ 1.42 & 18.91 $\pm$ 0.07 & 19.10 $\pm$ 0.07 & 9.0  & 0.02 \\
 6918 & 11:56:28.13 & 55:07:30.8 &  21.8 $\pm$ 2.3 & SABb$^3$  & 12.0 $\pm$ 0.2 &  8.79 $\pm$ 0.02 & 0.061 & 0.005 & 3.2 $\pm$ 0.2 & 10.96 $\pm$ 0.22 & 15.76 $\pm$ 0.04 & 16.03 $\pm$ 0.04 & 6.2  & 0.06 \\
 7244 & 12:14:18.08 & 59:36:55.6 &  65.4 $\pm$ 2.2 & SBcd      & 14.6 $\pm$ 0.4 & 11.81 $\pm$ 0.19 & 0.092 & 0.008 & 2.7 $\pm$ 0.4 & 12.19 $\pm$ 1.23 & 19.24 $\pm$ 0.18 & 19.32 $\pm$ 0.18 & --   & 0.00 \\
 7917 & 12:44:26.20 & 37:07:16.4 & 102.9 $\pm$ 2.3 & SBbc      & 13.8 $\pm$ 0.2 &  9.66 $\pm$ 0.04 & 0.093 & 0.008 & 4.1 $\pm$ 0.2 & 16.96 $\pm$ 0.79 & 17.64 $\pm$ 0.09 & 17.87 $\pm$ 0.09 & 10.3 & 0.14 \\
 8196 & 13:06:04.43 & 55:39:21.9 & 119.7 $\pm$ 2.3 & Sb        & 14.0 $\pm$ 0.2 &  9.97 $\pm$ 0.05 & 0.074 & 0.006 & 4.0 $\pm$ 0.2 &  8.35 $\pm$ 0.19 & 16.88 $\pm$ 0.04 & 16.99 $\pm$ 0.04 & 9.1  & 0.24 \\
 9177 & 14:20:30.49 & 10:25:55.5 & 132.4 $\pm$ 2.6 & Scd       & 14.0 $\pm$ 0.2 & 11.05 $\pm$ 0.12 & 0.141 & 0.012 & 2.8 $\pm$ 0.2 & 10.96 $\pm$ 0.44 & 18.04 $\pm$ 0.08 & 18.40 $\pm$ 0.08 & 7.0  & 0.08 \\
 9837 & 15:23:51.68 & 58:03:10.6 &  43.2 $\pm$ 2.3 & SABc      & 14.1 $\pm$ 0.3 & 10.45 $\pm$ 0.14 & 0.072 & 0.006 & 3.6 $\pm$ 0.3 & 27.83 $\pm$ 1.42 & 19.70 $\pm$ 0.08 & 19.89 $\pm$ 0.08 & 7.5  & 0.03 \\
 9965 & 15:40:06.76 & 20:40:50.2 &  70.6 $\pm$ 2.4 & Sc        & 14.0 $\pm$ 0.2 & 10.62 $\pm$ 0.08 & 0.266 & 0.023 & 3.1 $\pm$ 0.2 & 10.34 $\pm$ 0.29 & 17.61 $\pm$ 0.05 & 17.65 $\pm$ 0.05 & --   & 0.00 \\
11318 & 18:39:12.23 & 55:38:30.6 &  85.2 $\pm$ 2.2 & SBbc      & 13.8 $\pm$ 0.2 &  9.95 $\pm$ 0.06 & 0.210 & 0.018 & 3.7 $\pm$ 0.2 & 10.96 $\pm$ 0.22 & 17.18 $\pm$ 0.03 & 17.22 $\pm$ 0.03 & 7.0  & 0.12 \\
12391 & 23:08:57.17 & 12:02:52.7 &  66.8 $\pm$ 2.1 & SABc      & 14.0 $\pm$ 0.3 & 10.58 $\pm$ 0.10 & 0.371 & 0.032 & 3.1 $\pm$ 0.3 & 11.93 $\pm$ 0.52 & 17.70 $\pm$ 0.08 & 17.85 $\pm$ 0.08 & 5.1  & 0.04 \\
\hline
\end{tabular}
}
\tablefoot{
(1) UGC number; (2) Right Ascension; (3) Declination; (4) distance; (5) morphological
type; (6) $B$-band magnitude taken from \citetalias{bershady2010a}; (7) total 
extrapolated $K$-band magnitude (see text); (8) Galactic $B$-band extinction 
\citep{schlegel1998}; (9) Galactic $K$-band extinction \citep{schlegel1998}; 
(10) $B-K$ color corrected for Galactic extinction; (11) representative $K$-band disk
scale length; (12) representative disk central $K$-band surface brightness; (13) column
12, corrected for Galactic extinction and applied k-correction and face-on correction;
(14) the ``bulge radius'', where the light from the bulge contributes 10\% to the total
light; (15) bulge-disk ratio from $K$-band surface brightness profile decomposition.
Notes on morphologies: 1=peculiar, 2=starburst, 3=AGN.
}
\end{table*}

\section{Near-Infrared Photometry}
\label{sec:2MASSphot}

We retrieved the broadband $J$, $H$, and $K_s$ (hereafter $K$) imaging data for
each galaxy from the public 2MASS archive.\footnote{
Analysis of optical ($UBVRI$) and near-infrared ($JHK$) images obtained from the
2.1m telescope at Kitt Peak \citepalias{bershady2010a} are ongoing.
}
Where available, SDSS images were also retrieved from their respective archives
because these are used for source identification (see below). {\it Source
Extractor} \citep{Bertin1996} catalogs were constructed for all images to
identify neighbouring sources to be masked when producing the primary galaxy
photometry.  This masking is important both to ensure an accurate sky-level
determination and to inhibit foreground and background source contamination in
the target photometry apertures, since the targets are relatively large and we
need to reach low surface-brightness levels.
 
The so-called ``Kron'' apertures calculated by {\it Source Extractor} are of
adequate size for masking the majority of sources, which all are nearly stellar.
This aperture is defined to have a radius 2.5 times the first moment
of the light profile, $r_1$ (\citealt{kron1980}; cf.~Kron used $2r_1$ to define
a total aperture magnitude), and it has been shown that such schemes undercount
the total source flux by $10-25$\% for extended sources depending on the shape
of the light profile \citep{bershady1994}.  This appears to be particularly the
case for brighter sources with {\it Source Extractor}, likely because of the
isophotal limits used for determining $r_1$.\footnote{
For reference, $r_1 = 2 \hr$ for a purely exponential light profile observed at
infinite $S/N$. 
}  Empirically, we found that defining a masking aperture of $5 r_1 (1.625 -
0.03125 m_{\rm inst})$, where $m_{\rm inst}$ is the instrumental magnitude such
that 1 DN corresponds to 28.368 mag, resulted in a conservatively large mask at
bright magnitudes (including the primary source) without masking overly large
regions of the image from faint sources in all bands.

Additionally, {\it Source Extractor} has difficulty identifying the target
galaxies because of their relatively large size compared to the
point-spread-function (PSF), creating many smaller sources out of
non-axisymmetric substructure rather than finding a single source for all
galactic emission. For these reasons we hand-edited the $J$-band and (when
available) SDSS $r$-band catalogs, since these images were generally the
deepest, to increase the size of the contaminant sources, and to remove galactic
features erroneously identified as separate sources.  Some of the latter were
difficult to identify with confidence; for these sources we also consulted color
mosaics of SDSS data (when available) and when still in doubt chose to mask out
the source.  We also took this opportunity to identify any additional sources
not found by {\it Source Extractor} within $5r_1$ of the primary source.

An important aspect of the source masking is to force the masking apertures to
be identical in all bands. This ensures that the multi-aperture photometry
samples identical regions of the galaxy at all wavelengths in the case where
masked sources are within the target-source apertures. To accomplish this,
source catalogs from all bands were merged together to create one master
catalog. To avoid contamination near the galaxy, all sources within $7.5 r_1$
were copied to the master catalog verbatim from the cleaned catalog.  For all
other positions, sources were matched across all images, with a distance
threshold for matching of 0.5 times the semi-major axis of the ellipses (or the
closest match in the case of multiple matches), which we found resulted in
accurate matches without significant levels of false positives.  For all matched
sources, the source with the largest aperture was then chosen for the master
catalog.  This minimized contamination from the faint wings of foreground stars
and background galaxies.

After blanking all contaminant sources in the master catalog, and determining
the sky between $3.5 r_1$ and $6.5 r_1$ away from the primary galaxy, we
computed an initial surface-brightness profile of the galaxy using apertures
with constant ellipticity and position angle set to the isophotal value derived
from {\it Source Extractor}.  Next, we fit this profile between $1.25 r_1$ and
$2.5 r_1$ with an exponential function, and extrapolated this fit to $7.5 r_1$.
The extrapolated profile is then subtracted from the image to ensure that any
low-level galaxy flux outside of the masking region is not subtracted in the
subsequent surface fits that are used to correct the sky foreground.

\begin{figure}[t]
\centering
\includegraphics[width=0.25\textwidth,angle=270]{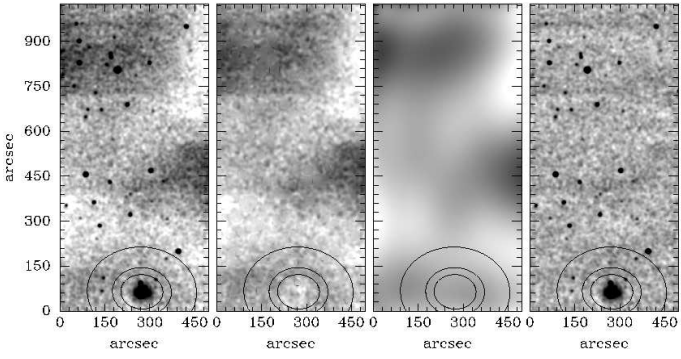}
\caption[]{
Illustration of our surface-fitting and source-blanking algorithms. {\bf Left panel:} 
$H$-band 2MASS image of UGC~4555. The galaxy itself is toward the bottom of the frame;
black ellipses mark radii of 2.5, 3.5, and 6.5 $r_1$. {\bf Second panel:} Same image but
with neighboring sources blanked and the estimated galaxy profile subtracted. {\bf Third
panel:} 9th-order surface fit using our iterative scheme, which was chosen to be the
``best'' fit to the image. {\bf Right panel:} Residual image after surface subtraction.
All images have the same dynamic range of 3DN, where 1DN corresponds to 
20.7~mag~arcsec$^{-2}$. They have been smoothed by a Gaussian with a FWHM of 9 pixels
($1\arcsec/{\rm pix}$) to enhance low-signal structure.
}
\label{fig:photometry}
\end{figure}

We fit and subtract the sky foreground separately in each passband using the
source masks and the iterative method described in \citetalias{westfall2011b}
and by \citet{Schechtman-Rook2012}.  At each iteration the order of the fit is
increased, up to order 9; the surface-brightness profile of the galaxy was
recomputed, re-fit, and re-subtracted for each iteration. The need for such a
rigorous effort is due both to the desire to obtain very high precision
photometry and the large-scale background fluctuations present in some of the
2MASS images.  Figure~\ref{fig:photometry} shows how well this correction works
for one of the worst cases.  While this surface-fitting method is generally
stable, we chose the fit order that minimized the standard deviation of the
pixels in the image as the ``best'' fit (typically orders between 5 and 9), and
use photometry from that order in our analysis.  The final surface photometry
was defined in elliptical apertures with constant ellipticity and position
angles defined from the kinematic analysis.

To improve the $S/N$ of the images, the $J$-, $H$- and $K$-band images were
combined into a single \jhk\ image, assuming that color gradients between these
passbands can be neglected; for 21 nearby S0 to Sbc galaxies,
\citet{peletier1997} found an average $J-K$ color gradient of -0.01 (with a
variance of 0.12) magnitudes per $K$-band disk scale length.  A \jhk\
curve-of-growth was determined for each galaxy within concentric elliptical
apertures with constant position angle and ellipticity as a function of radius,
and with a logarithmic radial sampling.  The apertures were centered on the
morphological centers of the galaxies (Sect.~\ref{sec:ContMaps}), with
position angles that correspond to the kinematic position angles and
ellipticities that reflect the inclinations as suggested by the inverted
Tully-Fisher relation (see Sect.~\ref{sec:orientation} for more details).

Based on the curves-of-growth, for each galaxy the radial range was determined in which
the residual sky foreground and its uncertainty were calculated. The radial range was
typically between $6\hr - 11\hr$. From this, the radial \jhk\ surface-brightness profiles
were derived. These \jhk\ profiles were re-scaled to the $K$-band surface-brightness 
profiles to provide a pseudo-$K$-band surface-brightness profile for each galaxy that
extends much further in radius than the $K$-band-only profile. The normalization was 
calculated using the photometric data within half the radius outside which the sky level
was determined. These pseudo-$K$-band surface-brightness profiles are presented in
Appendix~\ref{app:TheAtlas} (hereafter the Atlas).

For each galaxy, a representative photometric scale length ($\hr$) and central
surface brightness ($\mu_{0,K}$) of the disk was determined by fitting a
straight line to the pseudo-$K$-band luminosity profile in an iterative manner.
Starting with a radial range of $10\arcsec - 50\arcsec$, a straight line was fit
from which follows an initial estimate of $\hr$. Subsequently, the radial
fitting range was adjusted to $1\hr-4\hr$ to avoid the inner region where a
bulge or bar may dominate \citep{kormendy2004, erwin2005} and the outer region
where the exponential disk may be (anti-)truncated \citep{kregel2004,
pohlen2006}.  The fit was repeated to yield a new value of $\hr$, which defined
a new fitting range $1\hr-4\hr$.  This procedure was repeated until convergence
was achieved.  The corresponding values of $\mu_{0,K}$ and $\hr$ of the final
fit are listed in Table~\ref{tab:UGC}, and this fit is indicated in the Atlas
with a dotted line overlaid on the luminosity profile. This line is drawn as 
solid over the converged fitting range $1\hr-4\hr$.

The total extrapolated $K$-band magnitude is calculated by first integrating the
observed luminosity profile to the radius where the $K$-band surface brightness
plus its error falls below 25 mag/arcsec$^2$, yielding an aperture magnitude
$m_{{\rm aper},K}$.  Outside that radius, the luminosity profile is extended
with a modeled exponential decline that follows the scale length of the outer
disk, which can be steeper or shallower than the representative scale length
$\hr$ depending on the radial structure of the luminosity profile.  The extended
model profile is then integrated analytically from the aperture radius to
infinity to yield $m_{{\rm ext},K}$.  From this follows the total extrapolated
$K$-band magnitude as the proper sum of $m_{{\rm aper},K}$ and $m_{{\rm
ext},K}$.  These total extrapolated $K$-band magnitudes are also listed in
Table~\ref{tab:UGC}.

Finally, we performed bulge-disk decompositions using the one-dimensional
surface-brightness profiles.  For this purpose, the disk of a galaxy is modeled
with one or more exponential segments to properly subtract the inner disk, e.g.\
in the case of a type-II profile.  The modeled (multi-)exponential disk is
subtracted and the excess light in the inner region is modeled iteratively with
a general S{\'e}rsic profile that was convolved with a Gaussian to accommodate
seeing effects.  These convolved bulge models are also shown underneath the
inner luminosity profiles provided in the Atlas.  We produce bulge-to-disk
ratios by integrating the bulge model and comparing it to the total remaining
luminosity that is assumed to be from the disk component, i.e., $B/D = B/(m_K-B)$
where $B$ is the luminosity of the bulge, not the $B$-band magnitude.  A
proper, two-dimensional structural analysis of each galaxy based on our deep
2.1m imaging will be presented in a forthcoming paper.  These provisional $B/D$
measurements derived from the 2MASS photometry are listed in
Table~\ref{tab:UGC}.

Both the disk and the bulge have small Galactic extinction ($A_{K}^{g}=0.01-0.03$~mag;
Table~\ref{tab:UGC}) and k-corrections ($\kappa_{K}=0.01-0.10$~mag;
Table~\ref{tab:mK2iTF}). For the disk surface brightness, we also correct the
line-of-sight measurements to be the surface brightness as seen face-on by calculating
\begin{equation}
\mu_{K,\rm disk}^{i} = \mu_{K,\rm disk} - 2.5 C_{K}
\log\left[\cos\left(\itf\right)\right],
\label{eq:deproj_mu}
\end{equation}
where the coefficient $C_{K}=1$, assuming a transparent system with no internal
dust extinction. The face-on correction ranges from $0.01-0.38$~mag, depending
on the inclination ($\itf$; see Sect.~\ref{sec:Incl}) of the disk, with an
average and standard deviation of $0.14\pm0.09$~mag. The central surface brightness
with applied corrections for each galaxy can be found in Table~\ref{tab:UGC}.
The modeled bulge and resulting disk surface brightness can be seen in the Atlas.
For later analysis, we define the ``bulge radius'', $\rbulge$, to be the radius where
$\mu_{K,\rm bulge}-\mu_{K}=2.5$~mag, i.e.\ where the light from the bulge
contributes 10\% to the total light. These radii can also be found in Table~\ref{tab:UGC}.

\section{PPak Instrumental Setup}
\label{sec:Setup}

In this section, we provide a brief overview of the custom-built PPak fiber
bundle and the detailed setup used for our observations of the PPak sample.  We provide
a new astrometric table for the PPak fibers based on the post fabrication of the fiber
bundle \citep[as opposed to the nominal positions provided by][]{kelz2006}.  We
also discuss our analysis of the spectrum of the dome flood lamps, which we have
used for the wavelength calibration of our science spectra.

\subsection{The PPak fiber bundle}

\begin{figure}[t]
\centering
\includegraphics[width=0.49\textwidth]{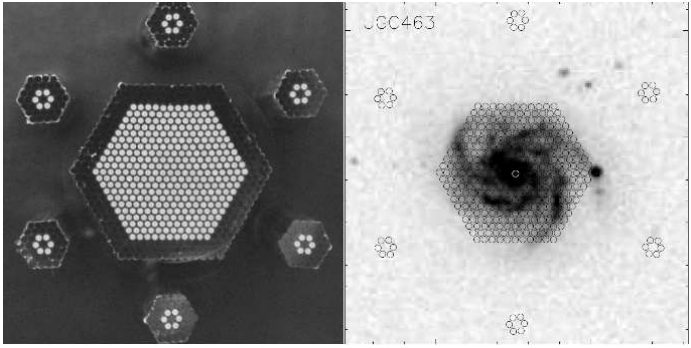}
\caption[]{
{\bf Left:} Photograph of the focal plane layout of PPak. The 331 science fibers
and 36 sky fibers are back-illuminated. {\bf Right:} The PPak fiber-footprint
overlaid on top of a direct image of UGC~463.
}
\label{fig:ppak}
\end{figure}

Galaxies in the PPak sample were observed using PPak, which is a bundle of 382
optical fibers that has been permanently integrated as an add-on module in the
pre-existing PMAS spectrograph at the 3.5-meter telescope at Calar Alto in
southern Spain.  Figure~\ref{fig:ppak} shows a direct image of the
back-illuminated PPak fiber-bundle in the focal plane, located behind a $f$/3.3
focal reducer lens.  The main fiber cluster consists of 331 science fibers, each
with a core diameter of $2\farcs7$, packed in a regular hexagonal grid with a
diagonal field-of-view of $64\arcsec\times74\arcsec$ and a filling factor of
60\%.  Thirty-six fibers, identical to the science fibers, are distributed over
6 mini-IFU's located on the circumference of a circle with a radius of
$72\arcsec$ relative to the main IFU head; these fibers nominally sample the
night sky and are used for sky subtraction of the science spectra.  The 367
active fibers are surrounded by short inactive fibers (dark in
Fig.~\ref{fig:ppak}) for protection during polishing and for a uniform
stress-load on the active fibers.  The remaining 15 fibers are arranged in a
similar mini-IFU configuration to receive light at the same focal ratio from the
integrating sphere of a separate calibration unit.  During the science
exposures, the 15 calibration fibers can be illuminated with a controllable
mixture of light from several arc lamps and a halogen lamp. The spots of the
emission lines in the calibration spectra are used to keep track of the flexure
in the Cassegrain-mounted PMAS spectrograph, as well as the effective spectral
resolution of a particular science exposure.  All 331$+$36$+$15 fibers are
uniformly distributed along a pseudo-slit in the spectrograph such that science,
sky and calibration spectra are recorded simultaneously in all areas of the
detector. A more detailed description of the PMAS spectrograph and PPak fiber
bundle is provided by \citet{roth2005} and \citet{kelz2006}.

\subsubsection{PPak fiber positions}
\label{sec:PPakFiber}

We have produced a post-fabrication astrometric fiber-position table of the PPak
IFU.  We converted the high-contrast direct image of PPak in
Fig.~\ref{fig:ppak} to FITS format and calculated the relative positions of
all 331 science fibers and 36 sky fibers using {\it Source Extractor}.  The
scale of the image (in pixels/arcsec) was calculated from the values given by
\citet{kelz2006}.  As can be seen in Fig.~\ref{fig:ppak}, the hexagonal grid
of fibers is not perfectly regular.  In particular, there is some appreciable
curvature in the most northern fiber rows.  The characterization of such
fabrication artifacts are important to the fidelity of our
reconstructed-continuum and kinematic maps (see the Atlas). The derived PPak
fiber-position table can be found in Appendix~\ref{sec:PPakPosTab}.

\subsection{Spectrograph settings}

All observations were taken with the J1200 grating, which has a blaze angle of
$46\fdg0$ and a blaze wavelength of 1200\AA\ in first order.  By mounting the
grating such that the grating normal is pointing toward the camera
(``backward''), one can increase the spectral resolution by a factor of two due
to the significant anamorphic demagnification.  We observed the second order of
the grating by properly setting the grating angle and using a stacked
combination of Schott's GG395 and BG40 glass filters.  Although meeting our
spectral resolution requirements \citepalias{bershady2010a}, this configuration
results in a geometric light loss of $\sim$15\% because the grating is
overfilled by the collimated beam.

Each fiber provides a spectrum covering approximately 430\AA\ dispersed along
rows of the CCD, as seen in Fig.~\ref{fig:rawdata_sci}.  We often refer to the
spectral and spatial (cross-dispersion) dimensions in the discussion of our PPak
spectroscopy using the variables $X$ and $Y$, respectively.  Due to the
different entrance angles of each fiber along the pseudo-slit, not all spectra
cover exactly the same spectral region: the spectra near the edges of the slit
are shifted $\sim$27\AA\ with respect to a spectrum near the center of the slit
(Fig.~\ref{fig:rawdata_sci}).  For each observation, the extracted spectral
range common to all spectra is typically 4975--5375\AA.

We minimized the effects of read-noise via on-chip binning of the CCD detector.
We binned by two pixels in the spatial (cross-dispersion) direction, but
performed no binning in the spectral direction.  The full-width at half maximum
(FWHM) of the PSF is $\sim$3 binned pixels in the cross-dispersion direction;
$\sim$5 pixels separate peaks of the PSF for adjacent fibers.  With this setup,
spatially adjacent spectra contributed less than 1\% scattered light to any
single aperture when defining an extraction aperture width of one FWHM centered
on the peak of the spatial PSF (Sect.~\ref{sec:dohydra}). In the spectral
direction, the FWHM of the resolution element is 0.66\AA\ and the linear
dispersion is $\sim$0.21\AA/pixel (or $\sim$12 \kms\ per pixel at 5150\AA).

Nominally, our spectrograph setup provides resolutions of
$\lambda/\Delta\lambda\approx7900$ at a central wavelength of
$\lambda_c=5200$\AA\ \citep{kelz2006}.  However, in practice the resolution was
often worse due to the combination of a less-than-perfect spectrograph focus,
flexure effects during the one-hour exposures, and/or the somewhat curved surface
of the CCD detector; we find that 97\% and 87\% of our observations have resolutions of
$\lambda/\Delta\lambda>6000$ and $\lambda/\Delta\lambda>7000$, respectively
(Sect.~\ref{sec:InstRes}).

\subsection{The dome flood lamps}

\begin{figure}[t]
\centering
\includegraphics[width=0.49\textwidth]{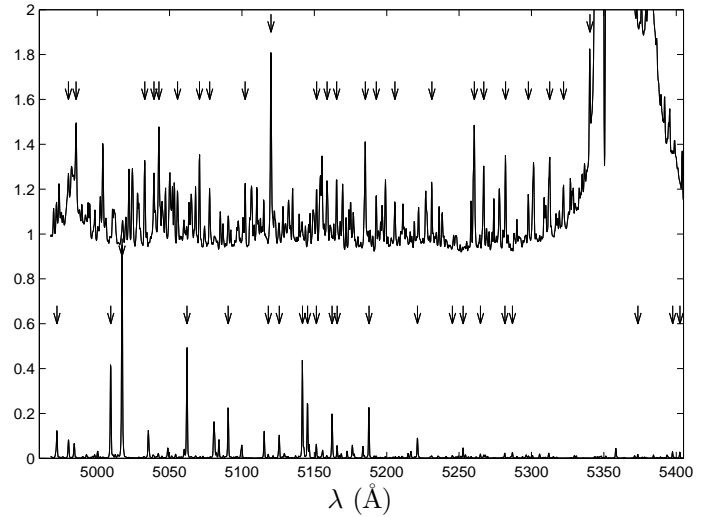}
\caption[]{
Spectrum of the dome flood lamps (top) and the ThAr calibration lamp (bottom);
the latter has been used to wavelength calibrate the former. The flux units are
arbitrary and offset for illustration purposes. The arrows mark the emission
lines in the ThAr-lamp spectrum that were used to calibrate the flood-lamp
spectrum (bottom), and the emission lines in the flood-lamp spectrum that were
used to wavelength calibrate our science spectra (top).
}
\label{fig:CalStra}
\end{figure}

The 15 calibration fibers collect light from the internal calibration lamps of
the PMAS spectrograph, but the (331) science and (36) sky fibers cannot.  Thus,
wavelength calibration of the science and sky fibers nominally requires
interpolation based on the calibration fibers \citep[e.g.][]{rosales-ortega2010},
which can limit the accuracy of the calibration.  Fortunately, the inside of the
dome can be illuminated by six bright flood lamps that are attached to a walkway
spanning the dome high above the telescope.  In the spectral region of our
interest, the spectrum of these flood lamps contains many more emission lines
than the internal Thorium-Argon (ThAr) lamps, as shown in Fig.~\ref{fig:CalStra}.
Therefore, we devised an observing strategy that takes advantage of these spectra
to increase the fidelity of our wavelength calibration.  We describe the wavelength
calibration of the flood-lamp spectrum below, and we describe its typical use in the
wavelength calibration of our science and sky spectra in Sect.~\ref{sec:Reduction}.

\subsubsection{Calibration of the flood-lamp spectrum}
\label{sec:CalCal}

For use in wavelength calibration of our science spectra, we first calibrated
the wavelengths of the flood-lamp emission lines; the wavelengths of these lines
were not provided by the manufacturer.  To do so, we partly dismantled the 
calibration unit that feeds the 15 calibration fibers of PPak such that they could
be illuminated by either the flood lamps via an external mirror or the normal
calibration lamps. Sets of five exposures each were taken of the internal ThAr lamp,
the internal halogen lamps, and the external flood lamps, respectively. Only the 15
calibration fibers were illuminated; the science and sky fibers were kept dark. The
telescope altitude and azimuth were fixed during these observations to prevent flexure
shifts.

The five images in each set were combined to increase $S/N$ and reject
cosmic-ray detections.  Using the {\it IRAF} task {\tt dohydra} in the {\it
HYDRA} package, we used the halogen image to trace and extract the 15 spectra
from both the ThAr lamp and flood lamps.  The wavelength scale of the flood-lamp
spectra were calibrated by the ThAr spectra by fitting a first-order
cubic-spline function to the $21-23$ isolated Argon lines in each ThAr spectrum.
None of the Thorium lines could be unambiguously identified from our library of
known Thorium emission lines.  The root-mean-square (RMS) scatter of the
wavelength calibration ranges from 0.009\AA\ to 0.015\AA.

Twenty-five isolated emission lines across the observed spectral range of the
flood-lamp spectrum were selected and their central wavelengths were determined
for each calibration fiber by fitting a single Gaussian to each line.  These
measurements made for the uppermost calibration-fiber aperture deviated
significantly from those of the corresponding emission lines in the other 14
apertures; this aperture is typically affected by the non-uniform focus of the
spectrograph across the detector, which results in platykurtic line profiles
(flatter than a nominal Gaussian) in the upper part of the frame.\footnote{
This is also the case for some of the science frames as discussed in 
Sect.~\ref{sec:dohydra}.
}  After excluding the measurements from the uppermost calibration-fiber
aperture, we calculated the average wavelength of each emission line in the
calibrated flood-lamp spectrum and find a standard deviation of less than
0.01\AA.  The error in these mean line centroids propagates in a systematic way
to the calibration of our science spectra; however, no relative error is
incurred among the calibrated science spectra.

Having calibrated the wavelength scale of the flood-lamp spectrum, we have found
that many of the lines seem to come from Dysprosium and possibly also from
Thorium by comparing the line centroids to literature line lists (e.g.\ in the
online atomic-line database provided by the National Institute of Standards and
Technology [NIST]\footnote{\tiny
\url{http://physics.nist.gov/PhysRefData/ASD/lines\_form.html}
}).
Unfortunately, most of the lines in the flood-lamp spectrum are blended such
that the measured centroids cannot be unambiguously correlated to the literature
reference values.  Instead, we choose to adopt the values found from our ThAr
calibration, which should have errors smaller than 0.01\AA.

\section{Observational Strategy}
\label{sec:Observations}

Our primary objective is to observe stellar absorption lines at the highest
spectral resolution achievable with PPak in the low-surface-brightness regions
($R\lesssim3\hr$) of stellar disks.  Therefore, we need to pay particular
attention to the quality of the calibration of the science exposures while
minimizing the amount of calibration overhead during the precious allocated dark
time.  To some extent, this notion has driven the design of the PPak IFU with
its 15 calibration fibers, and the observing and calibration strategy described
in this section has been devised to take advantage of these calibration fibers.
In the end, our observations span eight dark-night observing runs between
November 2004 and February 2008 with an overall success rate of $\sim$55\%; the
success rate is mainly due to adverse weather at Calar Alto.

\subsection{Calibration observations}

For calibration images (taken at the beginning of the evening and the end of the
morning of an observing night), the telescope was parked at a declination of
$\delta=+10\arcdeg$ and an hour angle of $-2$~hours.  This is the general
direction in which most of the science observations have been taken.  By parking
the telescope for several minutes before beginning observations, we minimize
flexure changes and, thereby, ensure that all calibration spectra were recorded
at the same location on the detector.  Moreover, this effort minimizes the
differential flexure between the calibration images and the science images.
Morning calibrations are taken in reverse order to those taken in the evening
(as described below) to maximize on-sky observing.

Our evening calibration sequence typically consisted of ten bias frames,
followed by five exposures each of the flood lamp, the continuum dome flat, and
the twilight sky flat.  Flood-lamp exposures were typically 30 seconds.  We
allowed several minutes between turning on the flood lamps and beginning the
sequence of exposures to minimize spectral variability in the flood-lamp spectra
due to temperature and pressure variations.  Simultaneous ThAr-lamp spectra via
the calibration fibers were taken during the flood-lamp exposures, which
provided spots on the detector that were used to correct for flexure differences
between the calibration and science observations.  Continuum dome-lamp exposures
were typically 20 seconds.  These data are used to flat-field and trace the
spectra in the science frames, as well as the other calibration frames.  During
dome-lamp exposures, the calibration fibers were illuminated by the halogen lamp
to allow for robust tracing of the ThAr spectra.  Sky-flat exposures ranged from
$1-360$ seconds, depending on the sky brightness.  The sky-flat data are used to
correct for variations in the fiber-to-fiber throughput.  Occasionally, a series
of hour-long dark-frame exposures were taken as well.  Figure~2.3 from
\citet{martinsson2011} provides examples of unprocessed bias, flood-lamp,
dome-flat, and sky-flat images.

\subsection{Galaxy observations}

\begin{figure}[t]
\centering
\includegraphics[width=0.49\textwidth]{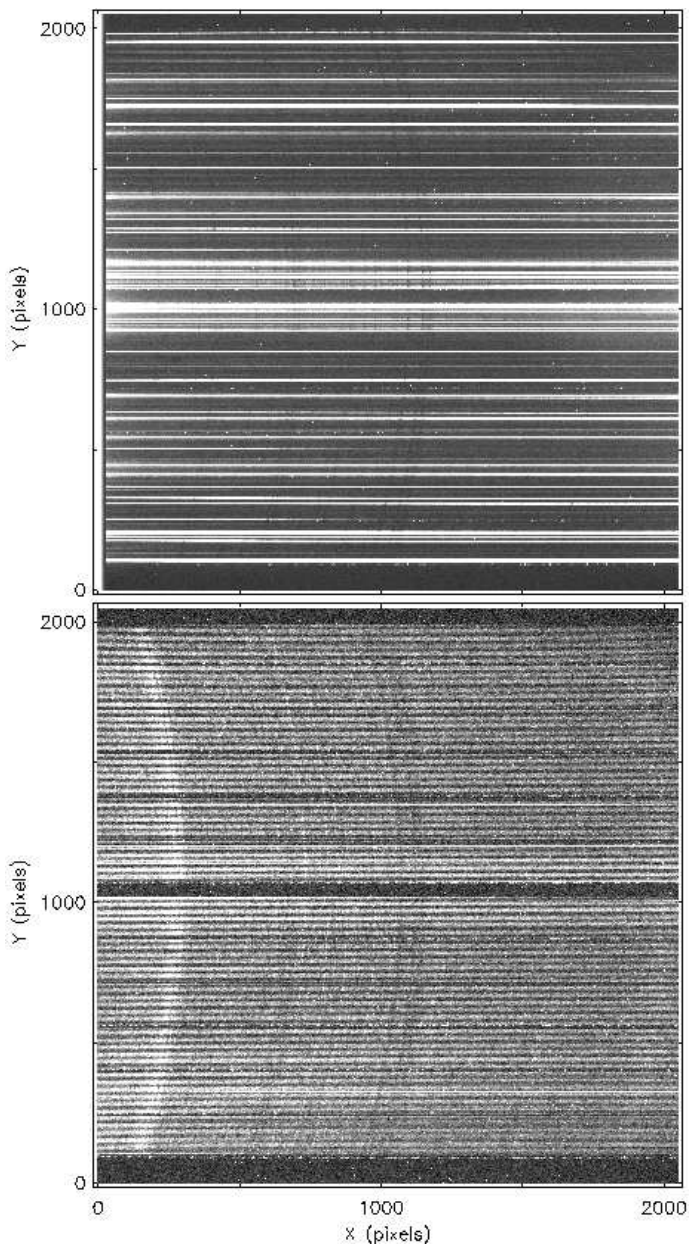}
\caption[]{
Examples of unprocessed PPak science exposures. {\bf Top:} The result of
drifting a template star through the central row of 21 fibers while the 15
calibration fibers are illuminated with the internal ThAr lamp. Spectra of the
template star are recorded all over the detector. {\bf Bottom:} A representative
one-hour exposure of the galaxy UGC~4368.  Again, the 15 calibration fibers are
illuminated with the internal ThAr lamp.  The curvature of the imaged
pseudo-slit is clearly visible.
}
\label{fig:rawdata_sci}
\end{figure}

Our galaxy observations are read-noise limited: the $S/N$ in the spectrum of
the dark night sky is limited by the read-noise of the detector, and the galaxy
surface-brightness levels at $R\lesssim3 \hr$ are typically well below the
surface brightness of the dark night sky.  Therefore, we need to maximize the
integration time for each science exposure.  In practice, we are limited by the
density of cosmic-ray hits in a single exposure.  We have chosen an integration
time of 3600 seconds for each science exposure, with the number of exposures per
galaxy depending on the surface brightness of the galaxy, the transparency of
the atmosphere and the darkness of the night sky. The number of exposures per
galaxy is presented in Table~\ref{tab:UGC_comb} and Table~\ref{tab:UGC_merge}.
Figure~\ref{fig:rawdata_sci} shows an example exposure of a target galaxy.

Since the PMAS spectrograph is mounted at the Cassegrain focus of the telescope,
its internal structure is subject to a changing gravity vector while the
telescope is moving during the night. This induces flexure between the various
components of the instrument due to a limited stiffness of the spectrograph
structure.  Such flexure may result in a deteriorating focus during the night, a
changing location of the spectra on the detector, and a smearing of the spectra
on the detector during the one-hour long exposure.  As part of the pilot and
feasibility study for the DMS with the PPak module, extensive flexure tests were
performed by pointing the telescope at different positions in hour angle and
declination, taking short arc-lamp exposures at each position to track shifts of
the emission-line spots on the detector.  The results are presented by
\citet{roth2005}, showing that the shifts are relatively small for negative hour
angles, i.e. on rising objects (see their Figure~21).  However, since the shift
of only one spot was considered, the test did not characterize any changes in
the plate scale or its rotation.  We present our measurements and corrections
for all such flexure distortions in Sect.~\ref{sec:Flexure}. To minimize the
detrimental effects of flexure during and in between our science exposures, we
decided to observe only rising targets east of the meridian.

%
\begin{table*}
\caption{\label{tab:TempStars}
Properties of the spectral template stars observed with PPak}
\centering
%
{\tiny
\renewcommand{\tabcolsep}{1.95mm}
\begin{tabular}{|r l l c c c r c || r l l c c c r c|}
\hline
\multicolumn{1}{|c}{HD}                      &
\multicolumn{1}{c}{SpT}                      &
\multicolumn{1}{c}{$\log g$}                 &
\multicolumn{1}{c}{[Fe/H]}                   &
\multicolumn{1}{c}{ref.}                     &
\multicolumn{1}{c}{Obs. Date}                &
\multicolumn{1}{c}{$S/N$}                    &
\multicolumn{1}{c||}{$\lambda/\Delta\lambda$}&
\multicolumn{1}{c}{HD}                       &
\multicolumn{1}{c}{SpT}                      &
\multicolumn{1}{c}{$\log g$}                 &
\multicolumn{1}{c}{[Fe/H]}                   &
\multicolumn{1}{c}{ref.}                     &
\multicolumn{1}{c}{Obs. Date}                &
\multicolumn{1}{c}{$S/N$}                    &
\multicolumn{1}{c|}{$\lambda/\Delta\lambda$} \\
\multicolumn{1}{|c}{(1)} &
\multicolumn{1}{c}{(2)}  &
\multicolumn{1}{c}{(3)}  &
\multicolumn{1}{c}{(4)}  &
\multicolumn{1}{c}{(5)}  &
\multicolumn{1}{c}{(6)}  &
\multicolumn{1}{c}{(7)}  &
\multicolumn{1}{c||}{(8)}&
\multicolumn{1}{c}{(1)}  &
\multicolumn{1}{c}{(2)}  &
\multicolumn{1}{c}{(3)}  &
\multicolumn{1}{c}{(4)}  &
\multicolumn{1}{c}{(5)}  &
\multicolumn{1}{c}{(6)}  &
\multicolumn{1}{c}{(7)}  &
\multicolumn{1}{c|}{(8)} \\
\hline
123299 & A0 III   & 3.30 & $-$0.56 & a & 2008-02-04 &  945 & 7293 &  63352 & K0 III   & 2.20 & $-$0.31 & a & 2004-11-19 &  632 & 7676 \\
  6695 & A3 V     & 4.30 & $-$     & g & 2004-11-19 & 1091 & 7712 &  74442 & K0 III   & 2.51 & $-$0.06 & a & 2005-11-02 &  527 & 7857 \\
159561 & A5 III   & 3.96 & $+$0.01 & g & 2008-02-04 & 1076 & 6905 &  85503 & K0 III   & 2.33 & $+$0.23 & a & 2005-11-06 & 1145 & 7647 \\
 13267 & B5 Ia    & 2.44 & $-$     & a & 2004-11-19 &  760 & 7729 & 102224 & K0 III   & 2.02 & $-$0.46 & a & 2006-10-14 &  472 & 6589 \\
 27295 & B9 IV    & 3.93 & $-$0.74 & a & 2004-11-19 & 1129 & 7792 & 145328 & K0 III   & 3.25 & $-$0.20 & b & 2005-05-02 &  618 & 8180 \\
 57118 & F0 Ia    & 1.7  & $-$     & g & 2004-11-19 &  612 & 7947 & 203344 & K0 III-IV& 2.62 & $-$0.24 & b & 2004-11-13 &  441 & 8322 \\
 29375 & F0 V     & 4.17 & $+$0.13 & g & 2004-11-19 &  743 & 7920 &   7010 & K0 IV    & 3.3  & $-$     & g & 2004-11-19 &  296 & 7988 \\
115604 & F3 III   & 3.4  & $+$0.18 & e & 2004-11-19 & 1257 & 7891 &  16160 & K1 III   & 1.88 & $-$0.25 & a & 2004-11-19 & 1388 & 7352 \\
 31236 & F3 IV    & 4.21 & $+$0.13 & g & 2004-11-19 &  579 & 7825 &  23841 & K1 III   & 1.30 & $-$0.95 & a & 2004-11-19 &  620 & 7766 \\
 43318 & F6 V     & 3.93 & $-$0.15 & a & 2004-11-19 &  687 & 7791 &\bf{162555}&\bf{K1 III}&\bf{2.72}&\bf{$-$0.21}&{\bf b}&\bf{2007-01-15}&\bf{1083}& \bf{8180} \\
136202 & F8 III-IV& 3.85 & $-$0.08 & a & 2008-02-04 & 1112 & 6601 & 176411 & K1 III   & 2.91 & $+$0.00 & b & 2005-05-02 &  927 & 8200 \\
 22879 & F9 V     & 4.29 & $-$0.83 & a & 2004-11-19 &  632 & 7632 & 205512 & K1 III   & 2.57 & $+$0.03 & a & 2006-10-15 &  732 & 7476 \\
101501 & G0-8 Vv  & 4.60 & $-$0.13 & a & 2004-11-19 & 1473 & 7745 &  51440 & K2 III   & 2.28 & $-$0.35 & a & 2004-11-19 &  993 & 7957 \\
 18391 & G0 Ia    & 0.00 & $-$0.28 & a & 2004-11-19 &  441 & 7975 &  72184 & K2 III   & 2.61 & $+$0.12 & a & 2005-11-06 &  871 & 7937 \\
  6903 & G0 III   & 2.9  & $-$     & g & 2004-11-19 &  851 & 7612 & 162211 & K2 III   & 2.45 & $+$0.05 & a & 2006-08-24 &  829 & 7661 \\
 30455 & G2 V     & 4.45 & $-$0.36 & g & 2004-11-19 &  612 & 7848 & 163588 & K2 III   & 2.61 & $-$0.09 & b & 2007-01-15 &  839 & 7842 \\
  6474 & G4 Ia    & 1.50 & $+$0.25 & a & 2004-11-19 &  301 & 7807 &  83618 & K3 III   & 1.74 & $-$0.08 & a & 2005-11-06 & 1046 & 7400 \\
 82210 & G4 III-IV& 3.19 & $-$0.28 & a & 2008-02-04 &  937 & 7155 &  94247 & K3 III   & 2.24 & $-$0.27 & b & 2004-11-19 & 1302 & 7881 \\
157910 & G5 III   & 1.83 & $-$0.32 & a & 2007-01-15 & 1306 & 8081 &  97907 & K3 III   & 2.07 & $-$0.10 & a & 2004-11-13 &  870 & 7572 \\
107950 & G6 III   & 2.61 & $-$0.16 & b & 2008-02-04 &  861 & 7124 & 102328 & K3 III   & 2.09 & $+$0.35 & a & 2005-05-03 & 1129 & 7647 \\
175535 & G7 IIIa  & 2.55 & $-$0.09 & a & 2007-01-12 & 1173 & 8296 & 127665 & K3 III   & 2.22 & $-$0.17 & b & 2007-01-15 & 1114 & 8223 \\
184492 & G8-9 III & 2.59 & $-$0.10 & b & 2005-05-02 &  487 & 8179 & 143107 & K3 III   & 2.34 & $-$0.32 & b & 2005-05-02 & 1419 & 8216 \\
  6833 & G8 III   & 1.25 & $-$0.99 & a & 2004-11-19 &  602 & 7911 & 184406 & K3 III   & 2.41 & $+$0.01 & a & 2005-05-02 &  836 & 8295 \\
 38656 & G8 III   & 2.52 & $-$0.22 & a & 2006-10-14 &  727 & 7405 &  48433 & K3 V     & 4.55 & $-$0.07 & d & 2004-11-19 &  710 & 7261 \\ 
 57264 & G8 III   & 2.72 & $-$0.33 & a & 2008-02-04 &  836 & 7549 & 149161 & K4 III   & 1.39 & $-$0.17 & a & 2005-05-02 &  770 & 8211 \\
 65714 & G8 III   & 1.50 & $+$0.27 & a & 2004-11-19 &  975 & 7644 &  49161 & K4 III   & 1.69 & $+$0.08 & a & 2004-11-19 & 1148 & 7364 \\
103736 & G8 III   & 2.3  & $-$     & g & 2004-11-16 &  756 & 7660 & 148513 & K4 IIIp  & 1.67 & $+$0.11 & a & 2005-05-02 &  635 & 8352 \\
141680 & G8 III   & 3.02 & $-$0.28 & b & 2005-05-02 &  789 & 8106 &  61603 & K5 III   & 1.50 & $+$0.24 & a & 2004-11-19 &  875 & 7523 \\
133640 & G8 IIIv? & $-$  & $-$     & g & 2007-01-14 &  661 & 7665 & 136028 & K5 III   & 1.90 & $+$0.19 & g & 2005-05-02 &  384 & 8154 \\
 73593 & G8 IV    & 2.25 & $-$0.12 & a & 2004-11-15 &  956 & 7733 & 146051 & M0.5 III & 1.40 & $+$0.32 & a & 2008-02-04 & 1023 & 6475 \\
 82885 & G8 IV-V  & 4.61 & $+$0.00 & a & 2005-11-06 &  488 & 7711 & 137471 & M1 III   & 1.10 & $+$0.07 & a & 2008-01-31 &  853 & 7642 \\
 41636 & G9 III   & 2.50 & $-$0.20 & a & 2004-11-19 &  804 & 7903 & 167006 & M3 III   & 0.70 & $+$0.00 & a & 2008-02-04 &  648 & 7106 \\
104985 & G9 III   & 2.52 & $-$0.26 & c & 2006-10-13 &  567 & 7153 & 123657 & M4 III   & 0.85 & $+$0.00 & a & 2008-02-04 &  739 & 7164 \\
180711 & G9 III   & 2.67 & $-$0.12 & a & 2007-01-13 & 1235 & 8519 & 148783 & M6 III   & 0.20 & $-$0.01 & f & 2008-02-04 &  839 & 6940 \\
 19476 & K0 III   & 3.08 & $+$0.04 & b & 2006-10-13 &  940 & 7358 &        &          &      &       &   &            &      &      \\
\hline
\end{tabular}
}
\tablefoot{
(1) HD~number; (2) spectral type; (3) surface gravitation; (4) metallicity;
(5) reference; (6) date of observation; (7) signal-to-noise ratio ($S/N$); 
(8) effective instrumental spectral resolution ($\lambda/\Delta\lambda$). 
The template star used for the stellar-kinematic analysis in this paper (HD 162555) has
been highlighted. The $S/N$ and $\lambda/\Delta\lambda$ are for the {\it combined} 
spectrum (Section~\ref{sec:CombTemp}). The spectral type, surface gravity and metallicity
have been taken from the following references:
(a) \cite{cenarro2007}; (b) \cite{mcwilliam1990}; (c) \cite{luck2007}; 
(d) \cite{soubiran2008}; (e) \cite{massarotti2008}; (f) \cite{ramirez2000}; 
(g) N. Cardiel (priv. comm.).
}
\end{table*}

Simultaneously with our science exposures, the 15 calibration fibers were
illuminated independently by the internal ThAr lamp. Typically, the shutter of
the internal lamp was opened $5\times8$ seconds, interspersed regularly
throughout each hour-long integration, producing emission-line spots in all
areas of the detector.  This procedure allowed us to trace any image shifts due
to the flexure (Sect.~\ref{sec:Flexure}) and to calculate the effective
resolution of every science exposure as a function of wavelength and aperture
number (Sect.~\ref{sec:InstRes}).

A particularly salient feature of the PMAS spectrograph and its built-in guide
CCD is its capability to save and reload the exact position of a guide star with
an accuracy of $0\farcs2$, the scale of a single pixel.  This allowed for an
accurate reproduction of the PPak pointing for each galaxy.  However, since
several telescope maintenance operations occurred over the course of our
observing campaign, including a removal of the PPak unit on 16~November~2006 and
a dismount of the Acquisition and Guide (A\&G) camera on 24~April~2006, the
relative offset between the guide camera and the PPak IFU was not consistent
between runs before and after these dates. Thus, although our reacquisition
strategy was successful for a given run, shifts in the PPak pointing between two
runs could be larger than one fiber diameter.  This has some consequences for
our ability to co-add multiple exposures of a single galaxy taken over more than
one run (Sect.~\ref{sec:Comb}).

Finally, the operating system of the PMAS spectrograph allows for each exposure
of the guide star, typically $15-120$ seconds long, to be archived. For each
galaxy exposure the series of guide-star exposures can be combined to yield an
average guide-star image from which the effective seeing for that particular
exposure can be determined (Sect.~\ref{sec:Seeing}).  The measured seeing will
be used in future papers to correct the data for the effects of beam smearing
caused by the relatively large on-sky fiber diameter.

\subsection{Observations of template stars}
\label{sec:ObsTempStars}

During nautical and astronomical twilight, we observed 69 stars with a wide
range in spectral type, luminosity class and metallicity. The stellar spectra,
obtained with the same instrumental setup as the science exposures, are used as
spectral templates in our cross-correlation approach for measuring the stellar
kinematics of the galaxies \citep[][hereafter
\citetalias{westfall2011a}]{westfall2011a}. A list of the observed template
stars can be found in Table~\ref{tab:TempStars}.  For the analysis in this paper
we have used the single stellar spectrum of the K1 III star HD162555 as a
template (highlighted in Table~\ref{tab:TempStars}).

To improve both the $S/N$ and efficiency of our template-star observations, we
observed the stars by drifting them from east to west over the central row of
fibers.  The drift speed depended on the apparent magnitude of the star,
resulting in effective integration times per fiber that varied from $3-27$
seconds.  In this way, 21 fibers could be illuminated in a single drift scan,
yielding at least 21 spectra that were later combined into a single template
spectrum with a very high $S/N$ as tabulated in Table~\ref{tab:TempStars}.
Figure~\ref{fig:rawdata_sci} shows an example exposure of a template star.

\section{Data Reduction}
\label{sec:Reduction}

This paper describes an extensive data set collected during eight different
observing runs over a period of more than three years.  We have made
substantial efforts to ensure that the reduction and analysis of these data have
been carried out in a robust, consistent, and homogeneous manner.  Due to the
need for high-precision kinematic measurements, we have been extra careful to
minimize the errors in the wavelength calibration.

The first basic reduction steps have been done in a standard way by subtracting
the bias and dark currents from all calibration and science images, and
subsequently dividing the resulting images by a flat-field image to correct for
pixel-to-pixel sensitivity variations. The distortions due to differential
flexure have been robustly determined, using the ThAr spots in the calibration
fibers as described in Sect.~\ref{sec:Flexure}.  Subsequently, the high $S/N$
calibration images have been reprojected to match the shift, rotation and plate
scale of each individual science exposure.  Using the {\tt hydra} package in
{\it IRAF},\footnote{
{\it IRAF} is distributed by the National Optical Astronomy Observatories, which
are operated by the Association of Universities for Research in Astronomy, Inc.,
under cooperative agreement with the National Science Foundation.
} the galaxy spectra have been extracted using dome-flat spectra to define the
extraction apertures, sky-flat spectra to correct for fiber-to-fiber throughput
variations, and calibrated spectra from the flood lamps to wavelength calibrate
the science spectra.  The quality of the wavelength calibrations has been
inspected extensively (Sect.~\ref{sec:dohydra}). Finally, we have subtracted
the simultaneously recorded sky spectra (Sect.~\ref{sec:Sky}) and combined the
exposures for each galaxy (Sect.~\ref{sec:Comb}) using various programs from
the {\it GIPSY} software package \citep{hulst1992, vogelaar2001}.

For subsequent analysis, it is important to determine the effective instrumental
spectral resolution in the extracted galaxy and template-star spectra. This can
vary significantly between different apertures, at different wavelengths, and in
different images.  Therefore, we have measured the spectral resolution of each
science exposure by fitting Gaussian profiles to the Argon lines in the
simultaneously recorded and extracted ThAr spectra from the 15 calibration
fibers. These measurements are then interpolated in both spatial and spectral
directions for every aperture and wavelength channel (Sect.~\ref{sec:InstRes}).

Below follows a more detailed description of these reduction steps.

\subsection{Bias and dark-current subtraction}

An overscan region is provided by the CCD controller, but we have not applied a
traditional overscan correction; charge from the exposed part of the detector,
especially in the areas exposed to light from bright science fibers, often
spills over into the overscan region.  Therefore, we simply subtracted a
master-bias image constructed from at least 14 individual bias read-outs.  There
is significant spatial structure in the bias with an increase in the bias level
of 7~ADU from the bottom to the top of the bias frames \citep[see Figure~2.3
from][]{martinsson2011}.  Depending on the temporal stability of the shape of
the bias pattern, master-bias images were constructed for each night, for a
series of consecutive nights, or for an entire observing run.  Maximizing the
number of individual bias read-outs combined in the master-bias minimizes the
contribution of the master-bias correction to the total error budget of the
extracted science spectra.

During each of 12 nights on 6 different runs, up to 5 dark exposures of 3600
seconds each were collected. After subtracting the associated master-bias
images, the statistical properties of the dark current, like its mean and RMS
noise, vary across the detector and among dark images from different observing
runs \citep{martinsson2011}.  However, the statistical variations within and
among the dark-current images is sufficiently small.  Therefore, we do not
subtract these noisy dark images from the science images but simply subtract a
representative dark current of 3.0~ADU/hour from all images.

\subsection{Pixel-to-pixel map and bad-pixel correction}
\label{sec:P2P}

Due to flexure, the location of the science spectra on the detector changes
slightly during the night such that the sensitivities sampled by our calibration
frames will not exactly correspond to those sampled by the science frames.  To
correct for this, we have created a normalized pixel-to-pixel sensitivity map
using unfocused and focused dome-flat frames over several different runs, which
homogenizes the illumination of the CCD frame.  To each row in this combined
dome-flat image we have fit a high-order (20) cubic-spline function in the
dispersion direction and have divided each row by this smooth function to obtain
a pixel-to-pixel sensitivity map normalized to unity \citep[see Figure~2.6
from][]{martinsson2011}.  The smooth variations in this normalized map are
typically much smaller than one percent, with the exception of a few small areas
where sensitivities drop to a few percent lower than the mean and one elliptical
``dark spot'' of $\sim100\times50$~pixels that is half as sensitive at its
center as compared to the rest of the frame \citep{martinsson2011}. This dark
spot typically affects 5--10 fibers in a $\sim$20\AA\ wide window in the red
(near $\lambda\sim5350$\AA).  We also identified defunct pixels and pixel
columns using the master-bias frames from both November 2004 and January 2007.
The values of these bad pixels in each image were replaced by a linear
interpolation of their neighboring pixels.

\subsection{Combining calibration spectra}

Dome-flat, sky-flat, and flood-lamp images were combined to increase their $S/N$
and to remove cosmic rays, typically using four or five images for each set of
exposures.  To both successfully reject cosmic rays and produce high $S/N$
images, the emission lines in these data must have a relatively stable flux; if
not, the statistical rejection scheme would have inadvertently modified the
shape, and thus the centroid, of the emission lines in the combined image
thereby degrading, e.g., the wavelength calibration.  This motivated our
decision to allow ample time for the temperature and pressure in the lamps to
settle.  In addition, we checked the emission-line stability of, specifically,
the flood-lamp spectra among a particular set by considering difference images.
If any of the images were different from the majority, they were excluded in the
combined image.  Often the first image in a series was affected and rejected,
indicating that our nominal settling time was insufficient for yielding a stable
flood-lamp temperature and pressure.  The sky-flat and dome-flat images were
also carefully checked to ensure their statistical similarity and co-location of
their spectra before combining them.

\subsection{Flexure corrections}
\label{sec:Flexure}

\begin{figure}[t]
\centering
\includegraphics[width=0.49\textwidth]{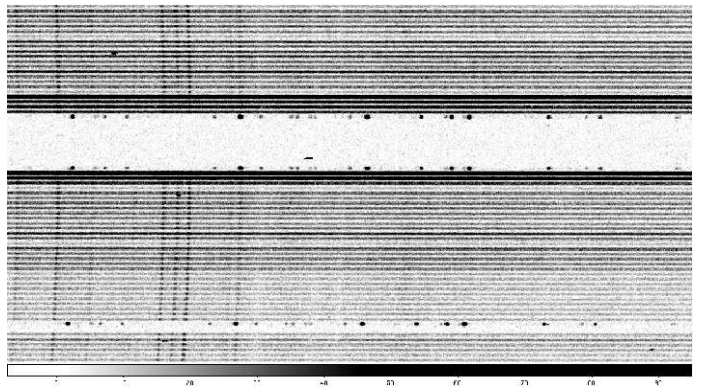}
\caption[]{
Central part of a raw image from a one-hour science exposure with inverted
color map, illustrating how the emission lines in the ThAr spectra create spots
all over the detector. These spots are used to track the effective spectral
resolution and flexure in the Cassegrain-mounted spectrograph.
}
\label{fig:ThArspots}
\end{figure}

Shifts in the location of the spectra on the detector due to flexure in the PMAS
spectrograph have been presented by \citet{roth2005}.  However, we also detect
non-negligible scale and rotation effects that can vary between the calibration
and science frames, not previously investigated by \citet{roth2005}.  Since the
dome-flat images are used to trace the spectra on the science frames, and the
flood-lamp images are used to wavelength calibrate the science spectra, it is
important to establish a robust method of measuring flexure in the spectrograph
and correct for its effects on our data.

The shift, rotation and scale differences were found by using several strong
ThAr emission lines in the 15 calibration spectra that are simultaneously
observed during the science and flood-lamp observations.  As can be seen in
Fig.~\ref{fig:ThArspots}, these emission lines create spots distributed over
the entire CCD frame; 60 of these spots have been used to calculate the
pixel-coordinate transformation matrix used to correct for flexure. By fitting
two-dimensional Gaussian profiles to these 60 spots, their positions in every
flood lamp and science frame were measured to an accuracy of 0.05~pixels or
better. Using these 60 positions, the mean shift, rotation and scale differences
between the flood-lamp and science frames could then be calculated with very
high accuracy.

\begin{figure}[t]
\centering
\includegraphics[width=0.49\textwidth]{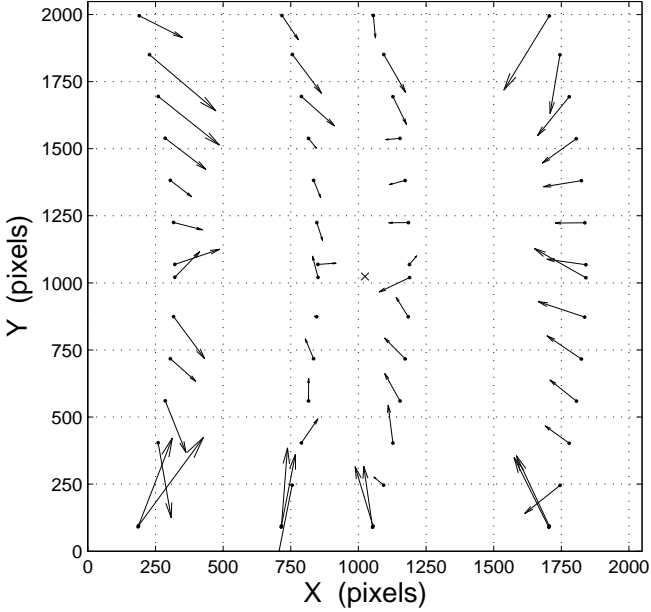}
\caption[]{
Measured shifts in the locations of 60 ThAr spots from a flood-lamp exposure and
from one of the science frames taken during the same night, illustrating a
severe case of flexure.  The shifts are indicated by arrows, calculated after
the mean shift ($\Delta X=0.21$ and $\Delta Y=0.46$ pixels) has been subtracted;
the arrows have been magnified 1000 times for illustration purposes. The change
in scale is clearly visible.
}
\label{fig:scalerot}
\end{figure}

We find that the shifts between the calibration and associated science frames,
due to the difference in telescope pointing between the observations, are
generally smaller than one pixel in both the $X$ (spectral) and $Y$ (spatial)
directions; however, 30\% of the frames have shifts of $1.0-3.6$ pixels in $X$
and/or $Y$, and 10\% of the frames are shifted more than one pixel in both
directions.  Figure~\ref{fig:scalerot} demonstrates the scale and rotation
effects between an example calibration frame and its associated science frame.
Over all data, we find that the scale differences are always less than 0.05\%,
and less than 0.01\% in more than 50\% of the images, resulting in shifts of,
respectively, 0.5 and 0.1 pixels at the edge of the detector.

After determining the coordinate transformation matrix, the calibration frames
were shifted, scaled and rotated with respect to every science frame. By
performing the transformations on the calibration frames, the interpolation
algorithms only affect the high $S/N$ frames, leaving the individual low $S/N$
science frames unaltered.

We assess the accuracy of our flexure corrections by determining the positions
of the ThAr lines in the reprojected flood-lamp frames using the method
described above.  The remaining differences between the centroids of the ThAr
spots in the science frames and in the reprojected flood-lamp frames are on
average 0.023$\pm$0.017 pixels, with an extreme case at 0.16 pixels. Comparing
the distribution of uncertainties to the non-corrected values \citep[see Figures
2.8 and 2.10 from][]{martinsson2011}, it is clear that the remaining deviations
are much smaller than the original offsets and generally negligible.  For frames
with large uncertainties in the reprojection matrix, however, additional checks
were made before further processing of the reprojected calibration frames.  In
particular, the precision of the wavelength calibration was assessed.  Overall,
the uncertainty in the wavelength calibration caused by these shifts is less
than 0.04 and 0.16 spectral pixels (0.5 \kms\ and 1.9 \kms\ at 5150\AA) for 90\%
and 100\% of our data.

\subsection{Error spectra}
\label{sec:Error}

We have calculated error spectra for our science observations as necessary for
our approach to measuring stellar kinematics (see Sect.~\ref{sec:extractkin}
and \citetalias{westfall2011a}).  We briefly describe the procedure used to
generate these error spectra here; a more detailed description can be found in
Appendix~B from \citet{westfall2009}.

Using the gain, $\eta$ (e$^-/$ADU), and read-noise, $\epsilon_{\rm RN}$ (e$^-$),
of the detector, we calculate the noise in each pixel of the raw image as
\begin{equation}
\epsilon(e^-) = \sqrt{\eta S + \epsilon_{RN}^2 + \epsilon_{b}^2},
\label{eq:errorspec}
\end{equation}
where $S$ (ADU) is the total signal and $\epsilon_{b}$ (e$^-$) is the added
noise due to subtraction of the master-bias images.  We neglect the noise that
is added when correcting for the pixel-to-pixel sensitivity variation
(Sect.~\ref{sec:P2P}).  We produced error spectra by processing the
two-dimensional error image in a way that complements the processing of our
science data, accounting for the appropriate propagation of the errors at every
step.  In particular, the spectral extraction, spectral resampling, and sky
subtraction are all properly accounted for in the final error spectra.

\subsection{Extraction and wavelength calibration of science spectra}
\label{sec:dohydra}

We have extracted and wavelength calibrated the science spectra using the {\it
HYDRA} package in {\it IRAF}.  For calibration, we selected the set of
calibration data (taken during the evening or morning) that are the most
complete (i.e., containing dome-flat, sky-flat, and flood-lamp images), have the
smallest flexure correction, and provide the most robust wavelength calibration.

Flexure solutions were calculated and applied to the dome-flat images for each
science and wavelength-calibration frame, and these corrected dome-flat images
were used to define the extraction aperture for each spectrum.  We used a 
fifth-order cubic-spline function to trace the spatial centroid of the PSF, and
we set the extraction width to the FWHM of the PSF near the central wavelength of
the spectrum.  This extraction width limits the scattered light between spatially
adjacent apertures to less than 1\%, as determined by considering the PSF of bright
stellar apertures adjacent to dark sky apertures in our template-star observations
\citep[see][for a detailed discussion of optimal extraction apertures]{bershady2005}.
Dome-flat spectra were also used to correct for spectral vignetting of the science
spectra \citep[Figure~2.3 from][]{martinsson2011}, whereas our sky-flat spectra were
used to correct for fiber-to-fiber throughput variations.

Wavelength solutions for each science spectrum were found by fitting a
first-order cubic-spline function to the pixel coordinates of the
wavelength-calibrated emission lines in the extracted flood-lamp spectra.
Low-quality wavelength solutions occurred for spectra near the top edge of the
PPak pseudo-slit for some frames (notably those taken on $2-3$ November 2005 and
$14-15$ October 2006).  This was due to a poor focus of the spectrograph that
resulted in strongly platykurtic (sometimes doubly peaked) emission lines,
particularly toward the blue part of the spectrum for high aperture numbers. For
the affected spectra, we found a high RMS scatter in our wavelength calibration;
we also flagged and omitted these spectra when combining the galaxy spectra from
multiple runs (Sect.~\ref{sec:Comb}).  After wavelength calibration, all
science spectra are resampled to a common (linear and log-linear) wavelength
scale for subsequent analysis; all spectra contain 2048 pixels and have a
wavelength range from $4980-5369$\AA\ (0.19\AA\ per pixel for the linear
wavelength scale).

We have performed several inspections of the wavelength-calibrated flood-lamp
and galaxy spectra to quantify and assess any adverse effects that our flexure
solutions may have had on the quality of the wavelength calibration.  First, we
have determined the RMS difference between the measured and tabulated
wavelengths of the $18-23$ flood-lamp emission lines as a measure of the
relative precision of the wavelength calibration.  The typical RMS difference is
$\sim$0.01\AA\ or $\sim$5\% of a spectral pixel; only 1.2\% of all calibrated
spectra have a wavelength-calibration RMS above 0.03\AA\ ($\sim$2\kms).  An
image with the RMS values for all observations is provided in Figure~2.11 from
\citet{martinsson2011}.

\begin{figure}[t]
\centering
\includegraphics[width=0.49\textwidth]{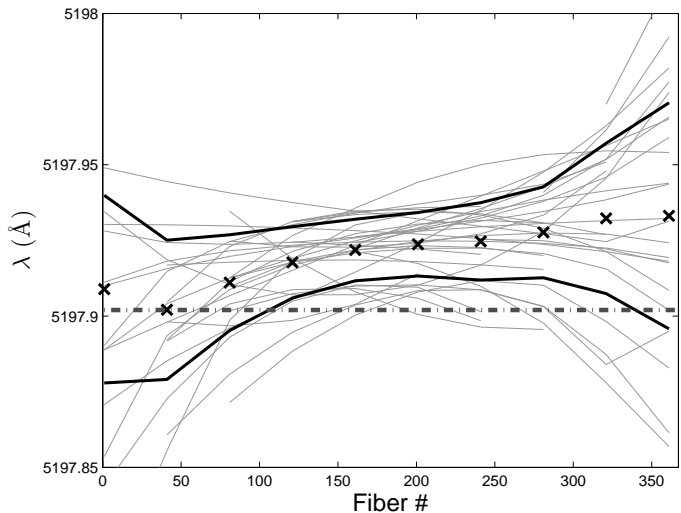}
\caption[]{
Measured wavelength of the \none\ sky-line at 5197.9\AA\ as a function of fiber
number.  The thin, grey lines connect measurements from 10 spectra in individual
science exposures.  Crosses show the averages of these measurements, and the
thick, black lines indicate the standard deviation from the mean. The
dash-dotted line depicts the tabulated value of \none\ in air, as given by NIST.
}
\label{fig:NIcheck}
\end{figure}

Second, we inspected the accuracy of the wavelength-calibration zero-point using
sky lines in the galaxy spectra.  Unfortunately, only the \none\ line at
5197.9\AA, near the center of the observed spectral range, was strong enough for
reliable centroid measurements.  When possible, we measured centroids for strong
\none\ lines (exceeding three times the RMS in the continuum level) in 10 galaxy
spectra spread equally along the pseudo-slit, which was possible for 39 science
exposures from all but one observing run. Figure~\ref{fig:NIcheck} compares
these measurements with the tabulated value of \none\ provided by NIST.

Figure~\ref{fig:NIcheck} demonstrates that the error in our wavelength
calibration is lowest for spectra near the middle of the pseudo-slit, increasing
toward both low and high aperture numbers.  However, the spread in \none\
centroids toward the edges of the pseudo-slit is also effected by the error in
the centroid measurements due to a decrease in the line $S/N$ according to the
vignetting function.  At the center of the pseudo-slit, we find a marginal
systematic shift of $\sim$0.022\AA\ ($\sim$0.12 spectral pixels or $\sim$1.3\kms)
toward the red, most likely introduced by the wavelength calibration of
the flood-lamp spectrum using the internal ThAr lamp.  We note that temperature
and/or pressure changes during calibration may have contributed to this effect:
an emission line may shift by 0.02\AA\ if the ambient temperature of the air
changes by 4 degrees.  Although not true of all frames, the \none\ line
generally shifts toward the red going from the bottom to the top of the
pseudo-slit, which may be due to a minute systematic residual in the
reprojection matrix from the flexure solutions (Sect.~\ref{sec:Flexure}).
In any event, the scatter and the trend is at the level of $1-2$ \kms\ and will
have no astrophysical implications when analyzing the galaxy spectra.

\subsection{Sky subtraction}
\label{sec:Sky}

The 36 dedicated sky fibers have been used to subtract the sky from our target
spectra for each science frame.  Given the change in spectral resolution across
the pseudo-slit, we do not subtract an average sky spectrum from our target
spectra.  Instead, we fit a second-order polynomial along the spatial direction
at every (wavelength-calibrated) spectral pixel.  Pixels in the sky spectra that
are affected by cosmic rays, or entire sky spectra that are contaminated by
nearby sources (e.g., stars), were rejected during this fit. We then subtract
the sky from the target spectra by interpolating the sky flux between sky
apertures as determined by this fit.  Our error spectra account for noise in
this sky subtraction by adding (in quadrature) the RMS difference in the fit to
errors calculated prior to sky subtraction; the sky-subtraction error is
typically 20\% of the noise in the outer, read-noise dominated galaxy spectra
before sky subtraction.

Sky subtraction of our template-star observations is generally more difficult
than for our galaxy spectra because the stellar spectra contaminate adjacent
spectra, including the sky spectra.  Therefore, we include any spectrum free of
stellar emission in the sky subtraction procedure, regardless of its designation
as a sky or science fiber.

\subsection{Combining science spectra}
\label{sec:Comb}

As explained in Sect.~\ref{sec:Observations}, we expect some galaxies observed
during several observing runs to have pointings that are not common to all
exposures.  In Table~\ref{tab:UGC_comb}, we group science exposures that have a
common pointing and provide some relevant information regarding the
observations.  In particular, this table provides the number of one-hour
exposures taken for each galaxy at a particular pointing.  This section details
how we have combined these exposures, typically $3-8$ per galaxy, to increase
$S/N$ and eliminate the effects of cosmic rays.

\begin{table}
\caption{\label{tab:UGC_comb}
Properties of the 41 combined galaxy exposures}
\centering
{\small
\begin{tabular}{|r c c c c c|}
\hline
  UGC  & Run   & N$_{\rm exp}$  & $\lambda/\Delta\lambda$ &  Seeing & ($\Delta$R.A. , $\Delta$Dec.) \\
  (1)  &    (2)     &  (3) &  (4)   &  (5)  &          (6)      \\
\hline
  448  &  Nov\hfill '05   &    7 &  7711  &   2.1 &                   \\
  463  &  Nov\hfill '04   &    6 &  7557  &   1.7 &                   \\
 1081  &  Aug\hfill '06   &    5 &  7726  &   1.4 &                   \\
       &  Oct\hfill '06   &    2 &  7208  &   1.5 & ($+0.68 , +0.50$) \\
 1087  &  Nov\hfill '04   &    7 &  7560  &   0.9 &                   \\
 1529  &  Oct\hfill '06   &    6 &  7748  &   1.6 &                   \\
 1635  &  Nov\hfill '04   &    6 &  7515  &   2.1 &                   \\
 1862  &  Nov\hfill '05   &    3 &  7478  &   1.0 &                   \\
       &  Oct\hfill '06   &    8 &  8102  &   1.3 & ($+0.43 , +0.30$) \\
 1908  &  Nov\hfill '05   &    8 &  6856  &   2.1 &                   \\
 3091  &  Nov\hfill '05   &    5 &  7310  &   1.6 &                   \\
 3140  &  Nov\hfill '04   &    7 &  7465  &   2.1 &                   \\
 3701  &  Nov\hfill '04   &    7 &  7635  &   1.0 &                   \\
       &  Jan\hfill '07   &    5 &  8125  &   1.7 & ($-2.46 , +0.82$) \\
       &  Jan\hfill '08   &    2 &  7319  &   1.7 & ($-4.27 , -2.81$) \\
       &  Jan\hfill '08   &    3 &  7639  &   1.2 & ($-2.68 , +0.13$) \\
 3997  &  Nov\hfill '05   &    6 &  7399  &   1.2 &                   \\
       &  Oct\hfill '06   &    2 &  6847  &   2.1 & ($+3.00 , +2.06$) \\
       &  Jan\hfill '07   &    8 &  8182  &   1.5 & ($+0.71 , +1.29$) \\
 4036  &  Nov\hfill '04   &    6 &  7582  &   1.0 &                   \\
       &  Jan\hfill '08   &    2 &  7238  &   1.7 & ($+4.88 , +1.75$) \\
 4107  &  Nov\hfill '05   &    5 &  6449  &   1.2 &                   \\
 4256  &  Oct\hfill '06   &    6 &  8136  &   1.5 &                   \\
 4368  &  Jan\hfill '07   &    6 &  7949  &   1.3 &                   \\
 4380  &  Nov\hfill '04   &    5 &  7539  &   0.9 &                   \\
 4458  &  Jan\hfill '07   &    7 &  8021  &   1.2 &                   \\
 4555  &  Oct\hfill '06   &    7 &  7385  &   1.6 &                   \\
       &  Jan\hfill '08   &    5 &  7467  &   1.5 & ($+0.97 , -2.99$) \\
 4622  &  Nov\hfill '04   &    5 &  7435  &   2.0 &                   \\
 6903  &  Jan\hfill '08   &    1 &  7428  &   1.7 &                   \\
 6918  &  Nov\hfill '04   &    2 &  7676  &   0.8 &                   \\
       &  Jan\hfill '07   &    5 &  8325  &   0.9 & ($-3.10 , +0.73$) \\
 7244  &  Jan\hfill '07   &    5 &  7773  &   1.5 &                   \\
 7917  &  Jan\hfill '08   &    4 &  7243  &   1.3 &                   \\
 8196  &  Jan\hfill '07   &    8 &  7909  &   1.5 &                   \\
 9177  &  Jan\hfill '07   &    6 &  7926  &   1.2 &                   \\
       &  Jan\hfill '08   &    4 &  7249  &   1.5 & ($-0.82 , +0.61$) \\
 9837  &  May\hfill '05   &    6 &  7447  &   1.6 &                   \\
 9965  &  May\hfill '05   &    5 &  7869  &   1.0 &                   \\
11318  &  May\hfill '05   &    6 &  7634  &   1.2 &                   \\
12391  &  Oct\hfill '06   &    6 &  7869  &   1.1 &                   \\
\hline
\end{tabular}
}
\tablefoot{
(1) UGC number; (2) when
the observations were taken; (3) number of exposures combined; (4) average effective
spectral resolution; (5) average effective seeing (arcsec); (6) pointing offset of
combined image with respect to the first combined image obtained (arcsec). Every single
exposure was 3600 seconds, except the two exposures of UGC~6918 from Nov '04 which had 2100
seconds each. UGC~3701 was observed with different pointings during two different nights
in Jan '08. The multiple combined images of eight galaxies were eventually merged as
described in Section~\ref{sec:merging} and listed in Table~\ref{tab:UGC_merge}.
}
\end{table}


We mask the cosmic rays by identifying statistically aberrant pixels in each
exposure.  We do this by subtracting the median of each group of exposures from
each exposure in the group.  Subsequently, we ran an iterative sigma-clipping
algorithm to create a cosmic-ray mask for each exposure.  This works
satisfactorily despite the small flexure shifts between the exposures within a
group and the varying intensity of the sky during a long series of exposures.
For galaxies with only one or two exposures, cosmic-ray masks were created
manually.  Masked pixels have been given a zero weight when averaging the images
together.  UGC~6903 is the only galaxy in our sample with a single exposure; the
masked pixels in this case were replaced by a linear interpolation in the
spectral dimension using neighboring, uncontaminated pixels.

Extracted spectra from individual frames within a group of exposures have been
combined on a fiber-by-fiber basis.  For the central fibers (fibers $148-183$;
all within $11\arcsec$ from the central fiber 164) the spectra were weighted by
the square of their mean $S/N$, times the mean spectral resolution of the
spectrum.  That is, the weight ($w_i$) of fiber $i$ is
\begin{equation}
w_i = (S/N)_i^2 (\lambda/\Delta\lambda)_i ,
\label{eq:combweight}
\end{equation}
where $(\lambda/\Delta\lambda)_i$ is the mean spectral resolution of the
spectrum.  The noise in fibers at larger distances from the center with fainter
continuum levels is dominated by the read-noise, and thus similar for all
spectra. Nevertheless, a varying transparency of the atmosphere motivates
assigning different weights to the different exposures.  Therefore, we use the
average $S/N$ of all recorded spectra as relative weights in
Eq.~(\ref{eq:combweight}).  Spectra with large errors on the wavelength
calibration, i.e., with a large RMS in the fit (Sect.~\ref{sec:dohydra}), are
completely masked in the weighted average. The error in each weighted-average
spectrum is calculated by propagating the errors on the individual spectra.

Weighted-average spectra for galaxies with multiple exposure groups (8 of 30
galaxies; see Table~\ref{tab:UGC_merge}) could not simply be combined on a
fiber-by-fiber basis because the pointing differences would result in an
unacceptable effective on-sky beam smearing.  These spectra were instead
combined at a later stage, after the difference in the pointing had been
determined (see Sect.~\ref{sec:merging}).

Possible wavelength shifts between different exposures of the same galaxy, e.g.,
due to significantly different flexure corrections or wavelength calibrations,
have been estimated by considering spectral shifts of the \none\ sky line
(Sect.~\ref{sec:dohydra}).  Only negligible wavelength shifts were detected.
Note that possibly larger wavelength shifts may exist between the spectra from
different apertures in the same exposure than between spectra from the same
aperture in different exposures.  However, line broadening due to slight
wavelength shifts between the spectra from different exposures is always much
smaller than the instrumental resolution, and therefore ignored.

\subsubsection{Merging galaxy spectra from different pointing centers}
\label{sec:merging}

Eight of our galaxies were observed during multiple runs. For these galaxies, we
combined the spectra from exposures with identical pointings, determined the
continuum levels in the combined spectra, and reconstructed the continuum image
as described in Sect.~\ref{sec:ContMaps}; this produced a reconstructed
continuum image for each exposure set. Subsequently, we determined the pointing
offsets between the $2-4$ reconstructed continuum images within an exposure set
(Sect.~\ref{sec:ContMaps}) and these offsets are listed in Table~\ref{tab:UGC_comb}.

With the pointing offsets known, we determined sets of fibers to combine that
were within 1.5 fiber radii of one another.  If the pointing offsets were small
enough, this procedure simply led to a larger beam profile for the 331 science
fibers.  However, in many cases the offsets were enough that some fibers near
the edge of the field-of-view were not combined with any other fiber.

\begin{table}[t]
\caption{\label{tab:UGC_merge}
Galaxies with multiple pointings}
\centering
\begin{tabular}{|r r c c c|}
\hline
  UGC  & N$_{\rm exp}$  & $\lambda/\Delta\lambda$ & Seeing & N$_{\rm spec}$ \\
  (1)  &  (2)   &  (3)   &   (4)    &  (5) \\
\hline
 1081  &      7 &  7590  &   1.4    &  331 \\
 1862  &     11 &  7807  &   1.2    &  331 \\
 3701  &     17 &  7762  &   1.4    &  373 \\
 3997  &     16 &  7928  &   1.4    &  357 \\
 4036  &      8 &  7517  &   1.1    &  372 \\
 4555  &     12 &  7431  &   1.5    &  352 \\
 6918  &      7 &  8267  &   0.9    &  352 \\
 9177  &     10 &  7719  &   1.3    &  331 \\
\hline
\end{tabular}
\tablefoot{
(1) UGC number; (2) number of exposures in the merged data; (3) 
effective spectral resolution; (4) effective seeing (arcsec); (5) number of galaxy spectra
after merging.
}
\end{table}


The groups of ``spatially overlapping'' fibers were combined using a weighted
average as described above when using Eq.~(\ref{eq:combweight}).  Note that
the final number of merged spectra may exceed the number of fibers in the PPak
IFU if the pointing offsets are sufficiently large. Because consecutive
observations of a galaxy may have been acquired after several months had passed,
the spectra to be combined were shifted to zero heliocentric velocity in order
to avoid broadening of the spectral lines due to differences in apparent
recession velocity. For each merged spectrum, we calculated the new effective
spectral resolution and effective seeing.  The effective sky position of each
fiber is taken to be the weighted fiber position, where the weight for each
fiber is $w_i$ from Eq.~(\ref{eq:combweight}) divided by the square of the
error in the pointing position. The effective fiber diameter for each merged
spectrum is taken to be the sum of the nominal fiber diameter and the quadrature
difference between the weighted standard deviation in the fiber position and the
propagated error in the position in each dimension. These attributes of the
merged spectra can be found in Table~\ref{tab:UGC_merge}.  
The combined and merged individual spectra are displayed in greyscale in the
upper-right panel of the accompanying Atlas, where the spectra have been reordered
based on the aperture distance to the center.
The \oiii\ emission lines are clearly
visible, as well as many stellar absorption lines. With a favourable north-south
orientation of the kinematic major axis, the rotation of the stellar disk can be
clearly discerned by an offset in wavelength of the absorption lines. This
effect is particularly visible for UGC~4036 and UGC~6918.

\subsubsection{Combining template-star spectra}
\label{sec:CombTemp}

For our observed template-star frames (Sect.~\ref{sec:ObsTempStars}), we
combined all extracted spectra with $S/N > 20$ and $\lambda/\Delta\lambda>5500$;
each spectrum was weighted according to $w_i=(S/N)_i^2$.  As provided in
Table~\ref{tab:TempStars}, our observational strategy has resulted in
template-star spectra with $S/N>300$ and $\lambda/\Delta\lambda>6500$
($\sinst\approx20$ \kms); the effective spectral resolution in each combined
spectrum is calculated using Eq.~(\ref{eq:sigmaInst}).

\subsection{The instrumental spectral resolution}
\label{sec:InstRes}

Accurate stellar-velocity-dispersion measurements require a detailed matching of
the instrumental dispersions ($\sinst$) in the galaxy and template spectra, or
for corrections to be made to the raw kinematic measurements (as we do here).
In either case, one requires measurements of $\sinst$ for both spectra.  For our
PPak observations, this has been achieved using the 15 ThAr spectra from the
simultaneously illuminated calibration fibers in the science frames.  These ThAr
spectra were extracted in an identical manner as the science spectra and
wavelength calibrated using the argon lines (see Fig.~\ref{fig:CalStra}).  Due
to the strong curvature of the imaged slit on the detector (see, e.g.,
Fig.~\ref{fig:rawdata_sci}), the algorithm used by {\tt dohydra} to automatically
shift the calibration to adjacent spectra failed, requiring an initial manual
shifting of the spectra to be done before running {\tt dohydra}.

\begin{figure}[t]
\centering
\includegraphics[width=0.49\textwidth]{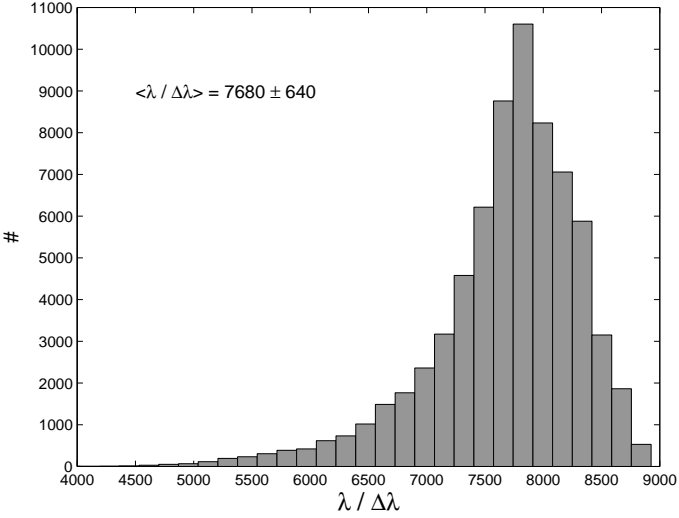}
\caption[]{
Distribution of the average instrumental spectral resolution in all science
exposures.  The average resolution is $\lambda/\Delta\lambda=7680\pm640$
($\sinst\approx16.6\pm1.4$ \kms), which is somewhat lower than the most common
resolution $\lambda/\Delta\lambda\approx7800$ due to the tail in the
distribution towards low $\lambda/\Delta\lambda$.
}
\label{fig:meanR}
\end{figure}

The intrinsic widths of the argon lines are much smaller than the instrumental
resolution ($\sigma_{\rm Ar}\approx1$ \kms\ $\ll\sinst$); therefore, we measured
$\sinst$ by fitting a Gaussian function to the Ar emission profiles.  We then
fit low-order-polynomial surfaces to these measurements to provide interpolated
$\sinst$ values at every pixel of every extracted spectrum.  Based on these
surface fits, we have calculated the average spectral resolution for every
galaxy spectrum extracted from all science frames; a histogram of these values
is shown in Fig.~\ref{fig:meanR}.  The tail of low-resolution values is
primarily due to spectra obtained from high aperture numbers in exposures taken
during only a few nights.  On the whole, 97\% and 87\% of all spectra have
resolutions of $\lambda/\Delta\lambda>6000$ ($\sinst\approx21$ \kms) and
$\lambda/\Delta\lambda>7000$ ($\sinst\approx18$ \kms), respectively. The average
instrumental dispersion for all spectra, $\sinst\approx16.6\pm1.4$ \kms, is
slightly better than reported by \citet[][hereafter
\citetalias{bershady2010b}]{bershady2010b}.

\begin{figure}[t]
\centering
\includegraphics[width=0.49\textwidth]{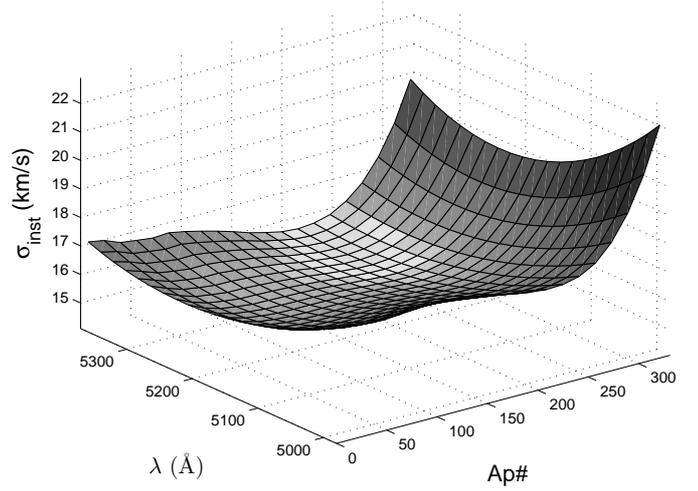}
\caption[]{
Average instrumental dispersion ($\sinst$) as a function of wavelength and aperture
number.
}
\label{fig:meanSigmaInst}
\end{figure}

For each galaxy, the spectral-resolution maps from each exposure were combined
to calculate the effective resolution of the combined galaxy spectra.  Each
measurement of $\sinst$ is assumed to describe a Gaussian instrumental profile,
such that $\sinst$ of the combined data is the second moment of a weighted sum
of Gaussian profiles with no offset in their centroid; i.e., we calculated
\begin{equation}
\sigma_{\rm comb} = \sqrt{ \frac{\sum{\left(w_i\,
\sigma_i^3\right)}}{\sum{\left(w_i\, \sigma_i\right)}} },
\label{eq:sigmaInst}
\end{equation}
where $w_i$ is the same weight as used in Eq.~(\ref{eq:combweight}) \citep{westfall2009}.
Figure~\ref{fig:meanSigmaInst} shows the average instrumental dispersion in all combined
images. In the center of the detector $\sinst$ is typically $15-17$ \kms\ and increases
toward high aperture numbers and shorter (bluer) wavelengths.  Except for the very highest
aperture numbers in some of the frames, the instrumental dispersion is always less than 20
\kms. The average spectral resolution in each of the 41 combined images is presented in
Table~\ref{tab:UGC_comb}.

The uncertainty in $\sinst$ is primarily due to the error in the Gaussian fit to
the argon lines. We estimate this uncertainty on the basis of the RMS about the
polynomial fit to the measured values, yielding one error value per spectrum.
Errors from single exposures are typically around 0.3 \kms\ ($0.4-1.0$ \kms\ for
the 30 highest aperture numbers where poor focus is a particular problem during
the nights mentioned earlier), assuming independent and random measurement
errors on the line fits. The errors on the instrumental dispersion for the
weighted-average spectra ($\delta\sigma_{\rm comb}$) are calculated by
propagating the errors from the individual spectra following
\begin{equation}
\delta\sigma_{\rm comb}= \frac{1}{2\sigma_{\rm comb}} \sqrt { \sum{ \left(
\frac{\partial\left(\sigma_{\rm comb}^2\right)}{\partial \sigma_i} \times
\epsilon_i \right)^2} }
\label{eq:sigmaInst_error}
\end{equation}
where $\epsilon_i$ is the error in a single spectrum and
\begin{equation}
\frac{\partial\left(\sigma_{\rm comb}^2\right)}{\partial \sigma_i} =
\frac{3w_i\sigma_i^2}{\sum{\left(w_i\sigma_i\right)}} - \frac{\sum{\left(
w_i\sigma_i^3\right)}\cdot w_i}{\sum{(w_i\sigma_i})^2}.
\label{eq:error_derivative}
\end{equation}
The typical formal error on $\sinst$ in the combined images
is $0.1-0.2$ \kms.

\subsection{Seeing}
\label{sec:Seeing}

On-sky seeing measurements are useful for our interpretation of, and correction
for, beam smearing of our data.  Therefore, we calculated the effective seeing
for every science exposure using guide-star images recorded by the A\&G camera
during our science observations.  These images are $50\times50$ pixels
($0\farcs2$ pixels) and have exposure times of $15-120$ seconds.  Each
guide-star image was ``bias-subtracted'' by subtracting the average value of the
upper six pixels in every column from all pixels in the same column.  This was
done to correct for differences of up to 30~ADU seen between neighbouring
columns in the images, and was especially important for seeing measurements of
faint guide stars that could have peak fluxes of less than 50~ADU above the bias
level.  We combined all guide-star images taken throughout each science exposure
(typically 100 or more) to produce a high-fidelity image of the on-sky PSF.

The {\it IRAF} task {\tt psfmeasure} was then used to measure the FWHM of the
stellar PSF; here, we define the effective seeing to be the FWHM of the Moffat
profile fitted to the average image of the guide star.  Our seeing measurements
range from $0\farcs6$ to $3\farcs1$ with an average and standard deviation of
$1\farcs4\pm0\farcs5$.  The effective seeing values in the combined science
exposures, calculated in the same way as the instrumental spectral dispersion
(Eq.~\ref{eq:sigmaInst}), are presented in Table~\ref{tab:UGC_comb}.

\subsection{Coordinate registration}
\label{sec:ContMaps}
Reconstructed continuum images from the PPak spectra were registered against blue
plates from the second Palomar Observatory Sky Survey (POSS-II) to accurately
determine the telescope pointing.  The PPak continuum images were constructed by
inserting sigma-clipped, mean continuum values for each spectrum into a blank
map at the relative locations of the corresponding fibers
(Sect.~\ref{sec:PPakFiber}); a $3\sigma$-clip level excludes the \oiii\ and
sky emission lines from the calculation of the mean continuum flux.
Interstitial pixels in this map were filled by convolving the discrete continuum
measurements with a Gaussian kernel having a FWHM that is 1.5 times the fiber
diameter.  The Atlas provides both the direct (POSS-II) and reconstructed
continuum images demonstrating that the detailed morphologies are easily visible
in the PPak data.  The pointing center of the PPak data is determined by
matching the pixel coordinates of two-dimensional Gaussian functions fitted to
the central parts of both the reconstructed (PPak) and direct (POSS-II)
continuum images using {\tt gaufit2d} in {\it GIPSY}.\footnote{
In the case of UGC~6918 the POSS-II image is saturated and we used the SDSS
image instead.
}  In this way, we have effectively attached a World Coordinate System (WCS) to
the PPak continuum images with an estimated accuracy of better than $1\arcsec$.

With a WCS attached, we can use the two-dimensional PPak images to identify the
locations of the morphological centers of the galaxies, which we define to
coincide with the peak flux in the inner regions.  For galaxies that have been
imaged by the Sloan Digital Sky Survey\footnote{
\url{http://www.sdss.org/}
} (SDSS), we have adopted the coordinates as reported by NED\footnote{
The NASA/IPAC Extragalactic Database, operated by the Jet Propulsion Laboratory,
California Institute of Technology, under contract with the National Aeronautics
and Space Administration.
}.  The correspondence between our fitted centers and the SDSS coordinates is
excellent: we find an RMS difference of $0\farcs4$ and an extreme case of
$0\farcs8$.  The coordinates reported by NED for galaxies without SDSS imaging
are mostly derived from 2MASS, where the agreement is often worse.  By comparing
with the centers provided by NED to the central peak in the maps, we find that
our Gaussian-fitted centers provided a better estimate for the morphological 
center, and we adopt our own center measurements for galaxies without SDSS imaging.
The coordinates of the adopted morphological centers of the galaxies can be found in
Table~\ref{tab:UGC}.

\subsection{[OIII] intensity maps}
\label{sec:OIIIint}

We have constructed two-dimensional maps of the relative \oiii\ intensities
using the same mapping algorithm as described in Sect.~\ref{sec:ContMaps}.
The \oiii\ intensities are calculated by integrating over a 20\AA\ region
surrounding the \oiii\ line after subtracting a baseline continuum flux.  The
continuum is set to be a straight line connecting the average flux in 20\AA\
wide spectral regions with centers separated by 20\AA\ on either side of the
\oiii\ line.  The resulting \oiii\ intensity maps are presented in the Atlas.
We find that the \oiii-emitting gas is patchier than the stellar continuum
light, characterized by small regions of strong emission.  In general, these
regions seem to be aligned with the spiral arms (e.g.\ UGC~7244 and UGC~9965)
and likely coincide with star-formation complexes.

\section{Kinematics of the Stars and Gas}
\label{sec:Analysis}

Using our properly reduced spectra, we measure the line-of-sight recession
velocity and velocity dispersion of both the stars and the \oiii\ gas within each
fiber aperture. We have used these data to produce the two-dimensional velocity
fields and velocity-dispersion maps presented in the Atlas.  We also determine
the orientation parameters of the galaxy disks, including their dynamical
centers, heliocentric systemic velocities, position angles and inclinations.
These orientation parameters allow us to derive the gas and stellar rotation
speed and velocity-dispersion as a function of radius.  As such, these
measurements are fundamental to many of the overall goals of our survey, such as
the calculation of asymmetric drift, estimation of the shape of the stellar
velocity ellipsoid, and determination of the dynamical disk mass surface
densities.

\subsection{Extracting stellar and gas kinematics}
\label{sec:extractkin}

The stellar kinematics in our target galaxies were derived using \textit{DC3}, a
program implementing the refined cross-correlation technique described fully in
\citetalias{westfall2011a}.  For our PPak spectra, this means that our stellar
kinematics (velocity, $V$, and velocity dispersion, $\slos$) are largely
determined by comparing the Mg and Fe absorption features present in both the
galaxy and template-star spectra.  Although we have observed many template stars
(Sect.~\ref{sec:ObsTempStars}), the stellar kinematics presented herein are
based on the K1 III template star HD162555 ($\log g=2.49$; [Fe/H]=$-0.15$).
In general, one should use a mix of stellar template spectra appropriately
matching the stellar populations represented within each fiber aperture;
however, this exercise is deferred to a future paper.  An analysis of the
anticipated effects of template mismatch is presented in
\citetalias{bershady2010b} (Section~3.4) and concludes that the median random
error on the derived velocity dispersion due to template mismatch is of order
4\%.

Stellar velocity measurements are corrected to the heliocentric reference frame
and for the velocity of the template star.  \citet{famaey2005} report a
heliocentric velocity of $-14.84 \pm 0.20$ \kms\ for HD162555.  By fitting
Gaussian profiles to the centroids of three MgIb and three Fe absorption lines,
we measure a less blue-shifted heliocentric velocity of $-13.09 \pm 0.80$ \kms.
The difference of $+1.75 \pm 0.82$ \kms\ is similar to the measured redshift of
the \none\ sky line as discussed in Sect.~\ref{sec:dohydra}.  For the
remainder of this paper, we will ignore this small apparent systematic offset in
velocity because it is irrelevant to our rotation-curve and velocity-dispersion
measurements.  We correct our stellar velocity-dispersion measurements for
instrumental broadening (Sect.~\ref{sec:InstRes}) following the procedure
outlined in the Appendix of \citetalias{westfall2011a}; see also
\citetalias{westfall2011b}.

The kinematics of the ionized gas in our target galaxies is derived by fitting
the \oiii$\lambda$5007\AA\ emission-line profile using software developed by
\citet{andersen2006, andersen2008}.  We fit both single- and double-Gaussian
profiles to each line within a 20\AA\ window centered at the redshifted
wavelength of the emission line. Emission-line profiles are considered more
aptly fit by a double-Gaussian function when the reduced chi-square decreases by
more than $10\%$ compared to that found for the single Gaussian function.  In
14\% of the fits, a double-Gaussian fit to the \oiii\ line was preferred.  For
each fitted line, the fitting procedure returns measures of the velocity and
velocity dispersion, and their associated errors as determined by a formal
covariance-matrix analysis.  In case of a double-Gaussian fit, the individual
Gaussian profile with the largest flux was taken to represent the recession
velocity and velocity dispersion of the ionized gas. The radial velocities of
the \oiii\ line have also been corrected to the heliocentric reference frame.
We determine the velocity dispersion of the ionized gas by subtracting $\sinst$
in quadrature from the velocity dispersion of the fitted line profile.

For illustration purposes, the heliocentric gas and stellar recession velocities
and velocity dispersions as measured in each spectrum are placed in a
two-dimensional map at the locations of the centers of the corresponding fibers.
Similar to the reconstructed continuum images, the pixels in between the fiber
locations have been filled by interpolation with a Gaussian convolution function
with a FWHM of 1.5 fiber diameter (see Appendix~A of
\citetalias[][]{westfall2011b} for details on the interpolation scheme).  This
yielded the two-dimensional velocity fields and velocity-dispersion maps as
presented in the Atlas.  We stress that the analysis of the kinematic data is
not based on these interpolated maps but rather on the individual measurements
from each fiber combined with the location of that fiber in the plane of the
sky. The right-most panels of the Atlas show the $S/N$ for each fiber footprint
in which the stellar or gas velocity dispersion has been measured.

\subsection{Orientation of the galaxy disks}
\label{sec:orientation}

To derive rotation curves and velocity-dispersion profiles for both dynamical
tracers (gas and stars), we first determine the location of the dynamical center
($x_0$,$y_0$), the systemic velocity ($\vsys$), the position angle ($\pa$), and
the inclination ($i$) of each galaxy using a step-wise approach.  Our procedure
assumes these geometric parameters are independent of radius; i.e., we
assume that the gas and stellar ensemble rotate on circular, concentric,
co-planar orbits around a common dynamical center.  Additionally, to minimize the
degrees of freedom when fitting the orientation of the rotating disks, we use a
hyperbolic-tangent (tanh) parameterization of the shape of the rotation curve
because it mimics a monotonically rising rotation curve with a near-linear rise
in the inner regions and a smooth turn-over at a scale radius ($\rs$) into a
flat part with an asymptotic maximum rotation speed ($\arot$); i.e.,
\begin{equation}
V(R) = \arot \times \tanh(R/\rs ).
\label{eq:tanh}
\end{equation}
At radius $R=\rs$ the rotation speed has reached $V(\rs)=0.76\arot$. It should
be noted, however, that this parameterization can not accommodate rotation
curves that are declining over a certain radial range (e.g.\ UGC~4458) or
rotation curves that show significant structure (e.g.\ UGC~4555).

In this way, we model the projected, two-dimensional velocity field of a rotating galaxy
disk using seven parameters: $x_0$, $y_0$, $\vsys$, $\pa$, $i$, $\arot$, and $\rs$. In
principle, all seven parameters can be fit simultaneously to the full two-dimensional
velocity field using a program developed by \citet{andersen2001phd} and 
\citet{andersen2003}. In practice, there is significant covariance between $i$ and $\arot$
for galaxies seen nearly face-on, and the dynamical center is poorly defined for a
velocity field in which the solid-body part of the rotation curve extends to a large
radius (large $\rs$) such that the inner isovelocity contours of the velocity field are
nearly parallel. To ameliorate these issues, we have adopted a few additional assumptions
as motivated in the following subsections.

\subsubsection{Dynamical center}
\label{sec:center}

\begin{figure}[t]
\centering
\includegraphics[width=0.49\textwidth]{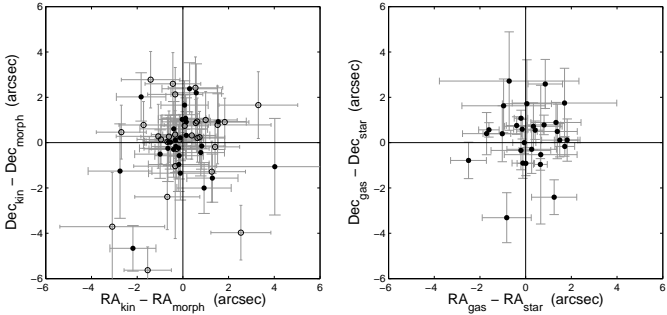}
\caption[]{
{\bf Left:} Offsets between the morphological centers of the galaxies as
determined from direct images, and the dynamical centers as determined from the
two-dimensional velocity fields.  Open and solid symbols correspond to the
dynamical centers derived for the \oiii\ and stellar velocity fields,
respectively. {\bf Right:} Offsets between the dynamical centers as derived for
the \oiii\ and stellar velocity fields.
}
\label{fig:CompCenters}
\end{figure}

We first determine the dynamical center of the rotating disk by simultaneously
fitting all seven velocity-field parameters.  The left panel of
Fig.~\ref{fig:CompCenters} demonstrates that we find no significant systematic
offset between the dynamical and morphological centers
(Sect.~\ref{sec:ContMaps}) for either tracer.  The offsets can be largely
explained by the measurement uncertainties on the locations of the dynamical
centers.  The distribution of offsets is similar for the dynamical centers
derived from the stellar and the gas velocity fields although the largest
outliers tend to come from the gas kinematics.  The right panel of
Fig.~\ref{fig:CompCenters} shows the differences between the dynamical centers
from the stellar and the gas kinematics.  Again, no systematic offset or trend
is detected and the width of the distribution is similar to that in the left
panel. 

Although Fig.~\ref{fig:CompCenters} shows that the dynamical centers lie
within half a fiber diameter from the morphological centers, which have a
precision of $0\farcs4$ (Sect.~\ref{sec:ContMaps}), there are still many
galaxies with significantly deviating dynamical centers.  We believe this to be
an artifact of the velocity-field modelling procedure, not a true reflection of
the difference between the morphological and dynamical centers.  Hereafter,
therefore, we have chosen to affix the dynamical centers to the more accurately
determined morphological centers for {\it all} galaxies in our modelling of the
velocity-field.  The underlying assumption is that the dynamical and
morphological centers are co-located astrophysically.

\subsubsection{Systemic velocity \& position angle}
\label{sec:PAVsys}

As the next step in our multi-stage fitting process, we have determined the
systemic velocities ($\vsys$) and kinematic position angles ($\pa$); these parameters 
are determined independently for the gas and stellar velocity fields.  In addition to
forcing the coincidence of the dynamical and morphological centers, we also fix other
velocity-field parameters that are largely independent of $\vsys$ and $\pa$; this 
strategy does not affect the accuracy of our $\vsys$ and $\pa$ measurements, even if the
fixed parameters are only approximations \citep{begeman1989}.  Therefore, while fitting
$\vsys$ and $\pa$, we have fixed $i$, $\arot \sin i$ and $\rs$ (see Eq.~\ref{eq:tanh})
in the following way.

For each galaxy, the projected asymptotic maximum rotation speed ($\arot \sin
i$) was kept fixed to $(V_{\rm max} - V_{\rm min})/2$, where $V_{\rm max}$ and
$V_{\rm min}$ are, respectively, the maximum and minimum observed recession
velocities after rejecting the 10\% highest and lowest values.  These
rejection percentiles were verified by inspecting the outliers in preliminary
made projected position-velocity diagrams.  The inclination of the galaxy disk
was kept fixed to the value calculated by inverting the Tully-Fisher (TF)
relation \citep{tullyfisher1977} using our fixed $\arot \sin i$ and the total
absolute $K$-band magnitude ($M_K$); see Sect.~\ref{sec:Incl} for further
details.  Finally, for the fitting of $\vsys$ and $\pa$, we fix $\rs = 0.42\hr$,
where $\hr$ is the $K$-band photometric radial scale length of the
exponential light distribution (Table~\ref{tab:UGC}).
The factor 0.42 follows from fitting a tanh function to the model
rotation curve of an exponential disk with scale length $\hr$ that was forced to remain
flat beyond 2.15$\hr$, where the model rotation curve of the exponential disk reaches its
maximum. 

\begin{figure}[t]
\centering
\includegraphics[width=0.49\textwidth]{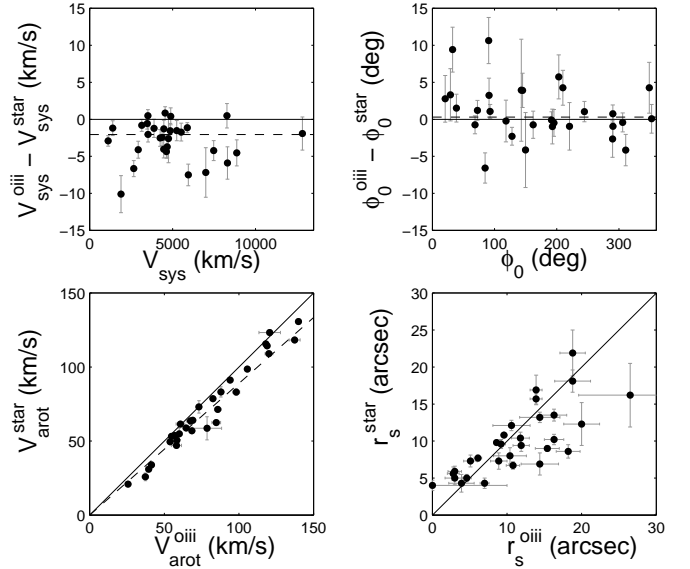}
\caption[]{
Differences between velocity-field model parameters for the stars and
\oiii-emitting gas.  The parameters compared are the systemic velocities ({\it
top left}), position angles ({\it top right}), projected rotation velocity ({\it
bottom left}) and rotation scale ({\it bottom right}).  The systematic offset
between the projected rotation speeds of the stars and gas is expected given the
significant asymmetric drift of the stars. The average asymmetric drift,
$\arot^{\rm star} = 0.89 \arot^{\rm OIII}$, is indicated with a dashed line
in the bottom left panel. Asymmetric drift should also result in a smaller scale
radius for the \oiii\ rotation curve; however this trend is generally not seen,
as discussed in the text.  In the bottom-right panel, the point located on the
vertical axis comes from UGC~4458, which has a declining gas rotation curve.
}
\label{fig:geometry}
\end{figure}

Provided with our independent fits using the two different tracers, we compare
the best-fitting velocity-field parameters for the gas and the stars in
Fig.~\ref{fig:geometry}. The upper-left panel shows $\Delta \vsys=\vsys^{\rm
OIII}-\vsys^{\rm star}$ versus the average value of $\vsys$ for each galaxy.
The weighted mean in $\Delta \vsys$ is $-2.06 \pm 0.20$ \kms\ (0.2 pixels),
indicated by the dashed line, such that the systemic velocity of the ionized gas
is slightly (but significantly) blue-shifted with respect to that of the stars.
In Sects.~\ref{sec:dohydra} and \ref{sec:extractkin}, we reported small
systematic offsets in the wavelength calibration that is consistent for,
respectively, the \none\ sky line and the template star. The insignificant
difference of $\sim0.5$ \kms\ between these shifts is, however, too small to
explain the observed $\Delta \vsys$.  This shift may, in fact, be physical: The
moderate blue-shift of the \oiii\ emission line relative to the stars may
reflect our biased view of ionized-gas expansion in star-forming regions, where
we preferentially see the near side due to extinction within these regions.
However, regardless of its physical and/or instrumental origin, this slight
difference is largely irrelevant to our measurements of the rotation curves.
Therefore, we bring the systemic velocities of the stars and the gas on par by
subtracting the observed average systematic offset of 2.1 \kms\ from all $\vsys$
measurements of the stars.  Velocity-field models fitted hereafter adopt the
weighted average of $\vsys$ from the two kinematic tracers for each galaxy,
after correcting for the offset of the stellar data.  These adopted average
values of $\vsys$ can be found in Table~\ref{tab:mK2iTF}. 

The upper-right panel of Fig.~\ref{fig:geometry} presents
$\Delta \pa$$=$$\pa^{\rm OIII}$$-$$\pa^{\rm star}$
for each galaxy.  Overall, the position angles agree very
well, with a weighted mean difference of $+0\fdg31\pm0\fdg33$ as indicated by
the dashed line.  We find $\Delta \pa$ to be significant for a few galaxies,
which is a result of velocity-field asymmetries combined with a patchy
distribution of the \oiii\ emission.  The errors in $\pa^{\rm OIII}$ are
typically larger than those in $\pa^{\rm star}$.  Velocity-field models fitted
hereafter adopt the error-weighted average of the gas and stellar $\pa$,
provided in Table~\ref{tab:mK2iTF}. 

\subsubsection{Inclination}
\label{sec:Incl}
Due to the nearly face-on orientation of the galaxies in our sample,
inclinations derived by kinematic modelling of the velocity fields can result in
large uncertainties \citep{andersen2001phd,martinsson2011}.  To the contrary,
inclinations inferred from the TF~relation ($\itf$) are generally more precise
in this inclination regime given (1) the well-defined $\arot \sin i$ of our gas
tracers (\oiii, \halp, and \hone); (2) the availability of well-measured $M_K$;
and (3) the small observed scatter in the $M_K$-$V_{\rm flat}$ TF~relation.
Given these considerations, we adopt $i=\itf$ for our analysis of the PPak data
in this paper.

\citet{verheyen2001b} determined the TF~relation for a sample of 21 spiral
galaxies in Ursa Major using deep $K$-band photometric images and the detailed
shapes of \hone\ rotation curves derived from 21cm aperture synthesis imaging.
The latter allowed for an accurate measurement of the amplitude of the flat part
of the extended rotation curves.  The total observed scatter in the
$M_K$-$V_{\rm flat}$ relation was 0.26 magnitudes.  The estimated depth of the
Ursa Major sample contributes 0.17 magnitudes to this total observed scatter,
and the remaining quadrature difference of 0.20 magnitudes can be attributed to
the measurement errors.  Here, we adopt the inverse TF~relation 
\begin{equation}
V_{\rm flat} = 0.5 \times 10^{(5.12 - M_K)/11.3}.
\label{eq:vflat}
\end{equation}
Our determination of $m_K$ (Sect.~\ref{sec:2MASSphot}) follows the same
procedure as used by \citet{verheyen2001b}, as appropriate for our use of
Eq.~(\ref{eq:vflat}).  We then calculate $M_K$ by applying the same
corrections for extinction and the distance modulus as listed in Table~2 of
\citetalias{bershady2010a}, and with k-corrections applied according to
\citet{bershady1995}.  Errors on $V_{\rm flat}$ include an adopted intrinsic
scatter in the TF~relation of 0.2~magnitudes, the photometric error on $m_K$,
and a distance error proportional to the recession velocity increased by 150
\kms\ to account for uncertainties in the peculiar motions of the galaxies. The
estimated values of $V_{\rm flat}$ and its error are listed in
Table~\ref{tab:mK2iTF}.

We use the TF-predicted $V_{\rm flat}$ and our measurements of $\arot \sin i$ to
calculate the inclination and its error.  Our measurements of $\arot \sin i$
take advantage of the multiple gas tracers provided by our full dataset (\oiii,
\halp, \hone; see forthcoming papers for a presentation of the latter two
tracers), but do not consider the stars such that we avoid the effects of
asymmetric drift.  Moreover, these measurements are made using the
position-velocity data for all three gas tracers, not the parameterized values
returned by our velocity-field models.  The observed $\arot \sin i$ determined
by a weighted average of data from our three tracers and the resulting $\itf$
are provided in Table~\ref{tab:mK2iTF}. 

The galaxies in the PPak sample are on average one magnitude brighter than the galaxies in
the Ursa Major sample, but otherwise span a similar range of $M_K$ magnitudes with a
similar scatter. The overlap of 3.5 magnitudes in $M_K$ is sufficient to limit any
systematic offset between the two samples. Furthermore, the morphological mix of galaxies
in the samples is quite similar, although the Ursa Major sample contains somewhat more
bar-dominated systems and low-surface-brightness galaxies. However, there is no
statistically significant offset in the $K$-band TF-relation between low- and
high-surface-brightness galaxies, or barred and non-barred galaxies 
\citep[e.g.][]{courteau2003}.
For the TF-relation in \cite{verheyen2001b}, a distance to Ursa Major of 18.6~Mpc
\citep{tully2000} was used. Based on recent results \citep{sorce2013}, a
distance of 18.0$\pm$0.9~Mpc seems more appropriate.  
Although the distance uncertainty for Ursa Major may result in a $\sim$0.5$\arcdeg$
systematic error on $\itf$ for the PPak galaxies, this is negligible and inconsequential
to the results presented in this paper.

It should be noted that we may refine the inclination estimates in forthcoming papers
using improved kinematic models, improved photometric measurements, and/or an improved
statistical treatment of the joint probability distributions for the various inclination
measurements. However, these improved values are not expected to alter the results presented here.

\begin{table*}
\caption{\label{tab:mK2iTF}
Parameterization of the orientation and kinematics of the galaxy disks}
\centering
%
{\tiny
\renewcommand{\tabcolsep}{1.05mm}
\begin{tabular}{|rrrccrrrrrrr|}
\hline
\multicolumn{1}{|c}{UGC}                        & 
\multicolumn{1}{c}{$V_{\rm sys}$}               & 
\multicolumn{1}{c}{$\pa$}                       & 
\multicolumn{1}{c}{$\kappa_{K}$}                &   
\multicolumn{1}{c}{$M_K$}                       &   
\multicolumn{1}{c}{$V_{\rm flat}^{\rm TF}$}     & 
\multicolumn{1}{c}{$\left<{V}_c{\rm sin}i\right>_w^{\rm gas}$} &
\multicolumn{1}{c}{$i_{\rm TF}$}                & 
\multicolumn{1}{c}{$V_{\rm arot}^{\rm stars}$}  & 
\multicolumn{1}{c}{$\rs^{\rm stars}$}           & 
\multicolumn{1}{c}{$V_{\rm arot}^{\rm [OIII]}$} &
\multicolumn{1}{c|}{$\rs^{\rm [OIII]}$}         \\ 
\multicolumn{1}{|c}{}                &   
\multicolumn{1}{c}{(km/s)}           & 
\multicolumn{1}{c}{(deg)}            & 
\multicolumn{1}{c}{(mag)}            & 
\multicolumn{1}{c}{(mag)}            & 
\multicolumn{1}{c}{(km/s)}           &   
\multicolumn{1}{c}{(km/s)}           & 
\multicolumn{1}{c}{(deg)}            &   
\multicolumn{1}{c}{(km/s)}           &   
\multicolumn{1}{c}{(arcsec)}         & 
\multicolumn{1}{c}{(km/s)}           &
\multicolumn{1}{c|}{(arcsec)}        \\
\multicolumn{1}{|c}{(1)}             &   
\multicolumn{1}{c}{(2)}              & 
\multicolumn{1}{c}{(3)}              & 
\multicolumn{1}{c}{(4)}              & 
\multicolumn{1}{c}{(5)}              &   
\multicolumn{1}{c}{(6)}              & 
\multicolumn{1}{c}{(7)}              &   
\multicolumn{1}{c}{(8)}              &   
\multicolumn{1}{c}{(9)}              & 
\multicolumn{1}{c}{(10)}             &
\multicolumn{1}{c}{(11)}             &
\multicolumn{1}{c|}{(12)}            \\
%
\hline
   448 &  4856.8 $\pm$ 0.5 & 306.1 $\pm$ 0.6 &{\it $-$0.04}&{\it $-$23.93 $\pm$ 0.10}&{\it 186 $\pm$           11}&{\it  83.3 $\pm$ 0.6}& 26.6 $\pm$ 1.6  &  79.0 $\pm$ 0.8 &  5.06 $\pm$ 0.45 &  82.1 $\pm$ \phantom{1}1.0 &  4.55 $\pm$ \phantom{1}0.62  \\
   463 &  4459.4 $\pm$ 0.6 &  68.5 $\pm$ 0.6 &{\it $-$0.04}&{\it $-$24.49 $\pm$ 0.09}&{\it 209 $\pm$           12}&{\it 106.2 $\pm$ 0.5}& 30.6 $\pm$ 1.9  &  99.7 $\pm$ 0.7 & 10.79 $\pm$ 0.33 & 107.5 $\pm$ \phantom{1}1.3 &  9.67 $\pm$ \phantom{1}0.53  \\
  1081 &  3135.2 $\pm$ 0.4 &  72.5 $\pm$ 0.6 &{\it $-$0.03}&{\it $-$23.05 $\pm$ 0.14}&{\it 156 $\pm$ \phantom{1}9}&{\it  62.7 $\pm$ 1.4}& 23.8 $\pm$ 1.6  &  58.7 $\pm$ 1.0 & 13.74 $\pm$ 0.80 &  63.8 $\pm$ \phantom{1}1.9 & 15.59 $\pm$ \phantom{1}1.69  \\
  1087 &  4480.9 $\pm$ 0.6 &  84.8 $\pm$ 1.0 &{\it $-$0.04}&{\it $-$23.18 $\pm$ 0.13}&{\it 160 $\pm$           10}&{\it  59.6 $\pm$ 0.5}& 21.9 $\pm$ 1.4  &  50.4 $\pm$ 1.6 &  7.07 $\pm$ 0.80 &  58.6 $\pm$ \phantom{1}1.0 &  5.09 $\pm$ \phantom{1}0.95  \\
  1529 &  4648.5 $\pm$ 0.7 & 162.1 $\pm$ 0.9 &{\it $-$0.04}&{\it $-$23.96 $\pm$ 0.11}&{\it 187 $\pm$           11}&{\it 118.8 $\pm$ 0.5}& 39.4 $\pm$ 2.6  & 118.4 $\pm$ 1.1 &  7.91 $\pm$ 0.40 & 121.1 $\pm$ \phantom{1}1.6 &  7.01 $\pm$ \phantom{1}0.58  \\
  1635 &  3516.5 $\pm$ 0.5 & 145.0 $\pm$ 1.0 &{\it $-$0.03}&{\it $-$22.93 $\pm$ 0.13}&{\it 152 $\pm$ \phantom{1}9}&{\it  53.4 $\pm$ 0.5}& 20.6 $\pm$ 1.3  &  50.5 $\pm$ 1.0 &  9.41 $\pm$ 0.71 &  53.7 $\pm$ \phantom{1}1.5 & 11.04 $\pm$ \phantom{1}1.30  \\
  1862 &  1388.1 $\pm$ 0.5 &  20.6 $\pm$ 1.4 &{\it $-$0.01}&{\it $-$20.96 $\pm$ 0.27}&{\it 102 $\pm$ \phantom{1}8}&{\it  57.1 $\pm$ 2.0}& 34.2 $\pm$ 3.3  &  46.6 $\pm$ 1.6 & 17.95 $\pm$ 1.51 &  56.6 $\pm$ \phantom{1}3.2 & 18.89 $\pm$ \phantom{1}2.33  \\
  1908 &  8257.8 $\pm$ 0.8 & 192.9 $\pm$ 1.2 &{\it $-$0.07}&{\it $-$25.11 $\pm$ 0.09}&{\it 237 $\pm$           14}&{\it  92.9 $\pm$ 0.8}& 23.1 $\pm$ 1.3  &  82.2 $\pm$ 2.0 &  9.35 $\pm$ 0.82 &  96.1 $\pm$ \phantom{1}1.9 & 11.67 $\pm$ \phantom{1}0.99  \\
  3091 &  5558.5 $\pm$ 0.6 & 209.9 $\pm$ 1.2 &{\it $-$0.04}&{\it $-$23.08 $\pm$ 0.18}&{\it 156 $\pm$           10}&{\it  67.1 $\pm$ 0.6}& 25.4 $\pm$ 1.7  &  58.3 $\pm$ 1.4 &  4.60 $\pm$ 0.66 &  68.8 $\pm$ \phantom{1}0.8 &  7.03 $\pm$ \phantom{1}0.65  \\
  3140 &  4622.4 $\pm$ 0.4 & 352.5 $\pm$ 0.9 &{\it $-$0.04}&{\it $-$24.51 $\pm$ 0.09}&{\it 209 $\pm$           12}&{\it  52.1 $\pm$ 0.4}& 14.4 $\pm$ 0.8  &  54.1 $\pm$ 0.9 &  6.14 $\pm$ 0.66 &  55.4 $\pm$ \phantom{1}1.1 &  3.03 $\pm$ \phantom{1}0.87  \\
  3701 &  2925.4 $\pm$ 0.6 &  91.0 $\pm$ 1.2 &{\it $-$0.03}&{\it $-$21.95 $\pm$ 0.22}&{\it 124 $\pm$ \phantom{1}9}&{\it  56.1 $\pm$ 0.7}& 26.9 $\pm$ 2.0  &  50.6 $\pm$ 3.3 & 19.82 $\pm$ 2.67 &  57.9 $\pm$ \phantom{1}1.7 & 17.45 $\pm$ \phantom{1}1.49  \\
  3997 &  5943.3 $\pm$ 0.7 &  32.5 $\pm$ 1.2 &{\it $-$0.05}&{\it $-$23.01 $\pm$ 0.22}&{\it 154 $\pm$           11}&{\it  68.2 $\pm$ 0.9}& 26.2 $\pm$ 1.9  &  63.8 $\pm$ 1.8 &  7.68 $\pm$ 0.87 &  71.0 $\pm$ \phantom{1}1.4 & 11.64 $\pm$ \phantom{1}1.02  \\
  4036 &  3466.2 $\pm$ 0.3 & 195.4 $\pm$ 0.8 &{\it $-$0.03}&{\it $-$23.96 $\pm$ 0.10}&{\it 187 $\pm$           11}&{\it  52.6 $\pm$ 0.5}& 16.3 $\pm$ 1.0  &  48.7 $\pm$ 0.7 & 12.90 $\pm$ 0.70 &  53.3 $\pm$ \phantom{1}1.6 & 13.59 $\pm$ \phantom{1}2.73  \\
  4107 &  3507.6 $\pm$ 0.4 & 290.1 $\pm$ 0.6 &{\it $-$0.03}&{\it $-$23.37 $\pm$ 0.12}&{\it 166 $\pm$           10}&{\it  67.7 $\pm$ 0.4}& 24.1 $\pm$ 1.5  &  63.8 $\pm$ 0.9 &  9.67 $\pm$ 0.58 &  68.4 $\pm$ \phantom{1}0.9 &  9.08 $\pm$ \phantom{1}0.62  \\
  4256 &  5248.2 $\pm$ 0.7 & 290.5 $\pm$ 1.0 &{\it $-$0.04}&{\it $-$24.81 $\pm$ 0.08}&{\it 223 $\pm$           13}&{\it  74.2 $\pm$ 1.1}& 19.5 $\pm$ 1.1  &  63.0 $\pm$ 1.4 &  9.27 $\pm$ 0.91 &  84.8 $\pm$ \phantom{1}2.8 & 17.94 $\pm$ \phantom{1}1.53  \\
  4368 &  3876.2 $\pm$ 0.6 & 128.1 $\pm$ 0.6 &{\it $-$0.03}&{\it $-$23.29 $\pm$ 0.12}&{\it 163 $\pm$           10}&{\it 116.1 $\pm$ 0.4}& 45.3 $\pm$ 3.3  & 113.1 $\pm$ 1.7 & 15.44 $\pm$ 0.68 & 117.0 $\pm$ \phantom{1}1.3 & 13.28 $\pm$ \phantom{1}0.72  \\
  4380 &  7481.3 $\pm$ 0.7 &  29.1 $\pm$ 1.8 &{\it $-$0.06}&{\it $-$23.98 $\pm$ 0.13}&{\it 188 $\pm$           12}&{\it  46.0 $\pm$ 0.5}& 14.2 $\pm$ 0.9  &  36.0 $\pm$ 2.0 &  4.89 $\pm$ 1.21 &  42.2 $\pm$ \phantom{1}1.5 &  4.04 $\pm$ \phantom{1}1.68  \\
  4458 &  4751.7 $\pm$ 1.1 & 289.7 $\pm$ 0.9 &{\it $-$0.04}&{\it $-$25.29 $\pm$ 0.08}&{\it 246 $\pm$           14}&{\it 140.1 $\pm$ 1.7}& 34.8 $\pm$ 2.2  & 117.9 $\pm$ 1.7 &  3.98 $\pm$ 0.68 & 138.0 $\pm$ \phantom{1}3.6 &            0\phantom{ 11.57} \\
  4555 &  4243.9 $\pm$ 0.5 &  92.9 $\pm$ 0.4 &{\it $-$0.04}&{\it $-$23.84 $\pm$ 0.10}&{\it 183 $\pm$           11}&{\it 115.9 $\pm$ 0.9}& 39.3 $\pm$ 2.6  & 109.9 $\pm$ 0.8 &  9.04 $\pm$ 0.40 & 114.5 $\pm$ \phantom{1}1.3 &  3.55 $\pm$ \phantom{1}0.84  \\
  4622 & 12831.9 $\pm$ 1.1 & 118.3 $\pm$ 1.4 &{\it $-$0.10}&{\it $-$24.95 $\pm$ 0.12}&{\it 229 $\pm$           14}&{\it  89.2 $\pm$ 0.7}& 22.9 $\pm$ 1.4  &  83.6 $\pm$ 3.5 &  5.68 $\pm$ 0.90 &  88.7 $\pm$ \phantom{1}1.2 &  6.77 $\pm$ \phantom{1}1.07  \\
  6903 &  1891.7 $\pm$ 1.2 & 143.0 $\pm$ 3.1 &{\it $-$0.02}&{\it $-$22.63 $\pm$ 0.20}&{\it 143 $\pm$           10}&{\it  76.6 $\pm$ 1.3}& 32.4 $\pm$ 2.5  &  62.1 $\pm$ 8.7 & 17.55 $\pm$ 4.62 &  87.7 $\pm$           18.2 & 34.26 $\pm$           11.57  \\
  6918 &  1112.2 $\pm$ 0.3 & 191.6 $\pm$ 0.5 &{\it $-$0.01}&{\it $-$22.90 $\pm$ 0.23}&{\it 151 $\pm$           11}&{\it  92.8 $\pm$ 0.4}& 38.0 $\pm$ 3.1  &  92.2 $\pm$ 0.3 & 10.01 $\pm$ 0.22 &  95.3 $\pm$ \phantom{1}0.8 &  8.87 $\pm$ \phantom{1}0.45  \\
  7244 &  4361.1 $\pm$ 0.5 & 149.8 $\pm$ 1.9 &{\it $-$0.04}&{\it $-$22.24 $\pm$ 0.21}&{\it 132 $\pm$ \phantom{1}9}&{\it  38.2 $\pm$ 0.6}& 16.8 $\pm$ 1.2  &  25.8 $\pm$ 2.6 & 12.35 $\pm$ 2.87 &  37.4 $\pm$ \phantom{1}2.1 & 20.02 $\pm$ \phantom{1}2.40  \\
  7917 &  6998.1 $\pm$ 0.9 & 220.5 $\pm$ 0.8 &{\it $-$0.06}&{\it $-$25.35 $\pm$ 0.06}&{\it 249 $\pm$           14}&{\it 129.8 $\pm$ 0.9}& 31.5 $\pm$ 1.9  & 120.7 $\pm$ 1.8 & 11.89 $\pm$ 0.72 & 118.1 $\pm$ \phantom{1}7.1 & 10.40 $\pm$ \phantom{1}2.49  \\
  8196 &  8336.9 $\pm$ 0.9 &  90.5 $\pm$ 1.1 &{\it $-$0.07}&{\it $-$25.36 $\pm$ 0.07}&{\it 249 $\pm$           14}&{\it  69.0 $\pm$ 0.8}& 16.1 $\pm$ 0.9  &  60.6 $\pm$ 1.5 &  4.71 $\pm$ 0.92 &  60.8 $\pm$ \phantom{1}2.9 &  2.99 $\pm$ \phantom{1}1.27  \\
  9177 &  8866.6 $\pm$ 0.9 & 244.5 $\pm$ 0.7 &{\it $-$0.08}&{\it $-$24.50 $\pm$ 0.13}&{\it 209 $\pm$           13}&{\it 135.4 $\pm$ 0.8}& 40.4 $\pm$ 2.7  & 134.6 $\pm$ 1.9 &  7.17 $\pm$ 0.57 & 143.0 $\pm$ \phantom{1}1.3 &  9.17 $\pm$ \phantom{1}0.89  \\
  9837 &  2654.9 $\pm$ 0.5 & 310.9 $\pm$ 1.0 &{\it $-$0.03}&{\it $-$22.71 $\pm$ 0.18}&{\it 145 $\pm$ \phantom{1}9}&{\it  75.4 $\pm$ 0.5}& 31.3 $\pm$ 2.2  &  70.1 $\pm$ 3.4 & 16.05 $\pm$ 1.67 &  75.0 $\pm$ \phantom{1}1.2 & 14.29 $\pm$ \phantom{1}0.77  \\
  9965 &  4527.1 $\pm$ 0.4 & 203.2 $\pm$ 1.5 &{\it $-$0.04}&{\it $-$23.61 $\pm$ 0.11}&{\it 174 $\pm$           10}&{\it  37.3 $\pm$ 0.6}& 12.4 $\pm$ 0.7  &  31.0 $\pm$ 1.2 &  7.32 $\pm$ 1.13 &  39.9 $\pm$ \phantom{1}0.9 &  9.33 $\pm$ \phantom{1}1.11  \\
 11318 &  5884.5 $\pm$ 0.4 & 348.7 $\pm$ 1.7 &{\it $-$0.05}&{\it $-$24.67 $\pm$ 0.08}&{\it 217 $\pm$           12}&{\it  21.9 $\pm$ 0.6}&  5.8 $\pm$ 0.4  &  20.9 $\pm$ 0.8 &  6.87 $\pm$ 1.48 &  25.8 $\pm$ \phantom{1}1.6 & 14.37 $\pm$ \phantom{1}2.52  \\
 12391 &  4880.5 $\pm$ 0.5 &  38.6 $\pm$ 0.9 &{\it $-$0.04}&{\it $-$23.54 $\pm$ 0.12}&{\it 172 $\pm$           10}&{\it  83.7 $\pm$ 0.7}& 29.2 $\pm$ 1.8  &  73.0 $\pm$ 1.6 & 11.20 $\pm$ 0.81 &  87.0 $\pm$ \phantom{1}2.2 & 16.55 $\pm$ \phantom{1}1.33  \\
\hline
\end{tabular}
}
\tablefoot{
Numbers in italics are used to calculate $i_{\rm TF}$. (1) UGC number; (2) 
weighted-average heliocentric systemic velocity; (3) position angle of the receding side 
of the kinematic major axis; (4) k-correction used to calculate $M_K$ following 
$\kappa_{K}=(1+\alpha)2.5{\rm log}(1+z)$ with $\alpha=1.25$ for the $K$-band according 
to \cite{bershady1995}; (5) absolute, total $K$-band magnitude based on Table~\ref{tab:UGC} 
with A$_{K}^{\rm g}$ and $\kappa_{K}$ applied; (6) amplitude of the flat part of 
the rotation curve as predicted by the Tully-Fisher relation; (7) weighted-average 
projected rotation velocity derived from the \halp, \oiii\ and HI position-velocity 
diagrams; (8) derived inclination based on the Tully-Fisher relation; (9) asymptotic 
maximum rotational velocity of the tanh-model fitted to the stellar data; (10) 
scale-radius of the tanh-model fitted to the stellar data; (11) and (12) identical to (9) 
and (10) but then for the \oiii\ gas.
}
\end{table*}

\subsection{Rotation curves and velocity-dispersion profiles}
\label{sec:ArotAsca}

Below, we use the orientation parameters and systemic velocities determined in
the previous section to calculate the rotation curves and the line-of-sight
velocity-dispersion profiles, for both the gas and stars, of the galaxy disks.

\subsubsection{Parameterized rotation curves}
\label{sec:tanhRC}

We derive $\arot$ and $\rs$ for both the gas and stellar kinematics in each
galaxy by fitting our kinematic model to their measured velocity fields while
keeping the geometric parameters and systemic velocities fixed; the results are
presented in the bottom row of Fig.~\ref{fig:geometry}.  The bottom-left panel
shows that, in general, the derived asymptotic rotation speed of the stars is
less than that of the ionized gas.  This is in accordance with the expectation
that the asymmetric drift (AD) of the stars is significantly larger than that of
the ionized gas. We find that 27 of the 30 galaxies show a $\geq 1\sigma$
signature of AD.  When deprojected to the in-plane rotation, for all 30 galaxies
we find $-5 \leq$ \kms\ $\arot^{\rm OIII}-\arot^{\rm star} \leq 65$ \kms, which
is between $-2$\% and $31$\% of the maximum gas rotation speed ($\vmaxoiii$).
The average AD is 11$\pm$8\% of $\vmaxoiii$, which is similar
to other measurements of local disk galaxies \citep[e.g.][]{ciardullo2004,
herrmann2009} and comparable to measurements in the solar neighborhood of the
Milky Way \citep[e.g.][]{ratnatunga1997, olling2003}.  A more detailed
investigation of AD, with an analysis of the data itself instead of
the parameterized maximum rotation speed, is deferred to a forthcoming paper.

The bottom-right panel of Fig.~\ref{fig:geometry} shows a significant scatter
in the correlation between the best-fitting $\rs$ for the gas and stellar
rotation curves.  Contrary to nominal expectations based on AD, the scale
lengths of the gas rotation curves are generally {\it larger} than those of the
stellar rotation curves.  However, we suspect that this is an artefact of the
kinematic model in combination with the relatively poor sampling of the gas
velocity field by the \oiii\ emitting gas near the dynamical center where the
rotation curve rises most steeply. Considering the formal errors from the fits,
it is noteworthy that the measurements with the smallest errors tend to indicate
that $\rs^{\rm star}>\rs^{\rm OIII}$ as would be expected for AD.  In general,
the errors correlate with the ratio $\rs^{\rm OIII} / \rs^{\rm star}$, a strong
indication for a measurement bias.  Therefore, we assign little astrophysical
meaning to any relation between $\rs^{\rm OIII}$ and $\rs^{\rm star}$ in
Fig.~\ref{fig:geometry}.

Velocity fields based on our kinematic model are presented in the Atlas.  These
maps project the rotation velocity of the model for each fiber to the sky plane
following the derived disk orientation and use an interpolation scheme identical
to that described for the observations (Sect.~\ref{sec:extractkin}).  The
irregular sampling of the velocity field and the finite size of the Gaussian
kernel introduce asymmetries into the interpolated model velocity fields despite
its intrinsic axisymmetry.  We have chosen this approach in an attempt to match
irregularities in the {\it observed} velocity field that may be introduced by
these same effects, as is particularly relevant to the residual maps shown in
the Atlas.  The residual maps help to identify large scale, non-circular or
streaming motions in the stellar and gas disks.  The particularly notable
residuals in the gas velocity field of UGC~4458 are primarily due to the
inability of our kinematic model to reproduce a declining rotation curve.  This
shortcoming motivates the non-parametric modelling of the rotation curve pursued
in the next section.

\subsubsection{Kinematics from constrained tilted-ring fits}
\label{sec:tiltedrings}

The parameterization of the rotation curves in Eq.~(\ref{eq:tanh}) helps to
stabilize the kinematic modelling, but this functional form is merely an
approximation to the actual shape of the rotation curve.  Therefore, we have
also derived non-parametric descriptions of the rotation curves by fitting
tilted rings to the two-dimensional kinematic data.

All rings are forced to have the same orientation, following the parameters
determined in Sect.~\ref{sec:orientation}.  Within the range $2.5\arcsec \leq
R \leq 37.5\arcsec$ (where $R$ is defined in the plane of the galaxy), rings are
defined by a continuous set of annuli with widths of $5\arcsec$, yielding seven
equally spaced rings.  Individual fiber measurements are considered within a
ring if the fiber {\it center} falls within the ring annulus.  Assuming circular
rotation and a marginal rotation-speed gradient within the ring, the
line-of-sight velocity as a function of azimuthal position within each ring is a
cosine function whose amplitude is the projected rotation speed of the ring
($\vrot \sin i$).  Thus, by plotting these amplitudes as a function of radius,
we create a non-parametric projected rotation curve for each tracer for each
galaxy at seven discrete radii.

Within the same seven annuli, we calculate the error-weighted mean of the
individual $\slos$ measurements for both the stars and the gas, excluding
measurements with errors in excess of 8 \kms.  We also determine the effective
radius of each annulus by calculating the weighted mean of each radius where the
weights are determined by the error in the velocity dispersion.

The azimuthally averaged $\vrot \sin i$ and $\slos$ profiles are presented in
the Atlas.  The seven radial points per profile are projected onto
position-velocity diagrams of measurements for individual spectra, separate for
each tracer.  In the Atlas, the radial profiles from both tracers are also
overplotted to facilitate a qualitative comparison.   In general, the
non-parametric rotation curves from the tilted-ring fits agree reasonably well
with the fitted tanh model; however, there are notable exceptions such as
UGC~4458.

\section{The Vertical Stellar Velocity Dispersion}
\label{sec:sigmaz}

\subsection{Exponential fits to $\slos$}
\label{sec:expsigma}

It is reasonable to expect that the velocity-dispersion profile of disk galaxies
closely follows an exponential in radius:
\begin{equation}
\slos(R) = \sigma_{\rm LOS,0} \cdot {\rm e}^{-R/\hsl},
\label{eq:sigmaexp}
\end{equation}
where $\hsl$ is the radial scale length of the velocity dispersion.  
This expectation follows from the following set of assumptions: (1) stars act as trace
particles of the disk mass distribution at all radii; (2) the scale height ($h_z$), disk
mass-to-light ratio, and axial ratios of the stellar velocity ellipsoid (SVE) are all
constant across the entire disk; and (3) the surface-brightness profile is
exponential in radius.  If these assumptions hold, one should find $\hsl = 2\hr$
\citep[e.g.][]{kruit1981}; see Sect.~\ref{sec:hrvshs}.

\begin{figure}[t]
\centering
\includegraphics[width=0.49\textwidth]{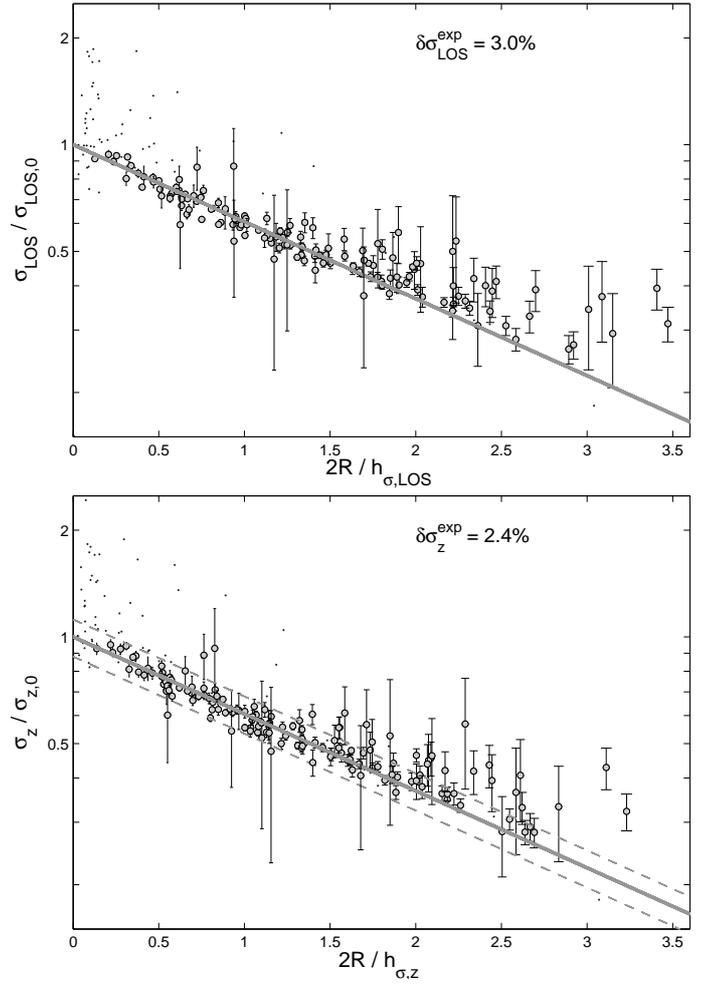}
\caption[]{
Velocity-dispersion profiles (top: $\slos$, bottom: $\sigz$) for all galaxies.
The plotted data are the seven azimuthally averaged values from
Sect.~\ref{sec:tiltedrings} normalized by the fitted central velocity
dispersion for each galaxy.  The radius of each datum is scaled by half of the
e-folding length of the exponential fit to its associated galaxy.  Data likely
affected by bulge kinematics, as well as all data from UGC~4458 and UGC~8196,
are marked with smaller symbols (and no errors); these data have been excluded
from the calculation of the scatter given near the top of each panel.  The
exponential model is plotted as a thick grey line with five times the scatter
about this line shown as dashed lines.  Points above the upper dashed line are
plotted in Fig.~\ref{fig:flaring}.  The models have not been fit directly to
the data shown, but to the individual fiber measurements for every galaxy
separately (see the Atlas).
}
\label{fig:sigmaexp}
\end{figure}

Thus, we have fit an exponential function to the radial distribution of the
individual $\slos$ measurements, incorporating the error in each measurement.
In this fit, we exclude $\slos$ measurements with errors that exceed 8 \kms\ and
those at $R\leq\rbulge$; for some galaxies, like UGC~7917, we also exclude
regions that are likely affected by a bar. The fitted parameters
($\sigma_{\rm LOS,0}$ and $\hsl$) are tabulated in Table~\ref{tab:expsig}.
Using the results from these fits,
the top panel in Fig.~\ref{fig:sigmaexp} provides $\slos/\sigma_{\rm LOS,0}$
versus $2R/\hsl$ for {\it all} galaxies, where the plotted
$\slos$ measurements are the {\it azimuthally averaged values} derived in
Sect.~\ref{sec:tiltedrings}.  Azimuthally averaged $\slos$ measurements from
designated bulge regions are represented by smaller symbols without errors, and
the solid line corresponds to the fitted exponential function. Excluding the
smaller symbols, the error-weighted scatter around the exponential fit is 3.0\%.

Several previous studies have reported a decrease in the observed stellar
velocity dispersion towards the centers of disk galaxies \citep{bottema1993,
emsellem2001, marquez2003, shapiro2003, ganda2006}.  It has been argued that
these ``dips'' are caused by a small, kinematically cold, central and young
stellar disk of high surface brightness that dominates the luminosity-weighted
kinematic measurements, outshining the hotter bulge component.  Indeed, some of
these studies focused on barred galaxies, Seyfert galaxies, or galaxies with a
significant bulge where one might expect to find such products of centrally
concentrated, recent star formation.  However, for edge-on galaxies,
\citet{kregel2005a} have shown that $\slos$ drops naturally towards the center
as a result of the sampling of the stellar kinematics along the sight line.
Moreover, disk-dominated galaxies may exhibit central dips in $\slos$ by
strictly adhering to the asymmetric-drift equation or having a disk with a
roughly constant stability \citep{westfall2009}. The galaxies in our PPak sample
are all close to face-on and were selected to, largely, avoid prominent bars,
bulges or nuclear activities. In addition, the galaxies also suffer from
beam smearing in their most inner regions. It may therefore not be surprising
that central dips in $\slos$ are not easily detected in our galaxies.

\subsection{Deriving $\sigz$ from $\slos$}

A fundamental goal of the DMS is to calculate dynamical mass surface densities of the
disks ($\sddisk$) for which the relevant kinematic measurement is not $\slos$ but the
vertical component ($\sigz$) of the SVE. We, therefore, deproject $\slos$ to
$\sigz$ following \citetalias{bershady2010b} (Equation~4): 
\begin{equation}
\sigz^2 = \frac{\slos^2}{\cos^2 i}\ \left[ 1 + \frac{\tan^2
i}{\alpha^2}\left(\sin^2\theta + \beta^2\cos^2\theta\right)\right]^{-1},
\label{eq:slos2sz}
\end{equation}
where $\alpha=\sigz/\sigr$ and $\beta=\sigp/\sigr$ describe the triaxial shape
of the SVE, $i$ is the inclination of the disk, and $\theta$ is the azimuthal
angle in the plane of the disk relative to the major axis.  As demonstrated for
UGC~463 in \citetalias{westfall2011b}, we can derive $\alpha$ and $\beta$ using
the combination of our gas and stellar kinematic data, assuming that the
epicycle approximation and the asymmetric-drift equation hold.  Similar analyses
for the PPak sample and the remainder of the DiskMass Phase-B sample is ongoing.
Here, we follow \citetalias{bershady2010b} and adopt $\alpha=0.6\pm0.15$ and
$\beta=0.7\pm0.04$ for all galaxies.

The conversion of $\slos$ to $\sigz$ is applied to every individual $\slos$ measurement.
Analogous to our fits to $\slos$ (Sect.~\ref{sec:expsigma}), we fit an exponential
function to these $\sigz$ measurements, yielding $\hsz$ and $\sigma_{\rm z,0}$
(Table~\ref{tab:expsig}). Due to the complexity in the error propagation, we do not
average the individual $\sigz$ measurement. Instead, the radially-binned $\sigz$
measurements are calculated from the error-weighted azimuthal averages of $\slos$ derived
in Sect.~\ref{sec:tiltedrings}.
The results for the radial profiles of $\sigz$ are shown in the bottom panel of
Fig.~\ref{fig:sigmaexp}.  Qualitatively, the behaviour of $\sigz$ matches
$\slos$, but the weighted RMS scatter is reduced slightly to 2.4\%, indicating
the merit of considering the SVE shape.

\begin{table}
\caption{\label{tab:expsig}
Exponential fits to the $\slos$ and $\sigz$ radial profiles}
\centering
{\small
\begin{tabular}{|r r r r r|}
\hline
\multicolumn{1}{|c}{UGC}                 & 
\multicolumn{1}{c}{$\sigma_{\rm LOS,0}$} &
\multicolumn{1}{c}{$\hsl$}               &
\multicolumn{1}{c}{$\sigma_{\rm z,0}$}   &
\multicolumn{1}{c|}{$\hsz$}              \\
\multicolumn{1}{|c}{}                    & 
\multicolumn{1}{c}{(km/s)}                 &
\multicolumn{1}{c}{(arcsec)}               &
\multicolumn{1}{c}{(km/s)}                 &
\multicolumn{1}{c|}{(arcsec)}              \\
\multicolumn{1}{|c}{(1)}             &   
\multicolumn{1}{c}{(2)}              & 
\multicolumn{1}{c}{(3)}              & 
\multicolumn{1}{c}{(4)}              & 
\multicolumn{1}{c|}{(5)}            \\
\hline
  448 &  57.3 $\pm$   4.8 & 22.2 $\pm$  2.8 &    47.8 $\pm$  2.7 & 24.6 $\pm$  2.4  \\
  463 &  79.3 $\pm$   2.5 & 27.7 $\pm$  1.4 &    69.5 $\pm$  2.5 & 27.5 $\pm$  1.5  \\
 1081 &  46.0 $\pm$   2.0 & 40.0 $\pm$  3.4 &    43.4 $\pm$  1.8 & 37.5 $\pm$  2.8  \\
 1087 &  50.9 $\pm$   6.6 & 16.2 $\pm$  3.7 &    42.5 $\pm$  4.6 & 19.2 $\pm$  3.8  \\
 1529 &  60.8 $\pm$   4.0 & 24.5 $\pm$  2.7 &    49.5 $\pm$  2.5 & 24.6 $\pm$  2.3  \\
 1635 &  33.4 $\pm$   1.7 & 31.6 $\pm$  4.4 &    29.0 $\pm$  1.6 & 37.9 $\pm$  6.3  \\
 1862 &  30.3 $\pm$   0.8 & 79.0 $\pm$  9.5 &    26.4 $\pm$  0.9 & 72.4 $\pm$  9.3  \\
 1908 &  85.4 $\pm$   7.4 & 17.3 $\pm$  2.1 &    74.0 $\pm$  6.6 & 18.6 $\pm$  2.4  \\
 3091 &  35.5 $\pm$   3.4 & 21.4 $\pm$  4.3 &    33.8 $\pm$  2.8 & 19.3 $\pm$  3.4  \\
 3140 &  76.8 $\pm$   4.3 & 22.5 $\pm$  2.0 &    73.4 $\pm$  3.9 & 22.9 $\pm$  1.9  \\
 3701 &  29.0 $\pm$   1.8 & 96.9 $\pm$ 28.2 &    25.9 $\pm$  1.1 & 91.7 $\pm$ 24.4  \\
 3997 &  47.2 $\pm$   3.3 & 16.6 $\pm$  2.3 &    38.5 $\pm$  3.2 & 19.4 $\pm$  3.3  \\
 4036 &  55.3 $\pm$   2.4 & 34.8 $\pm$  2.5 &    51.1 $\pm$  2.1 & 37.0 $\pm$  2.5  \\
 4107 &  46.5 $\pm$   1.7 & 31.5 $\pm$  2.3 &    41.8 $\pm$  1.7 & 32.2 $\pm$  2.5  \\
 4256 &  82.1 $\pm$   5.6 & 20.5 $\pm$  2.1 &    73.0 $\pm$  5.0 & 22.5 $\pm$  2.4  \\
 4368 &  62.3 $\pm$   3.0 & 30.6 $\pm$  2.3 &    43.6 $\pm$  2.3 & 35.4 $\pm$  2.9  \\
 4380 &  51.2 $\pm$   9.7 & 15.1 $\pm$  5.5 &    45.9 $\pm$  7.9 & 17.3 $\pm$  6.1  \\
 4458 & 113.6 $\pm$ 110.7 & 21.4 $\pm$  6.3 &    62.8 $\pm$ 35.3 & 33.7 $\pm$ 10.5  \\
 4555 &  49.0 $\pm$   2.2 & 49.6 $\pm$  5.0 &    39.5 $\pm$  1.2 & 52.3 $\pm$  3.9  \\
 4622 &  52.3 $\pm$   9.3 & 16.9 $\pm$  4.6 &    46.8 $\pm$  8.5 & 17.5 $\pm$  4.7  \\
 6903 &  32.9 $\pm$   4.1 & 32.0 $\pm$  9.1 &    28.2 $\pm$  3.8 & 36.3 $\pm$ 10.1  \\
 6918 &  56.9 $\pm$   2.4 & 35.3 $\pm$  2.6 &    38.5 $\pm$  0.9 & 54.6 $\pm$ 12.6  \\
 7244 &  30.5 $\pm$   1.7 & 42.6 $\pm$  9.5 &    29.1 $\pm$  1.6 & 43.3 $\pm$ 11.2  \\
 7917 &  80.1 $\pm$  12.7 & 27.1 $\pm$  4.3 &    69.9 $\pm$ 11.8 & 26.5 $\pm$  4.3  \\
 8196 & 112.2 $\pm$  68.3 & 16.5 $\pm$  5.2 &   108.5 $\pm$ 72.4 & 16.3 $\pm$  4.9  \\
 9177 &  56.5 $\pm$   6.0 & 21.2 $\pm$  3.5 &    47.7 $\pm$  4.3 & 20.0 $\pm$  2.7  \\
 9837 &  40.8 $\pm$   3.0 & 26.9 $\pm$  4.7 &    35.6 $\pm$  2.9 & 26.2 $\pm$  5.0  \\
 9965 &  40.7 $\pm$   1.9 & 29.6 $\pm$  4.1 &    39.6 $\pm$  1.9 & 28.7 $\pm$  3.7  \\
11318 &  59.7 $\pm$   3.8 & 29.9 $\pm$  3.3 &    59.4 $\pm$  3.8 & 29.9 $\pm$  3.3  \\
12391 &  50.8 $\pm$   3.3 & 35.4 $\pm$  5.7 &    43.6 $\pm$  3.3 & 35.8 $\pm$  6.3  \\
\hline
\end{tabular}
}
\tablefoot{
(1) UGC number; (2) central line-of-sight stellar velocity dispersion;
(3) the scale length of the exponential fit to $\slos$; (4) central vertical stellar
velocity dispersion; (5) the scale length of the exponential fit to $\sigz$.
}
\end{table}

\begin{figure}[t]
\centering
\includegraphics[width=0.49\textwidth]{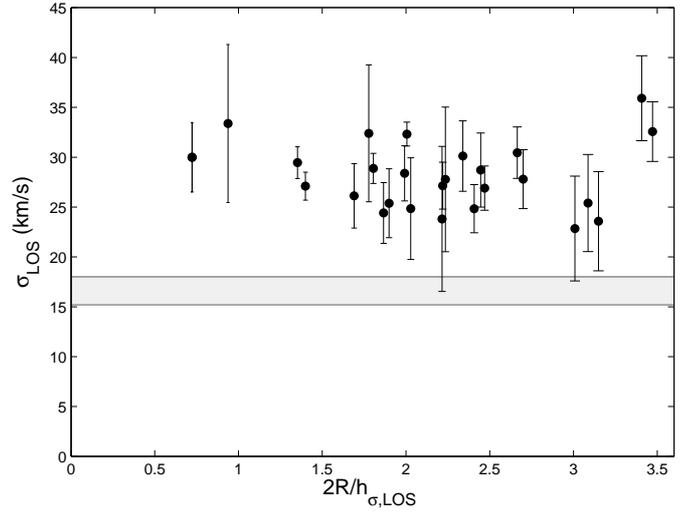}
\caption[]{
Measurements of $\slos$ that significantly deviates from the exponential
decline.  The grey band indicates the range in $\sinst$; the $\slos$
measurements presented in this figure have not been corrected for the
instrumental broadening to emphasize that these measurements are well above the
instrumental resolution limit.  One measurement from UGC~8196 at $\sim58$ \kms\
is not plotted.
}
\label{fig:flaring}
\end{figure}

Figure~\ref{fig:sigmaexp} shows that $\sigz$ (as well as $\slos$) decreases more
slowly than the fitted exponential function beyond a radius of $R \approx 1.5
(\hsz/2) \approx 1.5 \hr$ for several galaxies.  The data at these radii are
from lower surface-brightness regions of the disk where errors in $\sigz$ result
in a minor influence on the results of the exponential fit compared to the
higher surface brightness inner regions.  It is interesting to question whether
or not this difference between the inner and outer scale length of $\sigz$, or
``kinematic flaring'' of the disk, is astrophysical in origin.  In
Fig.~\ref{fig:flaring}, we plot the $\slos$ data from
Fig.~\ref{fig:sigmaexp} that lie five times the weighted scatter above the
exponential fit; 15 galaxies have at least one $\slos$ measurement outside of
$1.3\hr$ that satisfies this criterion.  The range of $\sinst$ is also shown to
demonstrate that most of the measured $\slos$ values are significantly larger 
than the instrumental dispersion, indicating that this ``kinematic flaring'' 
is likely real. A similar observation was made by \citet{herrmann2009}: Using
observations of planetary nebulae in five nearby spiral galaxies, they find that
$\sigz$ falls exponentially out to $\sim$3$\hr$, then stops declining and becomes
flat with radius.  This effect appears to occur at smaller radii for our galaxy
sample. Although a full discussion of this ``kinematic flaring'' is beyond the
scope of our paper, three example astrophysical scenarios that could lead to this
observation are (1) an increase in the disk mass-to-light ratio, (2) an increase
of the disk scale height (a flared disk), or (3) disk heating due to a massive
dark-matter halo.

\subsection{Photometric and dispersion scale lengths}
\label{sec:hrvshs}

\begin{figure}[t]
\centering
\includegraphics[width=0.49\textwidth]{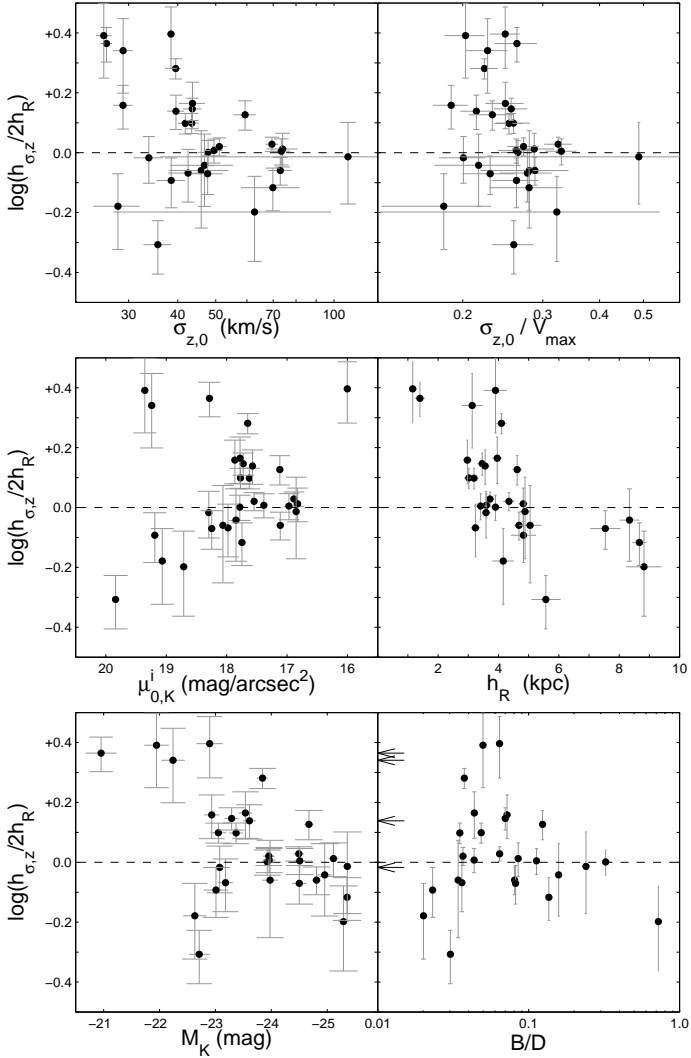}
\caption[]{
Ratio of the exponential scale length of $\sigz$ ($\hsz$) and twice the
photometric scale length ($2\hr$) as a function of several global properties of
the galaxies. This ratio is expected to be 1 ($\log[\hsz/2\hr]=0$) for an
isolated exponential stellar disk.
}
\label{fig:diffsigma}
\end{figure}

Using the parallel-plane approximation for a galaxy disk, \citet{kruit1988} has
shown that the mass surface density follows $\sddisk \propto \sigz^2$.  If mass
traces light and if the mass-to-light ratio of the disk is constant, we expect
$\sddisk \propto \mu$ such that $\sigz^2 \propto \mu$.  Thus, for an exponential
surface-brightness distribution with scale length $\hr$, we expect $\sigz$ to
have an exponential distribution with a scale length of $\hsz\equiv 2\hr$.  We
test this expectation using our direct measurements of $\hsz$ and the $K$-band
measurements of $\hr$ (Sect.~\ref{sec:2MASSphot}) by plotting the ratio of
these two quantities versus other salient galaxy properties in
Fig.~\ref{fig:diffsigma}.  As can be seen from the scatter, $\hsz/2\hr$ varies
by up to a factor of 2.5, which is substantially larger than expected given the
formal fitting errors. 
A weighted-average of all measurements yields $\log(\hsz/2\hr) = 0.069 \pm 0.092$ dex,
indicating that there is no significant systematic deviation from the expectation that
$\hsz\equiv 2\hr$. However, there is no correlation between those galaxies having 
$\hsz>2\hr$ and those that exhibit the ``kinematic flaring'' discussed in the previous
section. Disks that deviate from $\hsz\equiv 2\hr$ may have a radially varying
mass-to-light ratio, a radially varying scale height, and/or a change in the
vertical distribution of mass with radius.

Of all the abscissae in Fig.~\ref{fig:diffsigma}, $\hsz/2\hr$ appears to be
most strongly correlated with $\hr$.  We find that small disks yield $\sigz$
values that decline more slowly with radius than expected on the basis of the
light distribution, whereas $\sigz$ in larger disks declines more quickly.
We also find that the scatter in $\hsz/2\hr$ appears to decrease for disks with
higher $\sigma_{\rm z,0}$, $K$-band central surface brightness $\mu_{0,K}$, $M_K$,
and $B/D$ (Sect.~\ref{sec:2MASSphot}).  In conclusion, our theoretical expectation
that $\hsz=2\hr$ seems to hold best for stellar disks with a high central
velocity dispersion, high central surface brightness, in the brightest galaxies
with a significant bulge component.

\subsection{$\sigma_{\rm z,0}$ versus $V_{\rm max}$}
\label{sec:s2Vmax} 

As discussed in Sect.~\ref{sec:intro}, a key science objective of the DMS is to break the
disk-halo degeneracy in rotation-curve mass decompositions \citep{AlbadaSancisi1986}.
Using only our measurements of $\sigma_{\rm z,0}$ of the stellar disk and $\vmax$ of the
galaxy (based on our $\tanh$ fits to the \oiii\ gas from Sect.~\ref{sec:tanhRC}), we can
take a first step toward doing so by estimating the disk contribution to the rotational
velocity, without engaging in a full rotation curve decomposition analysis: For a radially
exponential disk with a certain oblateness ($q$), vertical density distribution ($k$) and
central vertical velocity dispersion of the disk stars ($\sigma_{\rm z,0}$), its dynamical
mass can be calculated analytically \citep{bottema1993}. From this follows the
gravitational potential that governs the corresponding maximum rotational velocity
$\vmax^{\rm disk}=f(q,k)\ \sigma_{\rm z,0}$ that occurs at a radius near 2.2$\hr$
\citep[][\citetalias{bershady2010b}]{casertano1983}.
Thus, we can define the maximality of the disk ($F_{\rm max}^{\rm disk}$) as the
fractional contribution of the stellar+gas disk to the total rotational velocity of the
galaxy at the radius where the rotation curve of the disk peaks: i.e., 
$F_{\rm max}^{\rm disk}\equiv\vmax^{\rm disk}/\vmax$.
Measurements of $F_{\rm max}^{\rm disk}$ made in this way for the PPak sample have been
presented by \citet[][hereafter \citetalias{bershady2011}]{bershady2011}, who also provide
the functional form of $f(q,k)$. Consequently, we limit ourselves here to investigating
the relation between the observables $\sigma_{\rm z,0}$ and the circular rotation speed of
the \oiii\ gas at 2.2$\hr$ ($\voiii$).

\begin{figure}[t]
\centering
\includegraphics[width=0.49\textwidth]{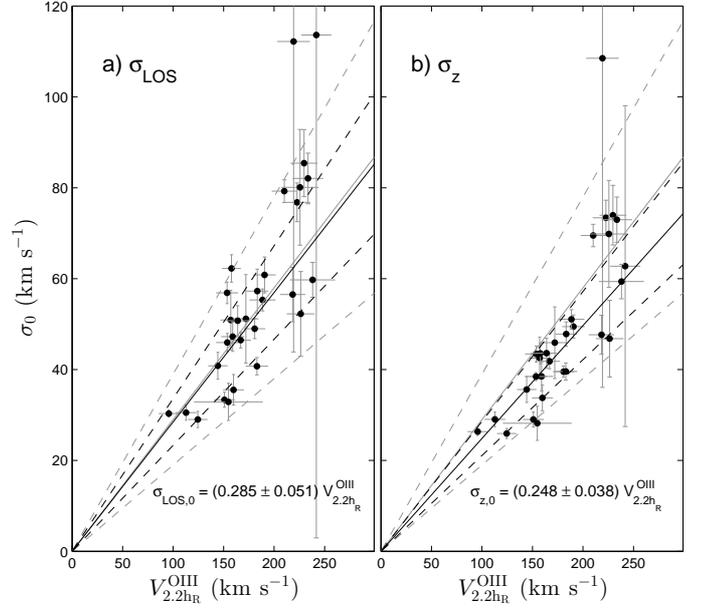}
\caption[]{
Relation between the central stellar velocity dispersion ($\sigma_{\rm LOS,0}$
to the left and $\sigma_{\rm z,0}$ to the right) and the circular speed of the
disks at 2.2$\hr$. The black solid line in each panel indicates the error-weighted
fit, and the black dashed lines are the error-weighted RMS scatter in the ratio.
The grey lines show the relation determined by \citet{bottema1993}.
}
\label{fig:sig2V}
\end{figure}

Figure~\ref{fig:sig2V} shows $\voiii$ versus $\sigma_{\rm LOS,0}$ and
$\sigma_{\rm z,0}$ for all galaxies in the PPak sample; as with
Fig.~\ref{fig:sigmaexp}, both $\slos$ and $\sigz$ are plotted to demonstrate
the merit of considering the shape of the SVE, even though the galaxies in our
sample are nearly face on.  Compared to the relation using $\sigma_{\rm LOS,0}$,
the relation between $\voiii$ and $\sigma_{\rm z,0}$ has a shallower
slope, as expected, and the scatter in the correlation decreases.  The slope is
also significantly shallower than what was found by \citet{bottema1993}.  The
fitted slope, $f(q,k)$, indicates that nearly all of the galaxy disks in our
sample are significantly submaximal, as discussed in \citetalias{bershady2011}.

\begin{figure}[t]
\centering
\includegraphics[width=0.49\textwidth]{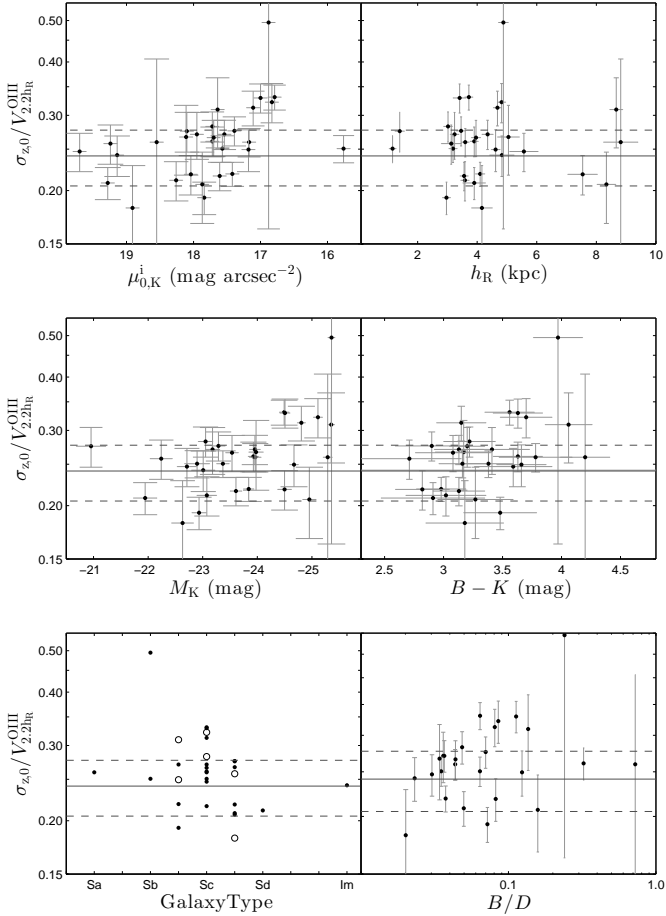}
\caption[]{
$\sigma_{\rm z,0}/\voiii$ versus global properties of the galaxies.
Solid and dashed lines indicate the mean and scatter in the ratio.
}
\label{fig:sig2Vtrends}
\end{figure}

Figure~\ref{fig:sig2Vtrends} shows the ratio
$\sigma_{\rm z,0}/\voiii=F_{\rm max}^{\rm disk}/f(q,k)$
versus various global properties of the galaxies. Weak correlations between
$\sigma_{\rm z,0}/\voiii$ and $\mu_{0,K}^i$, $M_K$, $B-K$, and morphological type can be
discerned. Although weak, these correlations are all consistent in terms of galaxy
properties along the Hubble sequence; later-type, fainter and bluer galaxies of lower 
surface brightness generally have kinematically colder disks for a fixed rotation
velocity. It should be noted that these global properties along the Hubble sequence also
correlate along the TF~relation; brighter galaxies have a larger $\voiii$. Therefore, a
significant covariance between $M_K$ and $\sigma_{\rm z,0}/\voiii$ would result in trends
that are opposite to those observed.

\section{Summary and Conclusions}
\label{sec:Summary}

We have presented gas and stellar kinematics obtained for 30 nearly face-on spiral
galaxies from an extensive observing campaign with PPak, a custom-built fiber-based IFU
module for the Cassegrain-mounted PMAS spectrograph on the 3.5m telescope at Calar Alto.
To measure the velocity dispersions of the stars in the outer low-surface-brightness
($\mu_B\approx25$ mag/arcsec$^2$) regions of the disks of spiral galaxies, multiple 
hour-long exposures per galaxy were carried out at the highest spectral resolution
($\lambda/\Delta\lambda=7680\pm640$, on average) achievable with PMAS, covering a
wavelength range of $4980-5369$\AA. During dusk and dawn, a total of 69 spectra of bright
stars were taken with the same spectrograph setting for use as template spectra in our
measurements of the stellar kinematics. The design of PPak in combination with the chosen
observing strategy allowed for zero calibration overheads during the precious dark nights.

We describe, in detail, the great care that was taken in our data reduction
procedure, as required to overcome the adverse effects of flexure and optical
aberrations in the instrument and ensure the highest possible quality of the
final data products.  The description of our data reduction procedure includes
spectral calibration of the external flood lamps, correction for shifts and a
changing plate scale on the detector due to flexure, pixel-to-pixel and
fiber-to-fiber flat-fielding, sky subtraction by interpolation along the slit to
account for optical aberrations in the spectrograph, and mapping the spectral
resolution that changes significantly across the detector.  After extraction and
wavelength calibration, spectra from the same fiber but from different exposures
were combined, weighting each spectrum by its $S/N$ and effective spectral
resolution. Observations of eight galaxies taken in multiple observing runs were
combined in groups of spatially overlapping fibers, taking into account the
differences in heliocentric velocities and in pointings of the instrument.

The kinematics of the ionized gas were obtained by fitting Gaussian functions to
the \oiii$\lambda$5007\AA\ emission line, provided the line was sufficiently
bright, yielding recession velocities and velocity dispersions of the \oiii\ gas
at each fiber position.  The stellar kinematics were measured using a robust
cross-correlation technique \citepalias[$DC3$;][]{westfall2011a} based on a
single K1 III template spectrum as a reference for all galaxies, yielding
recession velocities and velocity dispersions of the stars at each fiber
position.  For illustration purposes, two-dimensional maps of the stellar
continuum and \oiii\ line intensity were reconstructed for each galaxy, as well
as the velocity fields and velocity dispersion maps for both the stars and the
\oiii\ gas. The reconstructed continuum images were used to register a
coordinate system to the PPak fiber positions.

The measured recession velocities of the gas and stars in combination with the
fiber positions were used to derive the orientation of the galaxy disks.
Dynamical centers were assumed to coincide with the morphological centers and an
axisymmetric kinematic model with a hyperbolic-tangent prescription for the
shape of the rotation curve was fitted to the velocity fields.  A systematic
blueshift in the global recession velocities of the \oiii\ gas with respect to
the stars of $-2.1$ \kms\ was detected and preliminarily ascribed to the
expansion of star-forming regions from which the \oiii\ emission originates.
However, the offset is small, and it should be noted that a systematic
uncertainty of the same order exists between the gas and stellar velocities due
to an uncertainty in the heliocentric velocity of the used template star.  No
systematic offset is detected between the kinematic position angles of the
rotating gas and stellar disks.  The Tully-Fisher relation \citep{verheyen2001b}
was invoked to determine the inclination of the galaxy disks.  Given the
dynamical centers, systemic velocities and orientations of the galaxy disks, the
radial dependence of the rotational velocity and velocity dispersion of the gas
and stars was determined. 

The asymptotic maximum velocities of the fitted parameterized rotation curves
of the gas and stars clearly indicate the presence of asymmetric drift in the
stellar disks for nearly all galaxies in the sense that
$\vmaxoiii>\vmaxstar$. Using parameterized asymptotic rotation
speeds from our tanh-fit modeling we find this asymmetric drift to be on average
11$\pm$8\% of $\vmaxoiii$. However, several galaxies have large uncertainties
and a more detailed analysis using the actual data points instead of modeled rotation
curves will be presented in a forthcoming paper.

The observed line-of-sight velocity dispersions ($\slos$) of the stars were
corrected for an adopted shape of the stellar velocity ellipsoid ($\alpha$=0.6,
$\beta$=0.7), assumed to be the same within and among galaxies, yielding the
vertical velocity dispersions ($\sigz$) of the stars which have been measured
out to $2-3$ photometric disk scale lengths ($\hr$). The radial decline of the
stellar velocity dispersion has been fitted with an exponential decrease, yielding
central velocity dispersions and dispersion scale lengths for both $\slos$ and
$\sigz$. For several galaxies, the stellar velocity dispersion declines more slowly
than exponential beyond $\sim1.5\hr$, with 15 galaxies having at least one $\sigz$
value measured outside of $1.3\hr$ that is larger than five times the RMS above the
exponential fit.  The origin of this kinematic flaring is unclear, but may be related
to an increase in the disk mass-to-light ratio, an increase of the disk scale height
(a flared disk), or disk heating due to a massive dark-matter halo.

For the ensemble of 30 galaxies, the weighted-average ratio of the dispersion
scale lengths and two times the photometric scale length ($\hsz/2\hr$)
is consistent with the theoretically expected value of unity, but with a
variance of 0.09 dex; larger than can be explained on the basis of the
measurement errors.  More massive disks with a higher central surface brightness
show less scatter in this ratio and adhere closer to the theoretical
expectation. A weak trend with $\hr$ seems to suggest that $\hsz>2\hr$
in galaxies with $\hr<4$~kpc and $\hsz<2\hr$ in galaxies with
$\hr>4$~kpc.

The relation between the central stellar velocity dispersion of the disk and the
rotational velocity at $R$$=$2.2$\hr$ becomes tighter when $\sigz$ is used 
instead of $\slos$, indicating the merit of considering the shape of the stellar velocity
ellipsoid.  The ratio $\sigma_{\rm z,0}/\voiii$ seems to show trends
with the disk central surface brightness, absolute $K$-band magnitude, $B-K$
color and morphological types that are weak but consistent with the global
properties along the Hubble sequence; when compared to their rotational
velocities, later-type spiral galaxies have stellar disks that are kinematically
colder than earlier-type galaxies.

The main purpose of this paper has been to present the kinematic data of the gas
and stars as obtained with the custom-built PPak IFU for 30 galaxies in the
DiskMass Survey, and to present a first exploration of the kinematics. Any
further analysis of these data is deferred to forthcoming papers.

\acknowledgement{
We wish to thank the referee for useful comments and suggestions.
The staff at Calar Alto are thanked for their dedication and help during our
observations. M.A.W.V. and T.P.K.M. acknowledge financial support provided by NOVA, the
Netherlands Research School for Astronomy, and travel support from the Leids 
Kerkhoven-Bosscha Fonds. Support for this work has also been provided by the National
Science Foundation (NSF) via grants AST-0307417 and AST-0607516 (M.A.B., K.B.W., and 
A.S.-R.), OISE-0754437 (K.B.W.), and AST-1009491 (M.A.B. and A.S.-R.). K.B.W. is also
supported by grant 614.000.807 from  the Netherlands Organisation for Scientific Research
(NWO). R.A.S. and M.A.B. acknowledge support from NASA/Spitzer grant GO-30894.
This work has taken advantage of the online 
NED\footnote{{\tiny http://nedwww.ipac.caltech.edu/}}, 
SDSS\footnote{{\tiny http://www.sdss.org/collaboration/credits.html}}
DSS\footnote{{\tiny http://archive.stsci.edu/dss/acknowledging.html}} and 
2MASS\footnote{{\tiny http://www.ipac.caltech.edu/2mass/releases/allsky/faq.html\#reference}}
databases and data archives.
}

\bibliography{Martinsson_DMS_VI}  

\begin{thebibliography}{90}
\expandafter\ifx\csname natexlab\endcsname\relax\def\natexlab#1{#1}\fi

\bibitem[{{Andersen}(2001)}]{andersen2001phd}
{Andersen}, D.~R. 2001, PhD thesis, The Pennsylvania State University

\bibitem[{{Andersen} \& {Bershady}(2003)}]{andersen2003}
{Andersen}, D.~R. \& {Bershady}, M.~A. 2003, \apjl, 599, L79

\bibitem[{{Andersen} {et~al.}(2006){Andersen}, {Bershady}, {Sparke},
  {Gallagher}, {Wilcots}, {van Driel}, \& {Monnier-Ragaigne}}]{andersen2006}
{Andersen}, D.~R., {Bershady}, M.~A., {Sparke}, L.~S., {et~al.} 2006, \apjs,
  166, 505

\bibitem[{{Andersen} {et~al.}(2008){Andersen}, {Walcher}, {B{\"o}ker}, {Ho},
  {van der Marel}, {Rix}, \& {Shields}}]{andersen2008}
{Andersen}, D.~R., {Walcher}, C.~J., {B{\"o}ker}, T., {et~al.} 2008, \apj, 688,
  990

\bibitem[{{Begeman}(1989)}]{begeman1989}
{Begeman}, K.~G. 1989, \aap, 223, 47

\bibitem[{{Bell} \& {de Jong}(2001)}]{belldejong2001}
{Bell}, E.~F. \& {de Jong}, R.~S. 2001, \apj, 550, 212

\bibitem[{{Bershady}(1995)}]{bershady1995}
{Bershady}, M.~A. 1995, \aj, 109, 87

\bibitem[{{Bershady} {et~al.}(2004){Bershady}, {Andersen}, {Harker}, {Ramsey},
  \& {Verheijen}}]{bershady2004}
{Bershady}, M.~A., {Andersen}, D.~R., {Harker}, J., {Ramsey}, L.~W., \&
  {Verheijen}, M.~A.~W. 2004, \pasp, 116, 565

\bibitem[{{Bershady} {et~al.}(2005){Bershady}, {Andersen}, {Verheijen},
  {Westfall}, {Crawford}, \& {Swaters}}]{bershady2005}
{Bershady}, M.~A., {Andersen}, D.~R., {Verheijen}, M.~A.~W., {et~al.} 2005,
  \apjs, 156, 311

\bibitem[{{Bershady} {et~al.}(1994){Bershady}, {Hereld}, {Kron}, {Koo}, {Munn},
  \& {Majewski}}]{bershady1994}
{Bershady}, M.~A., {Hereld}, M., {Kron}, R.~G., {et~al.} 1994, \aj, 108, 870

\bibitem[{{Bershady} {et~al.}(2011){Bershady}, {Martinsson}, {Verheijen},
  {Westfall}, {Andersen}, \& {Swaters}}]{bershady2011}
{Bershady}, M.~A., {Martinsson}, T.~P.~K., {Verheijen}, M.~A.~W., {et~al.}
  2011, \apjl, 739, L47

\bibitem[{{Bershady} {et~al.}(2010{\natexlab{a}}){Bershady}, {Verheijen},
  {Swaters}, {Andersen}, {Westfall}, \& {Martinsson}}]{bershady2010a}
{Bershady}, M.~A., {Verheijen}, M.~A.~W., {Swaters}, R.~A., {et~al.}
  2010{\natexlab{a}}, \apj, 716, 198

\bibitem[{{Bershady} {et~al.}(2010{\natexlab{b}}){Bershady}, {Verheijen},
  {Westfall}, {Andersen}, {Swaters}, \& {Martinsson}}]{bershady2010b}
{Bershady}, M.~A., {Verheijen}, M.~A.~W., {Westfall}, K.~B., {et~al.}
  2010{\natexlab{b}}, \apj, 716, 234

\bibitem[{{Bertin} \& {Arnouts}(1996)}]{Bertin1996}
{Bertin}, E. \& {Arnouts}, S. 1996, \aaps, 117, 393

\bibitem[{{Bosma}(1978)}]{bosma1978}
{Bosma}, A. 1978, PhD thesis, Groningen Univ.

\bibitem[{{Bosma}(1981{\natexlab{a}})}]{bosma1981a}
{Bosma}, A. 1981{\natexlab{a}}, \aj, 86, 1791

\bibitem[{{Bosma}(1981{\natexlab{b}})}]{bosma1981b}
{Bosma}, A. 1981{\natexlab{b}}, \aj, 86, 1825

\bibitem[{{Bottema}(1993)}]{bottema1993}
{Bottema}, R. 1993, \aap, 275, 16

\bibitem[{{Casertano}(1983)}]{casertano1983}
{Casertano}, S. 1983, \mnras, 203, 735

\bibitem[{{Cenarro} {et~al.}(2007){Cenarro}, {Peletier},
  {S{\'a}nchez-Bl{\'a}zquez}, {Selam}, {Toloba}, {Cardiel},
  {Falc{\'o}n-Barroso}, {Gorgas}, {Jim{\'e}nez-Vicente}, \&
  {Vazdekis}}]{cenarro2007}
{Cenarro}, A.~J., {Peletier}, R.~F., {S{\'a}nchez-Bl{\'a}zquez}, P., {et~al.}
  2007, \mnras, 374, 664

\bibitem[{{Ciardullo} {et~al.}(2004){Ciardullo}, {Durrell}, {Laychak},
  {Herrmann}, {Moody}, {Jacoby}, \& {Feldmeier}}]{ciardullo2004}
{Ciardullo}, R., {Durrell}, P.~R., {Laychak}, M.~B., {et~al.} 2004, \apj, 614,
  167

\bibitem[{{Contini} {et~al.}(1998){Contini}, {Considere}, \&
  {Davoust}}]{contini1998}
{Contini}, T., {Considere}, S., \& {Davoust}, E. 1998, \aaps, 130, 285

\bibitem[{{Courteau} {et~al.}(2003){Courteau}, {Andersen}, {Bershady},
  {MacArthur}, \& {Rix}}]{courteau2003}
{Courteau}, S., {Andersen}, D.~R., {Bershady}, M.~A., {MacArthur}, L.~A., \&
  {Rix}, H.-W. 2003, \apj, 594, 208

\bibitem[{{Dutton} {et~al.}(2005){Dutton}, {Courteau}, {de Jong}, \&
  {Carignan}}]{dutton2005}
{Dutton}, A.~A., {Courteau}, S., {de Jong}, R., \& {Carignan}, C. 2005, \apj,
  619, 218

\bibitem[{{Emsellem} {et~al.}(2001){Emsellem}, {Greusard}, {Combes}, {Friedli},
  {Leon}, {P{\'e}contal}, \& {Wozniak}}]{emsellem2001}
{Emsellem}, E., {Greusard}, D., {Combes}, F., {et~al.} 2001, \aap, 368, 52

\bibitem[{{Erwin}(2005)}]{erwin2005}
{Erwin}, P. 2005, \mnras, 364, 283

\bibitem[{{Famaey} {et~al.}(2005){Famaey}, {Jorissen}, {Luri}, {Mayor}, {Udry},
  {Dejonghe}, \& {Turon}}]{famaey2005}
{Famaey}, B., {Jorissen}, A., {Luri}, X., {et~al.} 2005, \aap, 430, 165

\bibitem[{{Fraternali} {et~al.}(2011){Fraternali}, {Sancisi}, \&
  {Kamphuis}}]{fraternali2011}
{Fraternali}, F., {Sancisi}, R., \& {Kamphuis}, P. 2011, \aap, 531, 64

\bibitem[{{Ganda} {et~al.}(2006){Ganda}, {Falc{\'o}n-Barroso}, {Peletier},
  {Cappellari}, {Emsellem}, {McDermid}, {de Zeeuw}, \& {Carollo}}]{ganda2006}
{Ganda}, K., {Falc{\'o}n-Barroso}, J., {Peletier}, R.~F., {et~al.} 2006,
  \mnras, 367, 46

\bibitem[{{Herrmann} \& {Ciardullo}(2009)}]{herrmann2009}
{Herrmann}, K.~A. \& {Ciardullo}, R. 2009, \apj, 705, 1686

\bibitem[{{Hurst} {et~al.}(1998){Hurst}, {Armstrong}, \& {Arbour}}]{hurst1998}
{Hurst}, G.~M., {Armstrong}, M., \& {Arbour}, R. 1998, \iaucirc, 6875, 1

\bibitem[{{Kauffmann} {et~al.}(2003){Kauffmann}, {Heckman}, {White}, {Charlot},
  {Tremonti}, {Brinchmann}, {Bruzual}, {Peng}, {Seibert}, {Bernardi},
  {Blanton}, {Brinkmann}, {Castander}, {Cs{\'a}bai}, {Fukugita}, {Ivezic},
  {Munn}, {Nichol}, {Padmanabhan}, {Thakar}, {Weinberg}, \&
  {York}}]{kauffmann2003}
{Kauffmann}, G., {Heckman}, T.~M., {White}, S.~D.~M., {et~al.} 2003, \mnras,
  341, 33

\bibitem[{{Kelz} {et~al.}(2006){Kelz}, {Verheijen}, {Roth}, {Bauer}, {Becker},
  {Paschke}, {Popow}, {S{\'a}nchez}, \& {Laux}}]{kelz2006}
{Kelz}, A., {Verheijen}, M.~A.~W., {Roth}, M.~M., {et~al.} 2006, \pasp, 118,
  129

\bibitem[{{Kormendy} \& {Kennicutt}(2004)}]{kormendy2004}
{Kormendy}, J. \& {Kennicutt}, Jr., R.~C. 2004, \araa, 42, 603

\bibitem[{{Kregel} \& {van der Kruit}(2004)}]{kregel2004}
{Kregel}, M. \& {van der Kruit}, P.~C. 2004, \mnras, 355, 143

\bibitem[{{Kregel} \& {van der Kruit}(2005)}]{kregel2005a}
{Kregel}, M. \& {van der Kruit}, P.~C. 2005, \mnras, 358, 481

\bibitem[{{Kregel} {et~al.}(2005){Kregel}, {van der Kruit}, \&
  {Freeman}}]{kregel2005}
{Kregel}, M., {van der Kruit}, P.~C., \& {Freeman}, K.~C. 2005, \mnras, 358,
  503

\bibitem[{{Kron}(1980)}]{kron1980}
{Kron}, R.~G. 1980, \apjs, 43, 305

\bibitem[{{Kuijken} \& {Gilmore}(1991)}]{kuijken1991}
{Kuijken}, K. \& {Gilmore}, G. 1991, \apjl, 367, L9

\bibitem[{{Luck} \& {Heiter}(2007)}]{luck2007}
{Luck}, R.~E. \& {Heiter}, U. 2007, \aj, 133, 2464

\bibitem[{{M{\'a}rquez} {et~al.}(2003){M{\'a}rquez}, {Masegosa}, {Durret},
  {Gonz{\'a}lez Delgado}, {Moles}, {Maza}, {P{\'e}rez}, \&
  {Roth}}]{marquez2003}
{M{\'a}rquez}, I., {Masegosa}, J., {Durret}, F., {et~al.} 2003, \aap, 409, 459

\bibitem[{{Martinsson}(2011)}]{martinsson2011}
{Martinsson}, T.~P.~K. 2011, PhD thesis, Univ.~of~Groningen

\bibitem[{{Massarotti} {et~al.}(2008){Massarotti}, {Latham}, {Stefanik}, \&
  {Fogel}}]{massarotti2008}
{Massarotti}, A., {Latham}, D.~W., {Stefanik}, R.~P., \& {Fogel}, J. 2008, \aj,
  135, 209

\bibitem[{{McWilliam}(1990)}]{mcwilliam1990}
{McWilliam}, A. 1990, \apjs, 74, 1075

\bibitem[{{Modjaz} \& {Li}(1999)}]{modjaz1999}
{Modjaz}, M. \& {Li}, W.~D. 1999, \iaucirc, 7237, 1

\bibitem[{{Molaro} {et~al.}(1995){Molaro}, {Gabrielcic}, {Boehm}, \&
  {Tessicini}}]{molaro1995}
{Molaro}, P., {Gabrielcic}, A., {Boehm}, C., \& {Tessicini}, G. 1995, \iaucirc,
  6137, 1

\bibitem[{{Navarro} {et~al.}(1997){Navarro}, {Frenk}, \& {White}}]{NFW1997}
{Navarro}, J.~F., {Frenk}, C.~S., \& {White}, S.~D.~M. 1997, \apj, 490, 493

\bibitem[{{Nilson}(1973)}]{nilson1973}
{Nilson}, P. 1973, {Uppsala general catalogue of galaxies}, ed. {Nilson, P.}

\bibitem[{{Noordermeer} {et~al.}(2007){Noordermeer}, {van der Hulst},
  {Sancisi}, {Swaters}, \& {van Albada}}]{noordermeer2007}
{Noordermeer}, E., {van der Hulst}, J.~M., {Sancisi}, R., {Swaters}, R.~S., \&
  {van Albada}, T.~S. 2007, \mnras, 376, 1513

\bibitem[{{Olling} \& {Dehnen}(2003)}]{olling2003}
{Olling}, R.~P. \& {Dehnen}, W. 2003, \apj, 599, 275

\bibitem[{{Ostriker} \& {Caldwell}(1979)}]{ostriker1979}
{Ostriker}, J.~P. \& {Caldwell}, J.~A.~R. 1979, in IAU Symposium, Vol.~84, The
  Large-Scale Characteristics of the Galaxy, ed. {W.~B.~Burton}, 441--448

\bibitem[{{Peletier} \& {Balcells}(1997)}]{peletier1997}
{Peletier}, R.~F. \& {Balcells}, M. 1997, \na, 1, 349

\bibitem[{{Pohlen} \& {Trujillo}(2006)}]{pohlen2006}
{Pohlen}, M. \& {Trujillo}, I. 2006, \aap, 454, 759

\bibitem[{{Quinn} {et~al.}(2006){Quinn}, {Garnavich}, {Li}, {Panagia}, {Riess},
  {Schmidt}, \& {Della Valle}}]{quinn2006}
{Quinn}, J.~L., {Garnavich}, P.~M., {Li}, W., {et~al.} 2006, \apj, 652, 512

\bibitem[{{Ram{\'{\i}}rez} {et~al.}(2000){Ram{\'{\i}}rez}, {Sellgren}, {Carr},
  {Balachandran}, {Blum}, {Terndrup}, \& {Steed}}]{ramirez2000}
{Ram{\'{\i}}rez}, S.~V., {Sellgren}, K., {Carr}, J.~S., {et~al.} 2000, \apj,
  537, 205

\bibitem[{{Ratnatunga} \& {Upgren}(1997)}]{ratnatunga1997}
{Ratnatunga}, K.~U. \& {Upgren}, A.~R. 1997, \apj, 476, 811

\bibitem[{{Roberts}(1966)}]{roberts1966}
{Roberts}, M.~S. 1966, \apj, 144, 639

\bibitem[{{Rogstad} \& {Shostak}(1972)}]{rogstad1972}
{Rogstad}, D.~H. \& {Shostak}, G.~S. 1972, \apj, 176, 315

\bibitem[{{Rosales-Ortega} {et~al.}(2010){Rosales-Ortega}, {Kennicutt},
  {S{\'a}nchez}, {D{\'{\i}}az}, {Pasquali}, {Johnson}, \&
  {Hao}}]{rosales-ortega2010}
{Rosales-Ortega}, F.~F., {Kennicutt}, R.~C., {S{\'a}nchez}, S.~F., {et~al.}
  2010, \mnras, 405, 735

\bibitem[{{Roth} {et~al.}(2005){Roth}, {Kelz}, {Fechner}, {Hahn}, {Bauer},
  {Becker}, {B{\"o}hm}, {Christensen}, {Dionies}, {Paschke}, {Popow}, {Wolter},
  {Schmoll}, {Laux}, \& {Altmann}}]{roth2005}
{Roth}, M.~M., {Kelz}, A., {Fechner}, T., {et~al.} 2005, \pasp, 117, 620

\bibitem[{{Rubin} {et~al.}(1980){Rubin}, {Ford}, \& {.~Thonnard}}]{rubin1980}
{Rubin}, V.~C., {Ford}, W.~K.~J., \& {.~Thonnard}, N. 1980, \apj, 238, 471

\bibitem[{{Sackett}(1997)}]{sackett1997}
{Sackett}, P.~D. 1997, \apj, 483, 103

\bibitem[{{Saha} {et~al.}(2001){Saha}, {Sandage}, {Tammann}, {Dolphin},
  {Christensen}, {Panagia}, \& {Macchetto}}]{saha2001}
{Saha}, A., {Sandage}, A., {Tammann}, G.~A., {et~al.} 2001, \apj, 562, 314

\bibitem[{{S{\'a}nchez} {et~al.}(2012){S{\'a}nchez}, {Kennicutt}, {Gil de Paz},
  {van de Ven}, {V{\'{\i}}lchez}, {Wisotzki}, {Walcher}, {Mast}, {Aguerri},
  {Albiol-P{\'e}rez}, {Alonso-Herrero}, {Alves}, {Bakos}, {Bart{\'a}kov{\'a}},
  {Bland-Hawthorn}, {Boselli}, {Bomans}, {Castillo-Morales}, {Cortijo-Ferrero},
  {de Lorenzo-C{\'a}ceres}, {Del Olmo}, {Dettmar}, {D{\'{\i}}az}, {Ellis},
  {Falc{\'o}n-Barroso}, {Flores}, {Gallazzi}, {Garc{\'{\i}}a-Lorenzo},
  {Gonz{\'a}lez Delgado}, {Gruel}, {Haines}, {Hao}, {Husemann},
  {Igl{\'e}sias-P{\'a}ramo}, {Jahnke}, {Johnson}, {Jungwiert}, {Kalinova},
  {Kehrig}, {Kupko}, {L{\'o}pez-S{\'a}nchez}, {Lyubenova}, {Marino},
  {M{\'a}rmol-Queralt{\'o}}, {M{\'a}rquez}, {Masegosa}, {Meidt},
  {Mendez-Abreu}, {Monreal-Ibero}, {Montijo}, {Mour{\~a}o}, {Palacios-Navarro},
  {Papaderos}, {Pasquali}, {Peletier}, {P{\'e}rez}, {P{\'e}rez}, {Quirrenbach},
  {Rela{\~n}o}, {Rosales-Ortega}, {Roth}, {Ruiz-Lara},
  {S{\'a}nchez-Bl{\'a}zquez}, {Sengupta}, {Singh}, {Stanishev}, {Trager},
  {Vazdekis}, {Viironen}, {Wild}, {Zibetti}, \& {Ziegler}}]{sanchez2012}
{S{\'a}nchez}, S.~F., {Kennicutt}, R.~C., {Gil de Paz}, A., {et~al.} 2012,
  \aap, 538, A8

\bibitem[{{Sancisi}(2004)}]{sancisi2004}
{Sancisi}, R. 2004, in IAU Symposium, Vol. 220, Dark Matter in Galaxies, ed.
  {S.~Ryder, D.~Pisano, M.~Walker, \& K.~Freeman}, 233

\bibitem[{{Schechtman-Rook} \& {Hess}(2012)}]{Schechtman-Rook2012}
{Schechtman-Rook}, A. \& {Hess}, K.~M. 2012, \apj, 750, 171

\bibitem[{{Schlegel} {et~al.}(1998){Schlegel}, {Finkbeiner}, \&
  {Davis}}]{schlegel1998}
{Schlegel}, D.~J., {Finkbeiner}, D.~P., \& {Davis}, M. 1998, \apj, 500, 525

\bibitem[{{Shapiro} {et~al.}(2003){Shapiro}, {Gerssen}, \& {van der
  Marel}}]{shapiro2003}
{Shapiro}, K.~L., {Gerssen}, J., \& {van der Marel}, R.~P. 2003, \aj, 126, 2707

\bibitem[{{Sorce} {et~al.}(2013){Sorce}, {Courtois}, {Tully}, {Seibert},
  {Scowcroft}, {Freedman}, {Madore}, {Persson}, {Monson}, \&
  {Rigby}}]{sorce2013}
{Sorce}, J.~G., {Courtois}, H.~M., {Tully}, R.~B., {et~al.} 2013, \apj, 765, 94

\bibitem[{{Soubiran} {et~al.}(2008){Soubiran}, {Bienaym{\'e}}, {Mishenina}, \&
  {Kovtyukh}}]{soubiran2008}
{Soubiran}, C., {Bienaym{\'e}}, O., {Mishenina}, T.~V., \& {Kovtyukh}, V.~V.
  2008, \aap, 480, 91

\bibitem[{{Swaters} {et~al.}(2011){Swaters}, {Sancisi}, {van Albada}, \& {van
  der Hulst}}]{swaters2011}
{Swaters}, R.~A., {Sancisi}, R., {van Albada}, T.~S., \& {van der Hulst}, J.~M.
  2011, \apj, 729, 118

\bibitem[{{Tully} \& {Fisher}(1977)}]{tullyfisher1977}
{Tully}, R.~B. \& {Fisher}, J.~R. 1977, \aap, 54, 661

\bibitem[{{Tully} \& {Pierce}(2000)}]{tully2000}
{Tully}, R.~B. \& {Pierce}, M.~J. 2000, \apj, 533, 744

\bibitem[{{van Albada} {et~al.}(1985){van Albada}, {Bahcall}, {Begeman}, \&
  {Sancisi}}]{albada1985}
{van Albada}, T.~S., {Bahcall}, J.~N., {Begeman}, K., \& {Sancisi}, R. 1985,
  \apj, 295, 305

\bibitem[{{van Albada} \& {Sancisi}(1986)}]{AlbadaSancisi1986}
{van Albada}, T.~S. \& {Sancisi}, R. 1986, Royal Society of London
  Philosophical Transactions Series A, 320, 447

\bibitem[{{van der Hulst} {et~al.}(1992){van der Hulst}, {Terlouw}, {Begeman},
  {Zwitser}, \& {Roelfsema}}]{hulst1992}
{van der Hulst}, J.~M., {Terlouw}, J.~P., {Begeman}, K.~G., {Zwitser}, W., \&
  {Roelfsema}, P.~R. 1992, in Astronomical Society of the Pacific Conference
  Series, Vol.~25, Astronomical Data Analysis Software and Systems I, ed. D.~M.
  {Worrall}, C.~{Biemesderfer}, \& J.~{Barnes}, 131--136

\bibitem[{{van der Kruit}(1988)}]{kruit1988}
{van der Kruit}, P.~C. 1988, \aap, 192, 117

\bibitem[{{van der Kruit} \& {Freeman}(1984)}]{kruit1984}
{van der Kruit}, P.~C. \& {Freeman}, K.~C. 1984, \apj, 278, 81

\bibitem[{{van der Kruit} \& {Freeman}(1986)}]{kruit1986}
{van der Kruit}, P.~C. \& {Freeman}, K.~C. 1986, \apj, 303, 556

\bibitem[{{van der Kruit} \& {Searle}(1981)}]{kruit1981}
{van der Kruit}, P.~C. \& {Searle}, L. 1981, \aap, 95, 105

\bibitem[{{Verheijen}(1997)}]{verheyen1997}
{Verheijen}, M.~A.~W. 1997, PhD thesis, PhD thesis, Univ.~Groningen, The
  Netherlands , (1997)

\bibitem[{{Verheijen}(2001)}]{verheyen2001b}
{Verheijen}, M.~A.~W. 2001, \apj, 563, 694

\bibitem[{{Verheijen} {et~al.}(2004){Verheijen}, {Bershady}, {Andersen},
  {Swaters}, {Westfall}, {Kelz}, \& {Roth}}]{verheyen2004}
{Verheijen}, M.~A.~W., {Bershady}, M.~A., {Andersen}, D.~R., {et~al.} 2004,
  Astronomische Nachrichten, 325, 151

\bibitem[{{V{\'e}ron-Cetty} \& {V{\'e}ron}(2006)}]{veron2006}
{V{\'e}ron-Cetty}, M.-P. \& {V{\'e}ron}, P. 2006, \aap, 455, 773

\bibitem[{{Vogelaar} \& {Terlouw}(2001)}]{vogelaar2001}
{Vogelaar}, M.~G.~R. \& {Terlouw}, J.~P. 2001, in Astronomical Society of the
  Pacific Conference Series, Vol. 238, Astronomical Data Analysis Software and
  Systems X, ed. {F.~R.~Harnden Jr., F.~A.~Primini, \& H.~E.~Payne}, 358

\bibitem[{{Westfall}(2009)}]{westfall2009}
{Westfall}, K.~B. 2009, PhD thesis, Univ. of Wisconsin--Madison.

\bibitem[{{Westfall} {et~al.}(2011{\natexlab{a}}){Westfall}, {Bershady}, \&
  {Verheijen}}]{westfall2011a}
{Westfall}, K.~B., {Bershady}, M.~A., \& {Verheijen}, M.~A.~W.
  2011{\natexlab{a}}, \apjs, 193, 21

\bibitem[{{Westfall} {et~al.}(2011{\natexlab{b}}){Westfall}, {Bershady},
  {Verheijen}, {Andersen}, {Martinsson}, {Swaters}, \&
  {Schechtman-Rook}}]{westfall2011b}
{Westfall}, K.~B., {Bershady}, M.~A., {Verheijen}, M.~A.~W., {et~al.}
  2011{\natexlab{b}}, \apj, 742, 18

\bibitem[{{Zibetti} {et~al.}(2009){Zibetti}, {Charlot}, \& {Rix}}]{zibetti2009}
{Zibetti}, S., {Charlot}, S., \& {Rix}, H. 2009, \mnras, 400, 1181

\bibitem[{{Zwicky}(1933)}]{zwicky1933}
{Zwicky}, F. 1933, Helvetica Physica Acta, 6, 110

\end{thebibliography}
\bibliographystyle{aa_DMS_VI}     
\setlength{\bibsep}{1.3pt}

\appendix


\section{Notes on Individual Galaxies}
\label{app:gal}

\begin{list}{}
  {
  \settowidth{\labelwidth}{\bf UGC 00000:}
  \setlength{\labelsep}{1em}
  \setlength{\itemsep}{0.2em}
  \setlength{\parskip}{0.2em}
  \setlength{\leftmargin}{\labelwidth}
  \setlength{\rightmargin}{0pt}
  }
  \item[{\bf UGC   448:}] IC 43.  High quality kinematics.  Significant bulge
  with second highest bulge-to-disk ratio ($B/D = 0.32$).  Gas rotation curve
  rises quite sharply.  A small bar and significant spiral structure are visible
  morphologically but exhibit little kinematic influence.
  \item[{\bf UGC   463:}] NGC 234.  High quality kinematics.  PPak and SparsePak
  data studied in detail in \citetalias{westfall2011b}.  Strong, three-arm
  spiral structure with minor streaming motions.  Some \oiii\ spots are clearly
  associated with star formation in the south-west arm.
  \item[{\bf UGC  1081:}] NGC 575, IC 1710. Strongly barred galaxy. Bright field star
  within the PPak field-of-view.  A Type-II break exists in $\mu_K(R)$ at roughly
  $1\hr$ (approximately the same as the bar length). Very patchy \oiii\ emission.
  Stellar rotation curve shows interesting structure within $1\hr$.
  \item[{\bf UGC  1087:}] Very patchy \oiii\ emission, some of which is
  associated with the visible spiral arms.  ``Ringing'' present in $\mu_K(R)$
  associated with the azimuthal coherence of the tightly wound spiral arms.
  ``Kinematic flaring'' starts just beyond $1\hr$.  Host to Type Ia SN 1999dk
  \citep{modjaz1999}.
  \item[{\bf UGC  1529:}] IC 193.  High quality kinematics.  Sc galaxy, rather
  typical of our sample, apart from the high inclination ($\itf=39\arcdeg$).
  Rather weak \oiii\ emission.  A dip in the \oiii\ rotation curve occurs at
  $\sim1\hr$ without a morphological counterpart.
  \item[{\bf UGC  1635:}] IC 208. Poor \oiii\ data. A Type II break in
  $\mu_K(R)$ occurs at $\sim1\hr$ with a corresponding dip in the \oiii\
  rotation curve; any dynamical association between these features is unknown.
  ``Kinematic flaring'' begins at about $2\hr$.
  \item[{\bf UGC  1862:}] Unique among our sample:  It has the lowest
  luminosity, $M_K=-21.0$, one magnitude fainter than the second least luminous
  galaxy (UGC~3701).  Two Type II breaks exist in $\mu_K(R)$ (at $\sim
  23\arcsec$ and $\sim 60\arcsec$), but no indication of a bulge component;
  inner break is caused by the spiral arms.  Appears to have a rather large bar,
  as well.  Data within $R<2\farcs5$ (indicated with a dotted vertical line in
  the Atlas) are excluded in the analysis of $\slos$ and $\sigz$, to avoid
  beam-smearing affects particularly strong in the center.  Rather constant
  $\slos$ with radius such that $\hsz/2\hr = 2.5$, an very extreme value among
  our sample with only UGC~3701 having a larger ratio.
  \item[{\bf UGC  1908:}] NGC 927, Mrk 593.  Barred galaxy with a rather weak
  Type II break in $\mu_K(R)$ at $\sim 1\hr$.  Classified as a Starburst Nucleus
  Galaxy (SBNG), but its nucleus has an ambiguous activity classification
  between \ion{H}{2} and LINER \citep{contini1998}.  Strong \oiii\
  emission near the galaxy center and associated with the spiral arms.
  \item[{\bf UGC  3091:}] Since the bulge/disk fitting routine resulted in a
  non-existing bulge, the excess light in the central region is interpreted as
  an inner disk.  As with UGC~1862, $\slos$ and $\sigz$ data within $R<2\farcs5$
  are excluded from our analysis.  Strong \oiii\ emission associated with the
  spiral arms.  \oiii\ velocity dispersions are roughly the same as $\slos$.
  \item[{\bf UGC  3140:}] NGC 1642. Very close to face on with $\itf=14\arcdeg$.
  Nicely defined spiral structure but slightly lopsided.  \oiii\ rotation curve
  rises more steeply than the stellar rotation curve then declines to
  approximately the same value at $\sim1\hr$.  The $\mu_K$ profile breaks to a
  {\it more extended} disk (larger scale length) at $R\sim16\arcsec$.
  \item[{\bf UGC  3701:}] Second lowest surface brightness disk in our sample.
  Rather constant $\slos$ resulting in $\hsz/2\hr = 2.9$, an extreme value for
  our sample (see also UGC~1862).  Some ringing in the $\mu_K$ profile due to
  the spiral arms.  Rotation curves rise slowly.
  \item[{\bf UGC  3997:}] Classified as Im by RC3 with low surface brightness.
  Bright field star just to the west of the nucleus.  Stellar and \oiii\
  rotation curves dip at $\sim1\hr$ where the ``kinematic flaring'' begins.
  Rather low value of $\hsz/2\hr=0.6$ measured.
  \item[{\bf UGC  4036:}] NGC 2441.  Host of Type Ia SN 1995E
  \citep{molaro1995, quinn2006}.  Observations are dominated
  by ring-like structure, likely due to weak bar.  Isovelocity twisting exists
  in the stellar velocity field.  Streaming motions likely affect both the
  stellar and \oiii\ rotation curves.  Very strong \oiii\ emission at the center
  likely indicates an active nucleus.
  \item[{\bf UGC  4107:}] High-quality kinematics.  Well-defined three-arm
  spiral structure.  Type II break in $\mu_K(R)$ at $R\sim20\arcsec$ accompanied
  by a drop in the \oiii\ velocity dispersion.  Bright field star toward the
  north-east.
  \item[{\bf UGC  4256:}] NGC 2532.  Two close companions $\sim4\arcmin$ to the
  north connected by an \hone\ bridge \citep{martinsson2011}.  Interaction has
  likely produced the bright arm toward the east and the lopsidedness of the
  galaxy.  High star-formation rate with very bright \oiii\ emission associated
  with visible star-formation regions.  Stellar and \oiii\ rotation curves have
  dramatically different shapes; the \oiii\ gas is likely effected by streaming
  motions.
  \item[{\bf UGC  4368:}] NGC 2575. Highest inclination in the sample
  ($\itf=45\arcdeg$) with two bright field stars within the PPak field-of-view.
  High-quality stellar kinematics.  Two Type II breaks in $\mu_K(R)$, one at
  $R\sim16\arcsec$ and the other at $R\sim30\arcsec$.
  \item[{\bf UGC  4380:}]  Low-inclination galaxy ($\itf=13.8\arcdeg$) with a
  small apparent size and scale length ($\hr=10\arcsec$).  Stellar kinematic
  data have limited radial extent, and the stellar rotation curve declines
  dramatically at $R>1.5\hr$, likely erroneously.
  \item[{\bf UGC  4458:}]  NGC 2599, Mrk 389.  Earliest morphological type in
  our sample (Sa), with the largest bulge-to-disk ratio ($B/D = 0.72$).  Some
  spiral structure visible at large radii, but very smooth morphology otherwise.
  \hone\ rotation curve declines from 350 \kms\ to 250 \kms\
  \citep{martinsson2011}; \oiii\ rotation curve not well fit by a tanh model
  yielding large error on $r_{\rm s,OIII}$.  Bulge dominates all stellar
  kinematic data; however, we fit an exponential to $\slos$ beyond $R=27\arcsec$
  (the dotted line in the Atlas). We find an outlying value of $\hsz/2\hr=0.4$.
  \oiii\ velocity dispersion declines smoothly over all radii.
  \item[{\bf UGC  4555:}] NGC 2649.  Strong spiral structure affects $\mu_K(R)$.
  Bright field star visible to the north.  High-quality stellar kinematic data.
  Weak \oiii\ emission in the center (resulting in aberrant isovelocity
  contours); very bright emission associated with the arms at larger radii.
  Rather extreme value of $\hsz/2\hr=1.8$.
  \item[{\bf UGC  4622:}] The most distant galaxy in the sample ($\vsys=12830$
  \kms; $D=178$ Mpc).  Stellar kinematics limited to $R<15\arcsec$; \oiii\
  emission is rather extended.  Type II break in $\mu_K(R)$ at $R\sim15\arcsec$.
  Fourth highest bulge-to-disk ratio with $B/D=0.16$.
  \item[{\bf UGC  6903:}] Barred galaxy with rather low surface brightness.
  Poorest quality kinematic data in our sample (only one 3600s observation).
  Stellar-continuum $S/N$ map shows that only fibers in the bar and spiral-arm
  regions have $S/N>1$.  Strong dip in $\mu_K(R)$ at $R\sim20\arcsec$.
  Rather low value of $h_\sigma/2\hr\sim0.5$. Stars rotate faster than the
  \oiii\ gas, violating asymmetric drift equation.  Latter two comments should
  consider the quality of the data.
  \item[{\bf UGC  6918:}] NGC 3982.  High-surface-brightness member of the Ursa
  Major cluster.  Very high-quality kinematic data.  Host of Type Ia SN 1998aq
  \citep{hurst1998, saha2001}.  Classified as a Seyfert~1.9
  \citep{veron2006}.  Very strong \oiii\ emission associated with the nucleus
  and with the spiral arm to the south.  Warped and lopsided extension to the
  \hone\ gas \citep{martinsson2011}; PPak kinematics are regular.  Included in
  DMS pilot sample as presented in early publications
  \citep{verheyen2004,bershady2005,westfall2009}.  Type II break in $\mu_K(R)$
  at $R\sim1\hr$; $\slos$ transitions to a shallower slope at this radius.
  Large value of $\hsz/2\hr = 1.6$.
  \item[{\bf UGC  7244:}] NGC 4195. As with UGC~3091, this barred galaxy has
  been modeled assuming no bulge; inner excess in $\mu_K(R)$ ($R\leq8\arcsec$)
  interpreted as inner disk.  As with UGC~1862, we exclude $R<2\farcs5$ from our
  analysis of $\slos$ (vertical dotted line in Atlas).  Stellar kinematic
  measurements only reach $R\sim20\arcsec$.  Rather high value of $\hsz/2\hr =
  1.7$.
  \item[{\bf UGC  7917:}] NGC 4662.  Weakest \oiii\ emission in our sample,
  particularly within $R<10\arcsec$.  High-quality stellar kinematics.  Excluded
  $\slos$ data at $R<1\hr$ (vertical dotted line in the Atlas) to remove
  bar-associated regions.  Type II break in $\mu_K(R)$ at $R\sim25\arcsec$.
  Fifth highest bulge-to-disk ratio, with $B/D=0.14$.
  \item[{\bf UGC  8196:}]  NGC 4977.  Early-type spiral (SAb).  Third highest
  bulge-to-disk ratio with $B/D = 0.24$.  Low-surface-brightness, extended disk
  with strong spiral structure not probed by our kinematics.  Rather complex
  $\mu_K(R)$, transitions to a shallower slope at $R\sim10\arcsec$, which may be
  an extent of the bulge not accounted for in our bulge-disk decomposition.
  Data within $R<15\arcsec$ is excluded from our analysis of $\slos$.  Central
  velocity dispersion errors are large due to the few data points used in the
  fit.  Weak \oiii\ emission in the disk yields a poor rotation curve.  Strong
  \oiii\ emission in the center suggests an active nucleus.
  \item[{\bf UGC  9177:}]  Well-defined spiral structure affecting $\mu_K(R)$.
  High inclination ($\itf=40\arcdeg$) and high-quality rotation curves for both
  the gas and stars.  Measurements of $\slos$ limited to $R<25\arcsec$, whereas
  \oiii\ dispersions are largely beyond that radius.  The \oiii\ emission
  associated with some bright star-forming regions.
  \item[{\bf UGC  9837:}] High-quality \oiii\ data with many star-forming
  regions producing large EW \oiii\ emission.  Stellar data has a limited radial
  extent.  Weak morphological bar does not affect the kinematics.  Shallower
  slope in $\mu_K(R)$ beyond $\sim1\hr$.  Low value of $\hsz/2\hr=0.5$.
  \item[{\bf UGC  9965:}] IC 1132.  Very nearly face-on ($\itf=12\arcdeg$) with
  strong spiral structure visible in $\mu_K(R)$.  Bulgeless galaxy; as with
  UGC~1862, we exclude data within $R\leq2\farcs5$ in the analysis of $\slos$
  (vertical dotted line in Atlas).  Many \oiii-emitting star-formation regions
  associated with the spiral arms.
  \item[{\bf UGC 11318:}] NGC 6691.  Barred galaxy. Lowest inclination in the
  sample ($\itf=5.8\arcdeg$), yielding a very low-amplitude projected rotation
  curves.  Type II break in $\mu_K(R)$ at $\sim1\hr$ that matches a flattening
  of the $\slos$ profile.  Strong \oiii\ emission near the galaxy center that
  may indicate an active nucleus.
  \item[{\bf UGC 12391:}] NGC 7495.  Stellar and \oiii\ rotation curves dip at
  $\sim1\hr$ that may match a slight rise in $\mu_K(R)$, indicating streaming
  along a spiral arm.  Type II break in $\mu_K(R)$ at $R\sim25\arcsec$.  Bright
  star-forming knots in the extended parts of the disk.
\end{list}


\section{Atlas of Galaxy Kinematics}
\label{app:atlas}

This Atlas presents a collection of the relevant photometric and kinematic data
obtained for the 30 galaxies in the PPak sample, one galaxy per page.  For each
galaxy, the top half of the page displays two-dimensional images, whereas the
bottom half shows the derived radial profiles and contains tables with relevant
parameters.

\subsection{Images}

The two-dimensional images are organized in three rows:  The left-most panel in
the first row shows the optical image of the galaxy extracted from blue POSS-II
plates, illustrating its morphology and giving an impression of its surface
brightness.  SDSS multi-colour images are provided in
\citetalias{bershady2010a}.  The elongated panel to the right of the first row
shows the extracted, wavelength-calibrated, sky-subtracted and combined(+merged)
galaxy spectra organized along pixel rows.  Most images consist of 331 spectra
(331 rows); however, five galaxies have more than 331 spectra due to spatial
offsets between the different combined pointings (see Sect.~\ref{sec:merging}
and Table~\ref{tab:UGC_merge}). The spectra have been reordered in radial distance
from the galaxy center. Toward the blue (left) spectral regions in this
panel, the \oiii$\lambda$5007\AA\ emission line (position indicated above the panel)
is readily visible, as well as the \oiii$\lambda$4961\AA\ line for most galaxies
with recession velocities larger than 1500 \kms.  For distant galaxies (e.g.,
UGC~1908), the H$\beta$ line has also been redshifted into the observed spectral
range.  Toward the red (right) spectral regions, various stellar absorption lines
are visible including the MgIb triplet (position indicated above the panel) and
several Fe lines.

In the following two rows, we provide images based on information extracted from
the PPak fibers reconstructed in two dimensions.  The second row provides
information based on the stellar continuum, while the third row is for the
\oiii$\lambda$5007\AA\ emission line.  Each map is 90\arcsec\ on a side,
centered on the galaxy, and the small white cross inside a slightly larger black
cross indicates the morphological center of the galaxy. From left to right, the
panels provide:
\begin{enumerate}
\item {\bf Reconstructed intensity maps} of the stellar continuum (second row)
and the total flux in the \oiii$\lambda$5007\AA\ emission line (third row).  The
procedure used to create these images, including the interpolation scheme used
to fill interstitial regions is discussed in Sect.~\ref{sec:ContMaps}.  The
FWHM of the Gaussian kernel used to smooth the image (1.5 fiber diameter or
$\sim4\farcs0$) is shown as a hatched circle in the lower-right corner of the
panel.  These reconstructed continuum maps reproduce, in detail, the
morphologies of the galaxies as seen in the POSS II images provided in the first
row. The \oiii\ intensity maps are much patchier, concentrated toward the spiral
arms, and have holes in areas where no \oiii\ emission was detected.
\item {\bf Observed velocity fields} with data points weighted by their errors
and smoothed with a Gaussian kernel with FWHM of $\sim8\farcs0$.  The method
used to generate these maps is the same as described in Appendix A of
\citetalias{westfall2011b}.  Isovelocity contours are plotted to highlight the
structure in each velocity field; the contour levels are given in the table at
the bottom right of each page.  The thick black contour indicates the systemic
velocity while the thin black and white contours show the receding and
approaching sides of the disk, respectively.
\item {\bf Model velocity fields} based on the fitted tanh-model for the
rotation-curve shape as described in Sect.~\ref{sec:orientation}.  Although
the model provides a velocity for all fibers, only those fibers with measured
velocities are included.  These model velocities are smoothed in exactly the
same way as the observational data, except that all data have the same weight.
As discussed in Sect.~\ref{sec:tanhRC}, non-axisymmetric features in the model
velocity fields are the result of irregular sampling due to missing fibers
without observed data and the interpolation scheme.  Consequently, deviations
from axisymmetry occur mainly at the edges of the field-of-view for the stellar
data and, for the same reason, near the galaxy centers for the \oiii\ velocity
fields.  The isovelocity contour levels are the same for the model and observed
velocity fields.
\item {\bf Residual velocity fields} calculated simply as the observed minus the
modeled velocity fields. The contour levels are listed in the table, where black
contours indicate positive residuals and white contours negative residuals.
\item {\bf Line-of-sight velocity dispersion maps}, also interpolated and
smoothed using a Gaussian kernel with a FWHM of $\sim8\arcsec$, where the data
points with errors on $\slos$ exceeding 8 \kms\ have been rejected and the
remaining data is weighted by the errors. Contours are drawn in steps of 8 \kms;
the lowest contour (thick line) is at 16 \kms, which is similar to the
instrumental spectral dispersion.
\item {\bf Signal-to-noise maps} that use no interpolation to fill interstitial
regions; any point lying within a fiber aperture is given the value of the $S/N$
of the spectrum from that fiber.  In the second row, the plotted values are the
average $S/N$ in the spectral continuum; for the third row, they are the ratio
of the \oiii\ intensity to its error. No values are plotted for spectra that had
measurements with $S/N < 1$.
\end{enumerate}

\subsection{Radial profiles}

The bottom half of each page presents various quantities measured for each
galaxy as a function of in-plane radius.  These radii are determined using the
geometric projection parameters derived in Sect.~\ref{sec:orientation}.

Above the tables to the right, we provide the $K$-band surface-brightness
profile discussed in Sect.~\ref{sec:2MASSphot}.  The dotted line indicates the
result from the iterative fit of an exponential disk; the line is solid within
the converged radial fitting range of 1--4$\hr$, where 1$\hr$ is indicated by an
arrow.  The corresponding photometric scale lengths and central surface
brightness are considered as representative for the disk.  The curved solid line
in the inner region represents the best-fitting bulge profile, a Gaussian
smoothed S{\'e}rsic profile as described in Sect.~\ref{sec:2MASSphot}.  The
dashed vertical line indicates the radius at which the bulge contributes 10\% to
the azimuthally-averaged surface brightness ($\rbulge$).

To the left, we provide four groups of two panels.  Clockwise, starting from the
upper left, these groups provide:
\begin{enumerate}
\item {\bf Continuum and \oiii\ intensities}:  The grey points show the
intensities in individual spectra, while the black points show the
weighted-average data in radial bins.  The vertical dashed line marks $\rbulge$.
The scatter in the grey points at a certain radius indicates the azimuthal
surface-brightness fluctuations at that radius.  The very high \oiii\
intensities at the centers of some galaxies indicate the presence of an active
galactic nucleus (see, e.g., UGC~1908, UGC~4036 and UGC~6918).
\item {\bf Position-velocity diagrams} of the stellar and \oiii\ gas recession
velocities for data within $\pm$45$^{\circ}$ of the kinematic major axis; the
line-of-sight velocities have been corrected for the in-plane azimuth to match
the line-of-sight velocities along the major axis.  The solid and dashed curves
indicate the hyperbolic-tangent function fitted to the shape of the rotation
curve for the stars and \oiii\ gas, respectively; see Sect.~\ref{sec:tanhRC}.
The black points indicate the velocities as derived from the tilted-ring fitting
(Sect.~\ref{sec:tiltedrings}).  Both the parameterized rotation curve and the
tilted-ring fits are based on data from both the approaching and receding sides
of the galaxy.  Thus, the derived rotation curves are symmetric around $R=0$,
even though the data may show kinematic asymmetries.
\item {\bf Line-of-sight velocity dispersions} ($\slos$) of the stars and \oiii\
gas.  Grey error bars correspond to individual measurements while black filled
(stars) and open (\oiii) circles represent weighted-averages in radial bins
(Sect.~\ref{sec:tiltedrings}). The vertical dashed line indicates $\rbulge$.
The black solid line represents the exponential fit to $\slos$ of the stars
described in Sect.~\ref{sec:expsigma}; this fit excluded $\slos$ measurements
with errors larger than 8~\kms\ (also excluded from the Figure) and data within
$\rbulge$.
\item {\bf Projected rotation curves and velocity-dispersion profiles} for both
the stars (solid points) and \oiii\ gas (open circles) in the same panel.  The
solid lines indicate the best-fitting parameterized models for the stars; the
best-fitting model rotation curve for the \oiii\ gas is shown as a dashed line.
These curves are the same as those provided in the second and third panel
groups.
\end{enumerate}

\subsection{Tables}

In the lower right corner of each Atlas page, we provide two tables.  The upper
table lists the parameters of the derived orientation and kinematics of the
galaxy, including the coordinates ($\alpha$ and $\delta$;
Sect.~\ref{sec:center}) of the morphological center, the systemic velocity and
position angle of the receding side of the kinematic major axis ($\vsys$ and
$\pa$; Sect.~\ref{sec:PAVsys}), the Tully-Fisher-based inclination of the
galaxy disk ($\itf$; Sect.~\ref{sec:Incl}), the asymptotic rotational velocity
and the scale radius of the fitted parameterized rotation curve (stellar:
$V_{\rm arot,star}$ and $r_{\rm s,star}$; \oiii\ gas: $V_{\rm arot,OIII}$ and
$r_{\rm s,OIII}$); Sect.~\ref{sec:ArotAsca}), and the fitted central
line-of-sight velocity dispersion and
dispersion scale length of the stellar disk ($\sigma_{\rm 0,*}$ and $\hs {_*}$;
Sect.~\ref{sec:expsigma}).  The optical radial scale length ($\hr$) and the
radius ($\rbulge$) at which the bulge contributes 10\% to the surface brightness
is tabulated in the bottom.  The second table provides the contour levels and
intervals for the observed, modeled and residual velocity fields, as well as the
velocity-dispersion maps.

\onecolumn
\newpage
\section{The PPak Fiber Position Table}
\label{sec:PPakPosTab}
\begin{table*}[!h]
\setlength{\tabcolsep}{4pt}
\renewcommand{\arraystretch}{0.94}
\centering
{\tiny
\begin{tabular}{|r r r||r r r||r r r||r r r||r r r|}
\hline
Fiber  &  X  &  Y  & Fiber  &  X  &  Y  & Fiber  &  X  &  Y  & Fiber  &  X  &  Y  & Fiber  &  X  &  Y  \\
\hline
 1 & -14.31 &  0.16 &  75 &   0.15 & 24.93 & 149 &   1.87 &   9.32 & 223 &   3.27 & -24.72 & 297 &  -3.71 & -12.39 \\
 2 &  16.31 &  3.24 &  76 & -14.24 & 25.15 & 150 &  -8.92 &   3.16 & 224 & -23.43 & -21.46 & 298 &   1.54 & -21.64 \\
 3 &   0.10 & 12.46 &  77 &  18.09 & 25.15 & 151 &   5.22 &  -9.29 & 225 &  28.61 &   0.28 & 299 & -21.52 &  -5.89 \\
 4 & -14.23 & 12.64 &  78 & -25.06 & 12.87 & 152 &  -3.70 &  -6.21 & 226 & -14.55 & -30.90 & 300 & -10.91 & -18.51 \\
 5 &  14.71 & 12.58 &  79 &  -5.22 & 28.05 & 153 &   5.37 &   3.09 & 227 & -26.96 &  -2.67 & 301 &  17.89 &  -0.01 \\
 6 & -10.73 &  6.37 &  80 &  21.85 & 12.79 & 154 &  -3.53 &   6.26 & 228 &  19.41 & -15.50 & 302 &   8.77 & -15.52 \\
 7 &   7.34 & 18.74 &  81 & -28.58 &  0.40 & 155 &  -9.00 &  -3.02 & 229 &  12.39 & -27.82 & 303 &  21.29 &  -6.19 \\
 8 &  -6.99 & 18.75 &  82 &   7.31 & 24.99 & 156 &   5.46 &   9.35 & 230 &  31.99 &  -5.95 & 304 & -12.64 &  -9.16 \\
 9 & -17.88 &  6.42 &  83 &  32.45 &  6.83 & 157 &   8.87 &  -3.13 & 231 & -18.05 & -18.41 & 305 &  10.62 &  -6.22 \\
10 &   9.00 &  9.30 &  84 & -19.51 & 22.18 & 158 &  -5.29 &   9.40 & 232 &  -7.37 & -30.92 & 306 &  -0.20 & -18.59 \\
11 &  -7.08 & 12.46 &  85 & -28.62 &  6.60 & 159 &  -1.93 &  -9.31 & 233 & -30.55 &  -8.83 & 307 & -12.74 & -15.38 \\
12 &   0.14 & 18.77 &  86 &  21.69 & 19.00 & 160 &   3.61 &   6.22 & 234 &  -3.83 & -24.70 & 308 &  12.26 & -15.49 \\
13 &  18.16 &  6.37 &  87 &   3.76 & 31.17 & 161 & -10.70 &   0.09 & 235 &  24.92 & -18.54 & 309 &  -7.39 & -18.48 \\
14 & -23.23 &  3.41 &  88 & -12.38 & 28.38 & 162 &  -7.23 &  -6.13 & 236 &  17.73 & -30.92 & 310 &   3.39 & -12.41 \\
15 &   7.20 & 12.38 &  89 & -23.25 & 15.98 & 163 &   3.55 &  -0.04 & 237 & -19.93 & -27.71 & 311 &  17.71 &  -6.26 \\
16 & -12.39 & 15.69 &  90 &  14.49 & 25.05 & {\bf164} &   {\bf0.00} &   {\bf0.00} & 238 &   6.90 & -30.84 & 312 & -10.81 &  -6.04 \\
17 &  12.57 &  3.06 &  91 &  26.99 &  9.76 & 165 &  -3.60 &  -0.01 & 239 & -25.15 &  -5.89 & 313 &  14.07 & -12.41 \\
18 &  -1.64 & 15.59 &  92 & -35.75 &  0.53 & 166 &   1.70 &  -3.06 & 240 &  26.70 &  -9.21 & 314 &  -7.26 & -12.37 \\
19 &  12.68 & 15.67 &  93 &  25.28 & 19.19 & 167 &   1.77 &   3.12 & 241 & -12.70 & -27.81 & 315 &  10.54 & -18.58 \\
20 & -21.41 &  0.32 &  94 & -19.55 & 28.52 & 168 &  -1.79 &  -3.07 & 242 &  17.77 & -24.71 & 316 &   3.34 & -18.55 \\
21 &  23.53 &  3.35 &  95 &   3.76 & 24.91 & 169 &  -1.81 &   3.11 & 243 & -25.22 & -12.05 & 317 &  14.18 &  -6.28 \\
22 & -12.39 &  9.53 &  96 &  30.77 &  3.62 & 170 &   6.99 &  -6.24 & 244 &   6.92 & -24.73 & 318 &  -1.99 & -15.49 \\
23 &  12.77 & 21.85 &  97 &  18.06 & 31.61 & 171 &   7.22 &   6.16 & 245 &  26.71 &  -2.97 & 319 & -17.95 &  -6.00 \\ 
24 &  -5.21 & 15.62 &  98 & -23.18 &  9.72 & 172 &  -5.50 &  -9.28 & 246 &  -2.03 & -27.80 & 320 &   7.01 & -12.44 \\
25 & -16.02 & 15.80 &  99 &  -1.65 & 28.02 & 173 &  -5.31 &   3.12 & 247 &  21.22 & -12.40 & 321 &  -9.00 &  -9.22 \\
26 &  12.64 &  9.39 & 100 & -24.97 & 19.20 & 174 &   7.11 &  -0.03 & 248 & -16.35 & -21.56 & 322 &   8.78 &  -9.33 \\
27 & -16.09 &  3.26 & 101 &  28.98 & 13.04 & 175 &  -1.72 &   9.42 & 249 & -28.72 &  -5.74 & 323 &  -5.60 & -15.47 \\
28 &   5.51 & 15.62 & 102 & -15.93 & 22.02 & 176 &  -0.09 &  -6.19 & 250 &  15.86 & -21.65 & 324 & -16.16 &  -2.88 \\
29 &  -1.61 & 21.83 & 103 &  19.87 & 22.06 & 177 &  -5.45 &  -3.02 & 251 &  -7.41 & -24.70 & 325 &  14.27 &  -0.06 \\
30 &  19.90 &  9.51 & 104 & -14.16 & 31.54 & 178 &   5.28 &  -3.12 & 252 & -19.91 & -15.23 & 326 &   5.15 & -15.55 \\
31 & -19.62 &  9.64 & 105 &  12.71 & 28.17 & 179 &   0.06 &   6.26 & 253 &  28.41 &  -6.05 & 327 &  -0.17 & -12.37 \\
32 &  10.81 &  6.21 & 106 &  27.19 &  3.56 & 180 &   1.63 &  -9.32 & 254 &   8.75 & -27.80 & 328 & -10.93 & -12.28 \\
33 &   5.56 & 21.75 & 107 & -32.22 &  0.46 & 181 &  -7.12 &   6.26 & 255 & -14.44 & -24.68 & 329 &  12.37 &  -9.36 \\
34 &  -3.50 & 12.47 & 108 &  -6.99 & 31.48 & 182 &   8.94 &   3.09 & 256 &  23.04 & -15.49 & 330 &  12.43 &  -3.09 \\
35 &  -8.77 & 21.87 & 109 & -30.33 &  9.91 & 183 &  -7.15 &   0.05 & 257 &  -0.28 & -24.72 & 331 & -12.60 &  -3.02 \\
36 &   3.61 & 12.42 & 110 &  23.55 & 16.01 & 184 &   3.44 &  -6.25 & 258 & -23.41 &  -8.95 &     &        &        \\
37 &  16.38 & 15.77 & 111 &  21.65 & 25.40 & 185 &  35.76 &   0.55 & 259 &  15.94 & -27.79 &     &        &        \\
38 & -12.50 &  3.20 & 112 & -17.76 & 18.95 & 186 & -10.99 & -30.89 & 260 &  23.10 &  -9.23 &     &        &        \\
39 & -14.14 & 18.92 & 113 &  -3.43 & 24.93 & 187 &  21.34 & -24.72 & 261 & -21.70 & -18.37 & 501 &  -0.69 &  74.92 \\
40 &  10.82 & 12.36 & 114 &  10.89 & 31.27 & 188 & -23.45 & -15.17 & 262 &  -9.20 & -27.82 & 502 &  63.19 & -31.30 \\
41 & -19.64 &  3.35 & 115 & -26.78 &  3.48 & 189 &  24.83 & -12.33 & 263 &  10.51 & -24.72 & 503 & -63.17 & -31.45 \\
42 &  14.39 &  6.21 & 116 &  34.43 &  3.83 & 190 &   3.38 & -30.92 & 264 &  17.64 & -18.58 & 504 &  60.68 &  41.19 \\
43 & -10.61 & 12.57 & 117 &   1.96 & 28.06 & 191 & -21.80 & -24.58 & 265 &   1.56 & -27.82 & 505 &  -1.27 & -67.61 \\
44 &   3.73 & 18.68 & 118 & -17.76 & 25.31 & 192 &  30.29 &  -2.80 & 266 &  25.01 &   0.15 & 506 & -63.12 &  38.40 \\
45 &  18.27 & 12.66 & 119 &  23.46 &  9.65 & 193 & -32.28 &  -5.75 & 267 & -19.79 &  -9.04 & 507 &   2.69 &  68.59 \\
46 & -12.30 & 21.90 & 120 & -26.83 & 16.04 & 194 &  14.15 & -30.88 & 268 &  -9.16 & -21.67 & 508 &  66.09 & -37.84 \\
47 & -21.39 &  6.51 & 121 &  -8.77 & 28.39 & 195 &  -5.63 & -27.81 & 269 &   5.14 & -21.66 & 509 & -58.75 & -37.90 \\
48 &  10.88 & 18.79 & 122 &  19.96 & 15.86 & 196 &  21.27 & -18.61 & 270 &  17.67 & -12.42 & 510 &  64.51 &  34.89 \\
49 &  -8.82 & 15.66 & 123 &  16.22 & 28.16 & 197 & -16.28 & -27.80 & 271 & -16.29 & -15.30 & 511 &   2.53 & -73.74 \\
50 &  19.96 &  3.28 & 124 & -21.47 & 12.86 & 198 &  30.24 &  -9.12 & 272 &  19.59 &  -3.10 & 512 & -59.05 &  32.33 \\
51 &   2.00 & 15.59 & 125 & -32.17 &  6.73 & 199 & -27.09 & -15.14 & 273 & -23.33 &  -2.75 & 513 &   2.84 &  75.35 \\
52 &  -5.19 & 21.84 & 126 & -21.35 & 19.08 & 200 &  19.55 & -27.75 & 274 &  -3.77 & -18.60 & 514 &  66.75 & -31.56 \\
53 & -16.03 &  9.57 & 127 &  28.85 &  6.71 & 201 &   5.05 & -27.78 & 275 &  14.15 & -18.57 & 515 & -59.52 & -31.23 \\
54 &  14.52 & 18.82 & 128 & -15.96 & 28.47 & 202 & -34.06 &  -2.62 & 276 & -12.69 & -21.56 & 516 &  64.26 &  41.48 \\
55 &  16.23 &  9.40 & 129 &  23.48 & 22.26 & 203 &  26.61 & -15.38 & 277 &   6.97 & -18.56 & 517 &   2.32 & -67.25 \\
56 & -17.81 &  0.24 & 130 &   5.53 & 28.09 & 204 & -26.95 &  -8.95 & 278 & -16.28 &  -9.08 & 518 & -59.38 &  38.67 \\
57 &   1.96 & 21.76 & 131 & -26.73 &  9.76 & 205 &  33.85 &  -2.78 & 279 &  19.53 &  -9.33 & 519 &  -0.91 &  68.61 \\
58 & -10.58 & 18.80 & 132 & -21.39 & 25.45 & 206 &  -3.83 & -30.94 & 280 &  -2.05 & -21.68 & 520 &  62.50 & -37.56 \\
59 &  21.70 &  6.43 & 133 &  -3.37 & 31.15 & 207 & -19.92 & -21.53 & 281 &  -9.14 & -15.40 & 522 &  60.83 &  34.57 \\
60 &   9.11 & 15.57 & 134 &  30.67 &  9.92 & 208 & -18.10 & -30.90 & 282 &  21.48 &   0.04 & 521 & -62.41 & -37.76 \\
61 &  -8.87 &  9.50 & 135 & -33.99 &  3.60 & 209 &  10.53 & -30.89 & 283 & -19.66 &  -2.85 & 523 & -64.56 & -34.86 \\
62 & -25.00 &  0.37 & 136 & -10.58 & 25.18 & 210 &  19.47 & -21.61 & 284 &  12.35 & -21.65 & 524 & -62.77 &  32.12 \\
63 &  -3.42 & 18.74 & 137 &  19.82 & 28.50 & 211 & -30.47 &  -2.65 & 285 & -18.08 & -12.16 & 525 &   4.50 &  71.97 \\
64 & -17.86 & 12.72 & 138 & -28.66 & 12.97 & 212 & -25.21 & -18.30 & 286 &  10.56 & -12.36 & 526 &  68.09 & -34.87 \\
65 &   9.16 & 21.91 & 139 &   7.34 & 31.22 & 213 &  28.44 & -12.30 & 287 &   1.64 & -15.46 & 527 & -57.39 & -34.55 \\
66 & -14.29 &  6.44 & 140 & -23.09 & 22.25 & 214 & -28.81 & -11.99 & 288 &  15.99 &  -3.17 & 528 &  66.09 &  38.42 \\
67 &  25.33 &  6.57 & 141 & -17.67 & 31.62 & 215 & -18.09 & -24.64 & 289 & -14.47 & -18.46 & 529 &   4.13 & -70.30 \\
68 &  -6.94 & 24.92 & 142 &  10.93 & 25.03 & 216 &  32.17 &   0.37 & 290 & -14.38 &  -6.09 & 530 & -57.42 &  35.56 \\
69 & -24.97 &  6.56 & 143 &  27.12 & 16.09 & 217 &  -0.19 & -30.92 & 291 &  15.93 & -15.50 & 531 &  -2.65 &  71.77 \\
70 &  18.14 & 18.89 & 144 &   0.15 & 31.16 & 218 &  14.17 & -24.72 & 292 &  -5.55 & -21.61 & 532 &  60.95 & -34.37 \\
71 &   9.08 & 28.04 & 145 &  16.36 & 21.96 & 219 &  24.84 &  -6.13 & 293 &  23.18 &  -3.02 & 533 &  -1.08 & -73.99 \\
72 & -19.62 & 15.92 & 146 &  14.47 & 31.33 & 220 & -10.95 & -24.69 & 294 &   8.75 & -21.63 & 534 &  58.81 &  37.76 \\
73 & -30.39 &  3.52 & 147 & -10.55 & 31.56 & 221 &  23.08 & -21.53 & 295 & -14.46 & -12.28 & 535 &  -2.96 & -70.77 \\
74 &  25.40 & 12.91 & 148 &  10.73 & -0.05 & 222 & -21.64 & -12.12 & 296 &  15.96 &  -9.36 & 536 & -64.74 &  35.14 \\
\hline
\end{tabular}
}
\caption{\tiny On-sky positions (in arcsec) of the 331 science fibers (1-331) and 36 sky 
fibers (501-536), relative to the central fiber \#164. Positive X values represent an 
offset towards the west and positive Y values represent an offset towards the south.}
\label{tab:FibPos}
\end{table*}



\clearpage
\onecolumn

\section{The Atlas}
\label{app:TheAtlas}

 \begin{figure}[!b]
 \centering
 \includegraphics[width=0.95\textwidth]{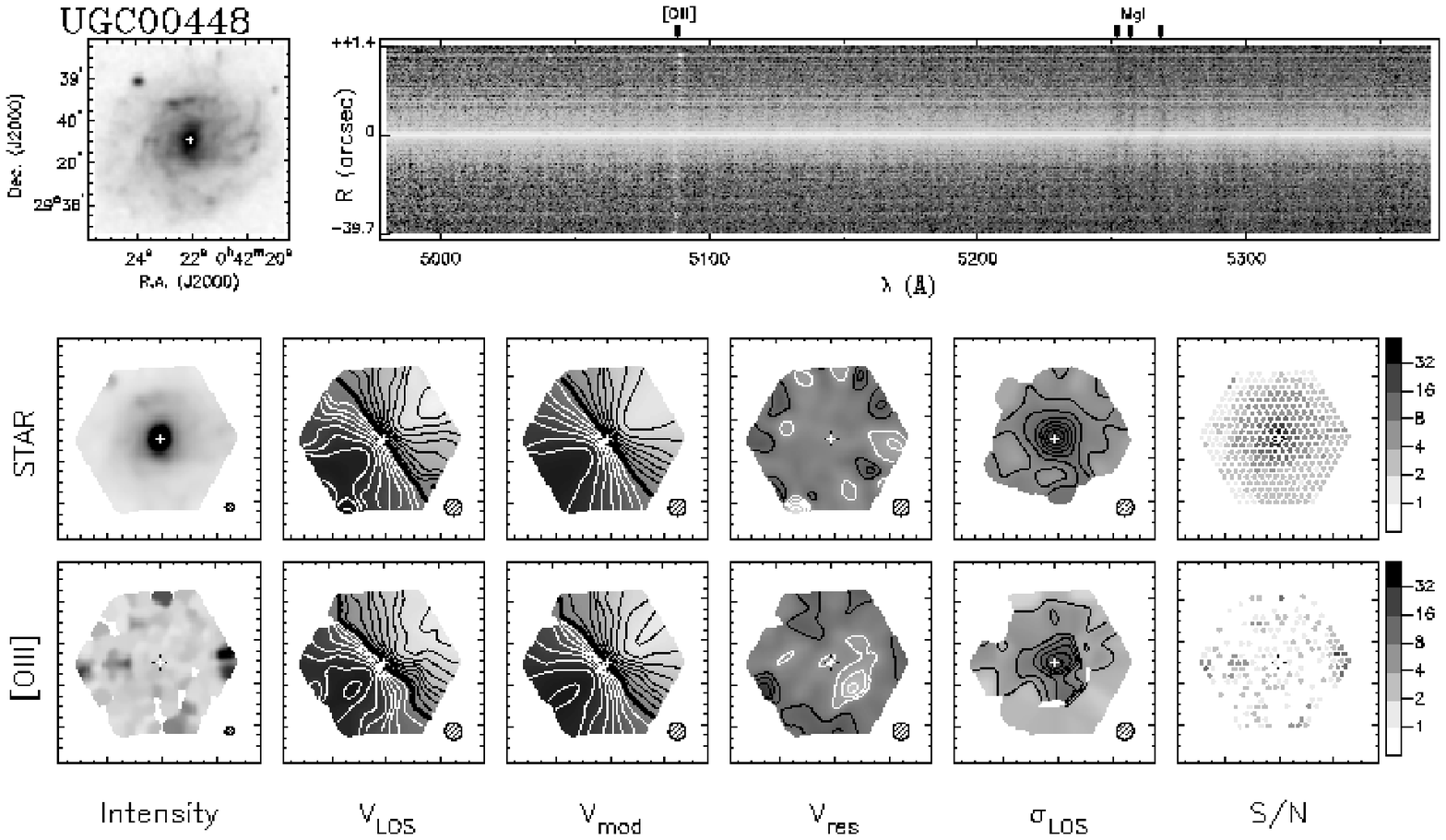}
 \end{figure}

 \begin{figure}[!b]
 \centering
 \includegraphics[width=0.95\textwidth]{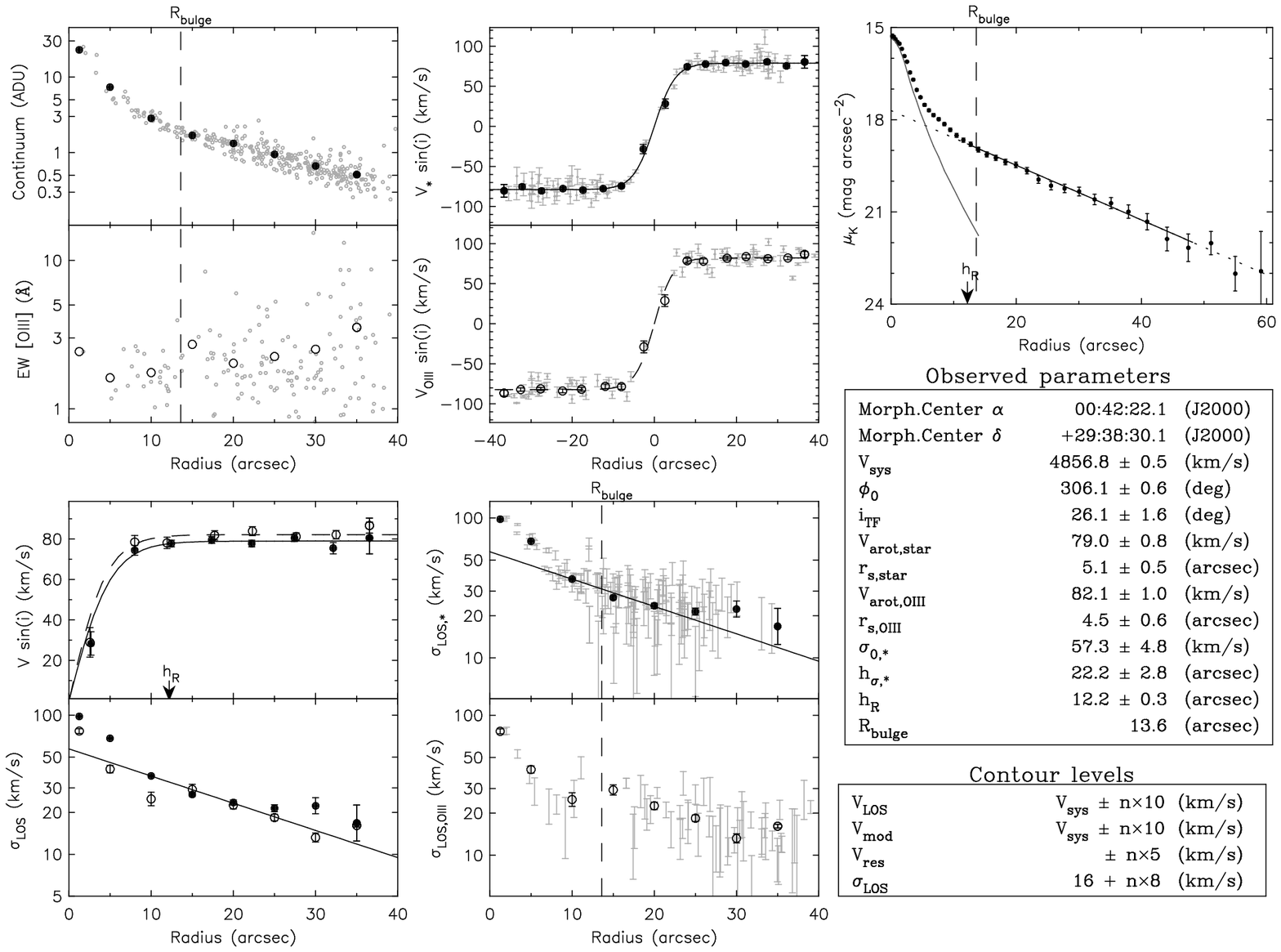}
 \end{figure}

\clearpage

 \begin{figure}
 \centering
 \includegraphics[width=0.95\textwidth]{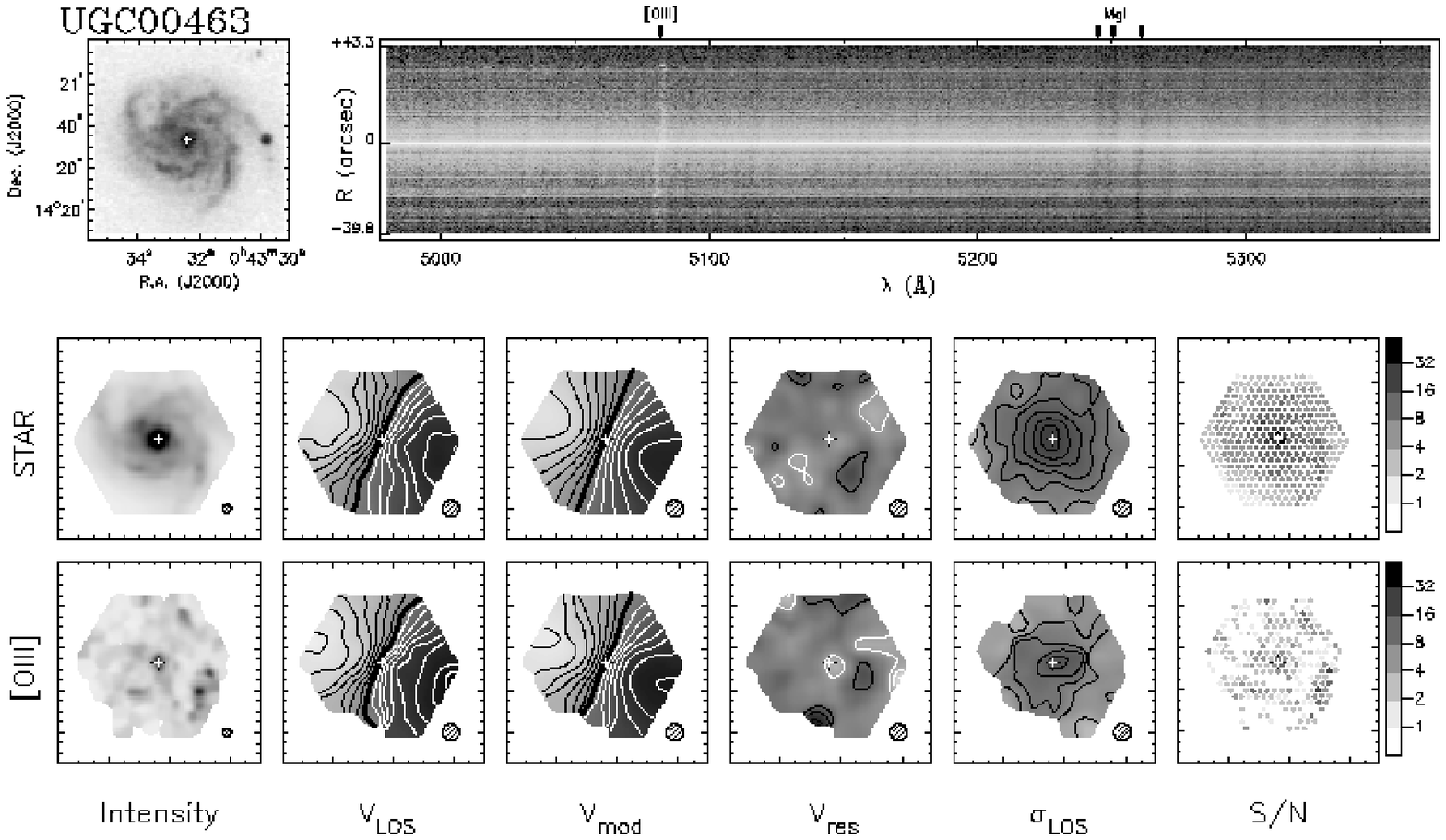}
 \end{figure}

 \begin{figure}
 \centering
 \includegraphics[width=0.95\textwidth]{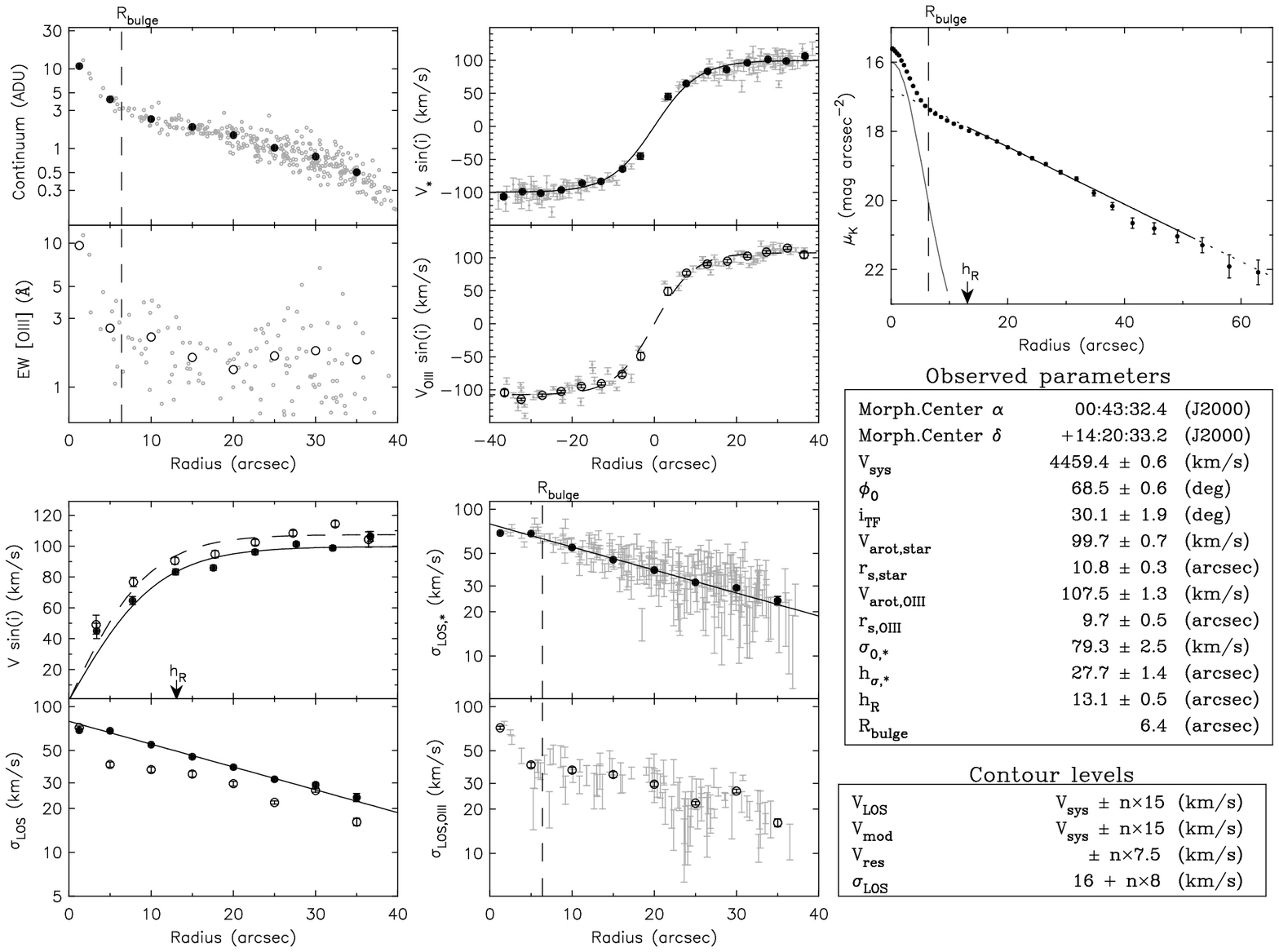}
 \end{figure}

\clearpage

 \begin{figure}
 \centering
 \includegraphics[width=0.95\textwidth]{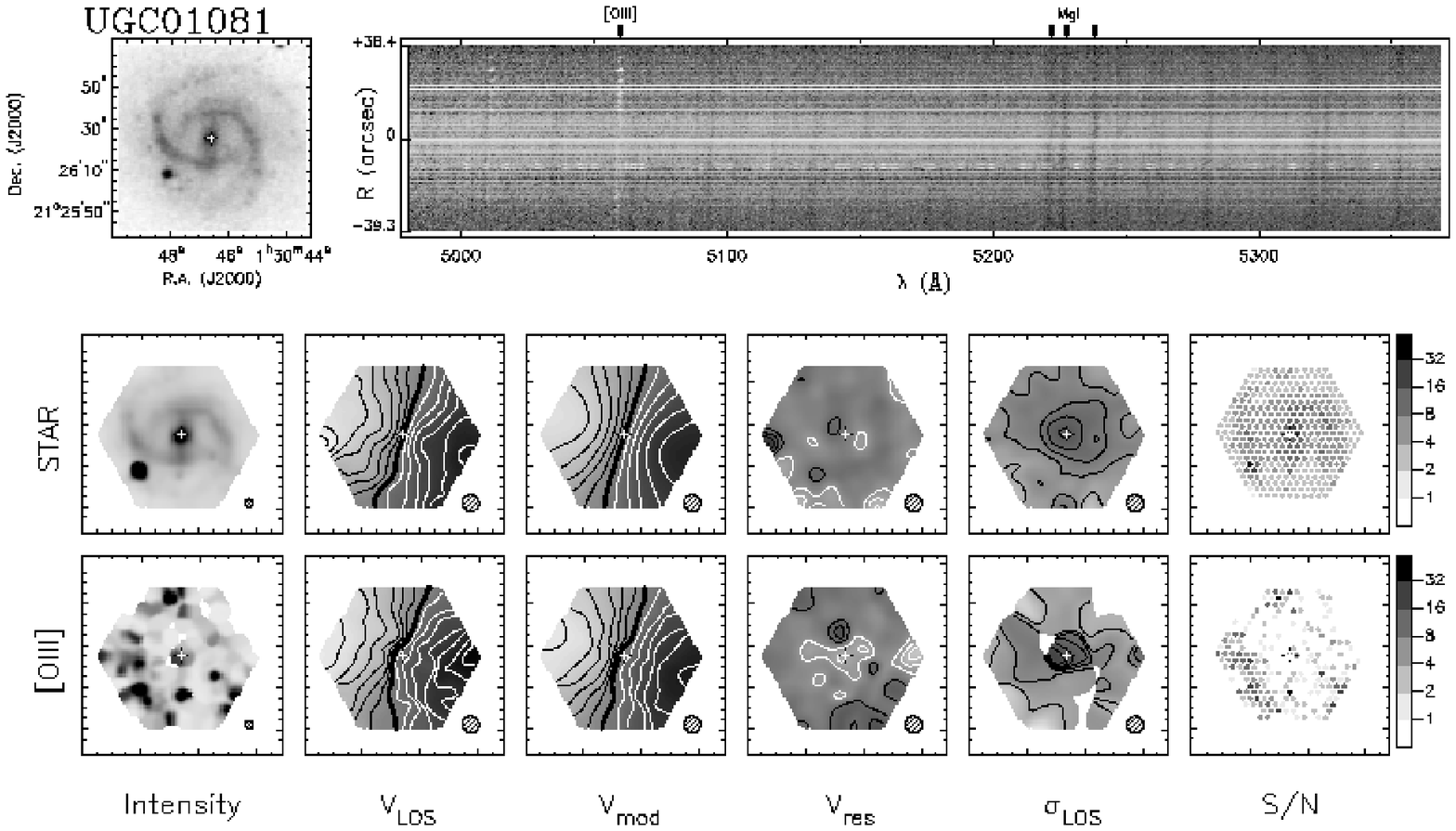}
 \end{figure}

 \begin{figure}
 \centering
 \includegraphics[width=0.95\textwidth]{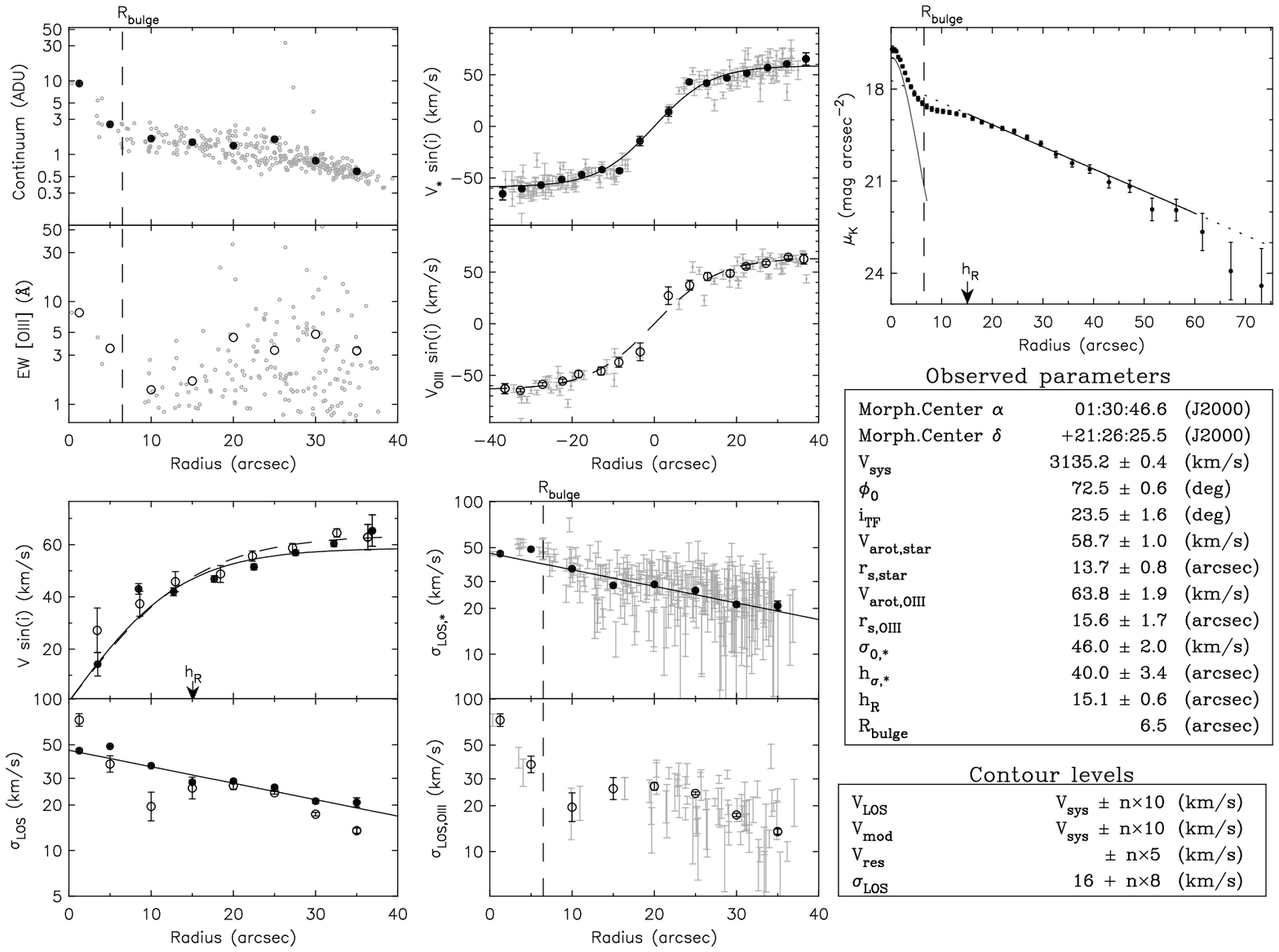}
 \end{figure}

\clearpage

 \begin{figure}
 \centering
 \includegraphics[width=0.95\textwidth]{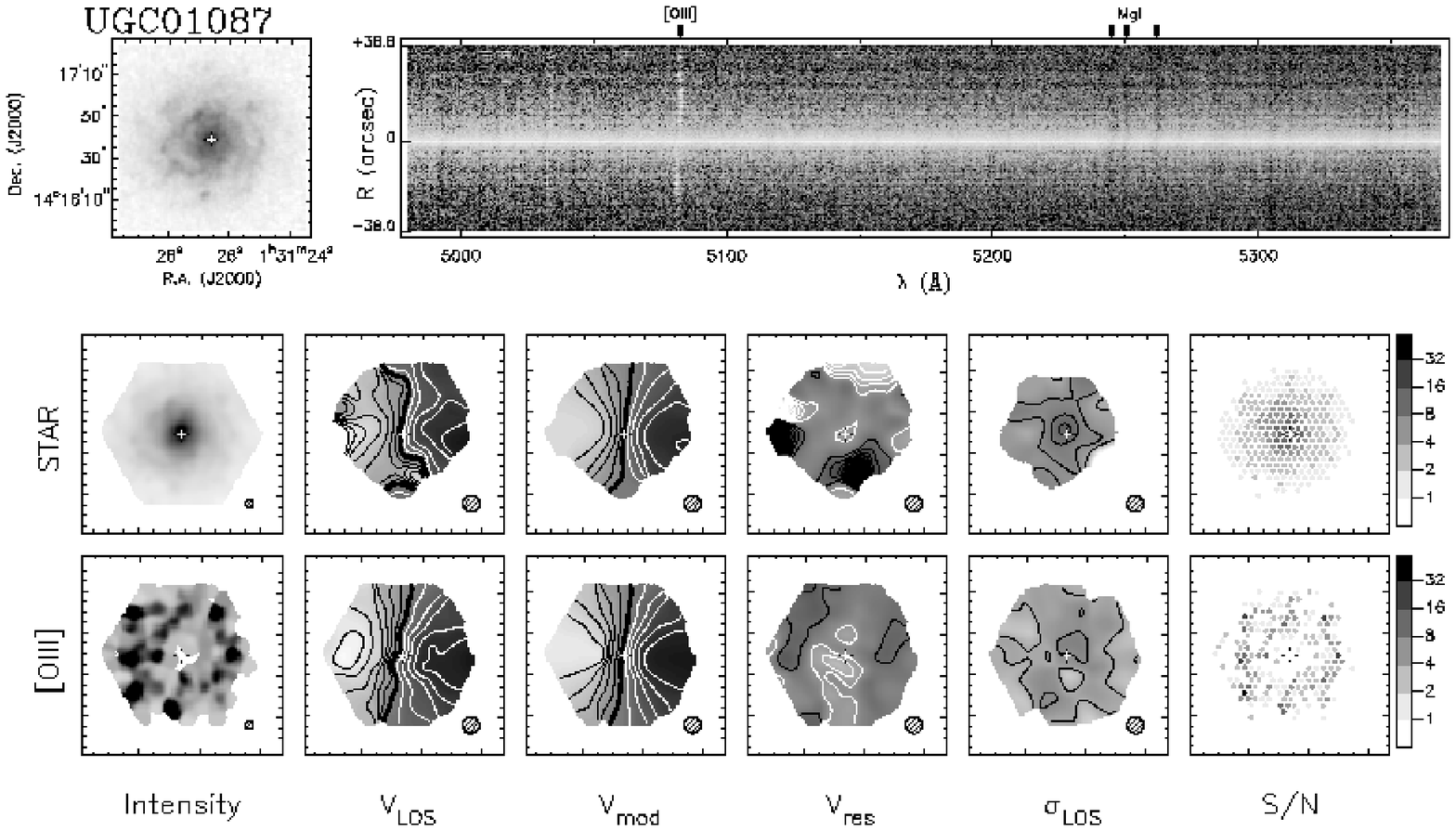}
 \end{figure}

 \begin{figure}
 \centering
 \includegraphics[width=0.95\textwidth]{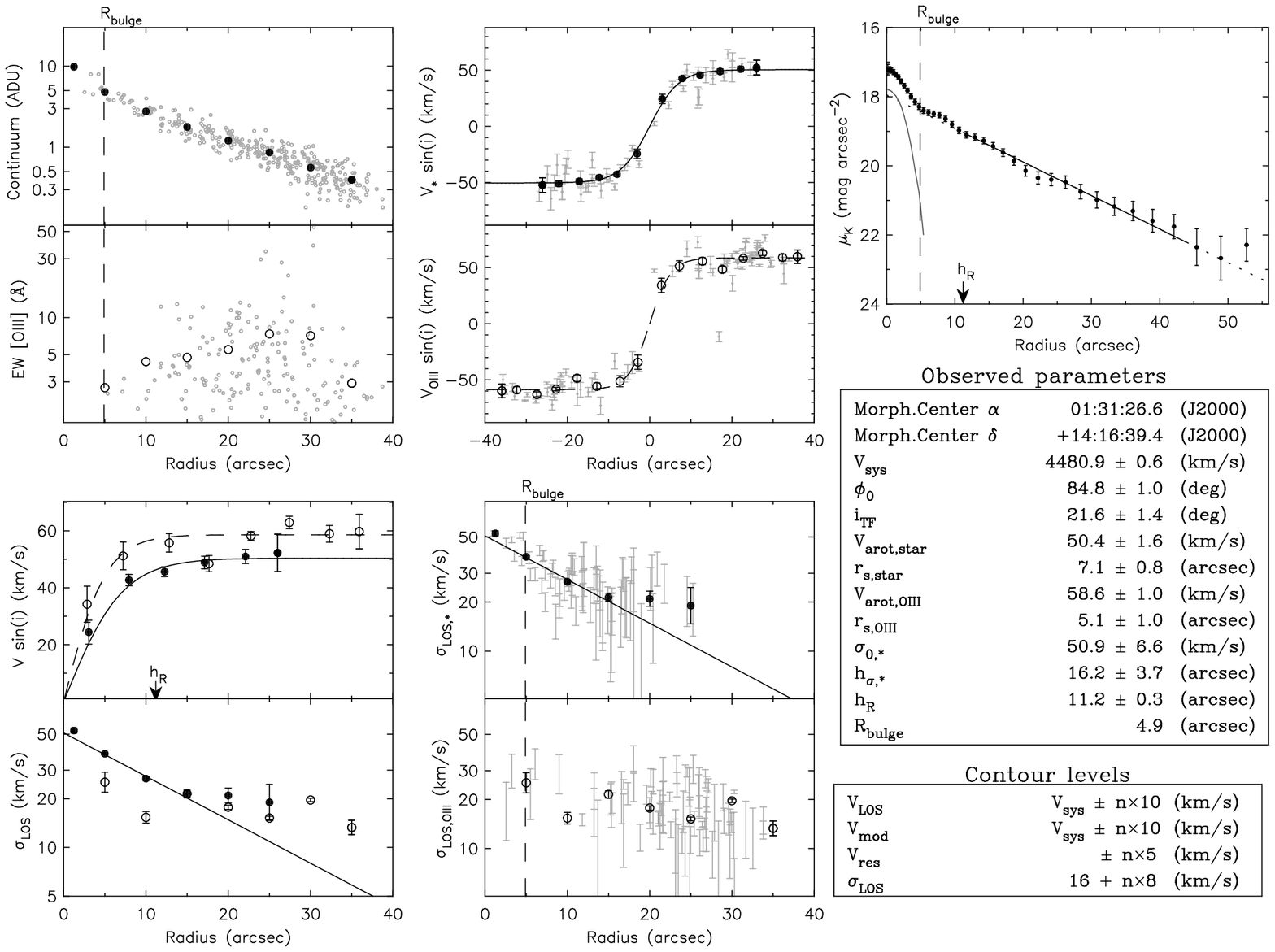}
 \end{figure}

\clearpage

 \begin{figure}
 \centering
 \includegraphics[width=0.95\textwidth]{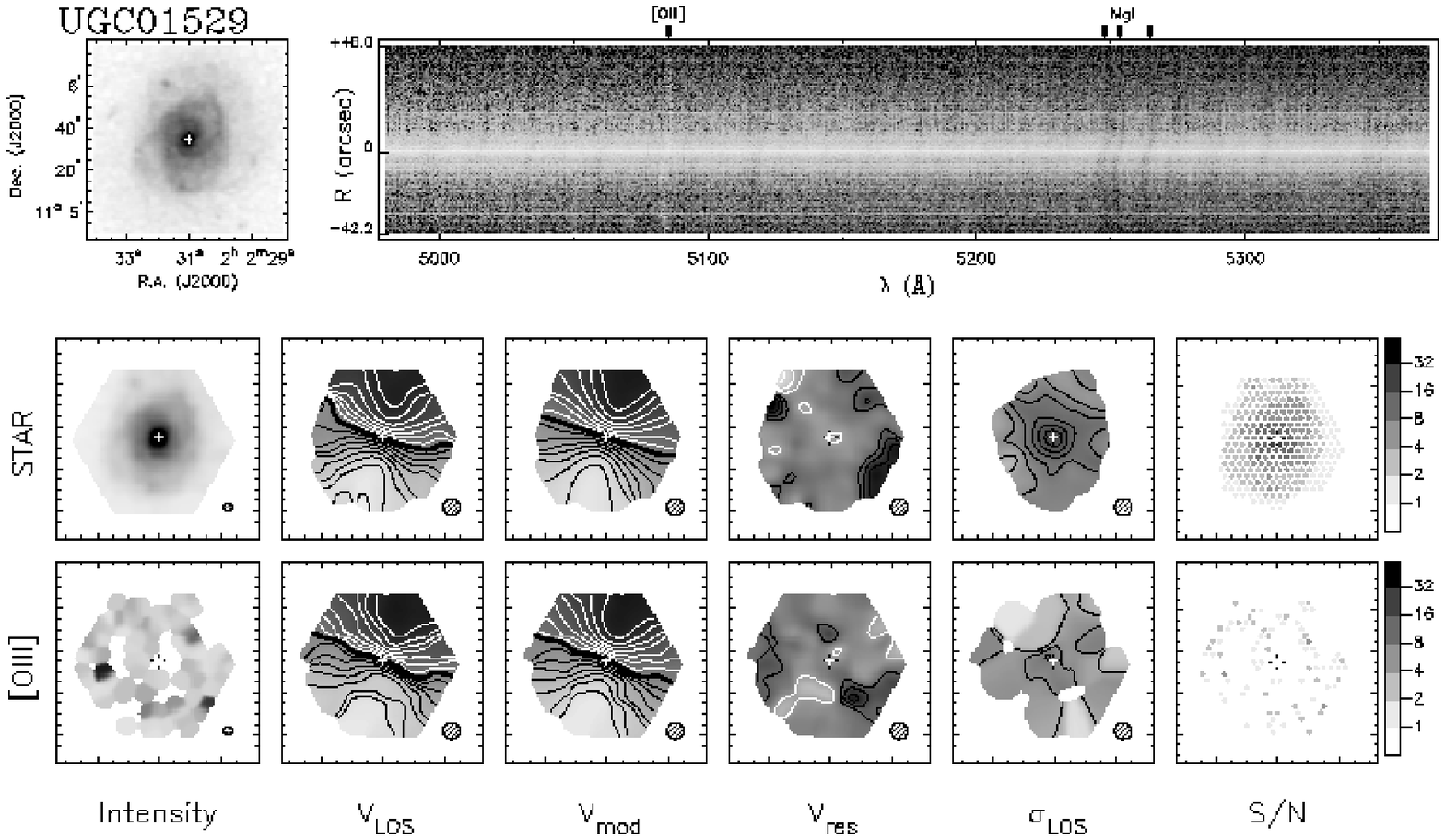}
 \end{figure}

 \begin{figure}
 \centering
 \includegraphics[width=0.95\textwidth]{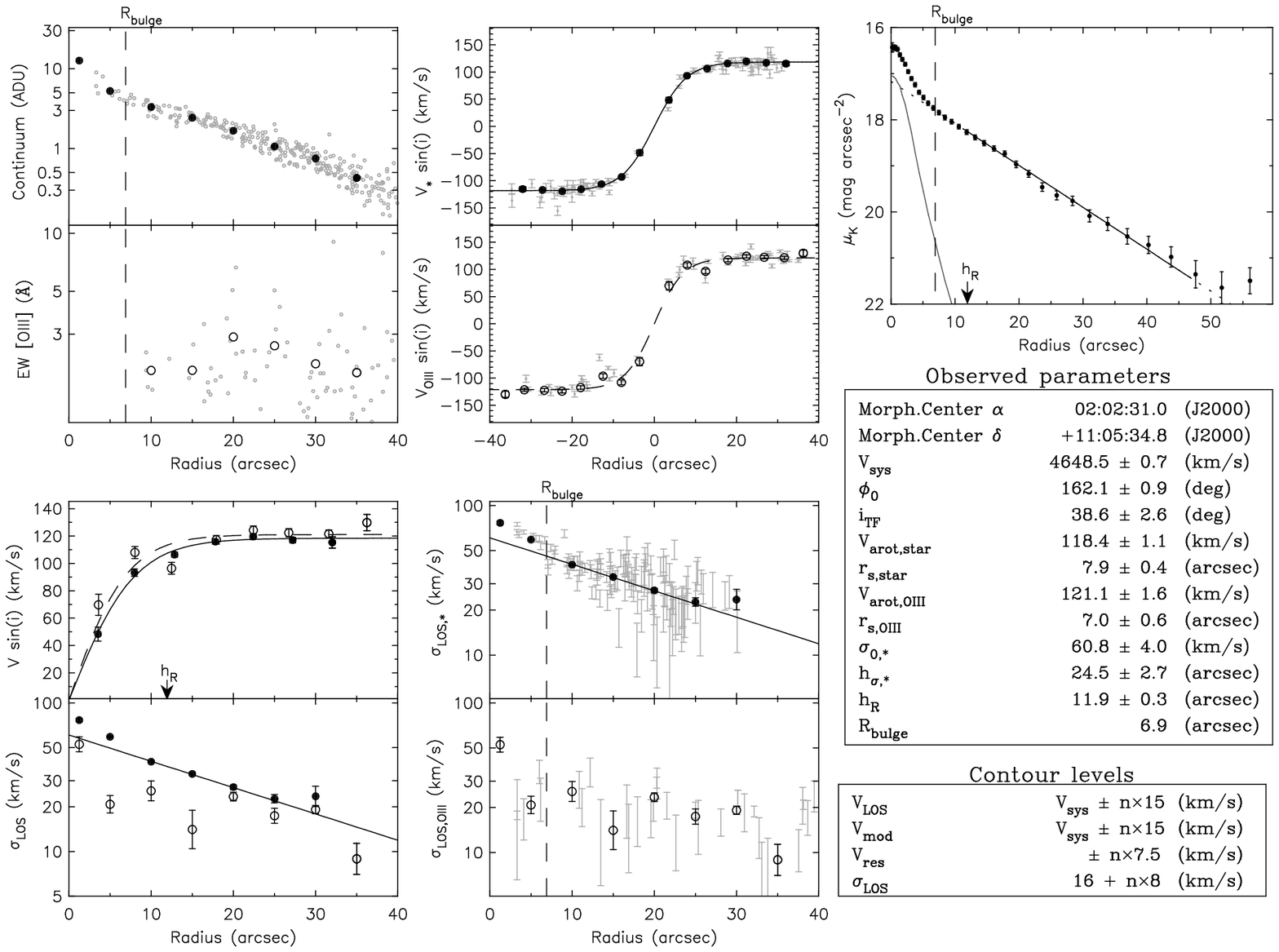}
 \end{figure}

\clearpage

 \begin{figure}
 \centering
 \includegraphics[width=0.95\textwidth]{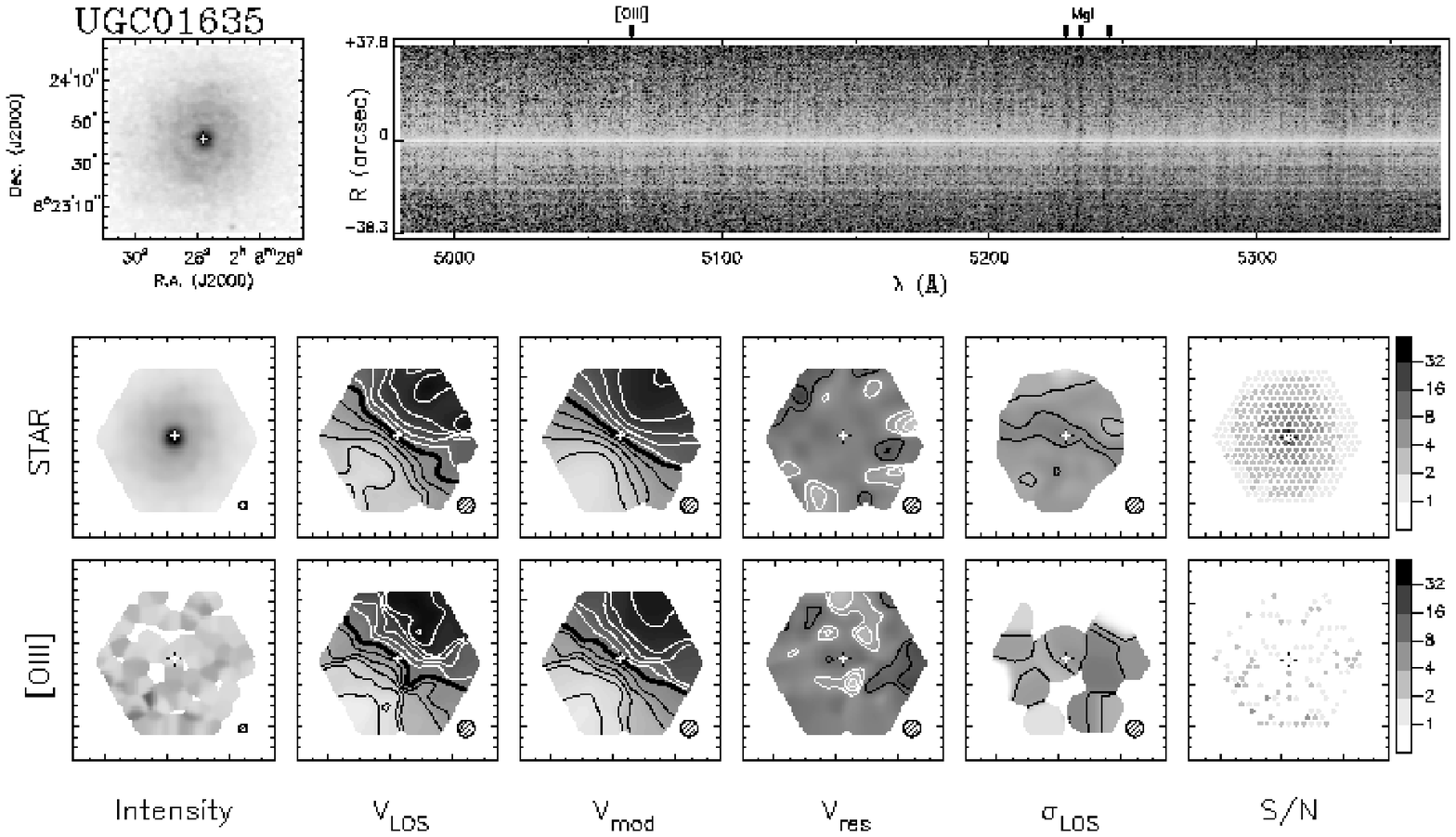}
 \end{figure}

 \begin{figure}
 \centering
 \includegraphics[width=0.95\textwidth]{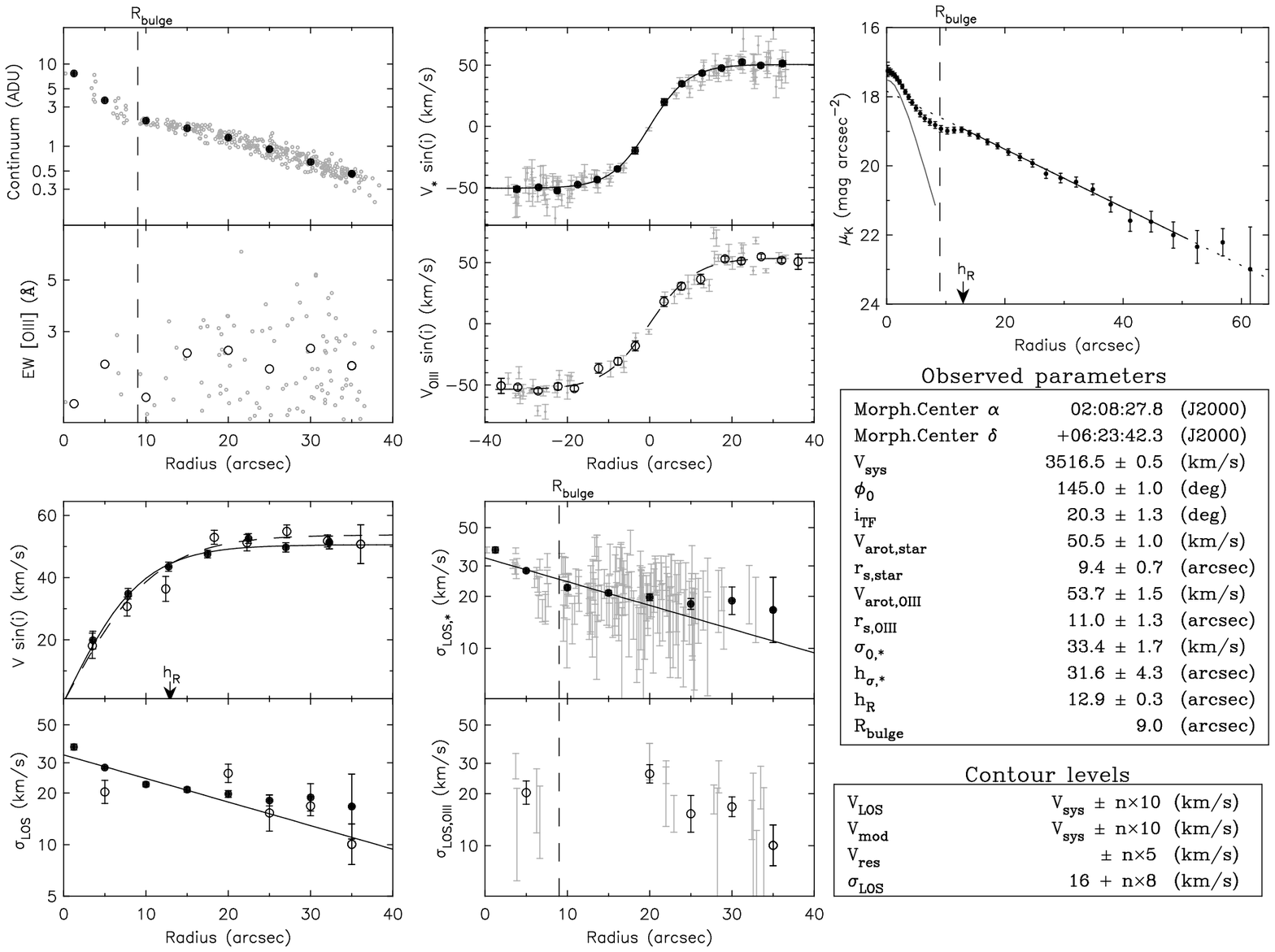}
 \end{figure}

\clearpage

 \begin{figure}
 \centering
 \includegraphics[width=0.95\textwidth]{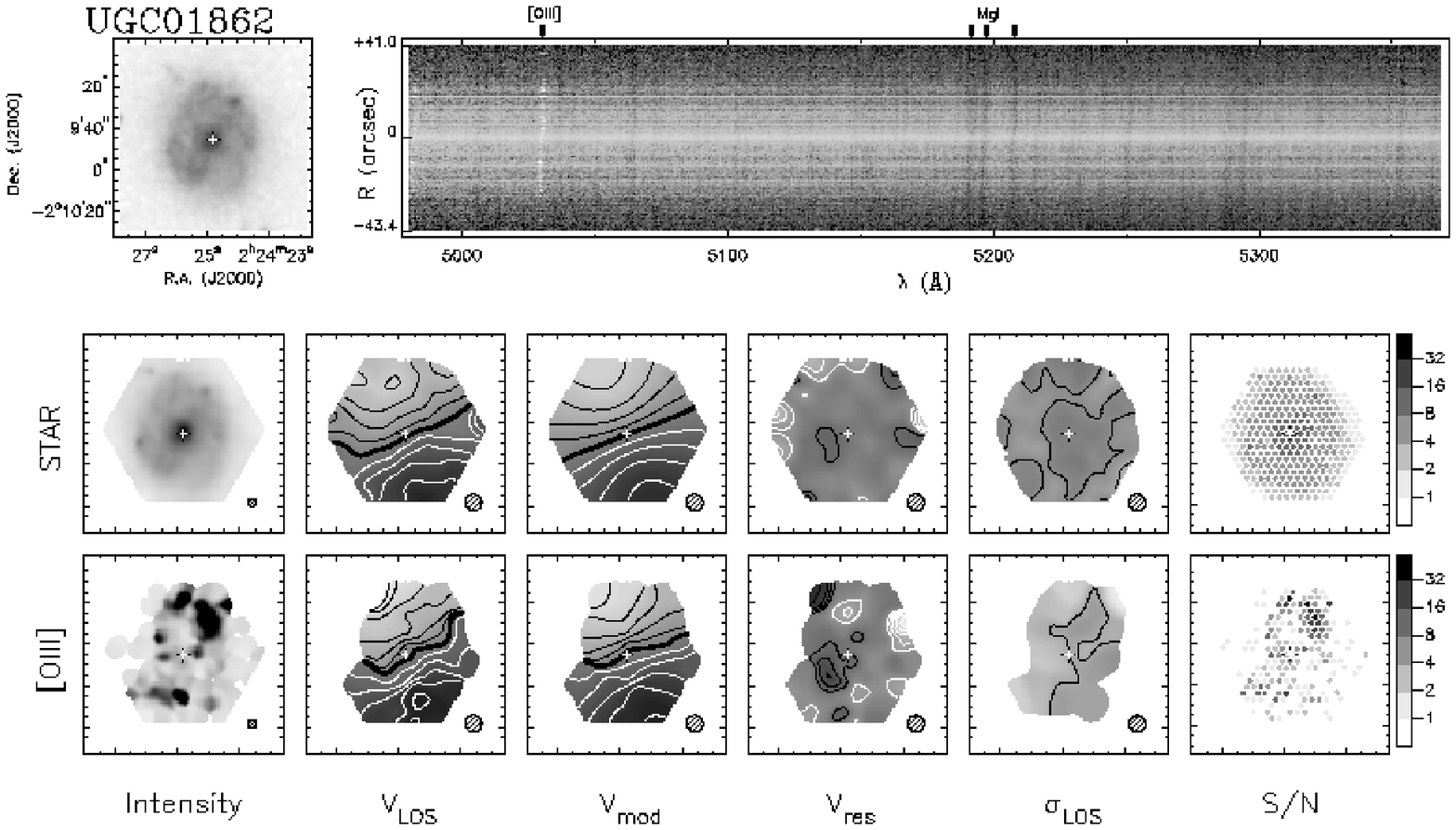}
 \end{figure}

 \begin{figure}
 \centering
 \includegraphics[width=0.95\textwidth]{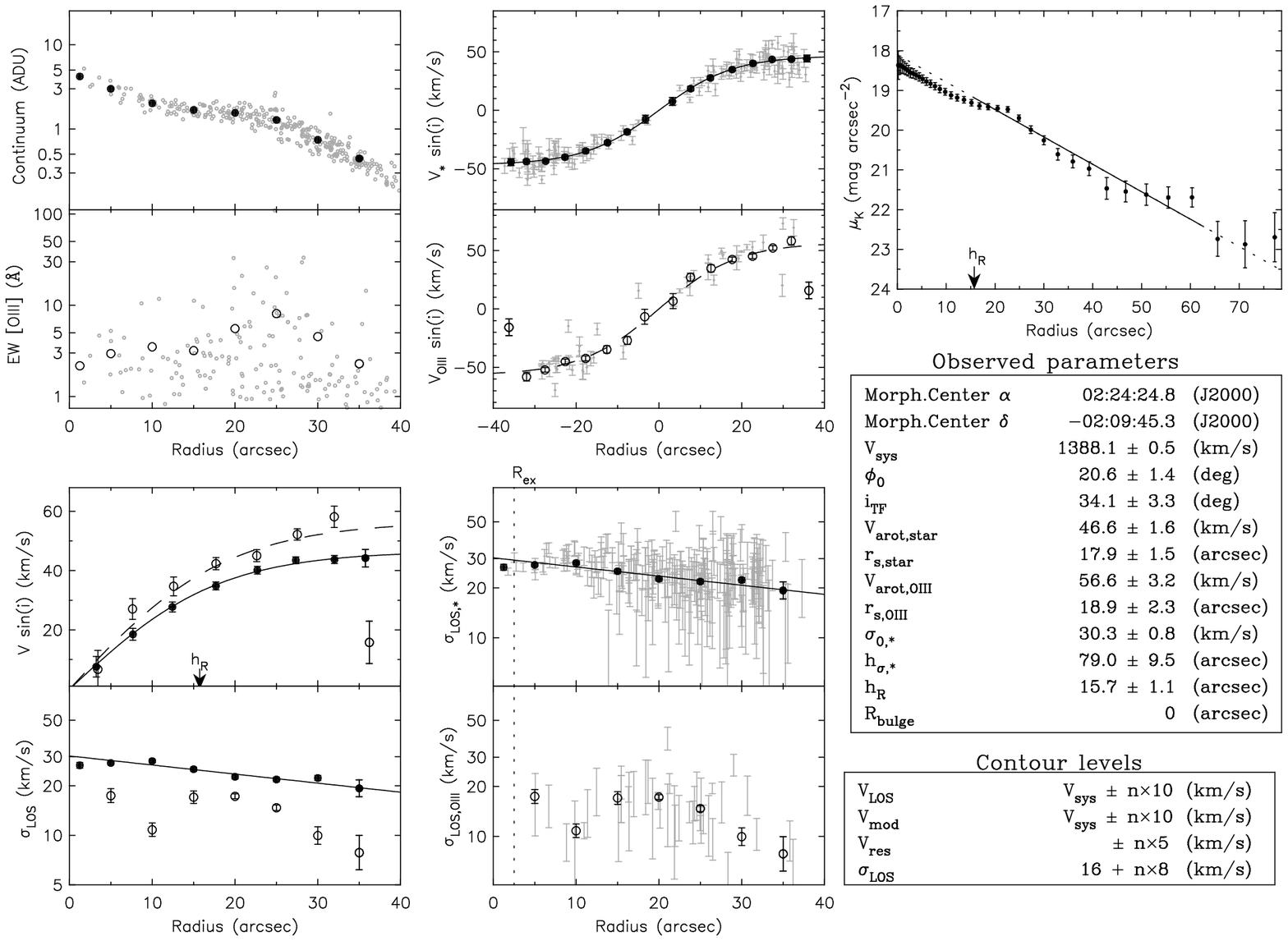}
 \end{figure}

\clearpage

 \begin{figure}
 \centering
 \includegraphics[width=0.95\textwidth]{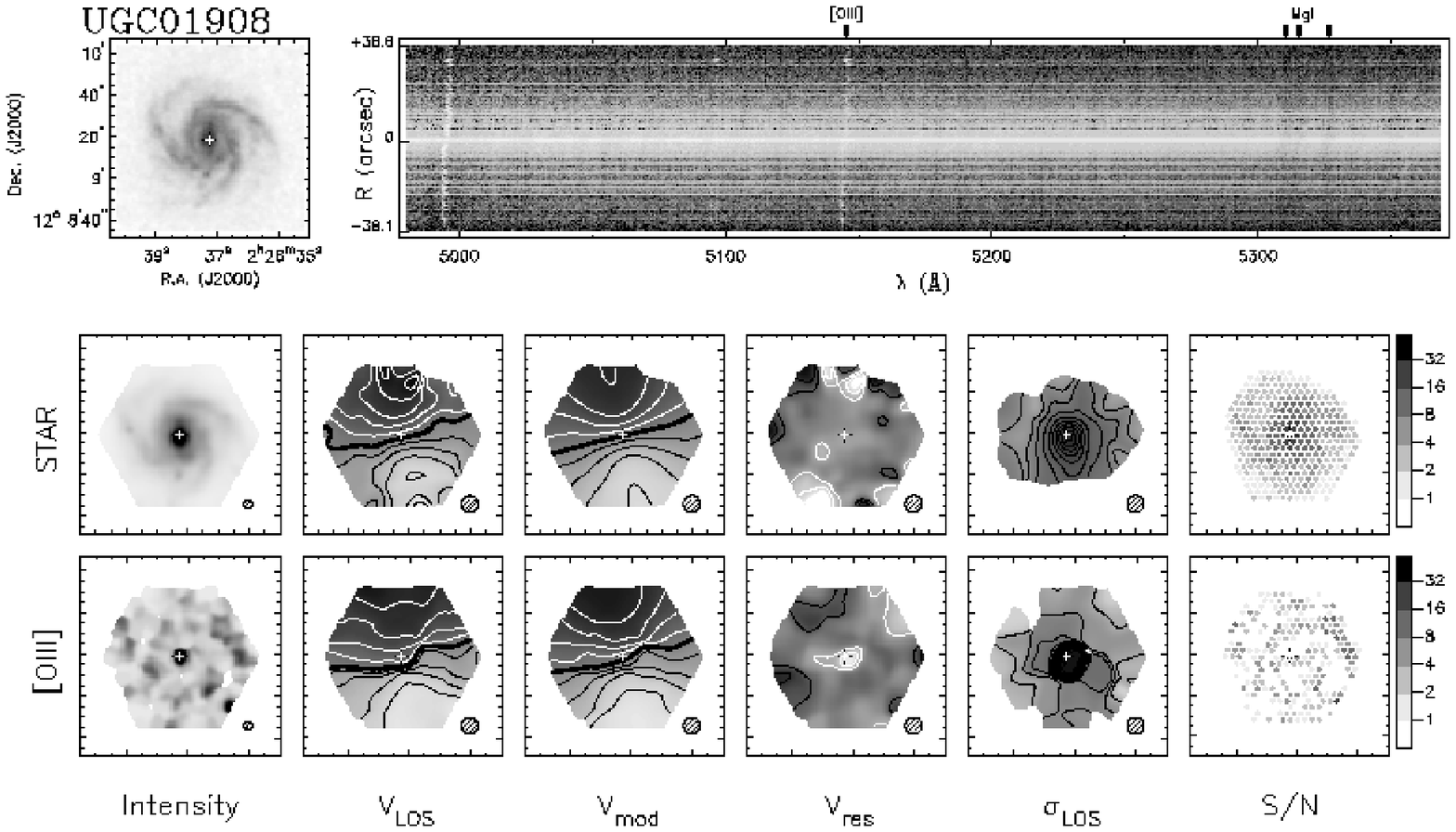}
 \end{figure}

 \begin{figure}
 \centering
 \includegraphics[width=0.95\textwidth]{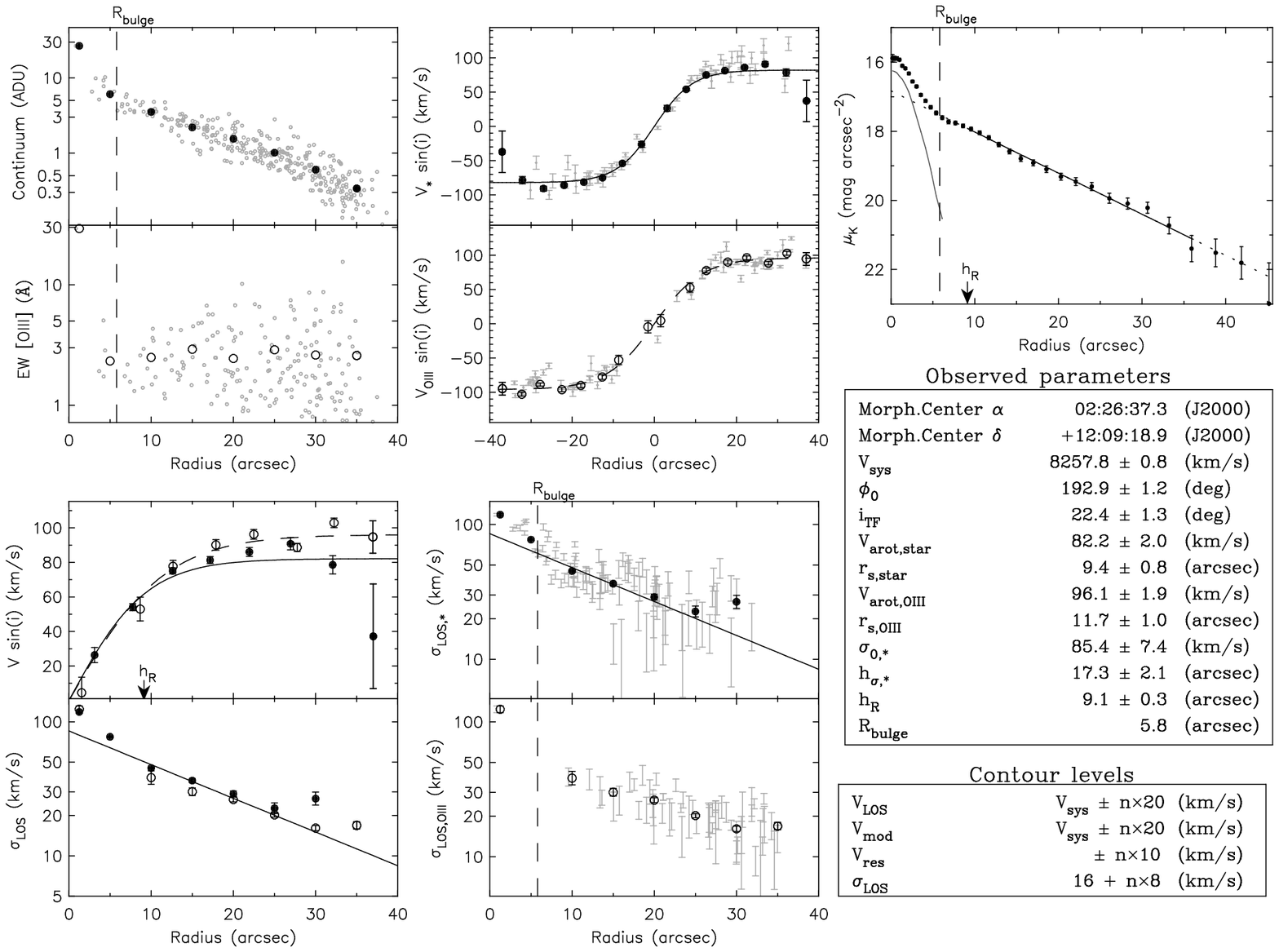}
 \end{figure}

\clearpage

 \begin{figure}
 \centering
 \includegraphics[width=0.95\textwidth]{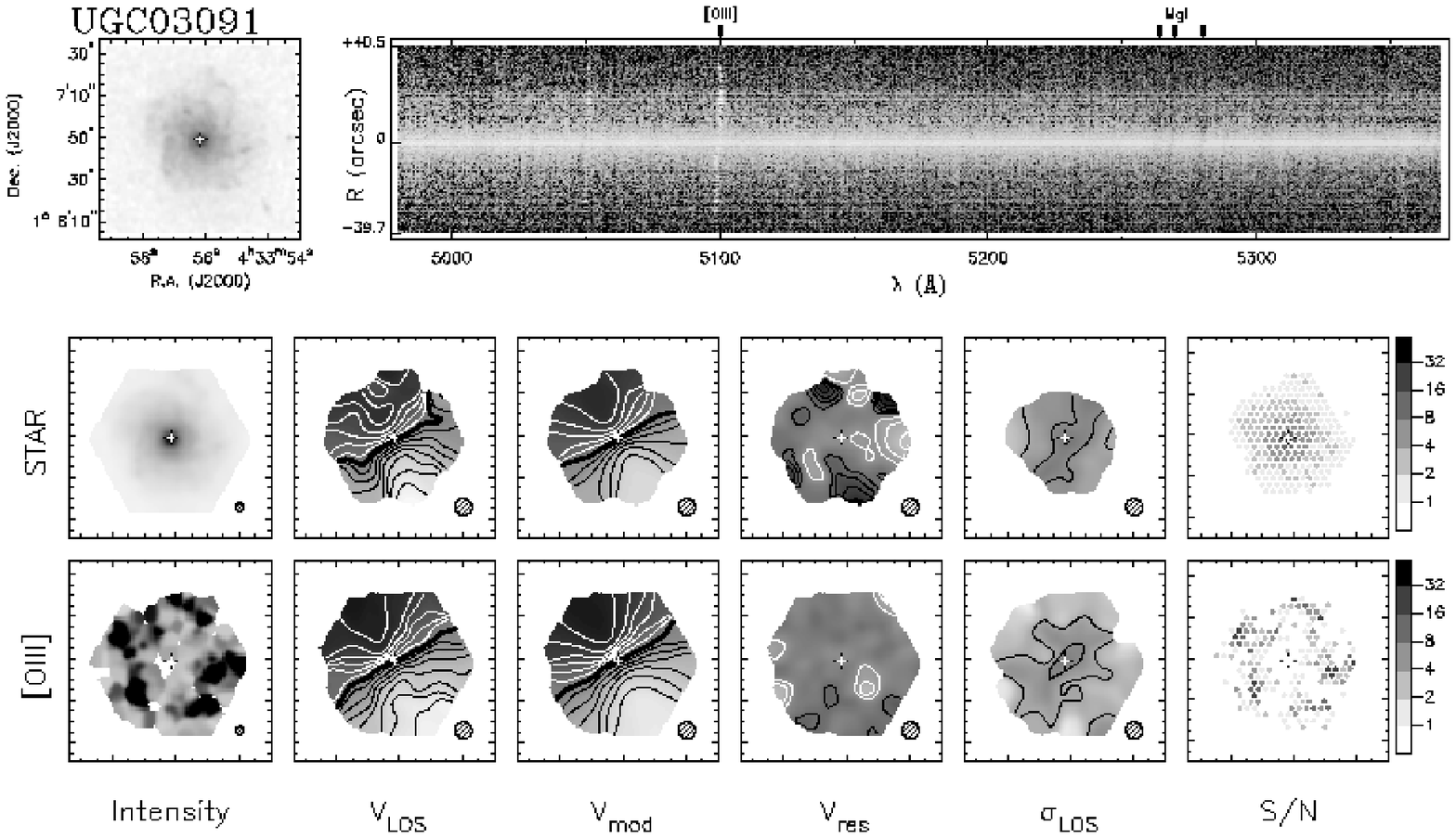}
 \end{figure}

 \begin{figure}
 \centering
 \includegraphics[width=0.95\textwidth]{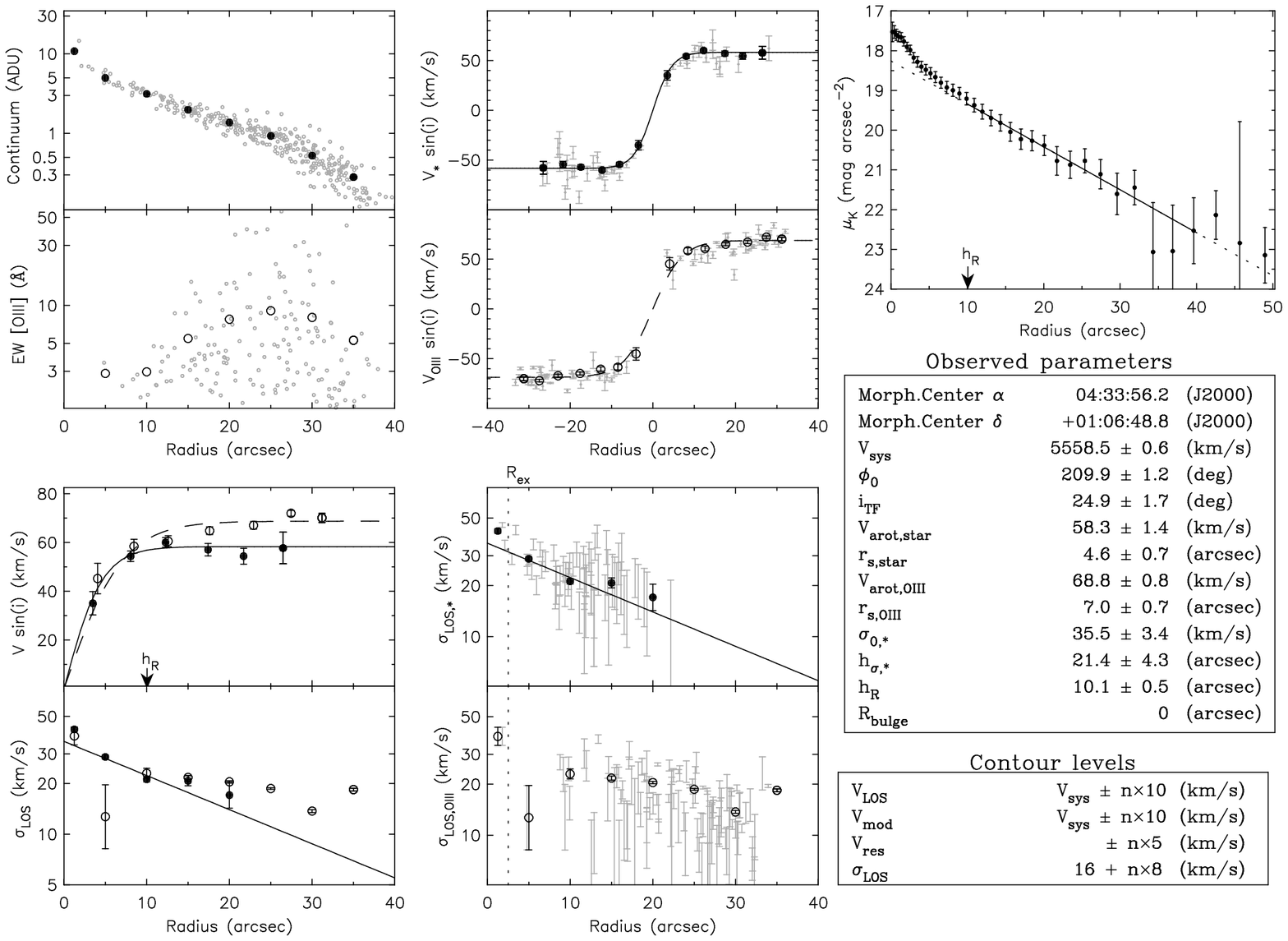}
 \end{figure}

\clearpage

 \begin{figure}
 \centering
 \includegraphics[width=0.95\textwidth]{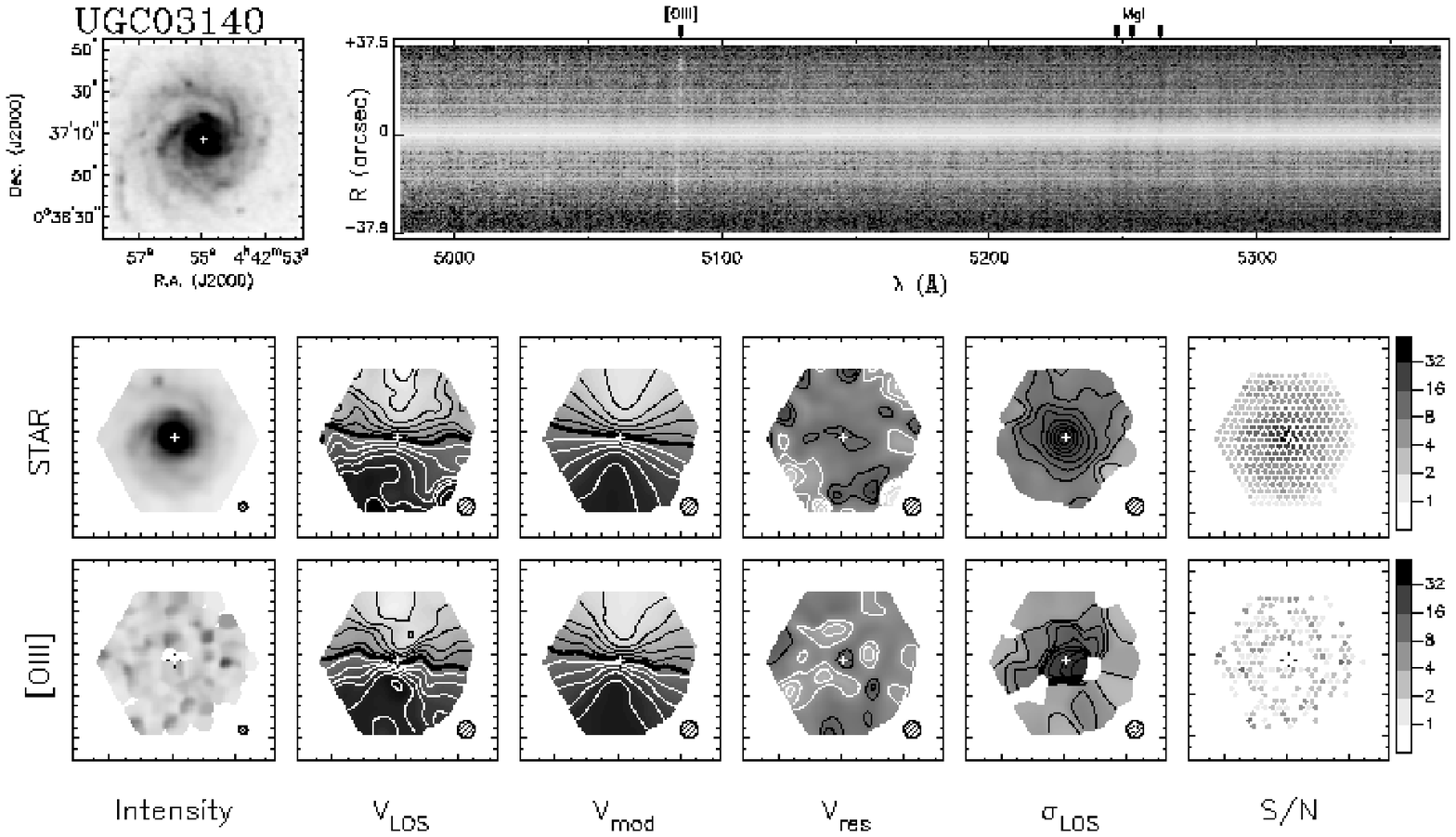}
 \end{figure}

 \begin{figure}
 \centering
 \includegraphics[width=0.95\textwidth]{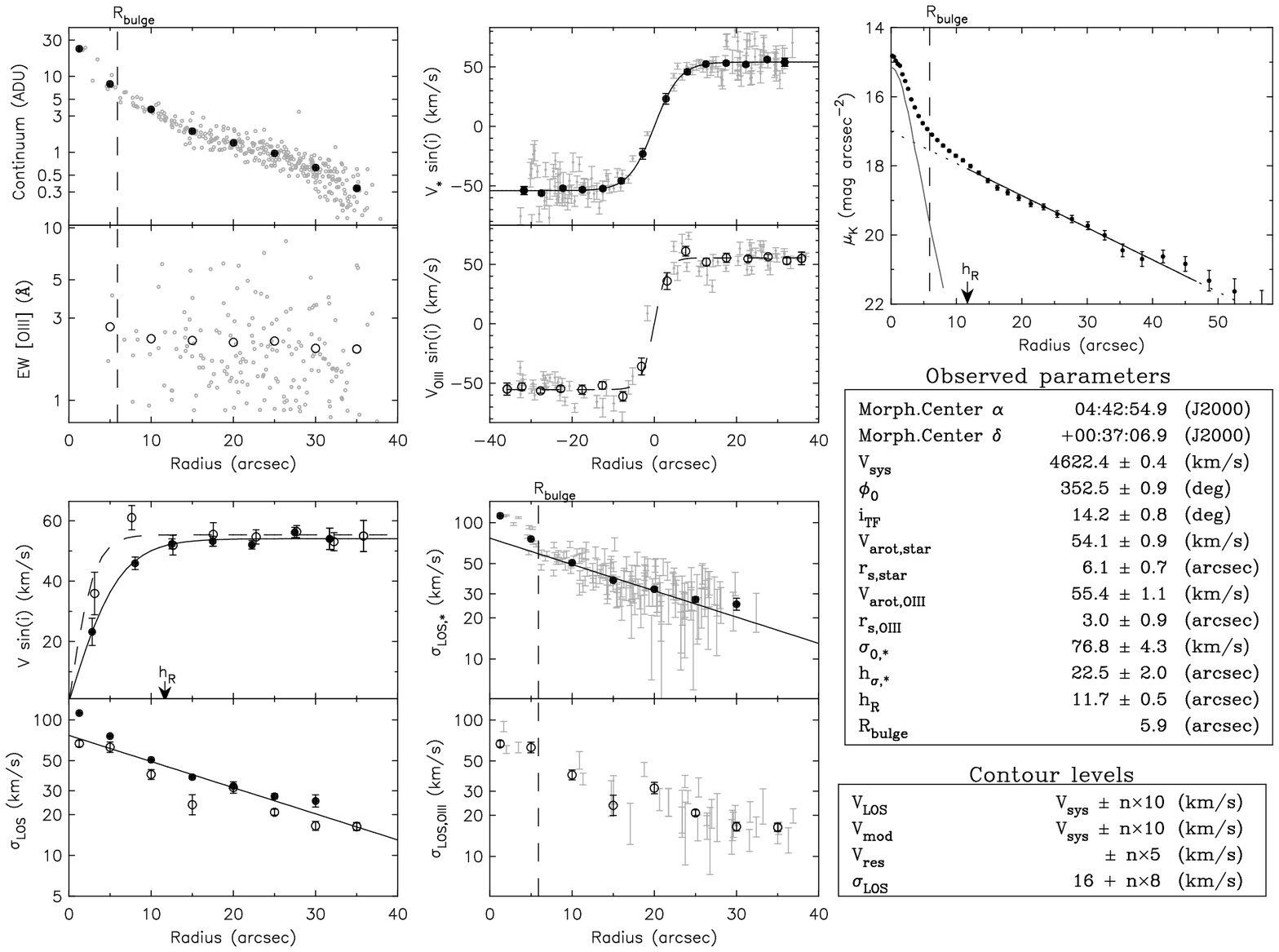}
 \end{figure}

\clearpage

 \begin{figure}
 \centering
 \includegraphics[width=0.95\textwidth]{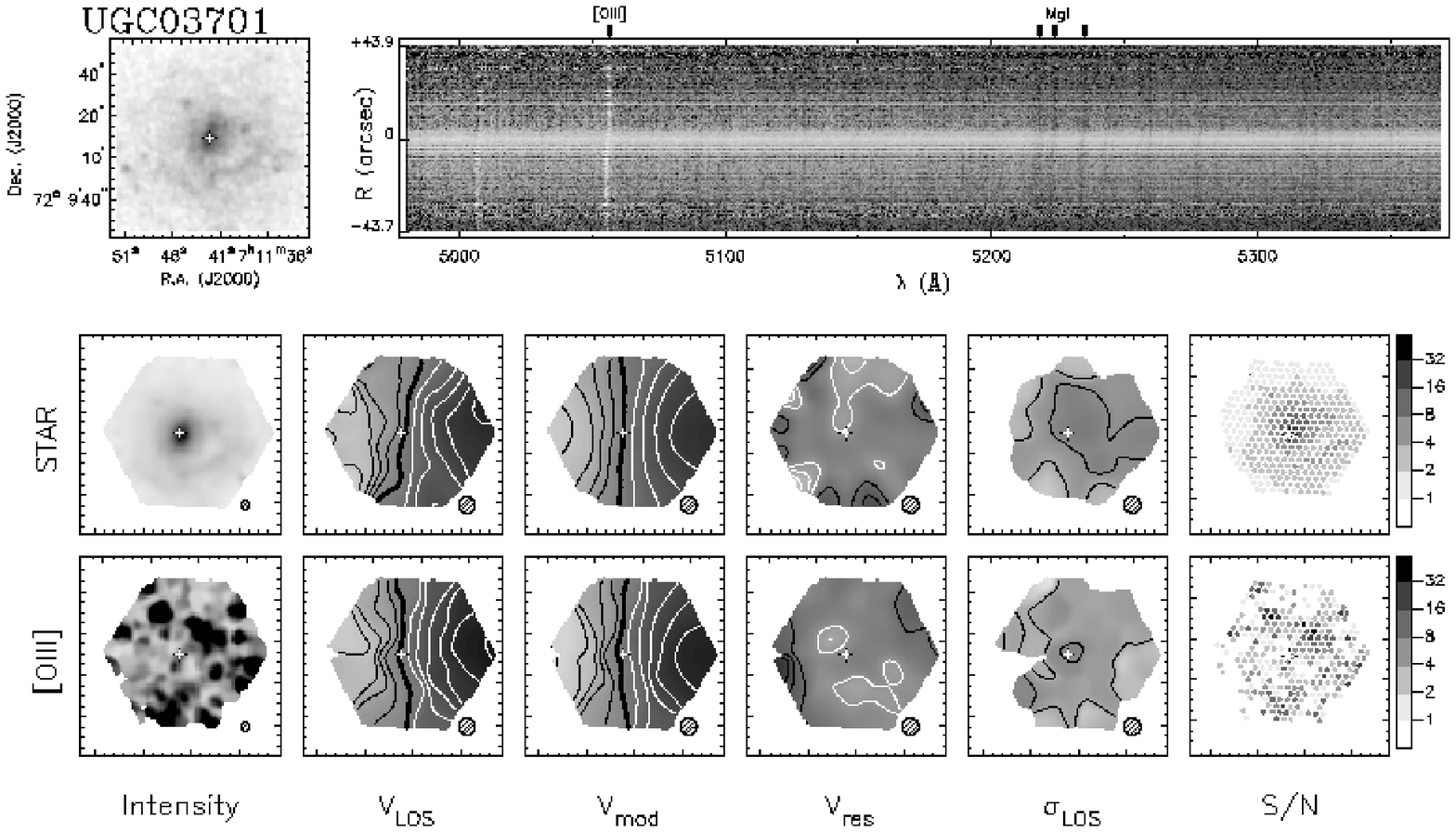}
 \end{figure}

 \begin{figure}
 \centering
 \includegraphics[width=0.95\textwidth]{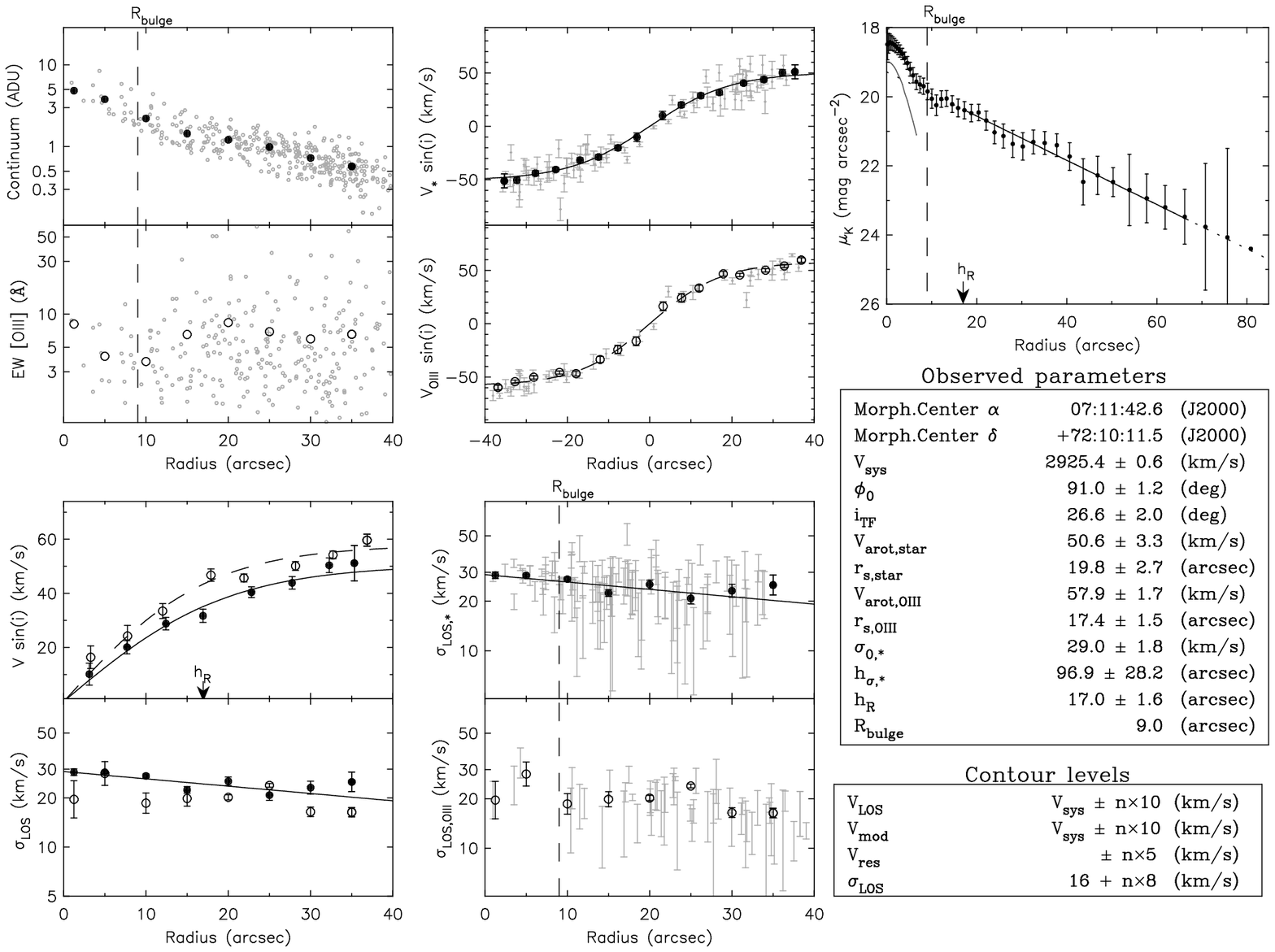}
 \end{figure}

\clearpage

 \begin{figure}
 \centering
 \includegraphics[width=0.95\textwidth]{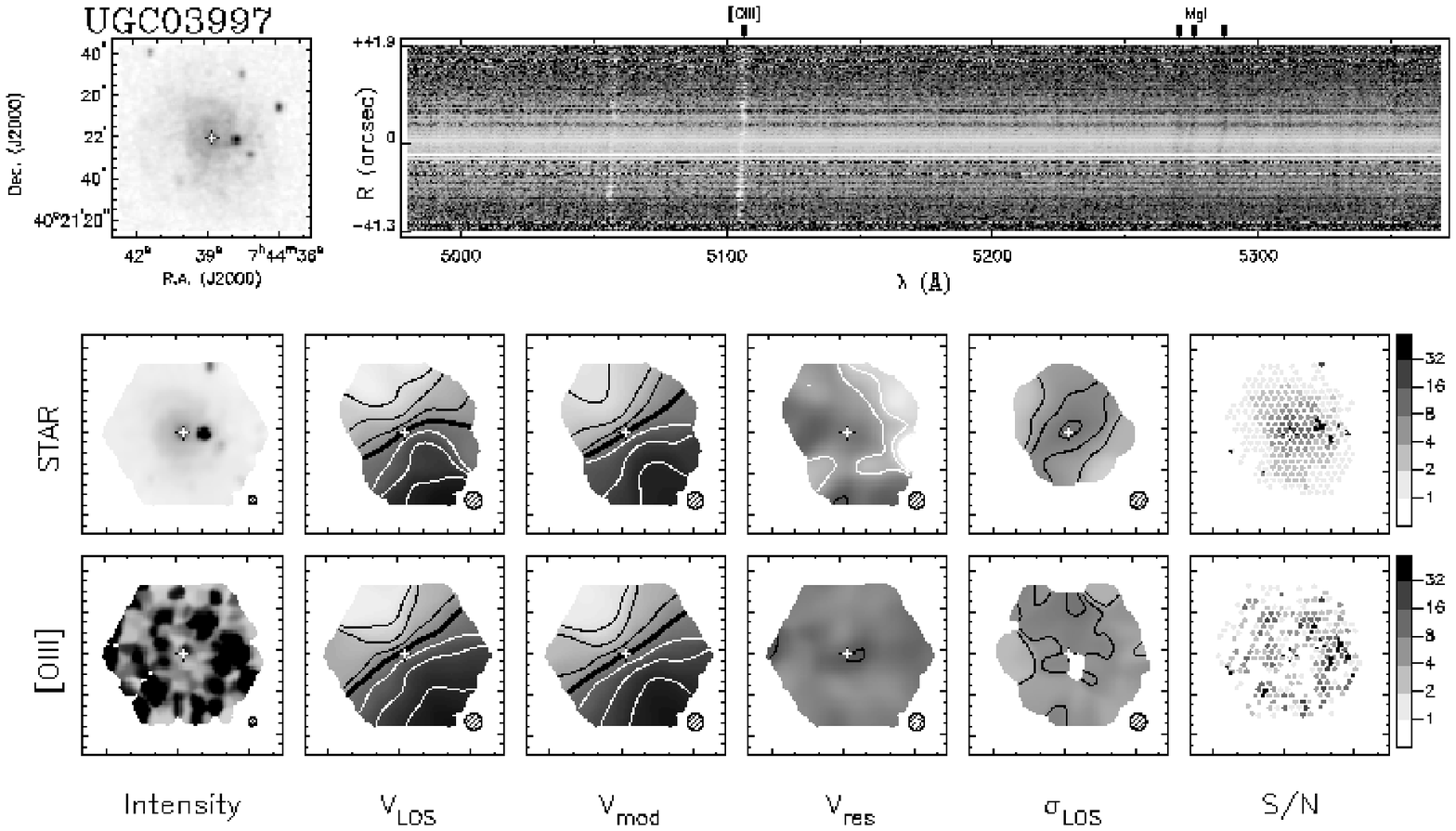}
 \end{figure}

 \begin{figure}
 \centering
 \includegraphics[width=0.95\textwidth]{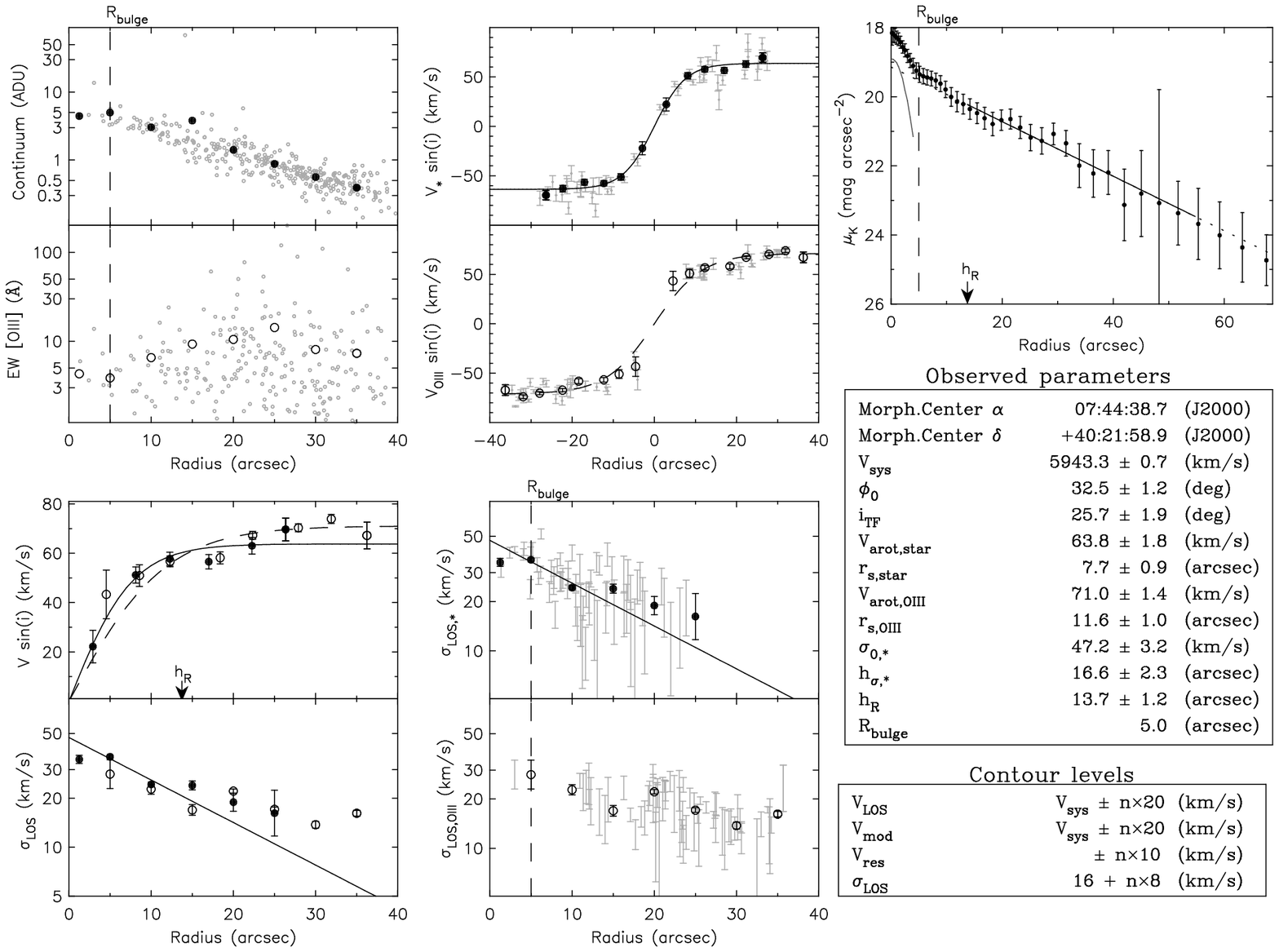}
 \end{figure}

\clearpage

 \begin{figure}
 \centering
 \includegraphics[width=0.95\textwidth]{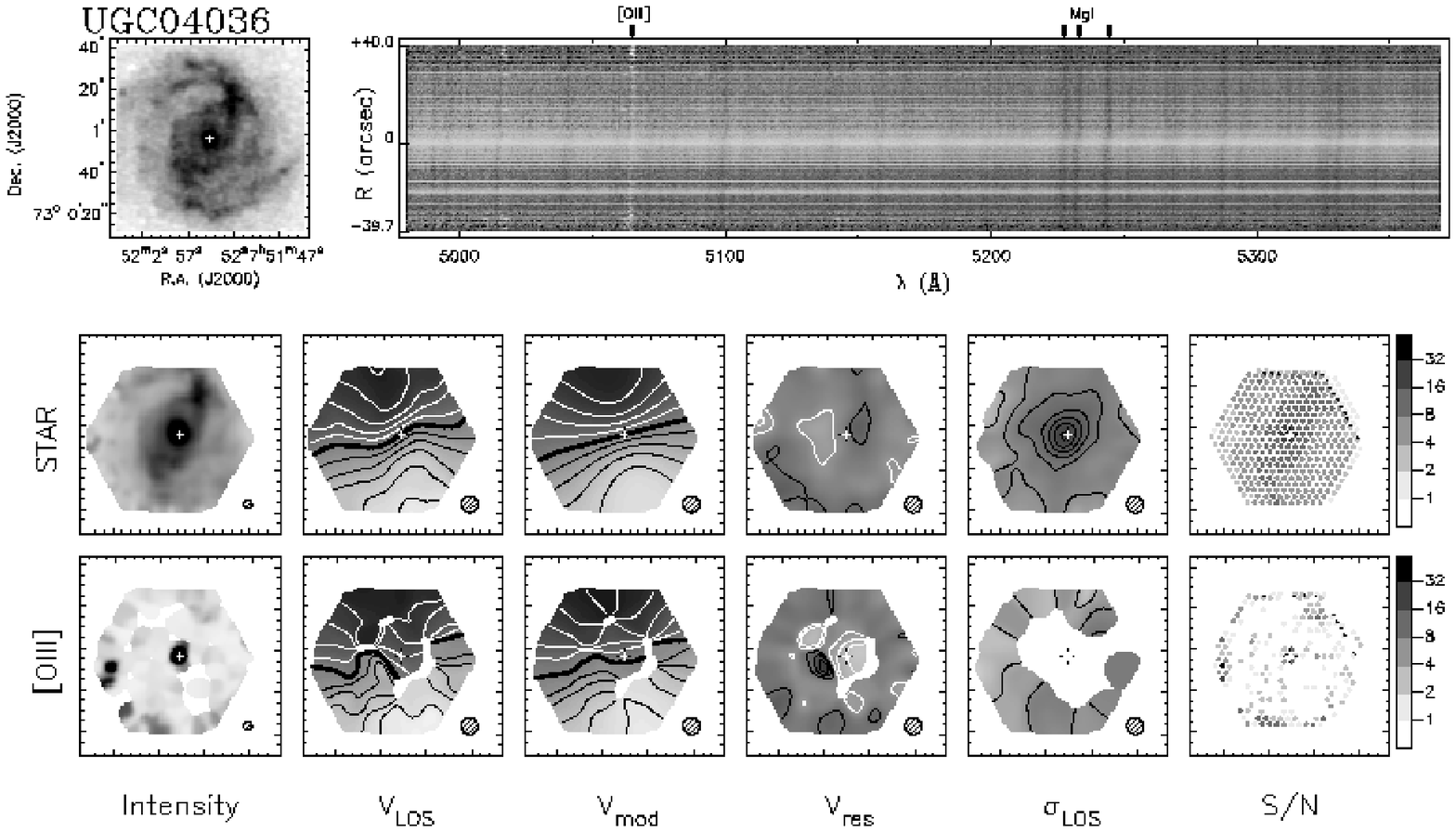}
 \end{figure}

 \begin{figure}
 \centering
 \includegraphics[width=0.95\textwidth]{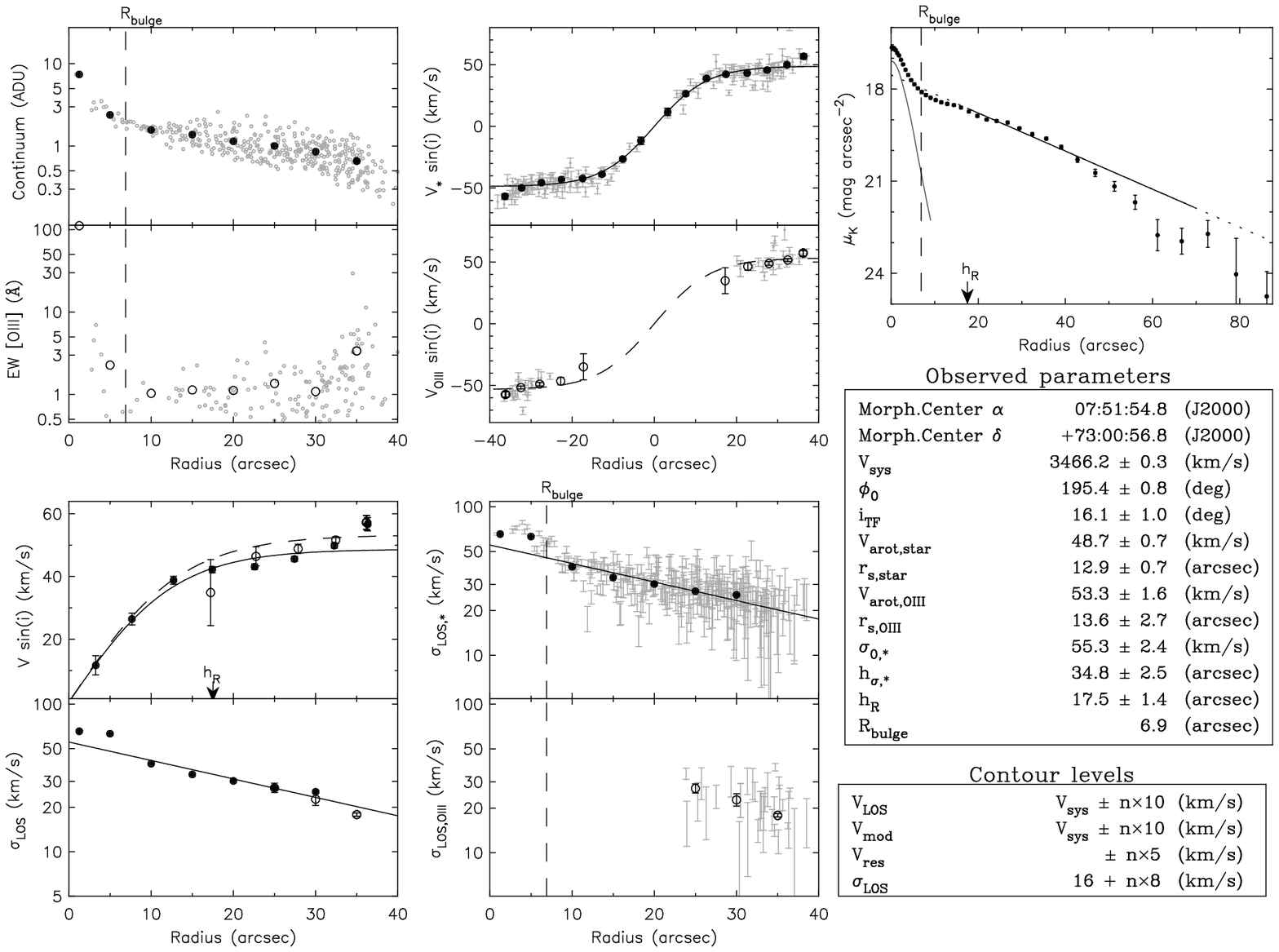}
 \end{figure}

\clearpage

 \begin{figure}
 \centering
 \includegraphics[width=0.95\textwidth]{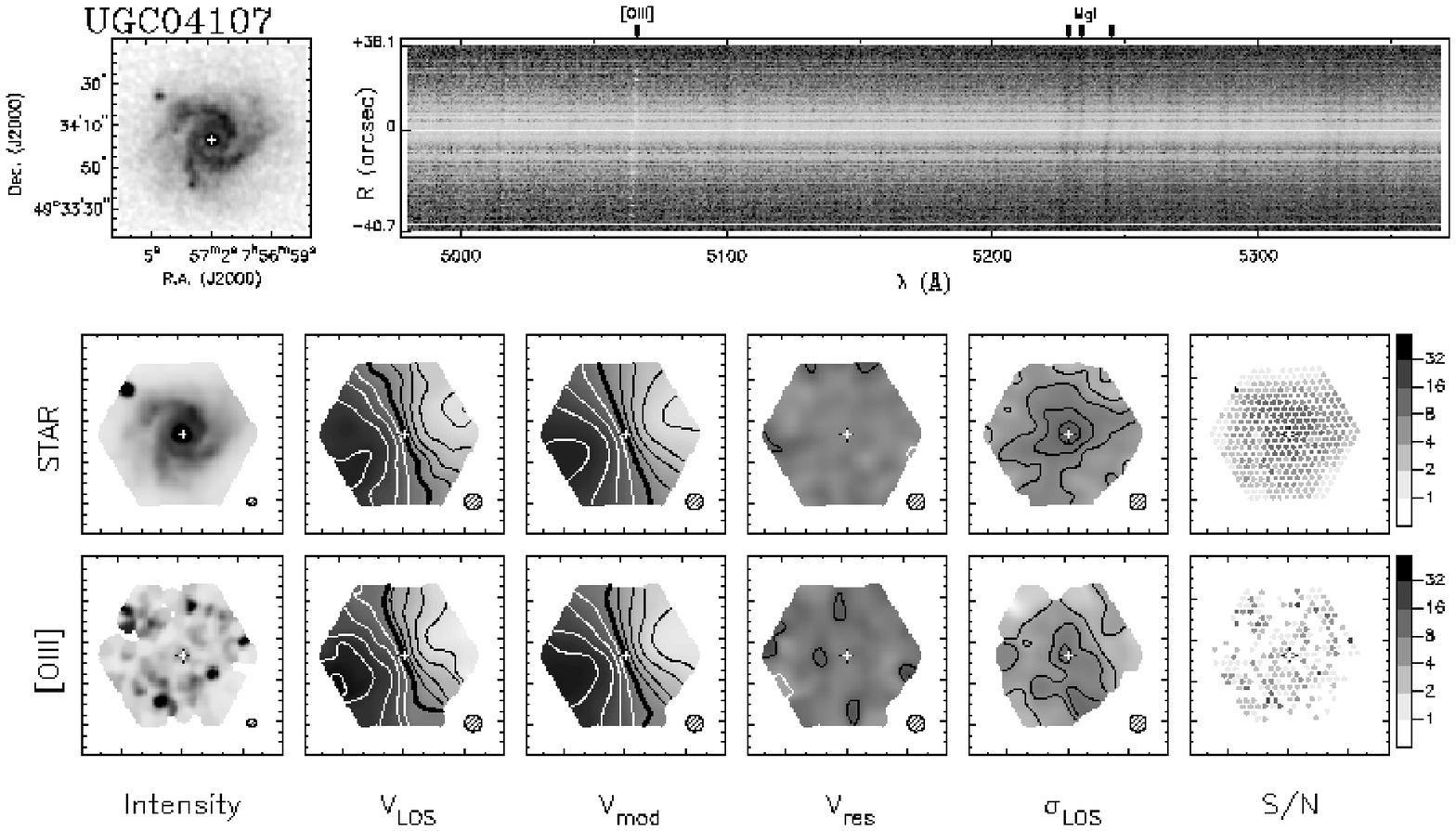}
 \end{figure}

 \begin{figure}
 \centering
 \includegraphics[width=0.95\textwidth]{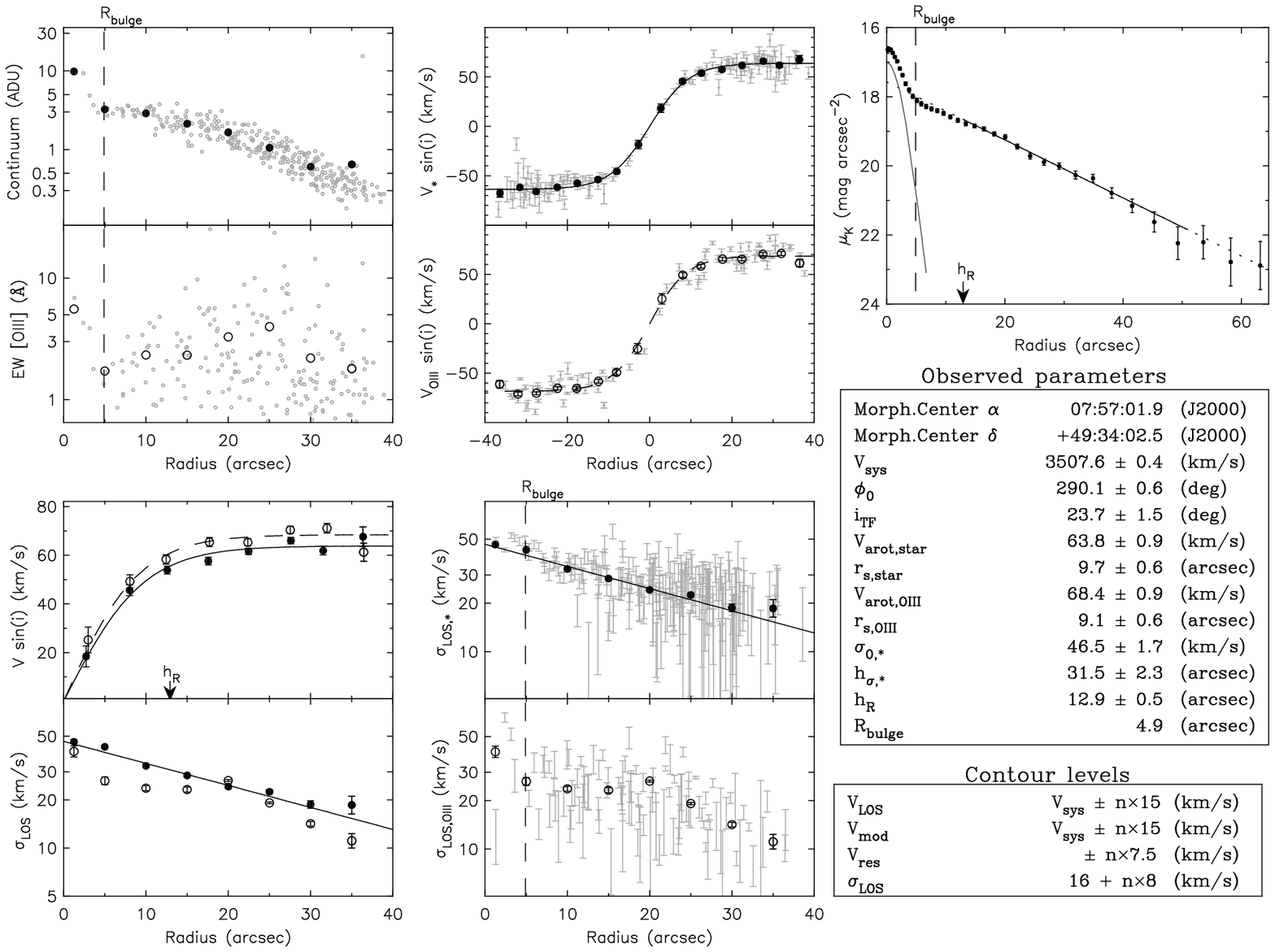}
 \end{figure}

\clearpage

 \begin{figure}
 \centering
 \includegraphics[width=0.95\textwidth]{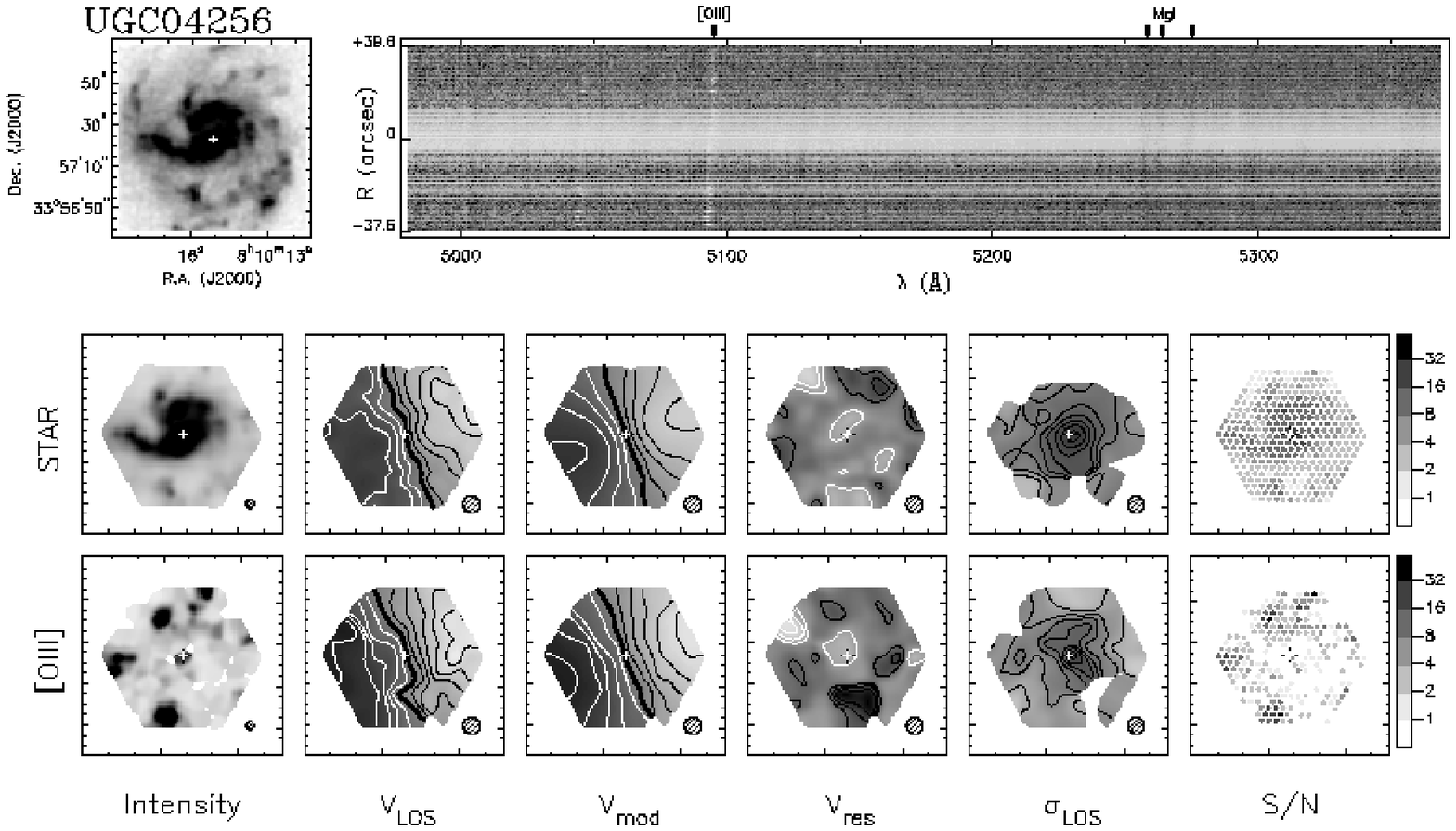}
 \end{figure}

 \begin{figure}
 \centering
 \includegraphics[width=0.95\textwidth]{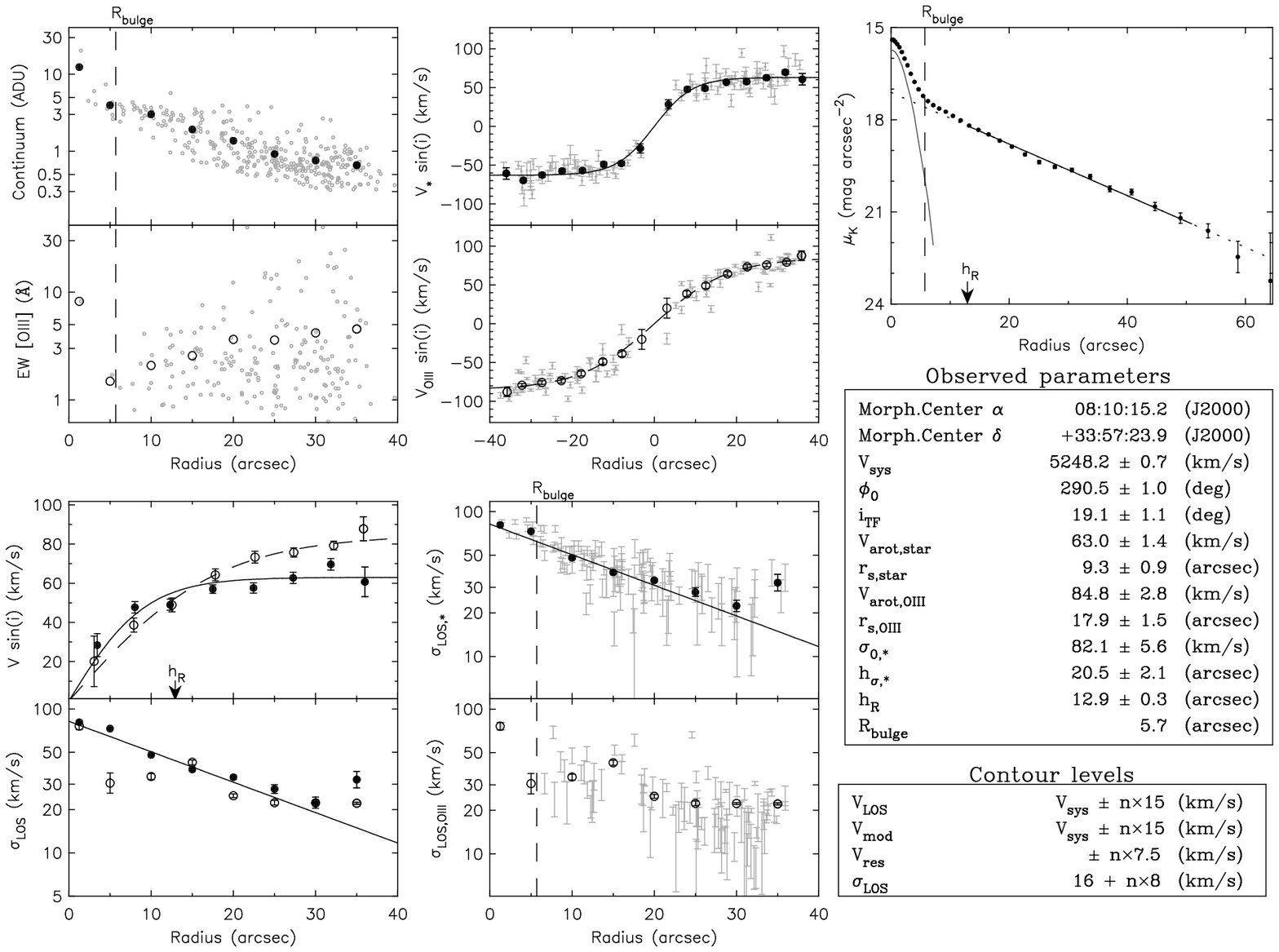}
 \end{figure}

\clearpage

 \begin{figure}
 \centering
 \includegraphics[width=0.95\textwidth]{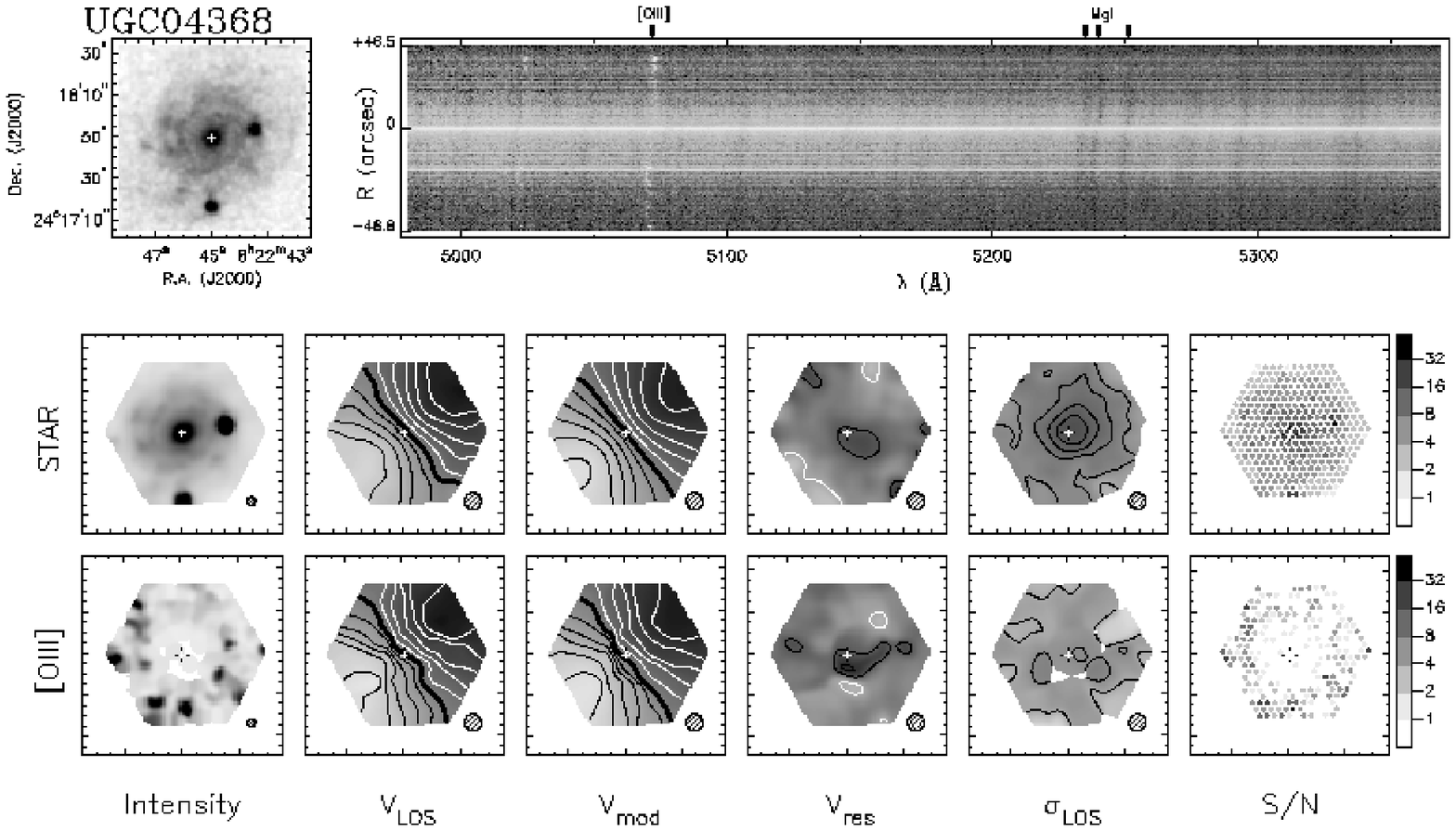}
 \end{figure}

 \begin{figure}
 \centering
 \includegraphics[width=0.95\textwidth]{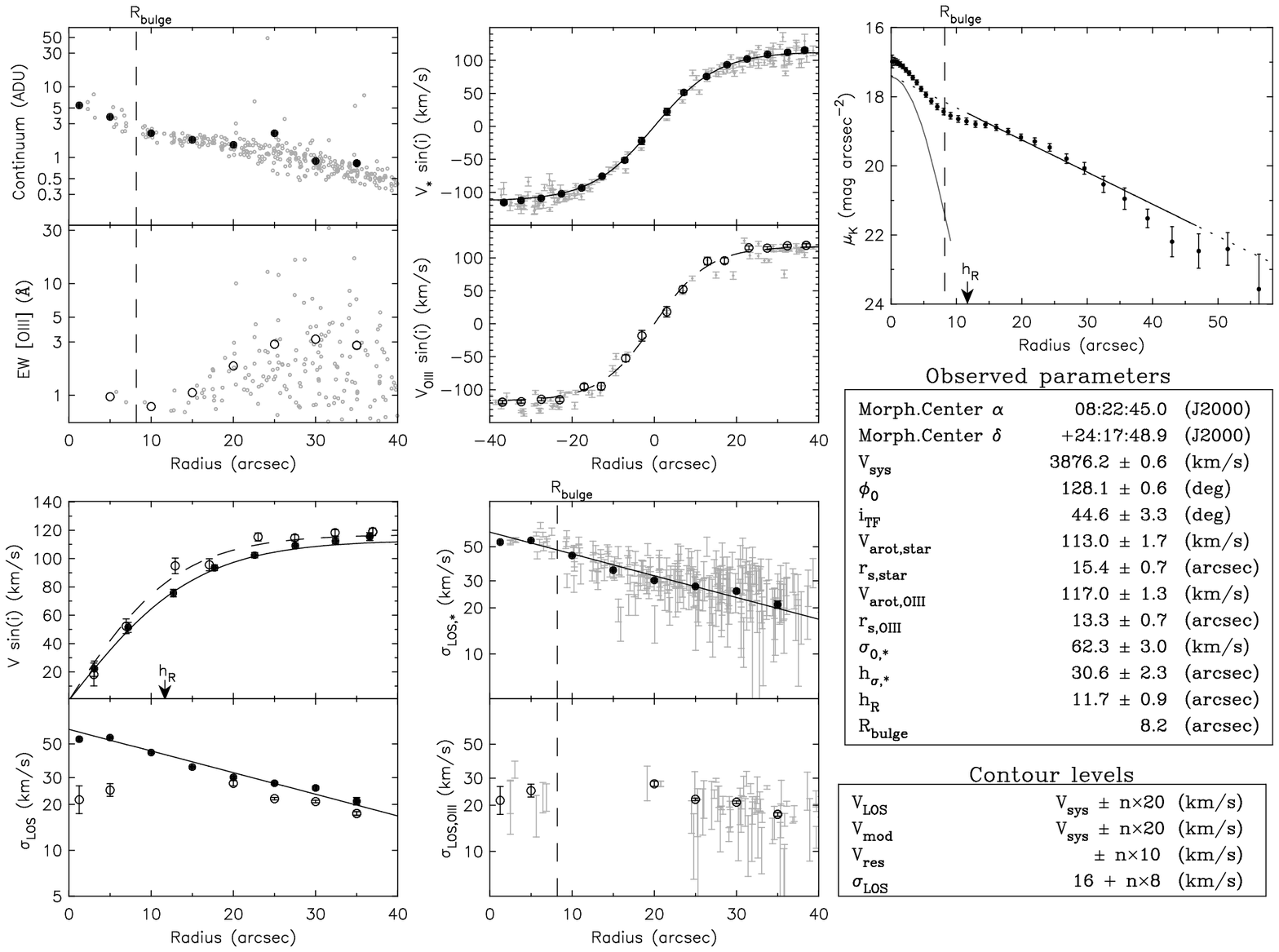}
 \end{figure}

\clearpage

 \begin{figure}
 \centering
 \includegraphics[width=0.95\textwidth]{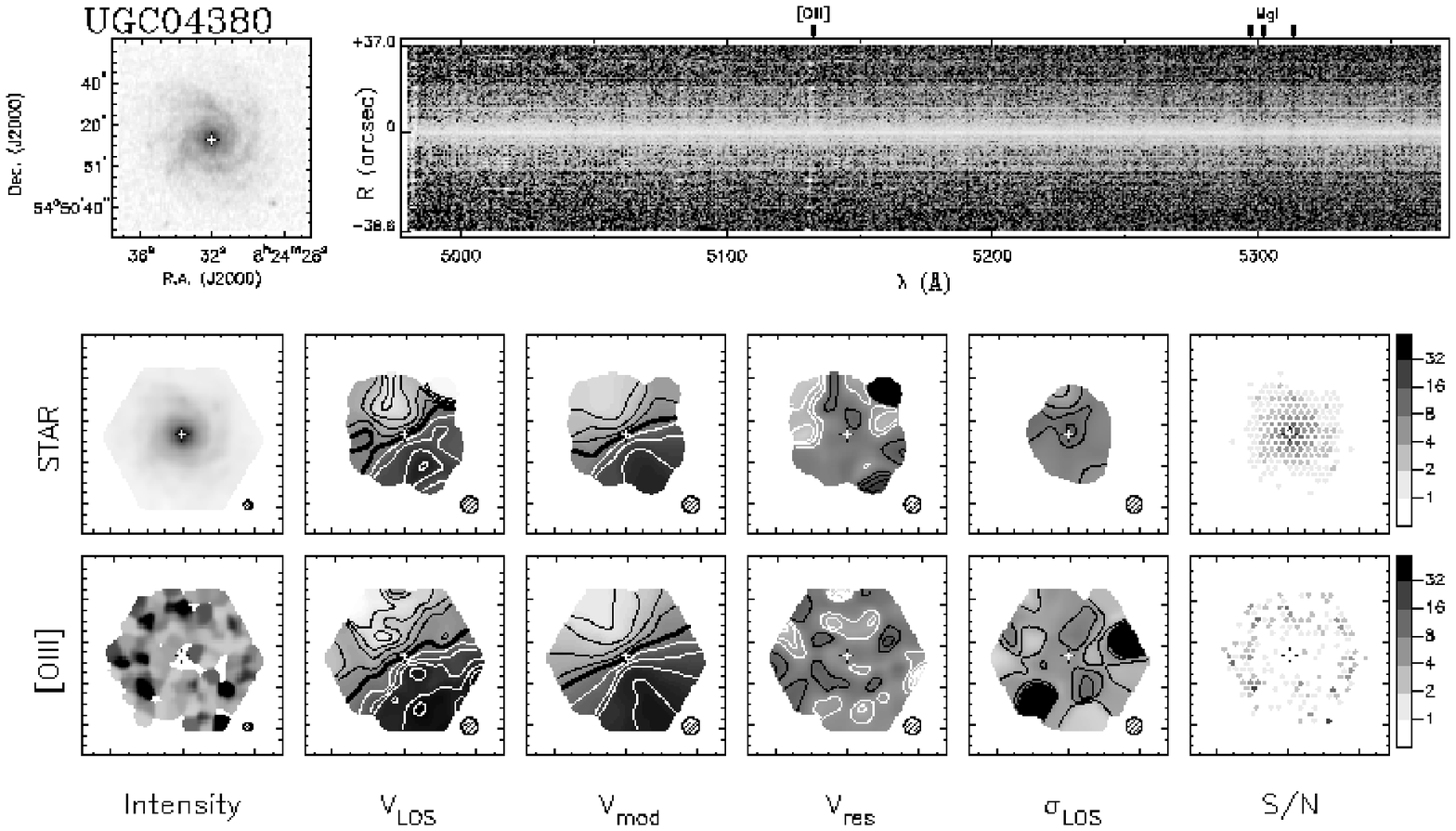}
 \end{figure}

 \begin{figure}
 \centering
 \includegraphics[width=0.95\textwidth]{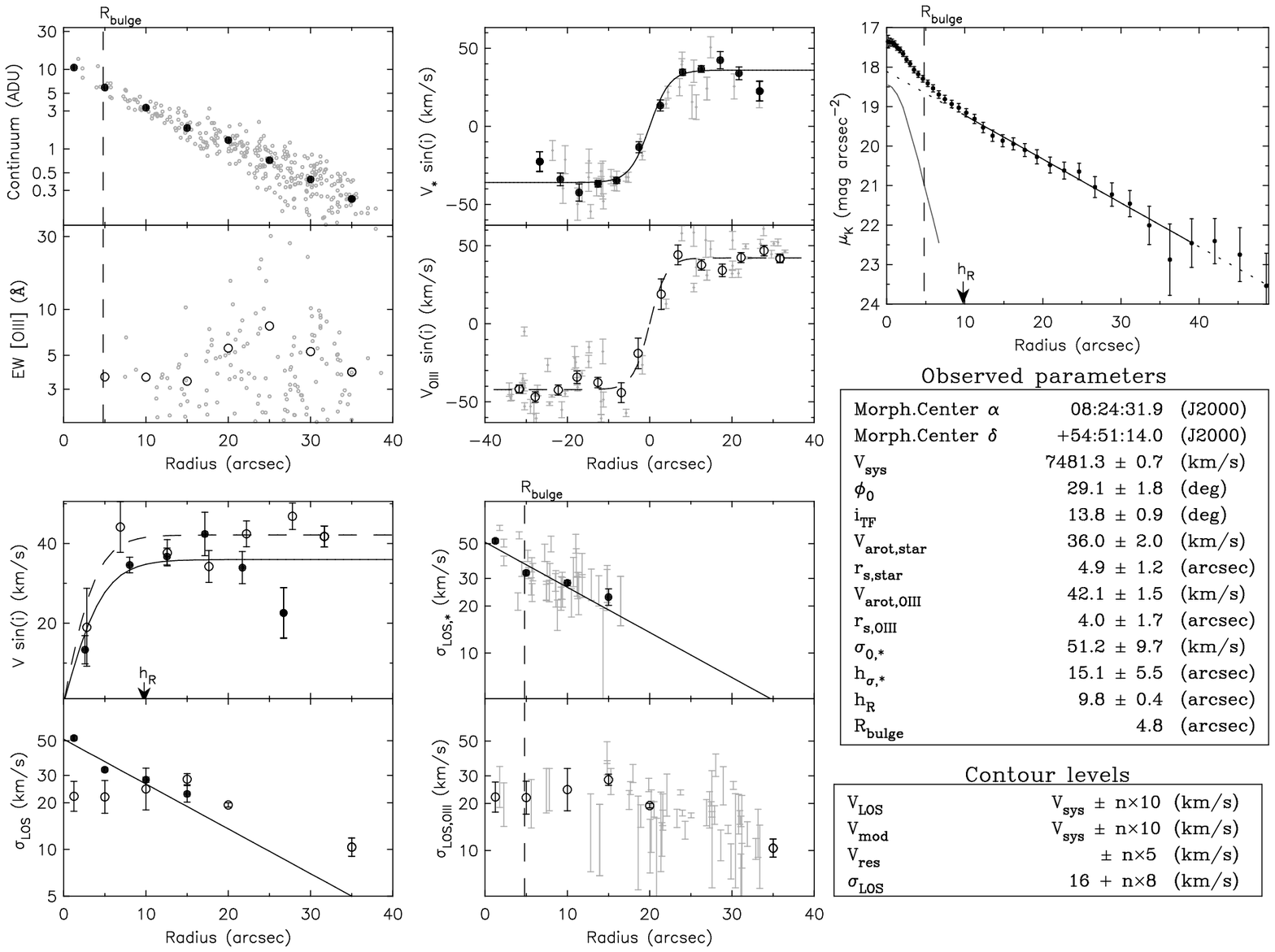}
 \end{figure}

\clearpage

 \begin{figure}
 \centering
 \includegraphics[width=0.95\textwidth]{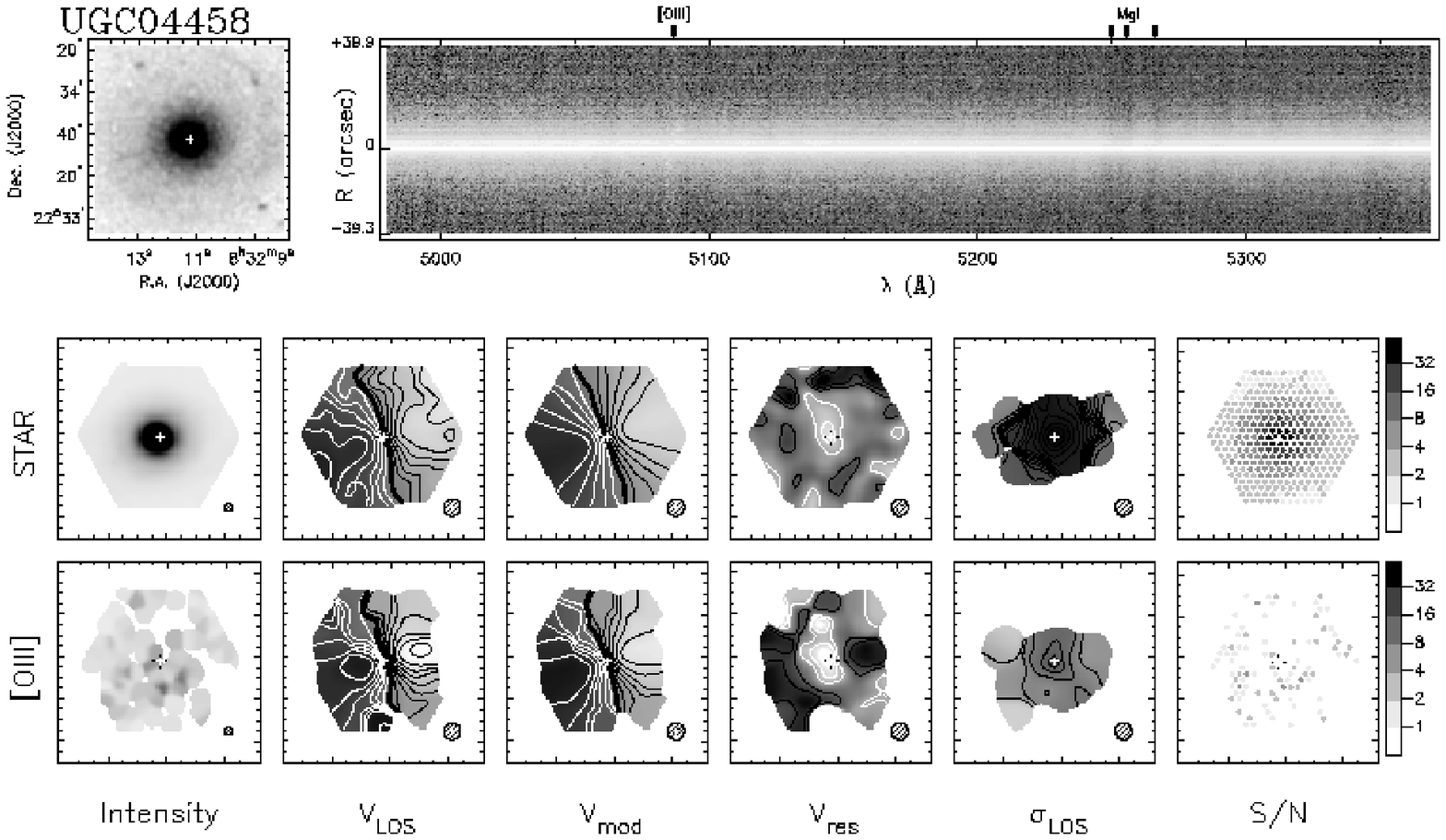}
 \end{figure}

 \begin{figure}
 \centering
 \includegraphics[width=0.95\textwidth]{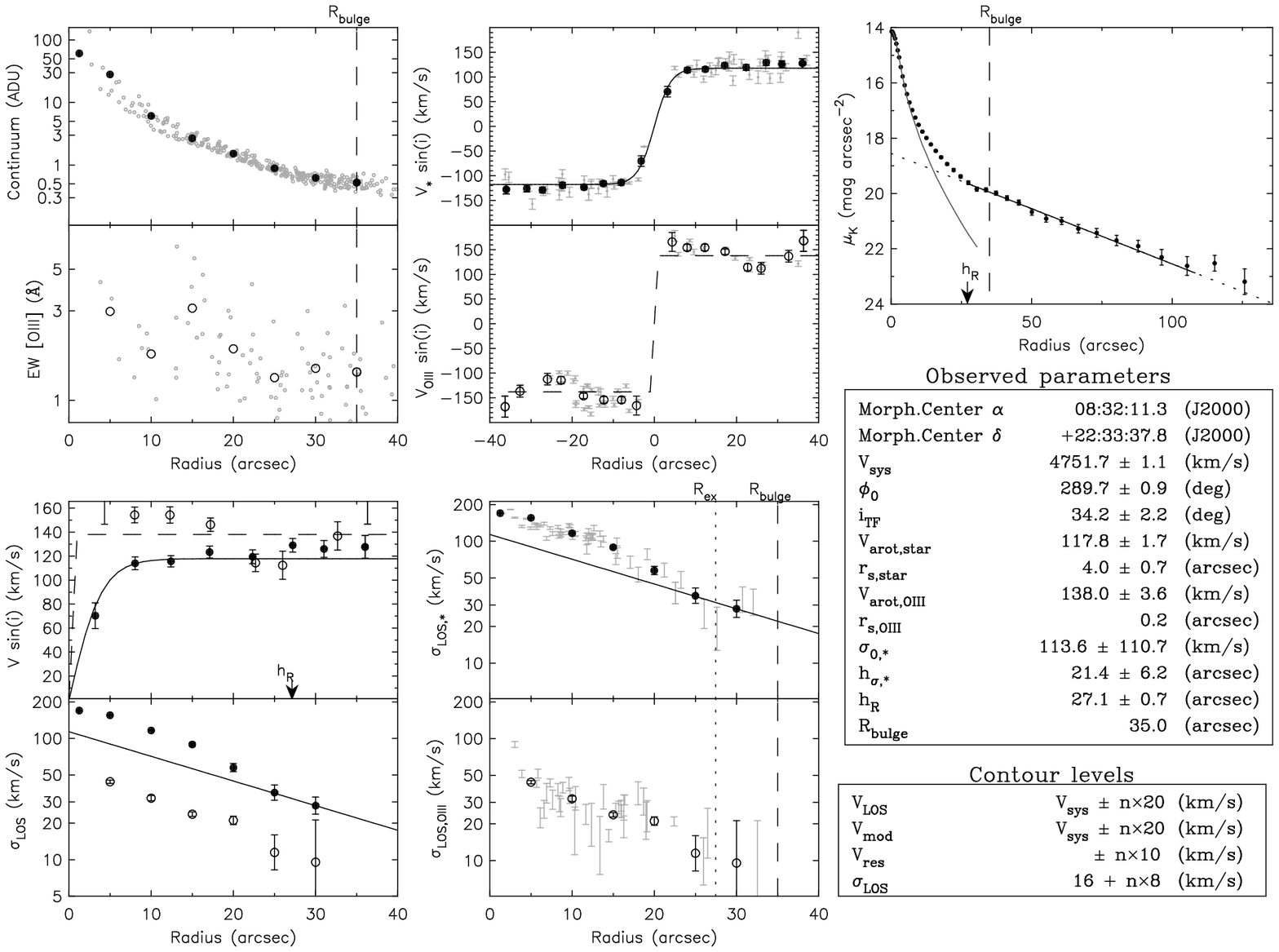}
 \end{figure}

\clearpage

 \begin{figure}
 \centering
 \includegraphics[width=0.95\textwidth]{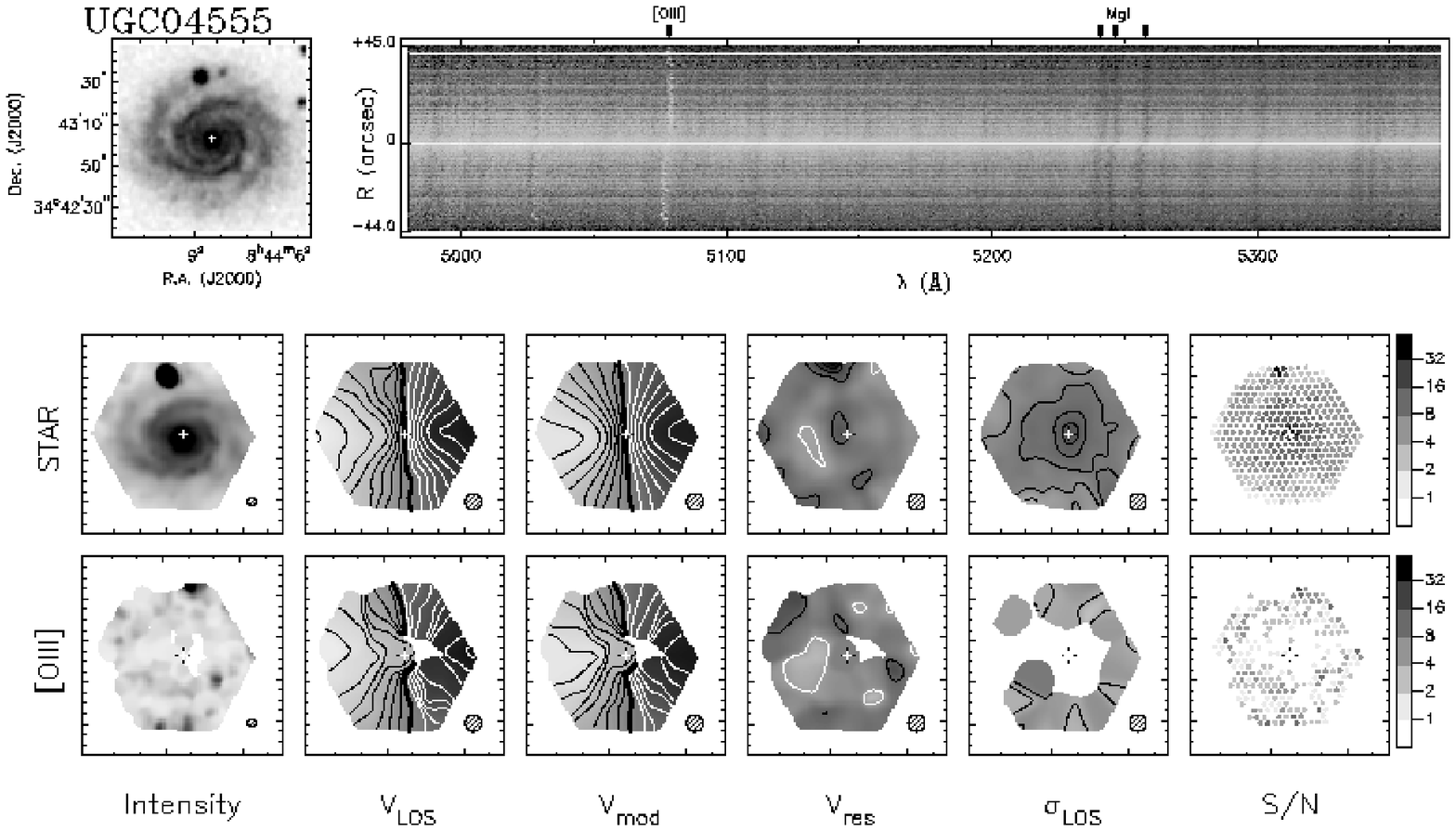}
 \end{figure}

 \begin{figure}
 \centering
 \includegraphics[width=0.95\textwidth]{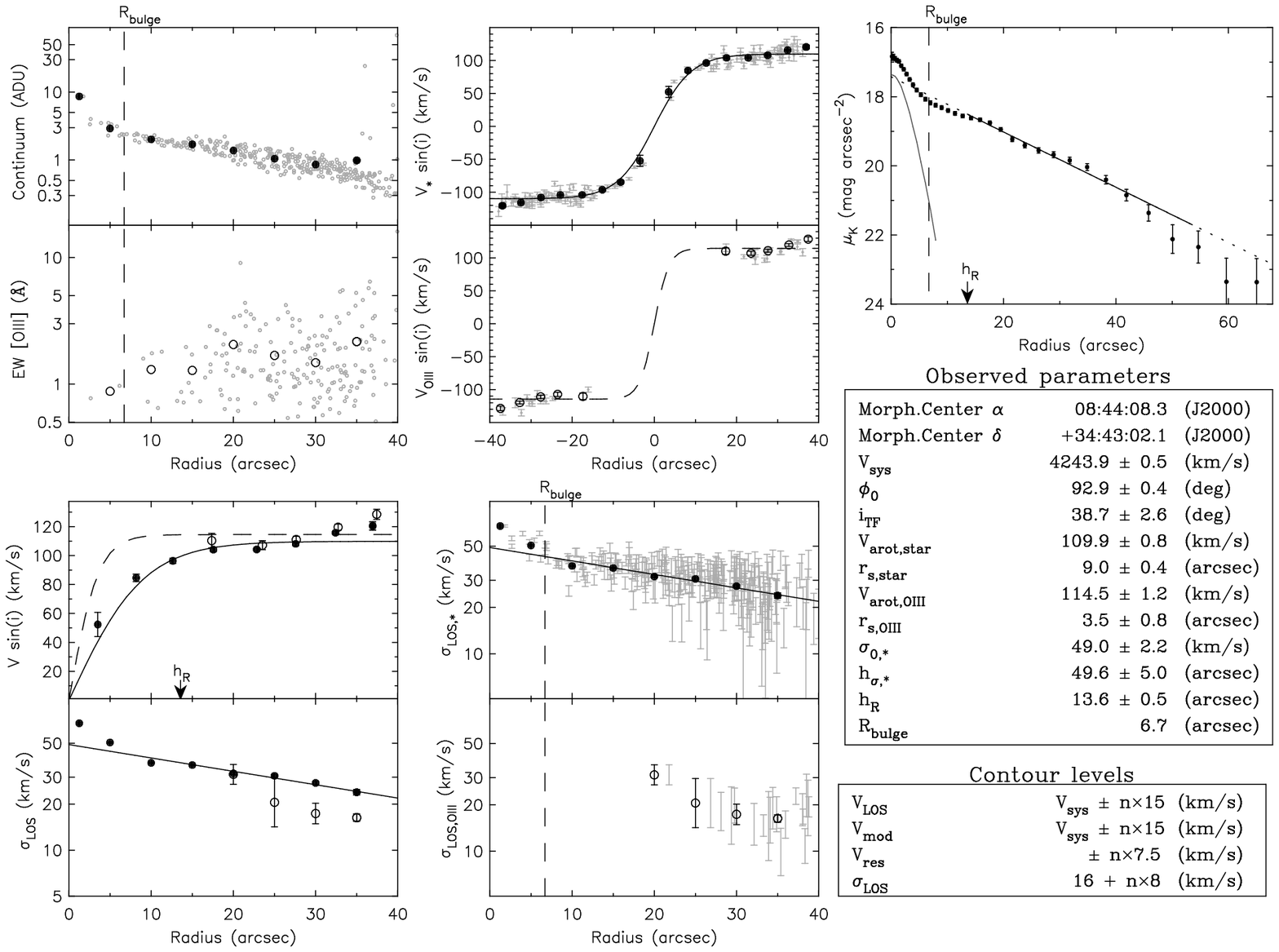}
 \end{figure}

\clearpage

 \begin{figure}
 \centering
 \includegraphics[width=0.95\textwidth]{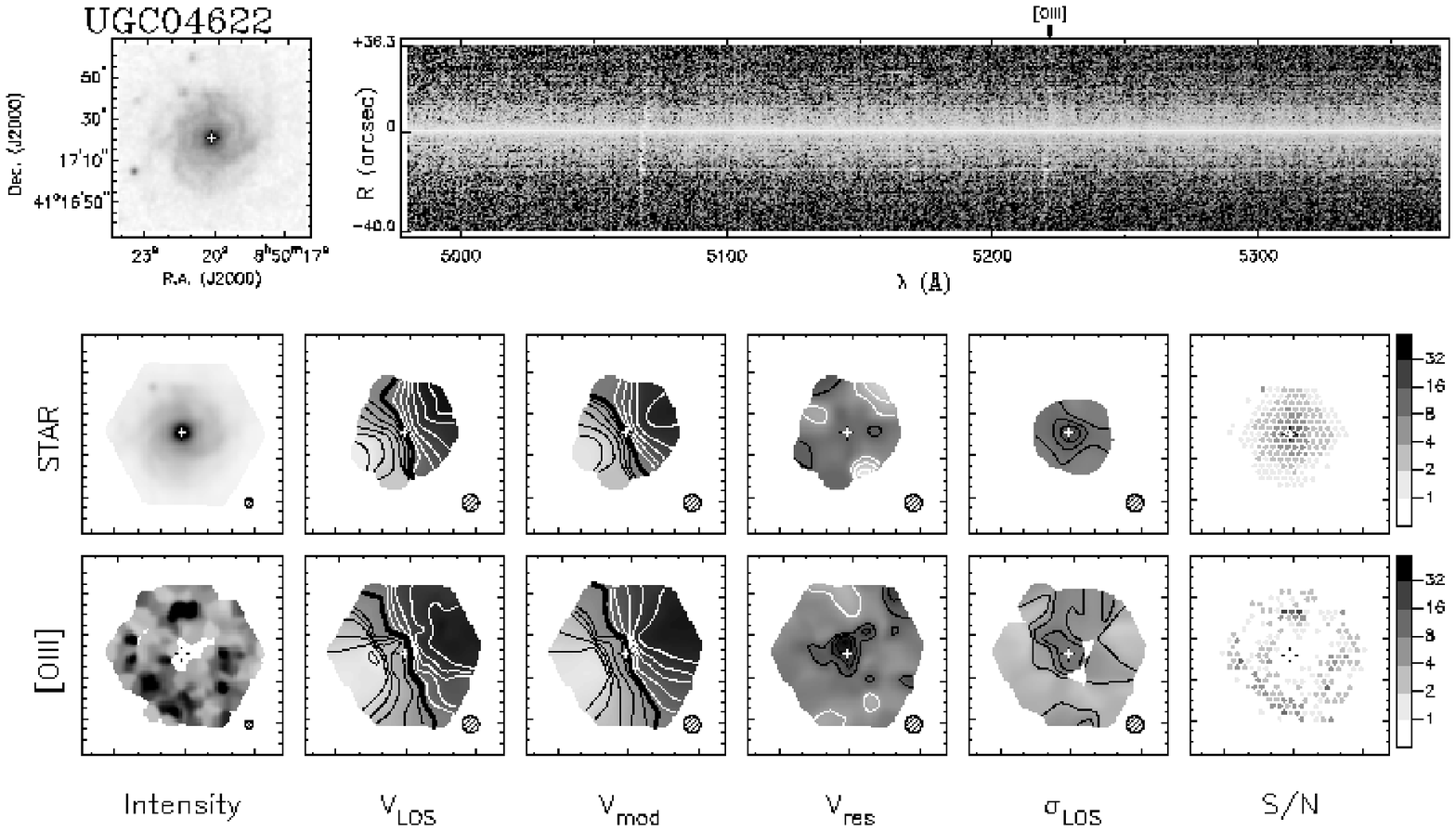}
 \end{figure}

 \begin{figure}
 \centering
 \includegraphics[width=0.95\textwidth]{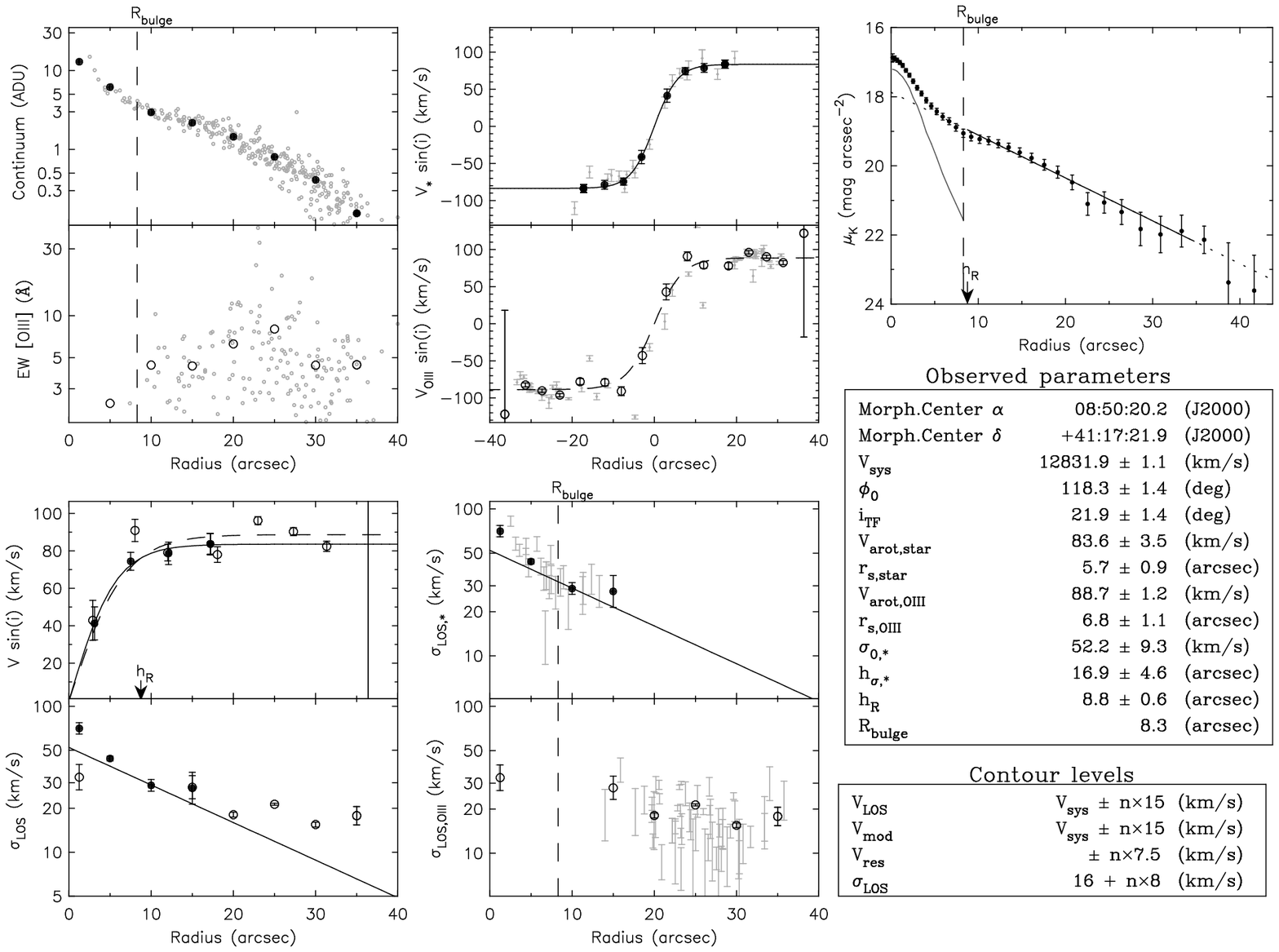}
 \end{figure}

\clearpage

 \begin{figure}
 \centering
 \includegraphics[width=0.95\textwidth]{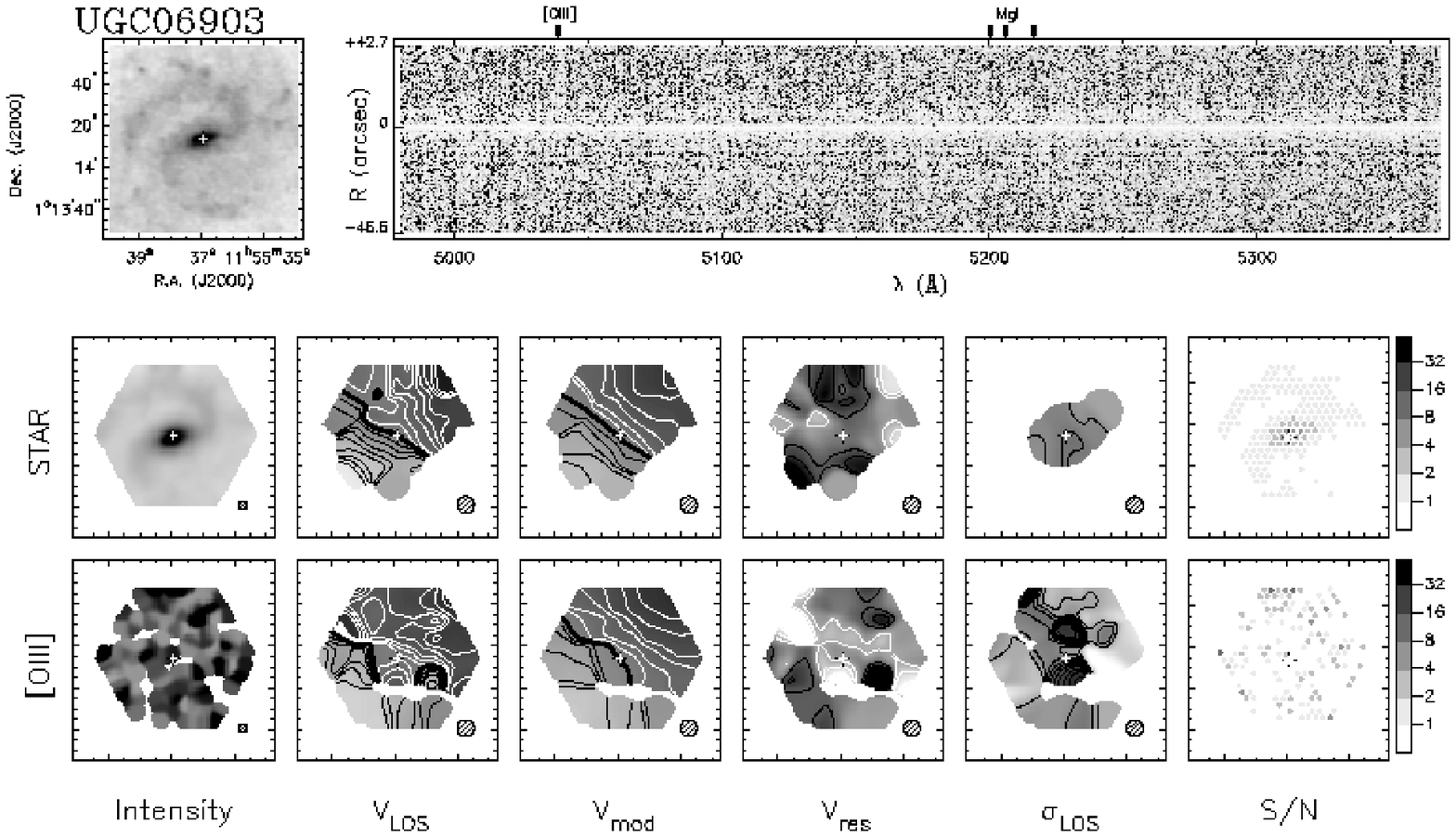}
 \end{figure}

 \begin{figure}
 \centering
 \includegraphics[width=0.95\textwidth]{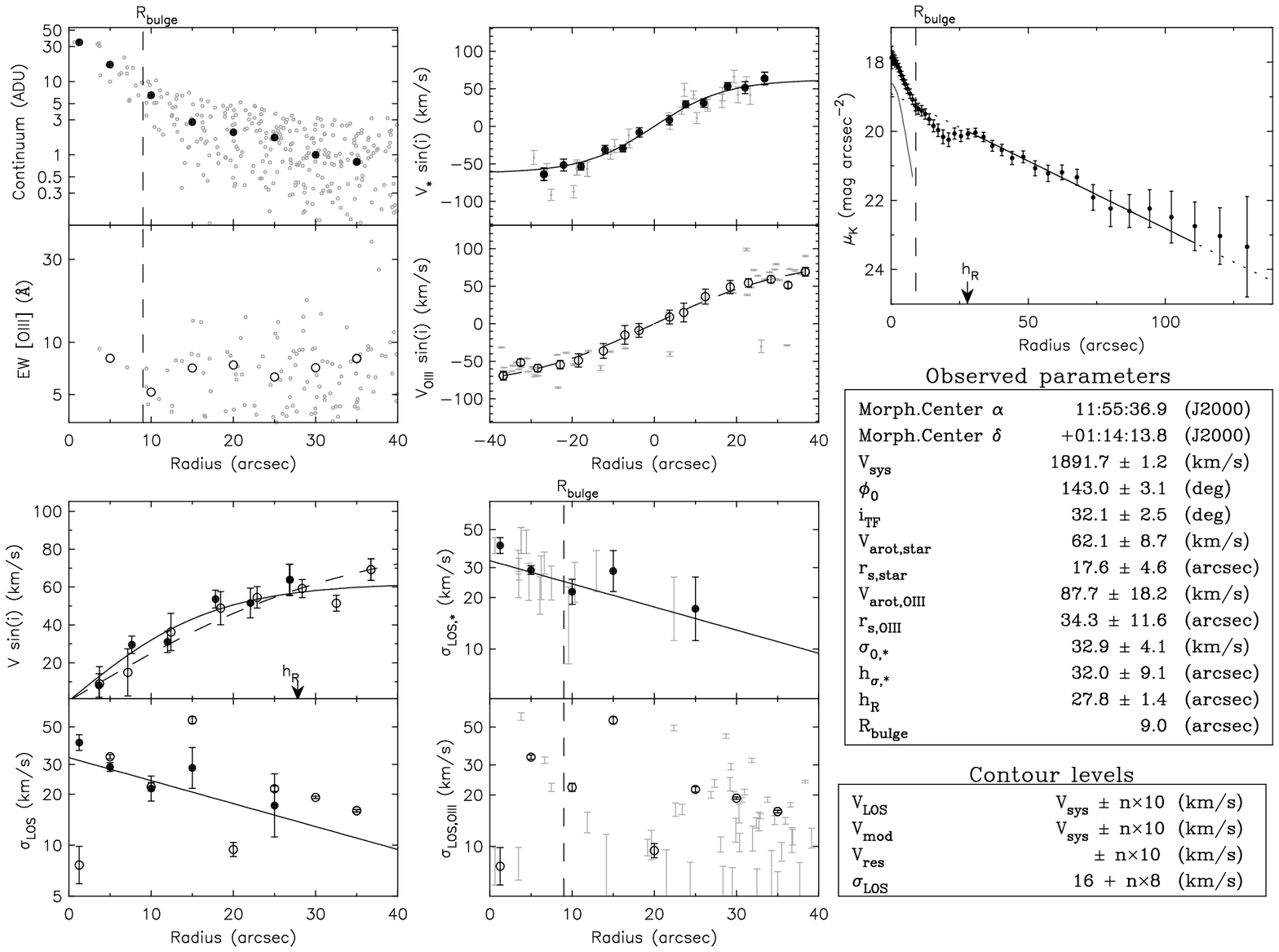}
 \end{figure}

\clearpage

 \begin{figure}
 \centering
 \includegraphics[width=0.95\textwidth]{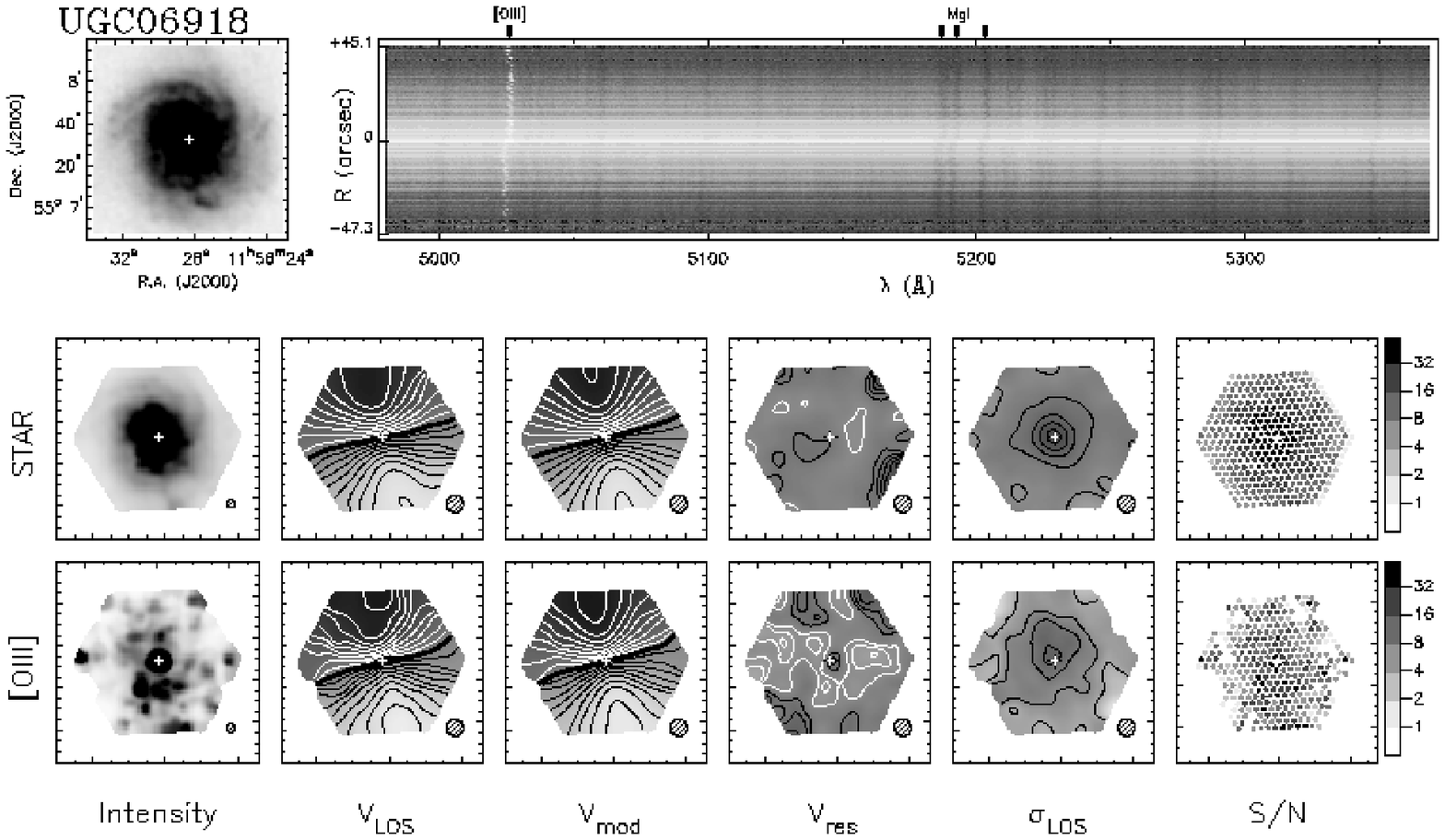}
 \end{figure}

 \begin{figure}
 \centering
 \includegraphics[width=0.95\textwidth]{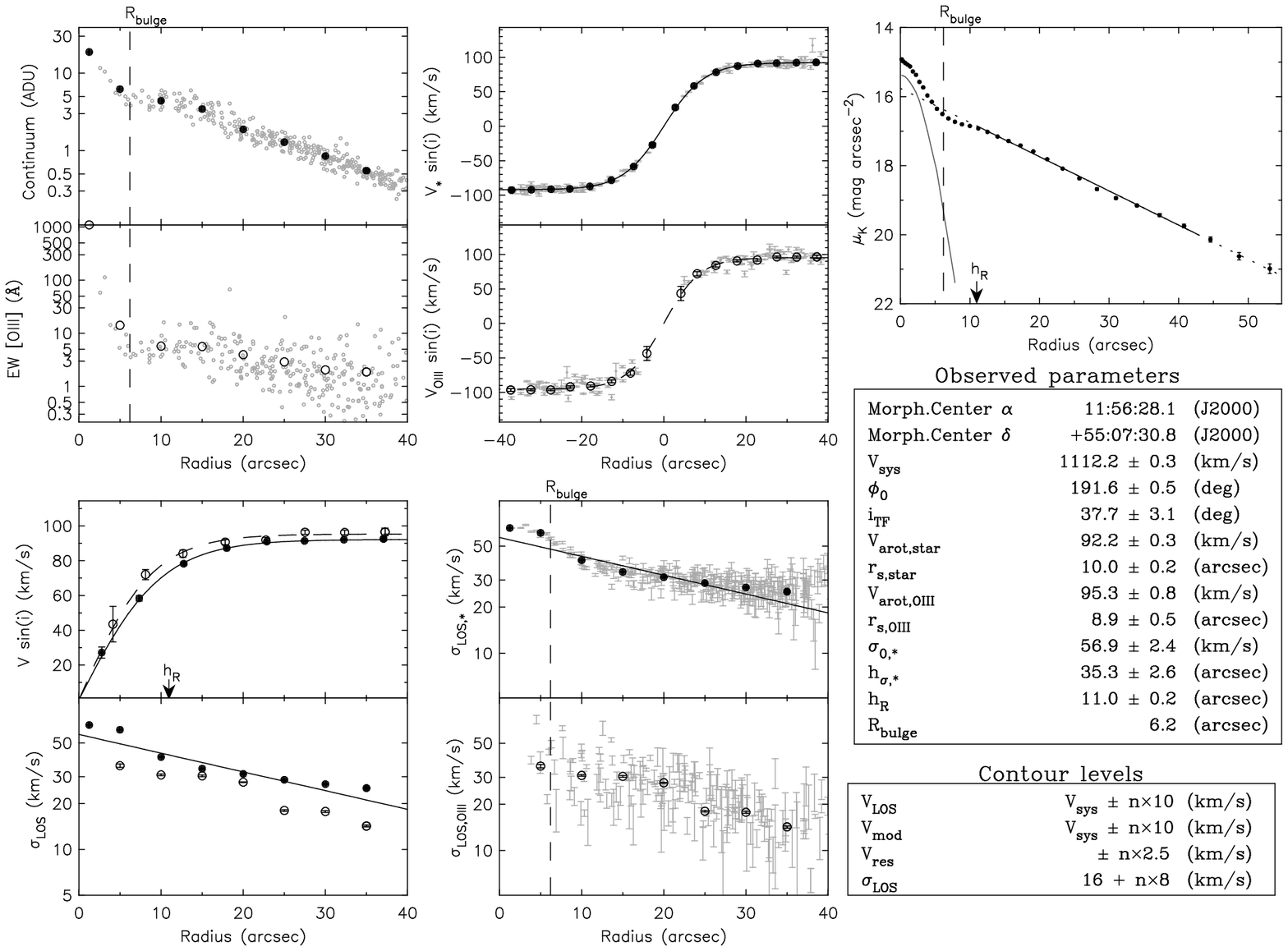}
 \end{figure}

\clearpage

 \begin{figure}
 \centering
 \includegraphics[width=0.95\textwidth]{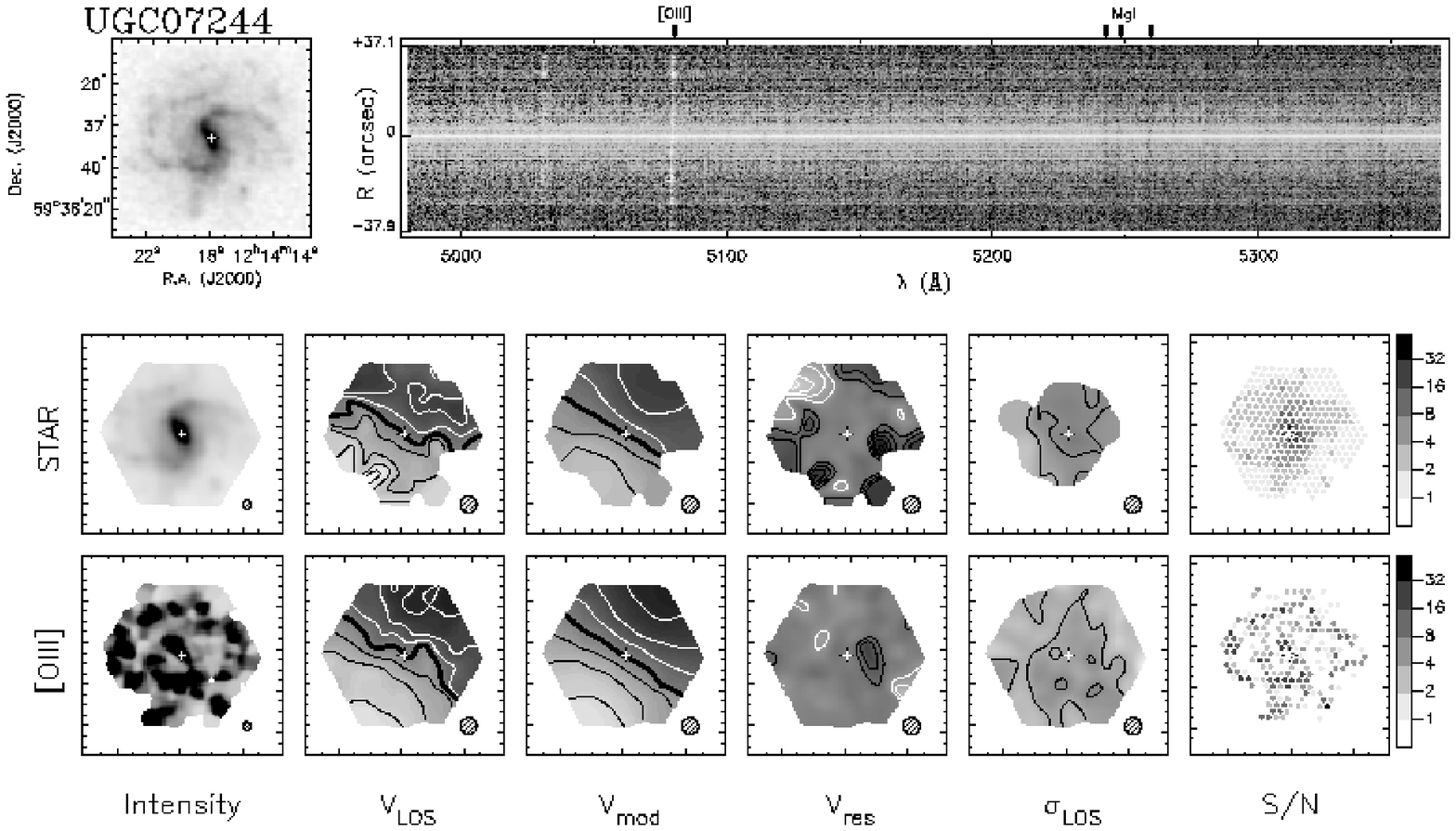}
 \end{figure}

 \begin{figure}
 \centering
 \includegraphics[width=0.95\textwidth]{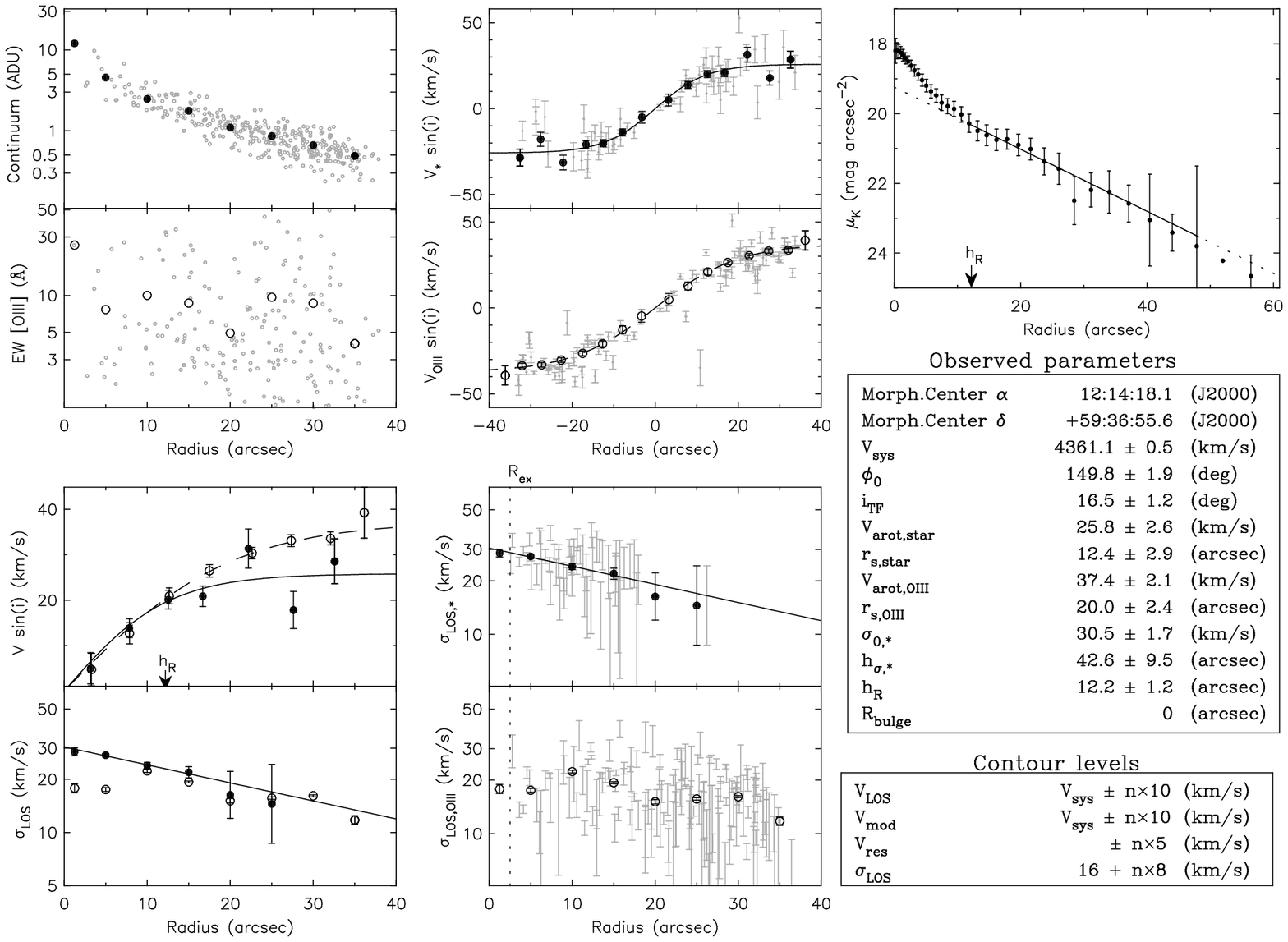}
 \end{figure}

\clearpage

 \begin{figure}
 \centering
 \includegraphics[width=0.95\textwidth]{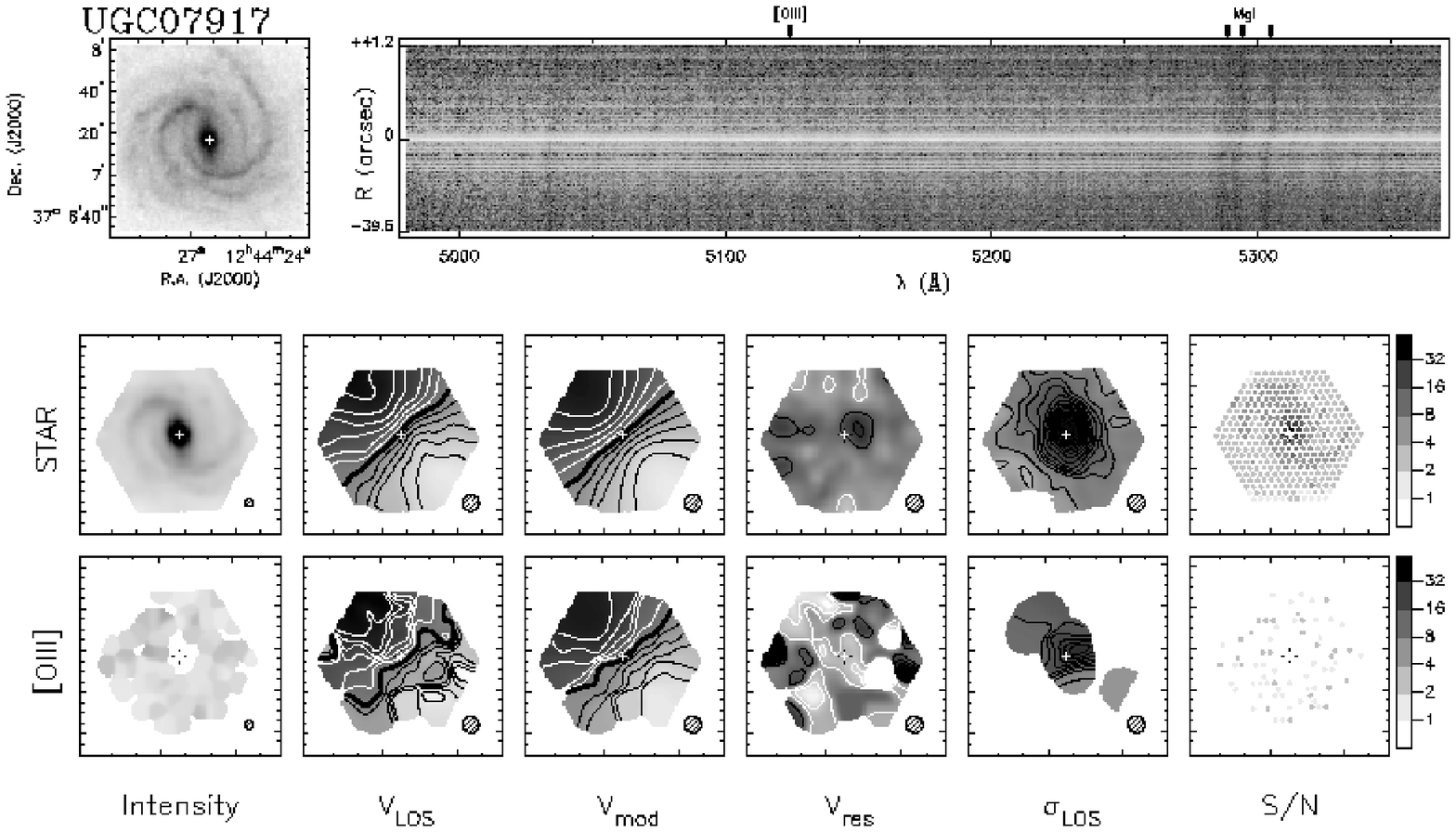}
 \end{figure}

 \begin{figure}
 \centering
 \includegraphics[width=0.95\textwidth]{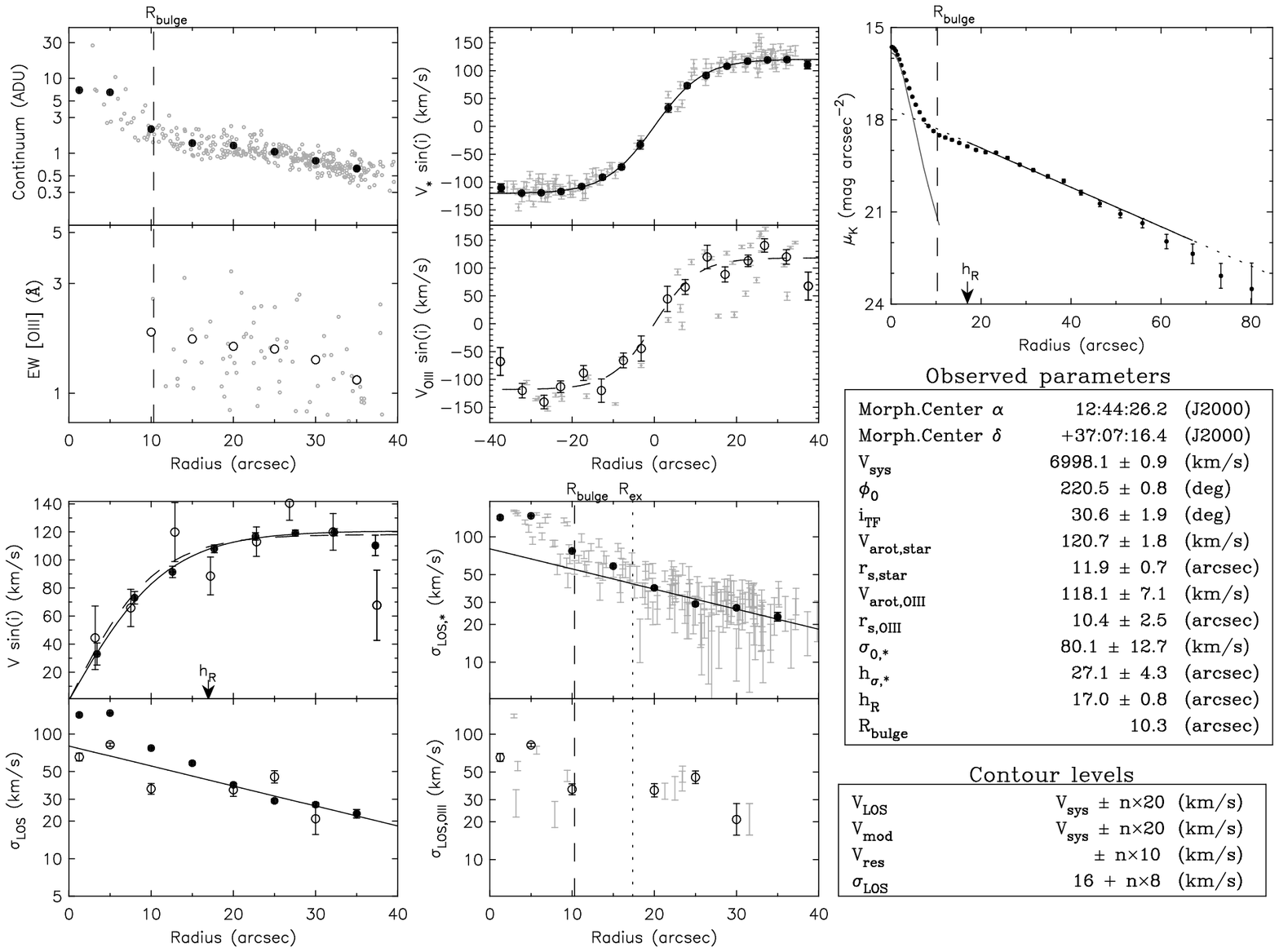}
 \end{figure}

\clearpage

 \begin{figure}
 \centering
 \includegraphics[width=0.95\textwidth]{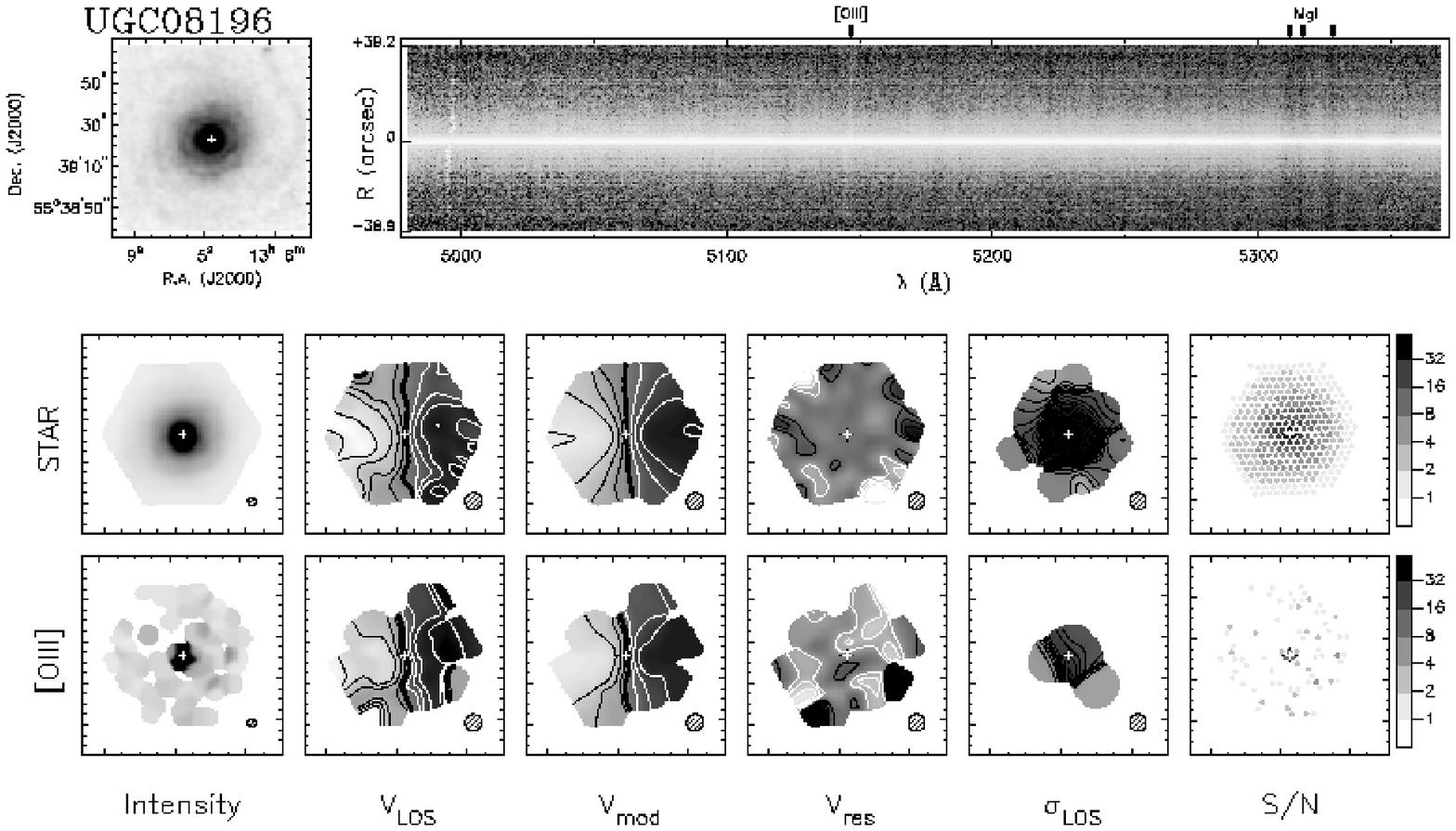}
 \end{figure}

 \begin{figure}
 \centering
 \includegraphics[width=0.95\textwidth]{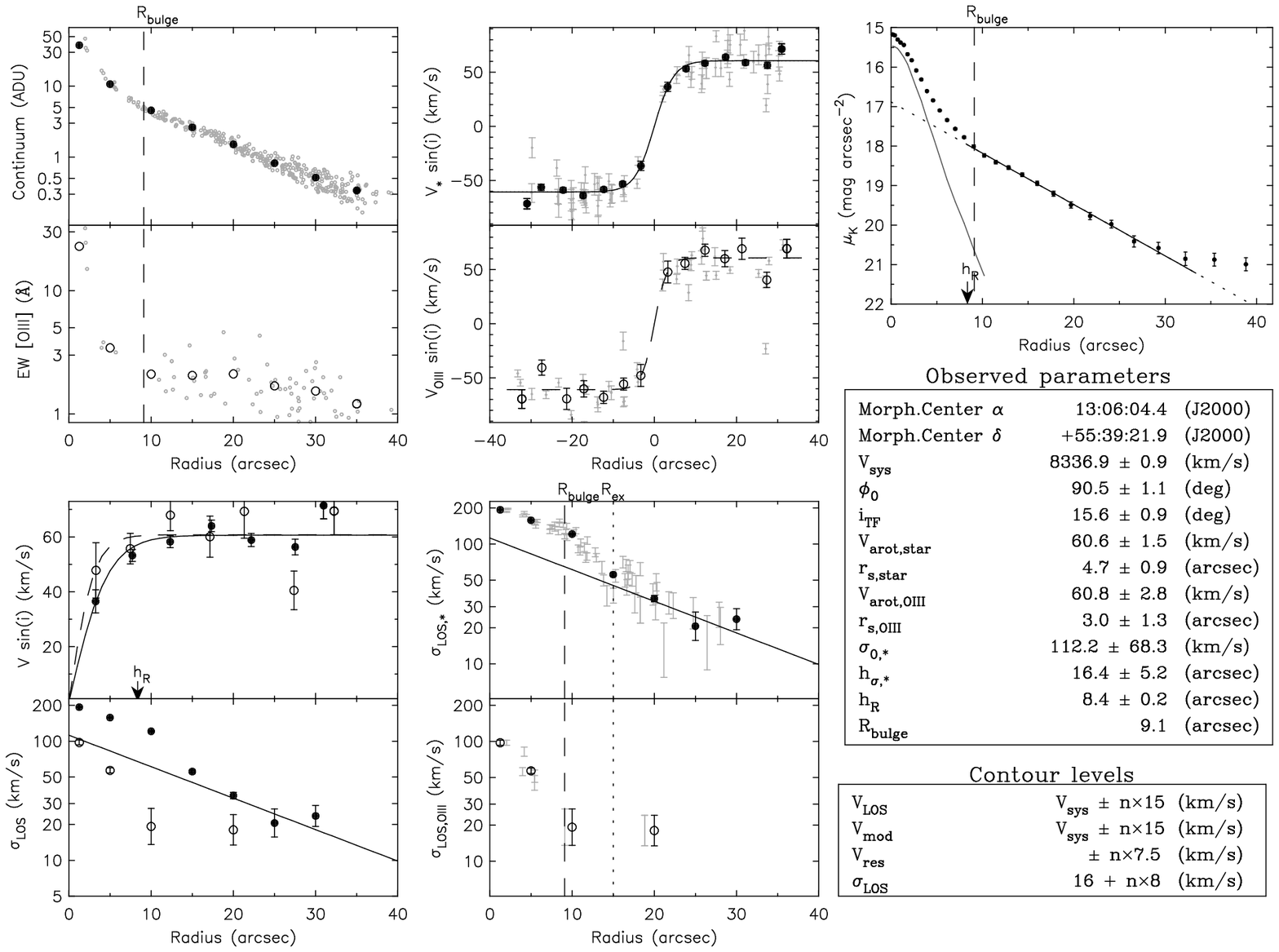}
 \end{figure}

\clearpage

 \begin{figure}
 \centering
 \includegraphics[width=0.95\textwidth]{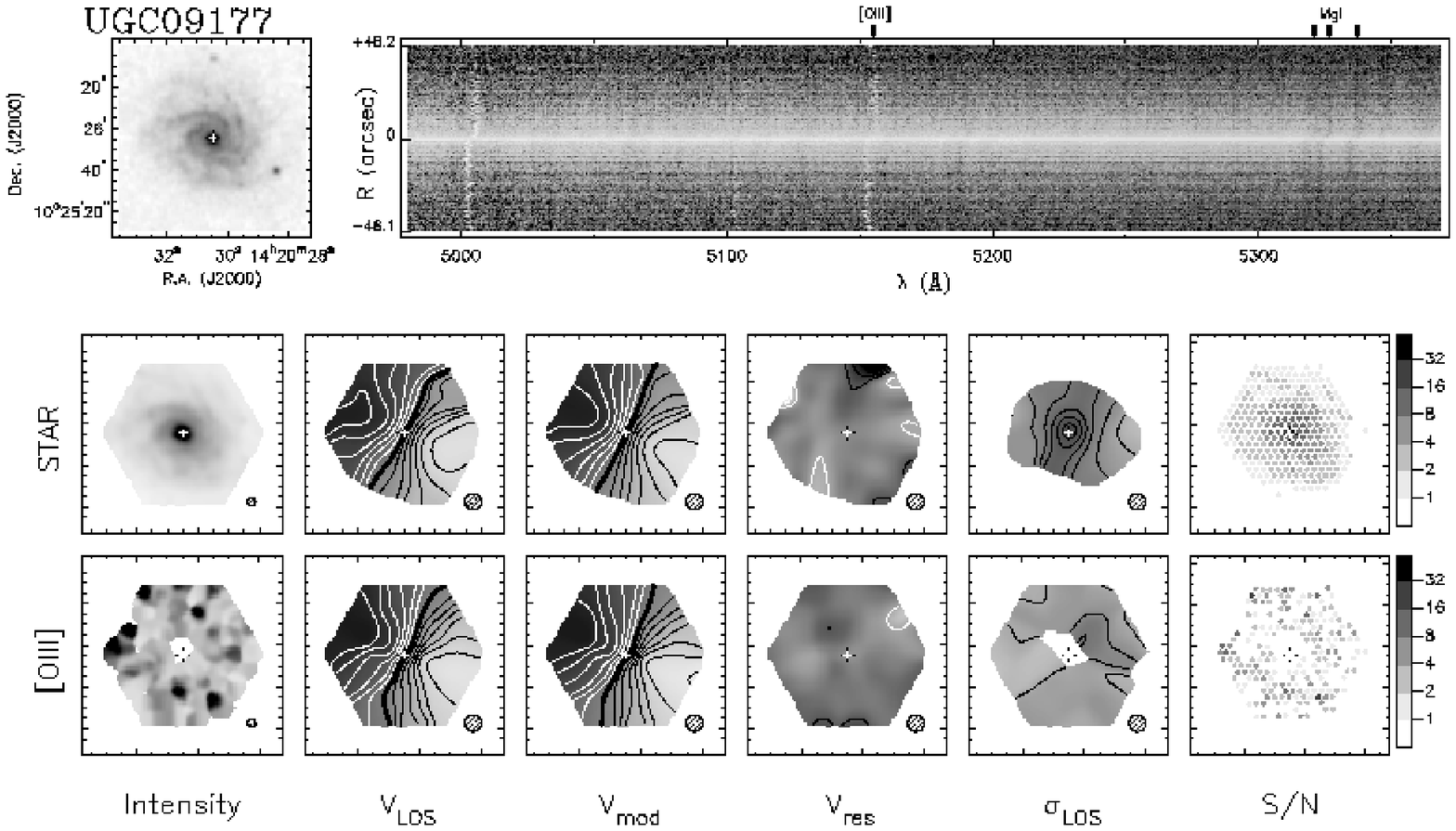}
 \end{figure}

 \begin{figure}
 \centering
 \includegraphics[width=0.95\textwidth]{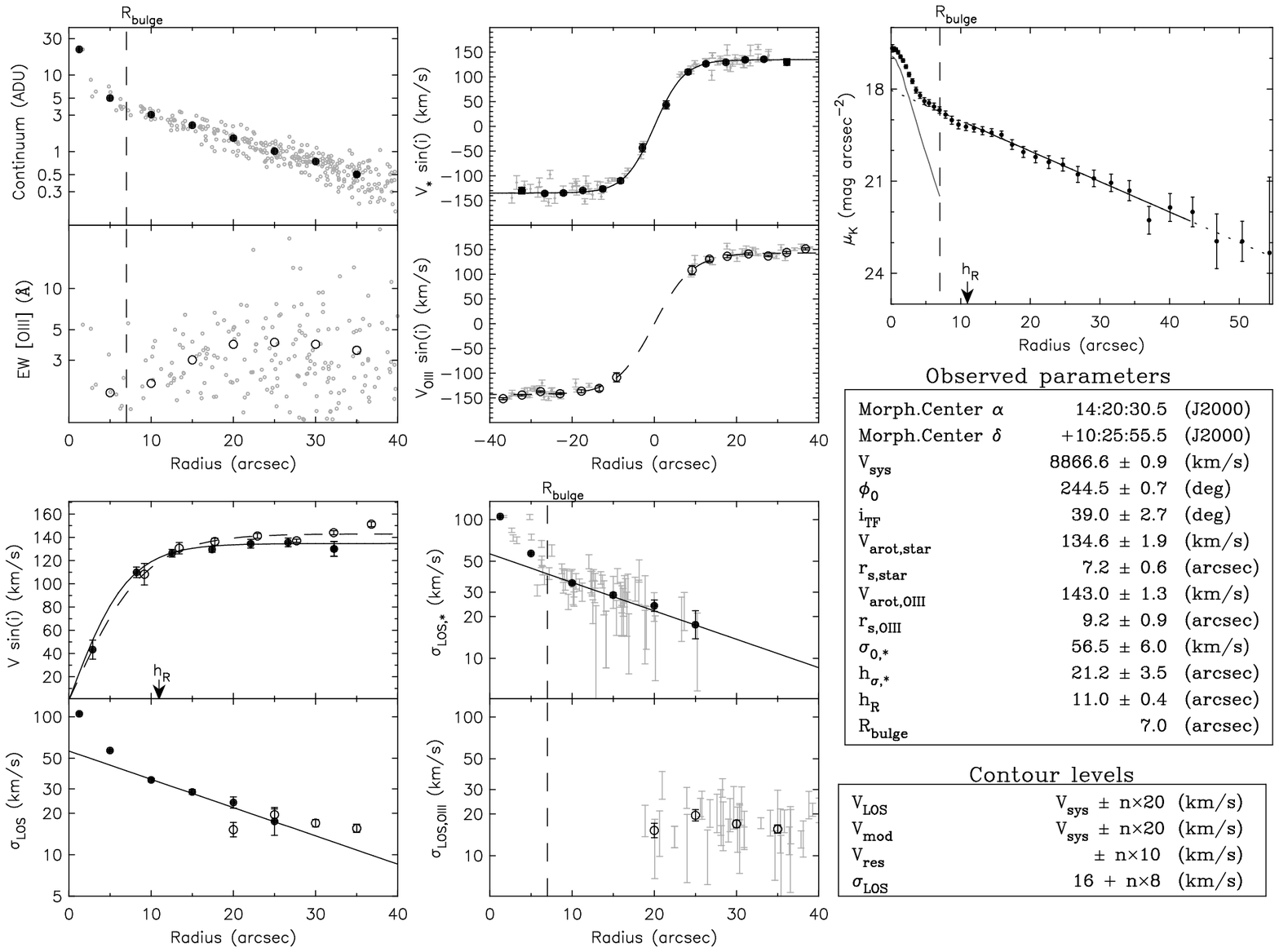}
 \end{figure}

\clearpage

 \begin{figure}
 \centering
 \includegraphics[width=0.95\textwidth]{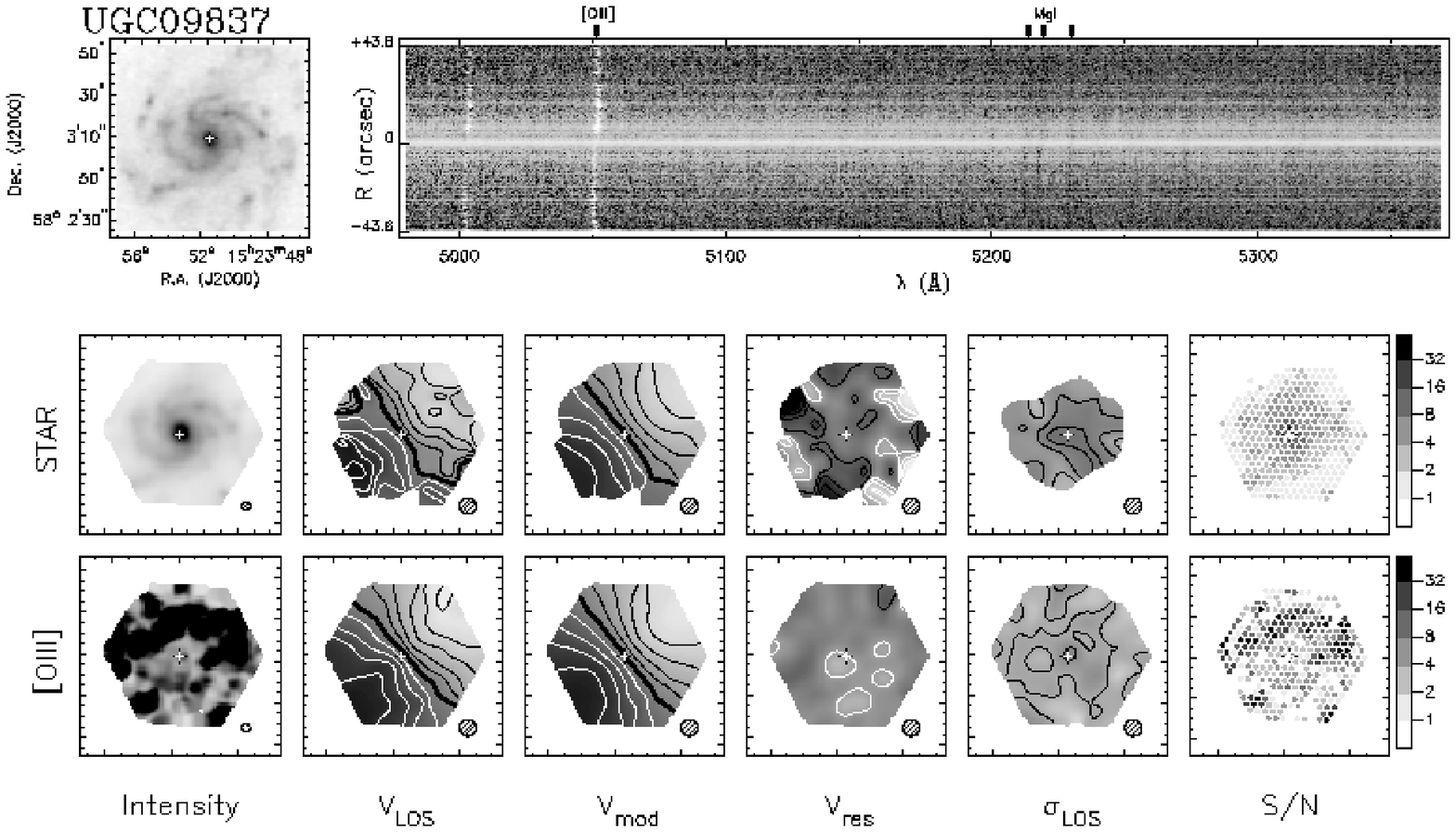}
 \end{figure}

 \begin{figure}
 \centering
 \includegraphics[width=0.95\textwidth]{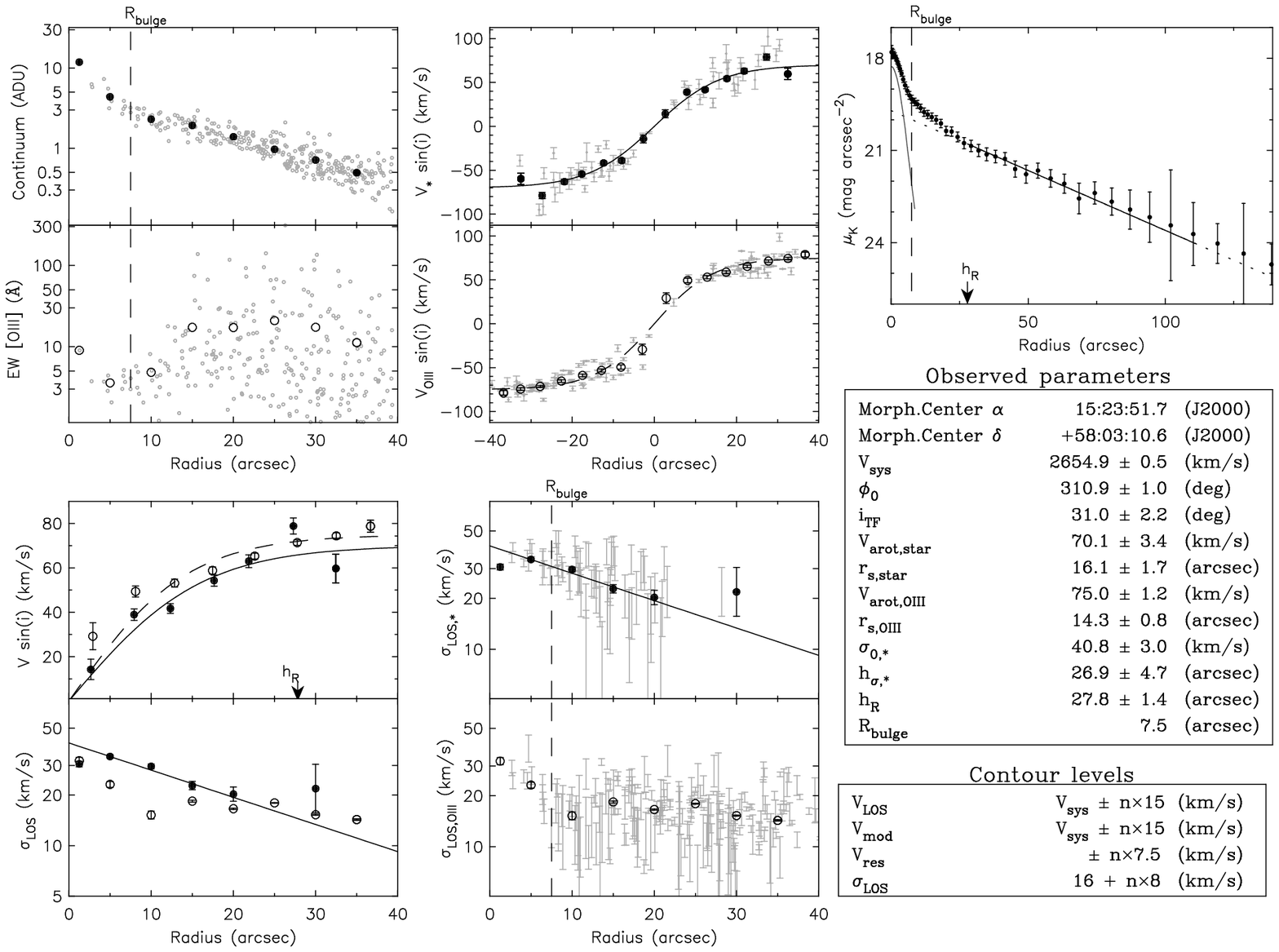}
 \end{figure}

\clearpage

 \begin{figure}
 \centering
 \includegraphics[width=0.95\textwidth]{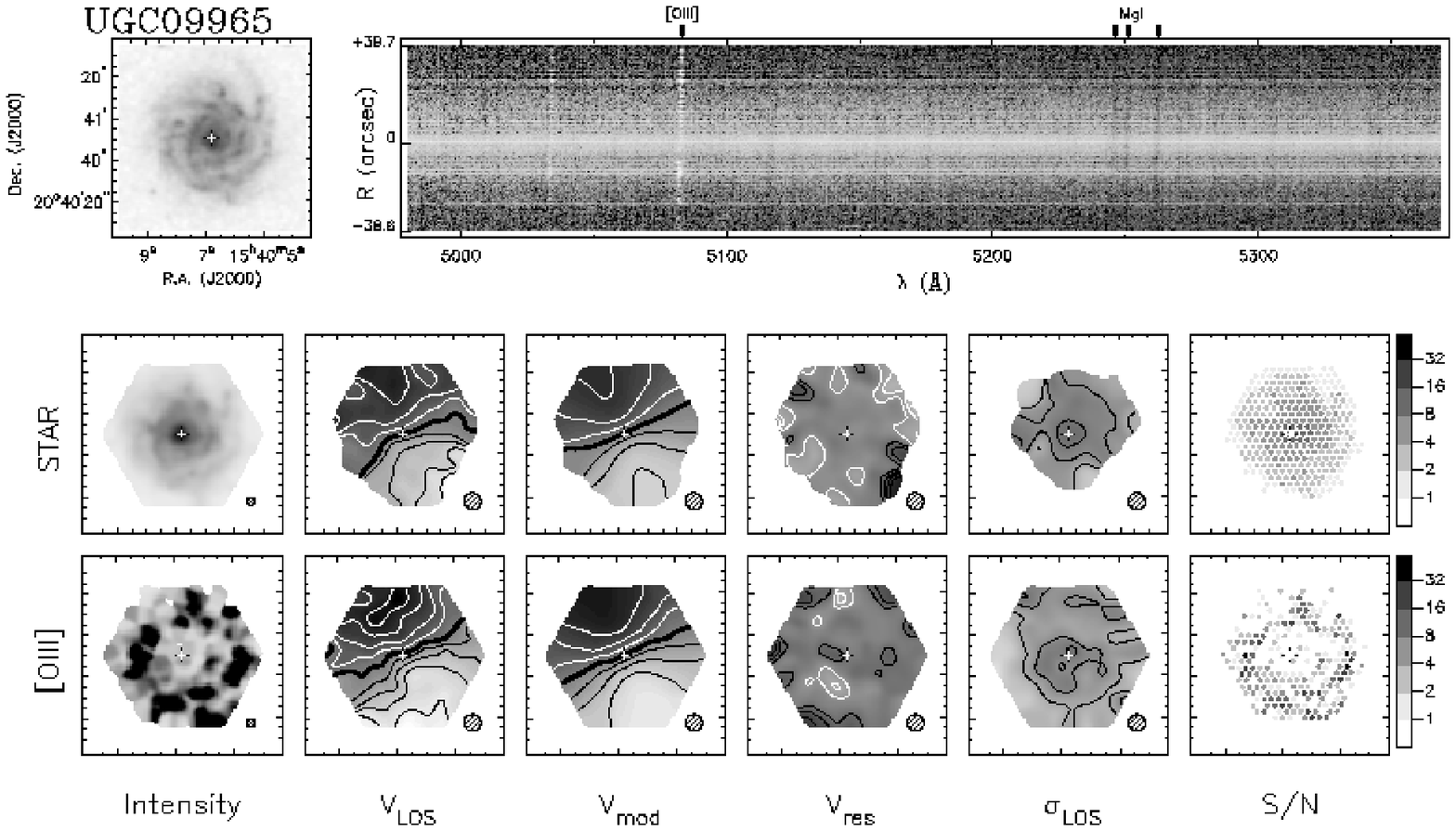}
 \end{figure}

 \begin{figure}
 \centering
 \includegraphics[width=0.95\textwidth]{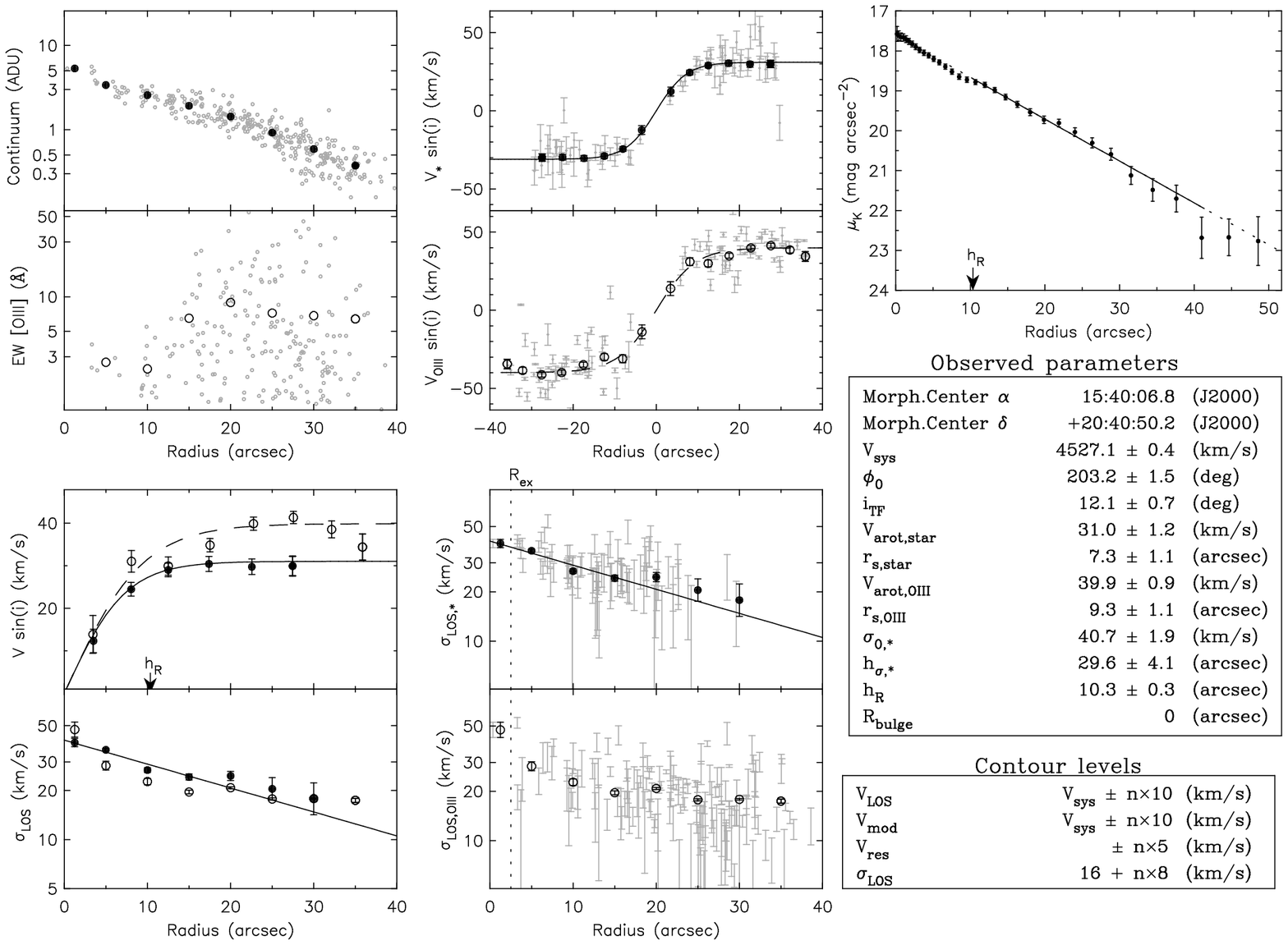}
 \end{figure}

\clearpage

 \begin{figure}
 \centering
 \includegraphics[width=0.95\textwidth]{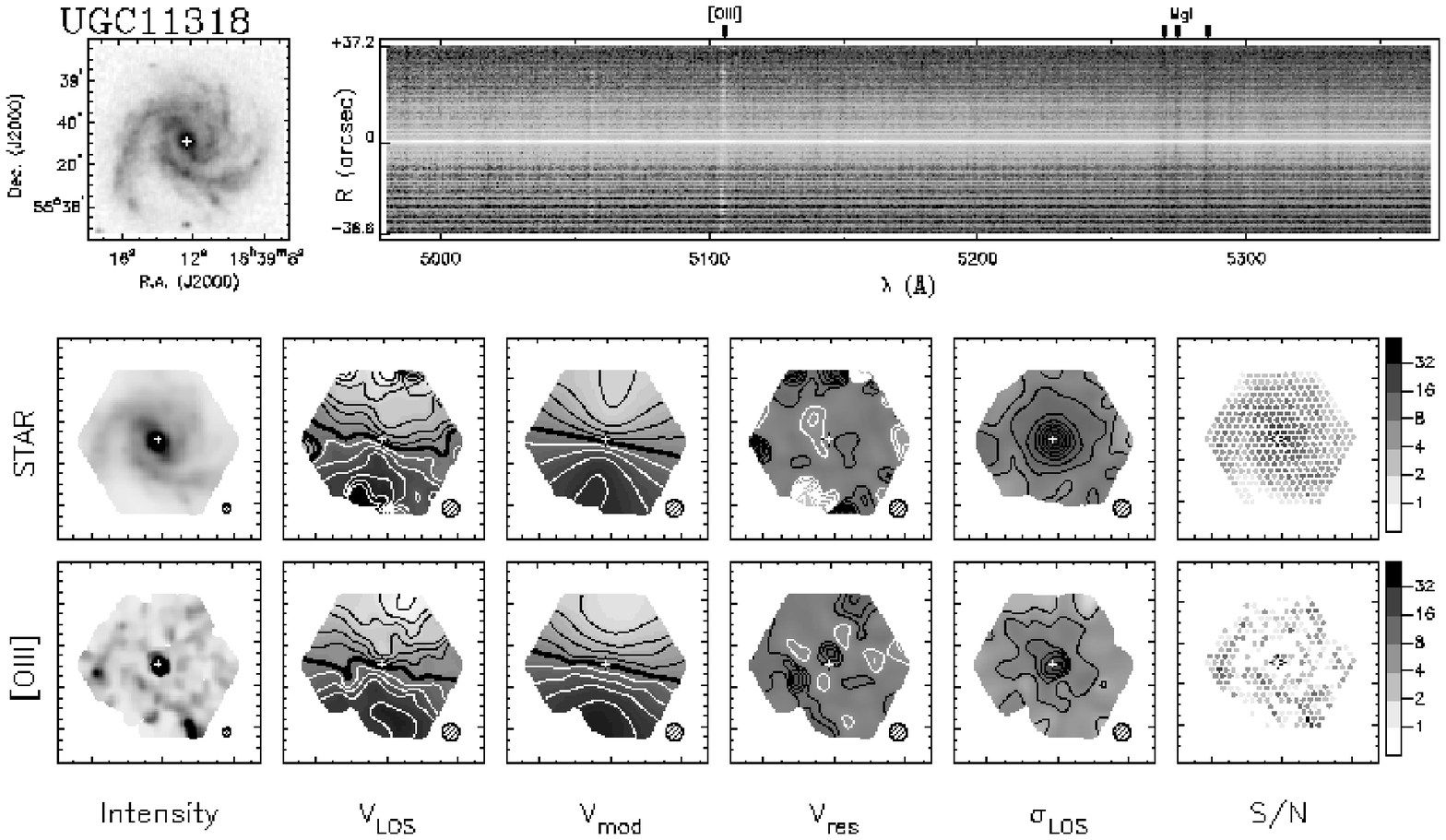}
 \end{figure}

 \begin{figure}
 \centering
 \includegraphics[width=0.95\textwidth]{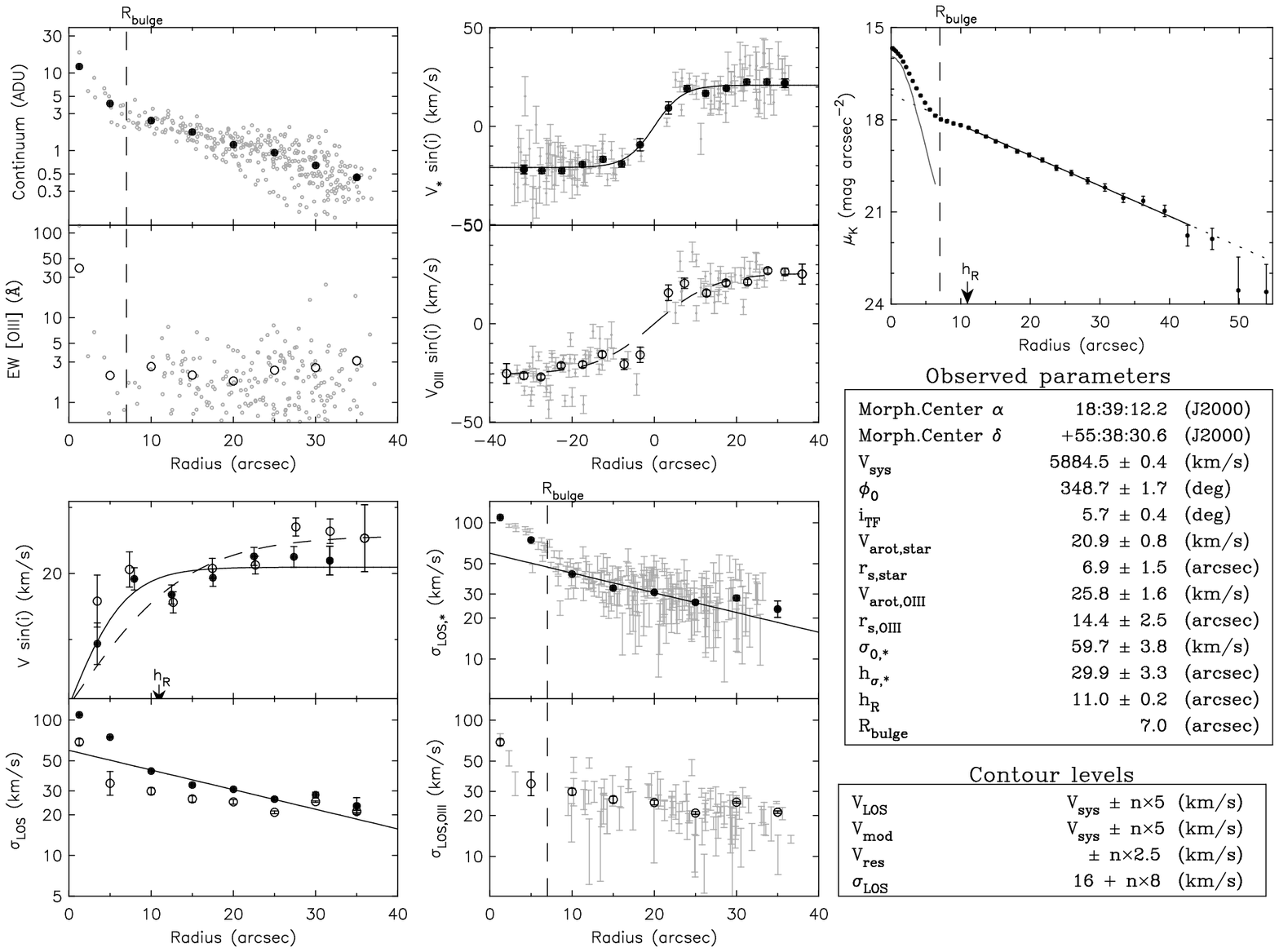}
 \end{figure}

\clearpage

 \begin{figure}
 \centering
 \includegraphics[width=0.95\textwidth]{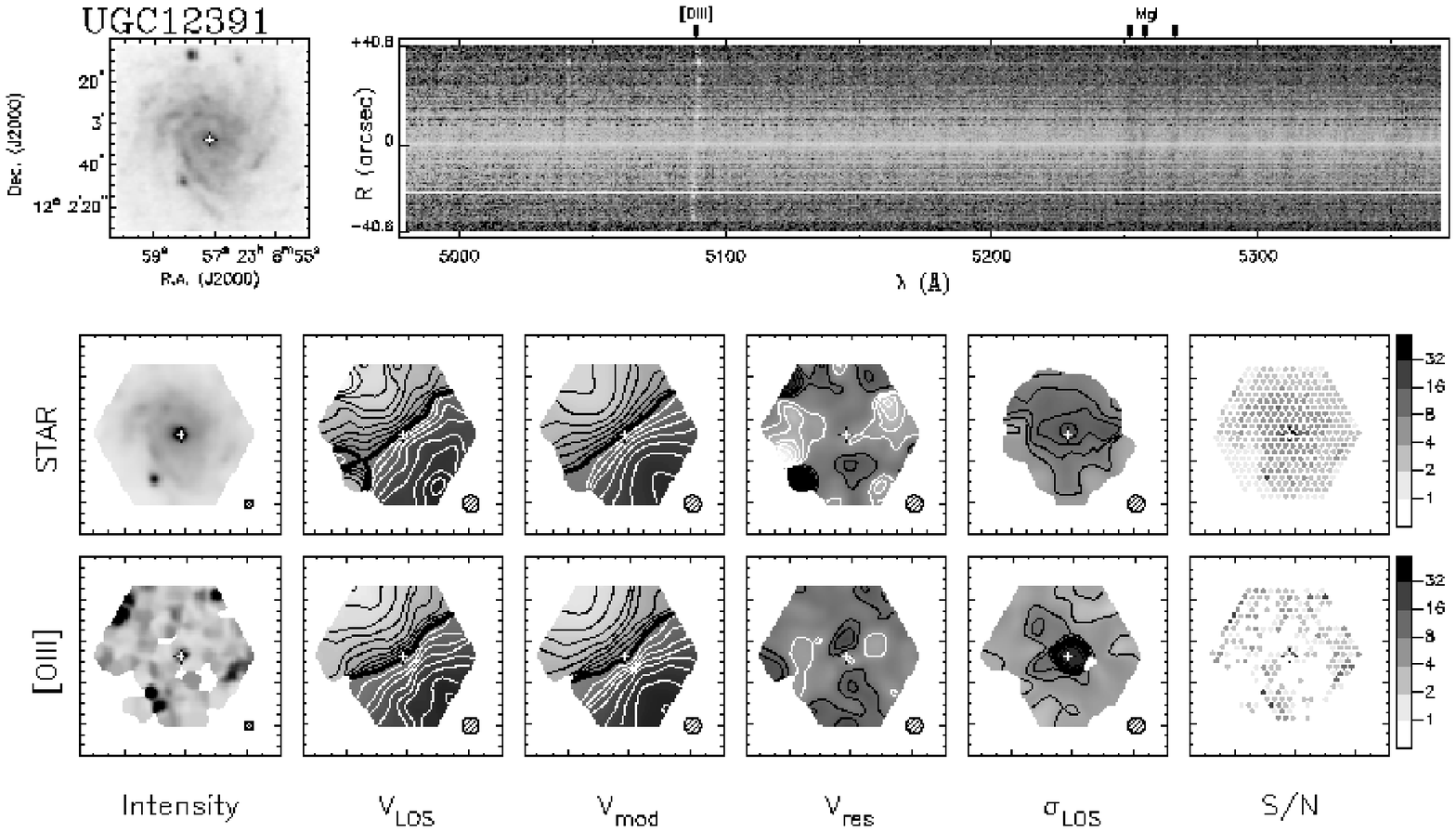}
 \end{figure}

 \begin{figure}
 \centering
 \includegraphics[width=0.95\textwidth]{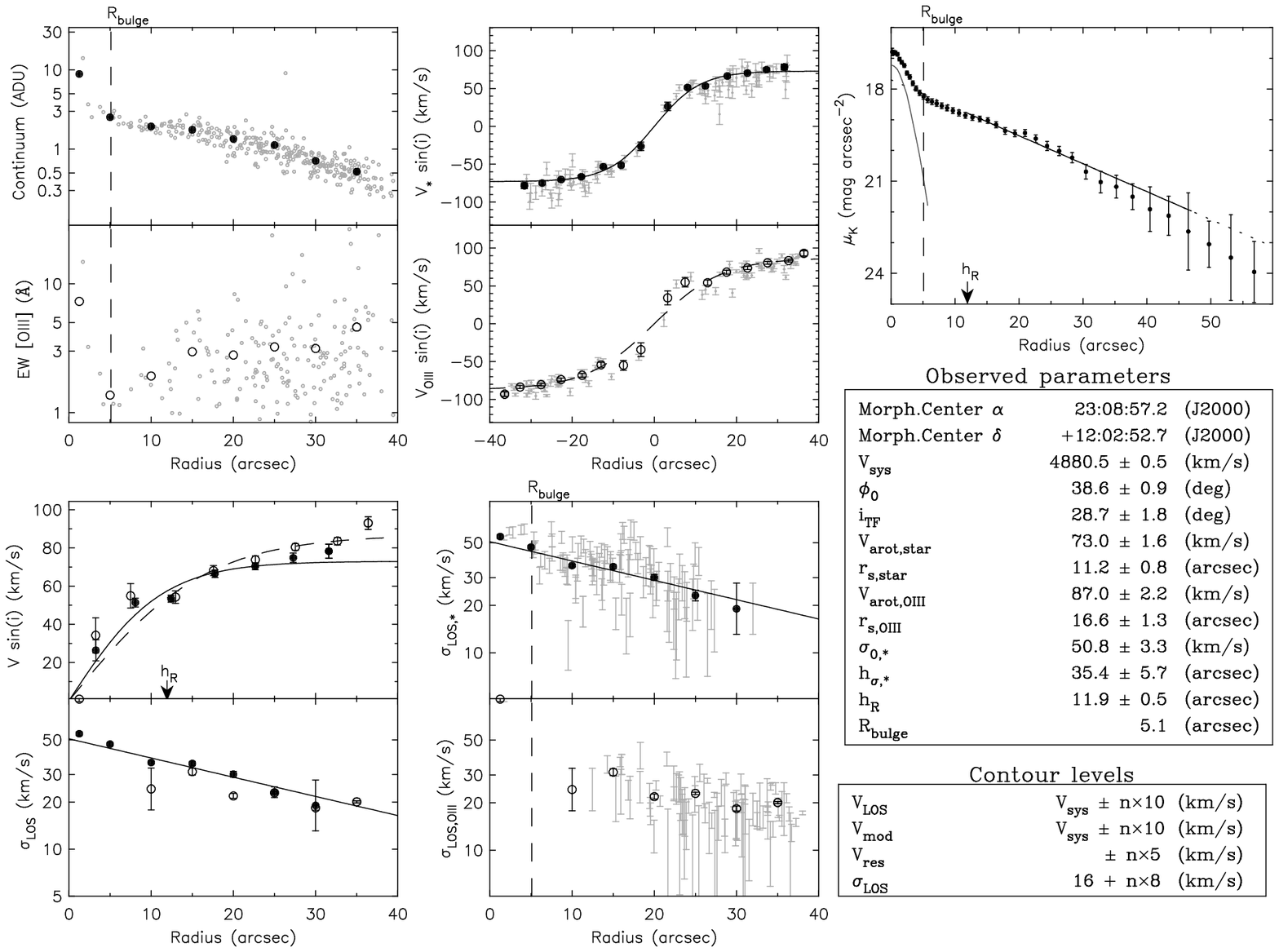}
 \end{figure}


\end{document}